\documentclass[12pt]{article}
\usepackage{graphicx,amsmath,setspace}
\usepackage{units}
\usepackage{color}

\newlength{\dinwidth}
\newlength{\dinmargin}
\newcommand{\dif}{\mathrm{d}}

\def\lapproxeq{\lower .7ex\hbox{$\;\stackrel{\textstyle                                                    
<}{\sim}\;$}}                                                    
\def\gapproxeq{\lower .7ex\hbox{$\;\stackrel{\textstyle                                                    
>}{\sim}\;$}}                                                    
\def\be{\begin{equation}}                                                    
\def\ee{\end{equation}}                                                    
\def\bea{\begin{eqnarray}}                      
\def\eea{\end{eqnarray}}

\def\GeV{\rm GeV}
\def\TeV{\rm TeV}

\def\sh{\hat s}
\def\sh2{{\hat s}^2}

\begin{document}

\begin{flushright}                                                    
LCTS/2014-47  \\
IPPP/14/97  \\
DCPT/14/194 \\                                                    
\today \\                                                    
\end{flushright} 

\vspace*{0.5cm}

\begin{center}
{\Large \bf Parton distributions in the LHC era:}\\ 
\vspace*{0.5cm}{\Large \bf MMHT 2014 PDFs}\\

\vspace*{1cm}
L. A. Harland-Lang$^{a}$, A. D. Martin$^b$, P. Motylinski$^a$ and R.S. Thorne$^a$\\                                               
\vspace*{0.5cm}                                                    
                                                  
$^a$ Department of Physics and Astronomy, University College London, London, WC1E 6BT, UK \\           
$^b$ Institute for Particle Physics Phenomenology, Durham University, Durham, DH1 3LE, UK                                                    
                                                    
\vspace*{1cm}

\begin{abstract} 
We present LO, NLO and NNLO sets of parton distribution functions (PDFs) of the 
proton determined from global analyses of the available hard scattering data. 
These MMHT2014 PDFs supersede the `MSTW2008' parton sets, but are obtained 
within the same basic framework. We include a variety of new data sets, from 
the LHC, 
updated Tevatron data and the HERA combined H1 and ZEUS data on the total and 
charm structure functions. We also improve the theoretical framework of the 
previous analysis. These new PDFs are compared to the `MSTW2008' parton sets. 
In most cases the PDFs, and the predictions, are within one standard 
deviation of those of MSTW2008. The major changes are the $u-d$ valence quark 
difference at small $x$ due to an improved parameterisation 
and, to a lesser extent, the strange quark PDF due to the effect of certain 
LHC data and a better treatment of the $D \to \mu$ branching ratio.
We compare our MMHT PDF sets with those of other collaborations; in particular with the NNPDF3.0 sets, which are contemporary with the present analysis.
\end{abstract}                                                        
\vspace*{0.5cm}                                                    
                                                    
\end{center}

\begin{spacing}{0.8}
\clearpage
\tableofcontents
\clearpage
\end{spacing}

\section{Introduction  \label{sec:1}} 

The parton distribution functions (PDFs) of the proton are determined from 
fits to the world data on deep inelastic and related hard scattering 
processes, see, for example, \cite{MSTW,CT10,NNPDF23,HERAPDF15,ABM14,JR14}. More than five years have 
elapsed since MSTW published \cite{MSTW} the results of their global PDF analysis entitled `Parton distributions for the LHC'.  Since then there have been significant improvements in the data, including especially the measurements made at the LHC.  It is therefore timely to present a new global PDF analysis within the MSTW framework, which we denote by MMHT2014\footnote{ We note that preliminary reports on these new PDFs have been presented in \cite{MMHTDIS, MMHTICHEP}.}.

In the intervening period, the predictions of the MSTW partons have been compared with the new data as they have become available.  The only significant shortcoming of these MSTW predictions was in the description of the lepton charge asymmetry from $W^\pm$ decays, as a function of the lepton rapidity. This was particularly clear in the asymmetry data measured at the LHC \cite{CMS-easym,ATLAS-WZ}.  This deficiency was investigated in detail in MMSTWW \cite{MMSTWW}.\footnote{The PDF sets in this article are often referred to as MSTWCPdeut, but we will use the nomenclature MMSTWW, i.e. the initials of the authors of the article, throughout this paper.} In that work, fits with extended `Chebyshev' parameterisations of the input distributions were carried out,  to exactly the same data set as was used in the original global MSTW PDF analysis.
To be specific, MMSTWW replaced the factors $(1+\epsilon x^{0.5}+\gamma x)$ in the MSTW valence, sea and gluon distributions by the Chebyshev polynomial forms $(1+\sum a_i T^{\rm Ch}_i(y))$ with 
$ y=1-2\sqrt{x}$, and $i=1 \ldots 4$.  The Chebyshev forms have the advantage that the parameters $a_i$ are well-behaved and, compared to the coefficients of the MSTW parameterisation, are rather small, with moduli usually $\le 1$.
At the same time, MMSTWW \cite{MMSTWW} investigated the effect of also extending, and making more flexible, the `nuclear' correction to the deuteron structure functions.
The extended  Chebyshev parameterisations resulted in an improved
 stability in the deuteron corrections.  The main changes in the PDFs found in the `Chebyshev' analysis, as compared to the MSTW fit, were in the valence up and down distributions, $u_V$ and $d_V$, for $x \lapproxeq 0.03$ at high $Q^2 \sim 10^4 ~\GeV^2$, or slightly higher $x$ at low $Q^2$; a region where there are weak constraints on the valence PDFs from the data used in these fits.
 These changes to the valence quark PDFs, essentially in the combination $u_V-d_V$, were sufficient to result in a good description of the data on lepton 
charge asymmetry from $W^\pm$ decays. Recall that the LHC data for the lepton 
asymmetry were not included in the MMSTWW \cite{MMSTWW} fit, but are predicted.
There were no other signs of significant changes in the PDFs, and for the 
overwhelming majority of processes at the LHC (and the Tevatron) the MSTW 
predictions were found to be satisfactory, see \cite{MMSTWW} (though the 
precise shape of the $W,Z$ rapidity data was not ideal, particularly at NNLO) 
and e.g. \cite{CMSW14,ATLASWW}.  

Nevertheless, it is time to take 
advantage of the new data in order to improve the precision of PDFs within the
same general framework of the MSTW analysis. This includes a fit to new data 
from HERA, the Tevatron and the LHC, where the data have all been published 
by the beginning of 2014, which was chosen as a suitable cut-off point.
It is worth noting at the beginning of the article that there are no very 
significant changes in the PDFs beyond those already in the MMSTWW set, and 
all predictions for LHC processes remain very similar to those for
MMSTWW and in nearly all cases to MSTW2008. Despite the inclusion of new data 
there is a slight increase of PDF uncertainty in general (particularly for
the strange quark) due to an improved understanding of the source of 
uncertainties. We also point out here that it is expected that there will 
be another update of the PDFs in the same framework with a time-scale 
consistent with the release of the final combination of HERA 
inclusive structure function data, more LHC data for a variety of
processes, and also the expected availability of the full NNLO calculation 
of inclusive jet production and of top quark pair production differential distributions.

The outline of the paper is as follows.
In Section \ref{sec:theory} we describe the improvements that we have in our theoretical procedures since the MSTW2008 analysis \cite{MSTW} was performed. In particular, we discuss the parameterisation of the input PDFs, as well as the improved treatments (i) of the deuteron and nuclear corrections, (ii)  of the heavy flavour PDFs, (iii) of the experimental errors of the data, and, (iv) in fitting the  neutrino-produced dimuon data.   In Section \ref{sec:preLHC}  we discuss the non-LHC data which have been added since the MSTW2008 analysis, while Section \ref{sec:4} describes the LHC data that are now included in the fit, where we determine these by imposing a cut-off date of publication by the beginning of 2014. The latter Section concentrates on the description of $W$ and $Z$ production data, together with a discussion of the inclusion of LHC jet production data.  

\begin{figure} 
\begin{center}
\includegraphics[height=7.5cm]{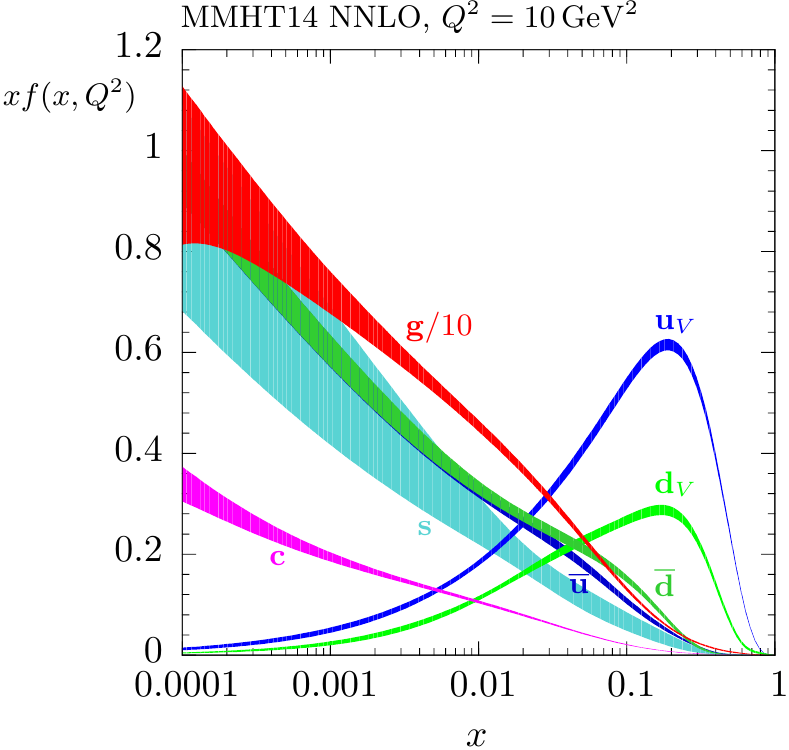}
\includegraphics[height=7.5cm]{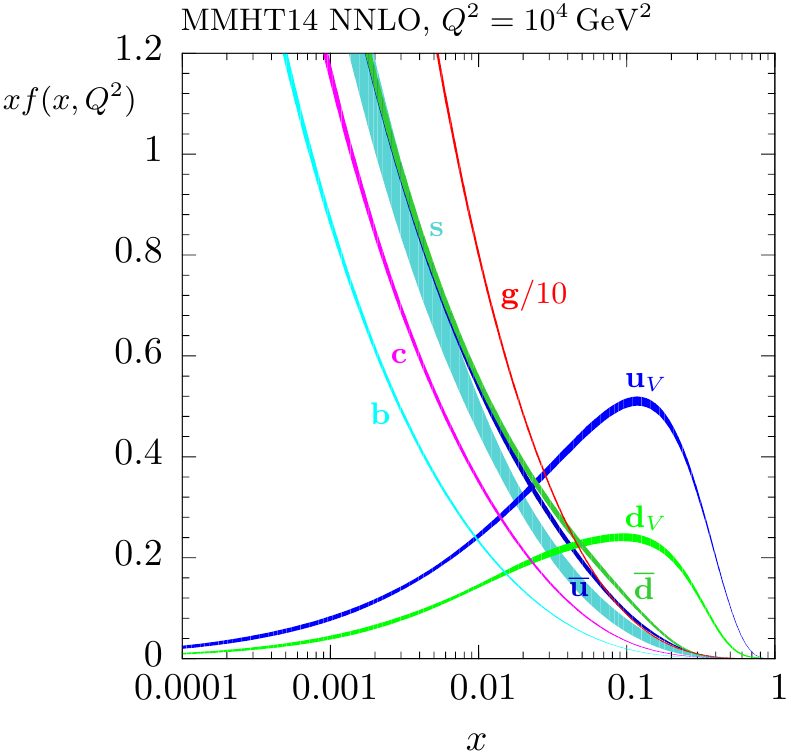}
\caption{\sf MMHT2014 NNLO PDFs at $Q^2=10 ~\GeV^2$ and $Q^2=10^4~\GeV^2$, with associated 68$\%$ confidence-level uncertainty bands. The corresponding plot of NLO PDFs is shown in Fig. \ref{fig:NLOpdfs}.}
\label{fig:NNLOpdfs}
\end{center}
\end{figure}

The results of the global analysis can be found in Section \ref{sec:5}. This section starts with a discussion of the treatment of the QCD coupling, and of whether or not to include $\alpha_S(M^2_Z)$ as a free parameter. We then present the LO, NLO and NNLO PDFs and their uncertainties, together with the values of the input parameters.  These sets of PDFs are the end products of the analysis -- the grids and interpolation code for the PDFs can be found at 
\cite{UCLsite} and will be available at \cite{LHAPDF} and 
a new HepForge \cite{hepforgesite} project site is foreseen. An example is given in Fig. \ref{fig:NNLOpdfs} which shows the NNLO PDFs at scales of $Q^2=10 ~\GeV^2$ and $Q^2=10^4 ~\GeV^2$, including the associated one-sigma (68$\%$) confidence-level uncertainty bands.

Section \ref{sec:5} also contains a comparison of the NLO and NNLO PDFs with those of MSTW2008 \cite{MSTW}. The quality of the fit to the data at LO is far worse than that at NLO and NNLO, and is included for completeness, and because of 
the potential use in LO Monte Carlo generators, though the use of generators with NLO matrix elements is becoming far more standard.
In Section \ref{sec:6} we make predictions for various benchmark processes at the LHC, and in Section \ref{sec:7} we discuss other data sets that are becoming available at the LHC which constrain the PDFs, but that are not included in the present global fit due to failure to satisfy our cut-off date; we refer to dijet and $W+c$ production and to the top quark differential distributions.   In Section \ref{sec:8} we compare our MMHT PDFs with those of the very recent NNPDF3.0 analysis \cite{NNPDF3}, and also with older sets of PDFs of other collaborations. In Section \ref{sec:9} we present our Conclusions.

\section{Changes in the theoretical procedures \label{sec:theory}}

In this Section, we list the changes in our theoretical description of the 
data, from that used in the MSTW analysis \cite{MSTW}.  We also glance 
ahead to mention some of the main effects on the resulting PDFs.

\subsection{Input distributions \label{sec:inputPDF}}
As is clear from the discussion in the Introduction, one improvement is to use parameterisations for the input distributions based on Chebyshev polynomials. Following the detailed study in \cite{MMSTWW}, we take for most PDFs a parameterisation of the form
\be
xf(x,Q_0^2)~=~A(1-x)^\eta x^\delta \left( 1+\sum^n_{i=1} a_i T^{\rm Ch}_i(y(x)) \right),
\label{eq:1}
\ee
where $Q_0^2=1~\GeV^2$ is the input scale, and $T^{\rm Ch}_i(y)$ are Chebyshev polynomials in $y$, with $y=1-2x^k$ where we take  $k=0.5$
and $n=4$. The global fit determines the values of the set of parameters  $A,~\delta,~\eta,~a_i$ for each PDF, namely for $f=u_V,~d_V,~ S,~ s_+$, where 
$S$ is the light-quark sea distribution
\be
S~\equiv~2(\bar{u}+\bar{d})+s+\bar{s}. 
\ee
For $s_+\equiv s+\bar{s}$ we set 
$\delta_+=\delta_S$. As argued in \cite{MSTW} the sea quarks at very low $x$ 
are governed almost entirely by perturbative evolution, which is flavour
independent, and any difference in the shape at very low $x$ is very quickly 
washed out. Hence, we choose to assume that this universality in the very low
$x$ shape is already evident at input. For $s_+$ we also set the 
third and fourth Chebyshev polynomials to be the same as for the light sea,
as there is not enough data which can constrain the strange quark, while
leaving all four parameters in the polynomial free leads to instabilities. 

We still have to specify the parameterisations of the gluon and of the differences $\bar{d}-\bar{u}$ and $s-\bar{s}$. For
 the parameterisation of $\Delta\equiv\bar{d}-\bar{u}$ we set 
$\eta_\Delta=\eta_S+2$, and use a parameterisation
\be
x\Delta(x,Q_0^2)~=~A_\Delta(1-x)^{\eta_{\Delta}}  x^{\delta_{\Delta}} \left( 1+ \gamma_\Delta x + \epsilon_\Delta x^2 \right).
\ee   
The (poorly determined) strange quark difference is taken to have a 
simpler input form than that in (\ref{eq:1}). That is
\be
s_- ~\equiv~ x(s-\bar{s}) ~=~A_- (1-x)^{\eta_{-}} x^{\delta_{-}} (1-x/x_0)
\label{eq:3}
\ee
where $A_-,~\delta_-$ and $\eta_-$ are treated as free parameters, and where the final factor in (\ref{eq:3}) allows us to satisfy the third number sum rule given in (\ref{eq:5}) below, i.e. $x_0$ is a crossing point.
Finally, it was found long ago \cite{MRST23}, that the global fit was considerably improved by allowing the gluon distribution to have a second term with a different small $x$ power
\be
xg(x,Q_0^2)=~A_g(1-x)^{\eta_g} x^{\delta_g} \left( 1+\sum^2_{i=1} a_{g,i} T^{\rm Ch}_i(y(x)) \right)~
+A_{g'}(1-x)^{\eta_{g'}} x^{\delta_{g'}},
\ee
where $\eta_{g'}$ is quite large, and concentrates the effect of this term 
towards small $x$. This means the gluon has 7 free parameters ($A_g$ being 
constrained by the momentum sum rule), which would be equivalent to using 
5 Chebyshev polynomials if the second term were absent.

The choice $k=0.5$, giving $y=1-2\sqrt{x}$ in (\ref{eq:1}), was found to be preferable in the detailed study presented in \cite{MMSTWW}. It has the feature that it is equivalent to a polynomial in $\sqrt{x}$, the same as the default MSTW parameterisation.  The half-integer separation of terms is consistent with the Regge motivation of the MSTW parameterisation.  The optimum order of the Chebyshev polynomials used for the various PDFs is explored in the fit. It generally turns out to be $n=4$ or 5.  The advantage of using a parameterisation based on Chebyshev polynomials is the stability and good convergence of the values found for the coefficients $a_i$.

The input PDFs are subject to three constraints from the number sum rules
\be
\int^1_0 dx ~u_V(x,Q^2_0)=2,~~~~~~\int^1_0 dx ~d_V(x,Q^2_0)=1,~~~~~~\int^1_0 dx ~(s(x,Q^2_0) - \bar s(x,Q^2_0)) 
=0,
\label{eq:5}
\ee
together with the momentum sum rule
\be
\int^1_0 dx ~x\left[u_V(x,Q^2_0)+d_V(x,Q_0^2)+S(x,Q_0^2)+g(x,Q^2_0)\right]~=~1.
\label{eq:6}
\ee
We use these four constraints to fix $A_g,~A_u,~A_d$ and $x_0$
 in terms of the other parameters.  In total there are 37 free (PDF) 
parameters in the optimum global fit, and there is also the strong coupling 
defined at the scale of the $Z$ boson mass $M_Z$, i.e.
$\alpha_s(M_Z^2)$, which we allow to be free when determining the best fit.  
Checks have been performed on our procedure which show that 
there is extremely little sensitivity to variation in $Q_0^2$ for either the fit quality 
or the PDFs extracted.

\subsection{Deuteron corrections   \label{sec:2.2}}
It is still the case that we need deep inelastic data on deuteron targets 
\cite{Benvenuti:1989fm,Arneodo:1996qe,Arneodo:1996kd,Adams:1996gu,Whitlow:1991uw,SLAC1990}
in order to fully separate the $u$ and $d$ distributions at moderate and large 
values of $x$.  Thus we should consider the correction factor $c(x)$ to be 
applied to the deuteron data
\be
F^d(x,Q^2)~=~c(x) \left[F^p(x,Q^2)+F^n(x,Q^2) \right]/2,
\ee
where we assume $c$ is independent of $Q^2$, and where $F^n$ is obtained 
from $F^p$ by swapping up and down quarks, and anti-quarks; that is, isospin 
asymmetry is assumed.  
In the MSTW analysis, motivated by \cite{BK}, despite the fact that the fit included all the 
deuteron data present in this analysis, the theory was only corrected 
for shadowing for small values of $x$, with a linear form for $c$ with 
$c=0.985$ at $x=0.01$ and $c=1$ just above $x=0.1$;  above this point it was 
assumed that $c=1$.  

In Ref. \cite{MMSTWW} we studied the deuteron correction factor in detail.  
We introduced the following flexible parameterisation of $c(x)$, which allowed 
for the theoretical expectations of shadowing (but which also allowed the 
deuteron correction factor to be determined by the data)
\bea
c(x)~&=&~(1+0.01N)~[1+0.01c_1 {\rm ln}^2(x_p/x)], ~~~~~~~~~~~~~~~~~~~~~~~~~~~~   x<x_p,\\
c(x)~&=&~(1+0.01N)~[1+0.01c_2 {\rm ln}^2(x/x_p)+0.01c_3{\rm ln}^{20}(x/x_p)], ~~~   x>x_p,
\eea
where $x_p$ is a `pivot point' at which the normalisation is $(1+0.01N)$. 
For $x<x_p$ there is freedom for $c(x)$ to increase or decrease smoothly 
depending on the sign of the parameter $c_1$. The same is true above $x=x_p$, 
but the very large power in the $c_3$ term is added
to allow for the expected rapid increase of $c(x)$ as $x \to 1$ due to Fermi 
motion.  If, as expected, there is shadowing at low $x$ and also a dip for 
high, but not too high, $x$ (that is if both $c_1$ and $c_2$ are found to be 
negative), then $x_p$ is where $c(x)$ will be a maximum, as expected from 
antishadowing (provided $N>0$).  If we fix the value of $x_p$, then the 
deuteron correction factor $c(x)$ is specified by the values of four 
parameters: the $c_i$ and $N$. In practice $x_p$ is chosen to be equal 
to $0.05$ at NLO, but a slightly smaller value of $x_p=0.03$ is marginally 
preferred at NNLO. 

\begin{figure} [t]
\begin{center}
\includegraphics[height=8cm]{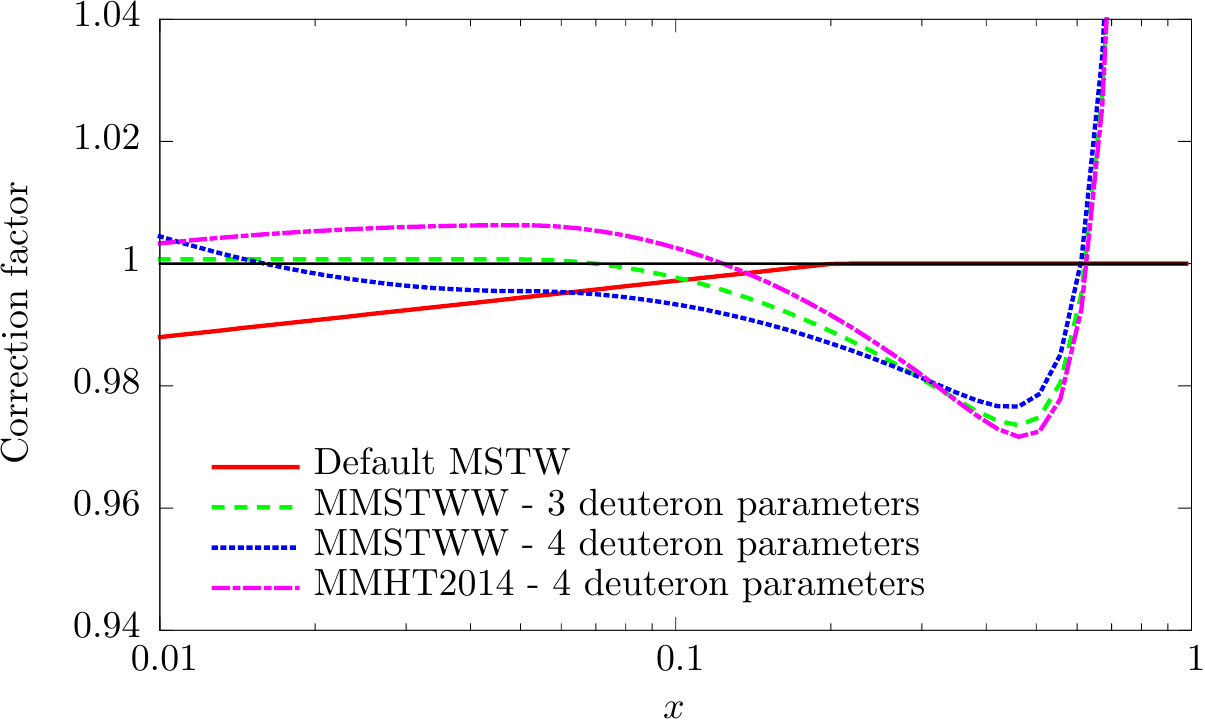}
\caption{\sf The deuteron correction factors versus $x$ at NLO shown for the fits listed in Table \ref{tab:1}.  The error corridor for the MMHT2014 curve is shown in Fig. \ref{fig:deutunc}, together with the result at NNLO.}
\label{fig:deut}
\end{center}
\end{figure}

As already emphasised, the introduction of a flexible parameterisation of the 
deuteron correction, $c(x)$, coupled with the extended Chebyshev 
parameterisation of the input PDFs was found \cite{MMSTWW}, unlike MSTW 
\cite{MSTW}, to describe the data for lepton charge asymmetry from $W^\pm$ 
decays well, and, moreover, to give a much better description of the same set 
of global data as used in the MSTW analysis.  The only blemish was that 
for the best possible fit the 4-parameter version of $c(x)$ had an unphysical 
form (with $c_1$ positive), so the preferred fit, even though it was of 
slightly lower quality, was taken to be the 
3-parameter form with $c_1=0$.  In the present analysis (which includes the 
post-MSTW data) this blemish does not occur, and the 4-parameter form of the 
deuteron correction factor turns out to be much as expected theoretically.  
The parameters are listed in Table \ref{tab:1} and the corresponding 
deuteron correction factors shown in Fig. \ref{fig:deut}. The fit quality for 
the deuteron structure function data for MMSTWW at NLO with 3 parameters was 
477/513, and was just a couple lower when 4 parameters were used. For 
MMHT2014 at NLO the value is 471/513 and at NNLO is slightly better at 
464/513. Hence, the new constraints on the flavour decomposition from the 
Tevatron and LHC are, if anything, slightly improving the fit to deuteron data,
though part of the slight improvement is due to a small change in the 
way in which NMC data is used -- see section 2.7.

\begin{table} [h]
\begin{center}
\begin{tabular}{|l|c|c|c|c|}\hline
 PDF fit &   $N$~~~ & $c_1$ &  $c_2$ & $c_3\times 10^8 $ \\ \hline
  MMSTWW -  3 pars. & 0.070 & 0  & $-0.608$  & 3.36\\
  MMSTWW - 4 pars. & $-0.490$ &  0.349 & $-0.444$ & 3.40\\
  MMHT2014 NLO  & $0.630 \pm 0.831$  & $-0.116 \pm 0.507$ & $-0.758 \pm 0.324$ & $3.44 \pm 1.89$\\
  MMHT2014 NNLO  & $0.589 \pm 0.738$ & $-0.116 \pm 0.996$ & $-0.384 \pm 0.182$ & $0.0489\pm 0.0056$\\
 \hline
\end{tabular}
\end{center}
\caption{\sf The values of the parameters for the deuteron correction factor  found in the MMSTWW \cite{MMSTWW} and the present (MMHT)  global fits.}
\label{tab:1}
\end{table}

The uncertainties for the parameters in the MMHT2014 PDF fits are also shown in 
Table \ref{tab:1}. These values are quoted as three times the uncertainty
obtained using the standard $\Delta \chi^2=1$ rule. In practice we use the so-called
``dynamic tolerance'' procedure to determine $\Delta \chi^2$ for each of our
eigenvectors, as explained in Section 6 of \cite{MSTW}, and also discussed in Section 
5 of this article, and a precise determination of the deuteron correction uncertainty
is only obtained from the similar scan over $\chi^2$ as used to determine eigenvector 
uncertainties. However, a typical value is three times the $\Delta \chi^2=1$ 
uncertainty, and this should give a fairly accurate representation of the 
deuterium correction uncertainty.\footnote{This choice works well for PDF 
uncertainties, as discussed in \cite{WattThorne}.} The correlation matrices for the 
deuteron parameters for the NLO and NNLO analyses are, respectively, 
\be
c_{ij}^{\rm NLO} = 
\begin{pmatrix}  
1.000  & -0.604 & -0.693 &  0.177 \\
-0.604 &  1.000 &  0.426 & -0.116 \\
-0.693 &  0.426 &  1.000 & -0.360 \\
 0.177 & -0.116 & -0.360 &  1.000 \\
\end{pmatrix},
\ee

\be
c_{ij}^{\rm NNLO} = 
\begin{pmatrix}  
1.000  & -0.540 & -0.692 &  0.179 \\
-0.540 &  1.000 &  0.371 & -0.118 \\
-0.692 &  0.371 &  1.000 & -0.341 \\
 0.179 & -0.118 & -0.341 &  1.000 \\
\end{pmatrix}.
\ee

\begin{figure}
\begin{center}
\includegraphics[height=8cm]{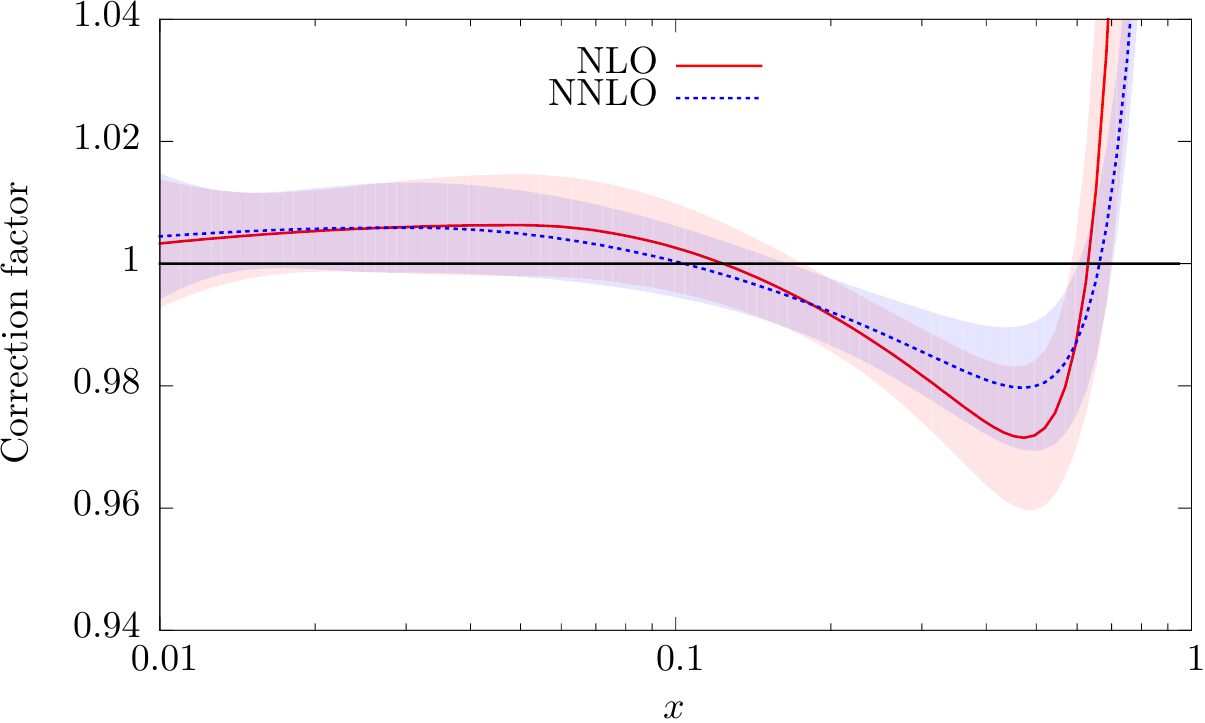}
\includegraphics[height=8cm]{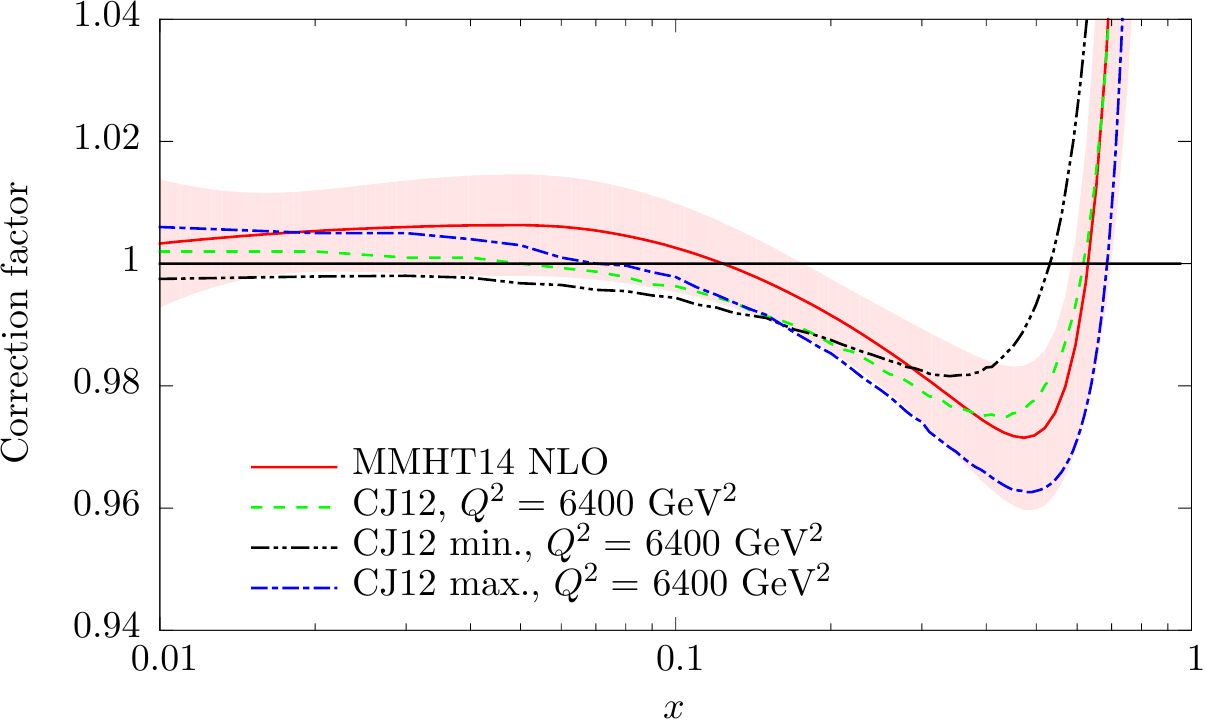}
\vspace{0cm}
\caption{\sf The deuteron correction factors versus $x$ at NLO and NNLO with uncertainties (top) and at NLO compared to the CJ12 corrections (bottom).}
\label{fig:deutunc}
\end{center}
\end{figure}

We plot the central values and uncertainties of the deuteron corrections 
at NLO and at NNLO in the higher plot of Fig.~\ref{fig:deutunc}. One can see that the uncertainty is 
of order $1\%$ in the region $0.01\lapproxeq x\lapproxeq 0.4$ well constrained by deuteron 
data. Although the best fits now correspond to a decrease as $x$ becomes
very small this is not determined within even a one standard deviation uncertainty band.  The lack 
of deuteron data at high $x$, $x\gapproxeq 0.75$, mean that the correction factor is not really 
well determined in this region, and the uncertainty is limited by the form of the parameterisation.
However, the sharp upturn at $x \sim 0.6$ is driven by data.   

Until recently, most of the other groups that have performed global PDF 
analyses do not include deuteron corrections. An exception is the analysis of 
Ref. \cite{OAM}.  In the present work, and in MMSTWW \cite{MMSTWW}, we have 
allowed the data to determine what the deuteron correction should be, with an 
uncertainty determined by the quality of the fit.  The CTEQ-Jefferson 
Lab collaboration \cite{OAM} have performed three NLO global analyses which 
differ in the size of the deuteron corrections. They are denoted CJ12min, 
CJ12med and CJ12max, depending on whether they have mild, medium or strong 
deuteron corrections. We plot the comparison of these to our NLO deuteron 
corrections in the lower plot of Fig.~\ref{fig:deutunc}. The 
CJ12 corrections are $Q^2$-dependent due to target mass and higher twist 
contributions, as discussed in \cite{Accardi}. 
These contributions die away asymptotically, so we compare to the CJ12
deuteron corrections quoted at a very high $Q^2$ value of $6400~\GeV^2$. 
In the present analysis it turns out that the data 
select deuteron corrections that are in very good agreement for $x>0.2$ with 
those given by the central CJ set, CJ12med.  
The behaviour at smaller values of $x$ is sensitive to the lepton charge 
asymmetry data from $W^\pm$ decays at the Tevatron and LHC, the latter of 
which are not included in the CJ12 fits.

\subsection{Nuclear corrections for neutrino data}

The neutrino structure function data are obtained by scattering on a heavy 
nuclear target. The NuTeV experiment \cite{NuTeV} uses an iron target, and 
the CHORUS experiment \cite{CHORUS} scatters on lead. Additionally the dimuon 
data from CCFR/NuTeV \cite{Dimuon} is also obtained from (anti)neutrino 
scattering from an iron target. 
In the MSTW analysis \cite{MSTW} we applied the nuclear corrections $R_f$, 
defined as
\be
f^A(x,Q^2)~=~R_f(x,Q^2,A)~f(x,Q^2),
\ee
separately for each parton flavour $f$ using the results of a NLO fit by de 
Florian and Sassot \cite{deFS}. The $f^A$ are defined to be the PDFs of a {\it proton} bound in a nucleus of mass number $A$. 
In the present analysis we use the updated results of de Florian et al., 
which are shown in Fig.14 of \cite{deF}. The nuclear corrections for the 
heavy flavour quarks are assumed to be the same as that found for strange quarks,
though the contribution from heavy quarks is very small.
The updated nuclear corrections are quite similar, except for the 
strange quark for $x<0.1$, though this does not significantly affect 
the extracted values of the strange quark. The new corrections 
improve the quality of the fit by $\sim 25$ units in $\chi^2$, spread over
a variety of data sets, including obvious candidates such as NuTeV 
$F_2(x,Q^2)$, but also HERA structure function data and CDF jet data which are 
only indirectly affected by nuclear corrections. 

As in \cite{MSTW} we multiply the nuclear corrections by a $3$-parameter 
modification function, eq.(73) in \cite{MSTW}, which allows a penalty-free
change in the details of the normalisation and shape. As in \cite{MSTW} the 
free parameters choose values $\lapproxeq 1$, i.e. they chose modification of 
only a couple of percent at most away from the default values. Hence, for 
both deuteron and heavy nuclear corrections, we allow the fit to choose 
the final corrections with no penalty; but in both cases the corrections are
fully consistent with expectation, i.e. any penalty applied would 
have very little effect.

\subsection{General Mass - Variable Flavour Number Scheme (GM-VFNS)}

The treatment of heavy flavours -- charm, bottom -- has an important impact 
on the PDFs extracted from the global analysis due to the data available for 
$F_2^h(x,Q^2)$ with $h=c,~b$, and also on the heavy flavour contribution to 
the total structure function at small $x$.  Recall that there are two 
distinct regions where heavy quark production can be readily described.  For 
$Q^2\sim m^2_h$ the massive quark may be regarded as being only produced in 
the final state, while for $Q^2 \gg m^2_h$ the quark can be treated as 
massless, with the ln$(Q^2/m^2_h)$ contributions being summed via the 
evolution equations. The GM-VFNS is the appropriate way to interpolate 
between these two regions, and as shown recently 
\cite{Thorne,NNPDFgmvfns,ThorneFFNS}, the use of the fixed flavour number 
scheme (FFNS) leads to significantly different results in a PDF fit 
to the GM-VFNS, even at NNLO.  However, there is freedom to define different 
definitions of a GM-VFNS, which has resulted in the existence of various 
prescriptions, each with a particular reason for its choice. Well known 
examples are the original  Aivazis-Collins-Olness-Tung (ACOT) \cite{ACOT} 
and Thorne-Roberts (TR) \cite{TR} schemes, and 
their more recent refinements \cite{Chu,Tung,TR1}. 
The MSTW analysis \cite{MSTW} adopted the more recent TR' prescription in \cite{TR1}.

Ideally one would like any GM-VFNS to reduce exactly to the correct fixed 
flavour number scheme at low $Q^2$ and to the correct zero-mass VFNS as $Q^2 \to \infty$.  
This has been accomplished in \cite{Thorne}, by introducing a 
new `optimal' scheme which improves the smoothness of the transition region 
where the number of active flavours is increased by one. The optimal scheme 
is adopted in the present global analysis.\footnote{We do not treat the top 
quark as a parton, i.e. even at high scale we remain in a 5 flavour scheme. 
Even at LHC energies the mass of the top quark is quite large compared
to any other scale in the process, and the expressions for the cross sections 
for top production are all available in the scheme where the top appears in 
the final state.}  

In general, at NLO, the PDFs, and the predictions using them 
can vary by as much as 2$\%$ from the mean value due to the ambiguity in the 
choice of the GM-VFNS, and a similar size variation feeds into predictions for
e.g. $W,Z$ and Higgs boson production at colliders. At NNLO there is far 
more stability to varying the GM-VFNS definition. Typical changes are less 
than 1$\%$, and then only at very small $x$ values. This is illustrated well 
by the plots shown in Fig. 6 of \cite{Thorne}. Similarly predictions for 
standard cross sections vary at the sub-percent level at NNLO.

\subsection{Treatment of the Uncertainties}

All data sets which are common to the MSTW2008 and the present analysis
are treated in the same manner in both, except that the 
multiplicative, rather than additive, definition of correlated uncertainties 
is used, 
as discussed in more detail below. All new data sets use the full treatment of 
correlated
uncertainties, if these are available. For some data sets these are provided 
as a set of individual sources of correlated uncertainty, while for others 
only the final correlation matrix is provided. 
 
If only the final correlation matrix is provided, then 
we use the expression 
\be 
\chi^2 = \sum_{i=1}^{N_{\rm pts}} \sum_{i=j}^{N_{\rm pts}}
(D_i-T_i) (C^{-1})_{ij} (D_j-T_j),
\ee
where $D_i$ are the data 
values $T_i$ are the parametrised\footnote{The parameters are those of the input PDFs, the QCD coupling $\alpha_s(M_Z^2)$ and the nuclear corrections.}
predictions, and $C_{ij}$ is the covariance matrix.

In the case where the individual sources of correlated errors are provided
the goodness-of-fit, $\chi^2$, including the full correlated error 
information, is defined as 
\be
\chi^2=\sum_{i=1}^{N_{\rm pts}}\left(\frac{D_i+\sum_{k=1}^{N_{\rm corr}}
r_k\sigma_{k,i}^{\rm corr}-T_i}{\sigma_i^{\rm uncorr}}\right)^2+\sum_{k=1}^{N_{\rm corr}}r_k^2,
\ee 
where $D_i+\sum_{k=1}^{N_{\rm corr}}r_k\sigma_{k,i}^{\rm corr}$ are the data 
values allowed to shift by some multiple $r_k$ of the systematic error
$\sigma_{k,i}^{\rm corr}$ in order to give the best 
fit, and where $T_i$ are the parametrised predictions. 
The last term on the right is the penalty for the shifts of 
data relative to theory for each source of correlated uncertainty. 
The errors are combined multiplicatively, that is 
$\sigma_{k,i}^{\rm corr}= \beta_{k,i}^{\rm corr}T_i$,  where
 $\beta_{k,i}^{\rm corr}$ are the percentage errors. Previously, in 
MSTW \cite{MSTW}, the additive definition was employed for all but the 
normalisation uncertainty. That is, $\sigma_{k,i}^{\rm corr}
= \beta_{k,i}^{\rm corr}D_i$ was used. 

To appreciate the consequence of the change we can 
think of the shift of data relative to theory as being approximately
given by
\be
\sum_{k=1}^{N_{\rm corr}} r_k\sigma_{k,i}^{\rm corr} = 
\sum_{k=1}^{N_{\rm corr}}\beta_{k,i}^{\rm corr}D_i(T_i) \approx \delta f 
D_i(T_i),
\ee
where $\delta f$ is the fractional shift in the data -- this is exactly correct 
for a normalisation uncertainty.

Defining $1+\delta f = f$, effectively the difference between the additive 
and multiplicative use of errors is that  
\be
D_i + \sum_{k=1}^{N_{\rm corr}}\beta_{k,i}^{\rm corr}D_i \sim
f*D_i \qquad \hbox{or} \qquad T_i - \sum_{k=1}^{N_{\rm corr}}\beta_{k,i}^{\rm corr}T_i \sim T_i/f.
\ee
So for the additive definition the data are effectively rescaled by $f$ while for 
the multiplicative definition the theory is rescaled by $1/f$.
This means that in the two cases the $\chi^2$ definition behaves like
\be
\chi^2\sim \left(\frac{f*D_i-T_i}{\sigma_i^{\rm uncorr}}\right)^2
\qquad \hbox{or} \qquad 
\chi^2\sim \left(\frac{D_i-T_i/f}{\sigma_i^{\rm uncorr}}\right)^2 = \left(\frac{f*D_i-T_i}{f*\sigma_i^{\rm uncorr}}\right)^2.
\ee
Hence, with our new choice, the uncorrelated errors 
effectively scale with the data, 
whereas with the previous additive definition the uncorrelated uncertainties 
remain constant as the data are rescaled. The additive definition can therefore 
lead to
a tendency for the data to choose a small scaling $f$ to bring the data closer 
together and hence reduce the $\chi^2$, as pointed out in 
\cite{D'Agostini} and discussed in \cite{NNPDFnorm}. Our previous treatment 
of uncertainties guarded against this for the most obvious case of normalisation 
uncertainty by using the multiplicative definition for this particular source. 
However, the same type of effect is possible in any relatively large systematic
uncertainty which affects all data points with the same sign, e.g. jet 
energy scale uncertainty, so the multiplicative definition is the safer choice,
and is recommended by many experiments.

The other change we make in our treatment of correlated uncertainties is 
that we now use the standard quadratic penalty in $\chi^2$ for normalisation 
shifts, rather than the quartic penalty adopted in MSTW \cite{MSTW}.  It is 
checked explicitly that this makes essentially no difference in NLO and NNLO
fits, but there is a tendency for some data to normalise down in a LO fit.
In some cases the quality of the fit at LO would be very poor without this 
freedom, though it could often be largely compensated by a change in 
renormalisation and/or factorisation scale away from the standard values.

\subsection{Fit to Dimuon data  \label{sec:2.6}}

\begin{figure} 
\begin{center}
\vspace{-4.0cm}
\includegraphics[height=12cm]{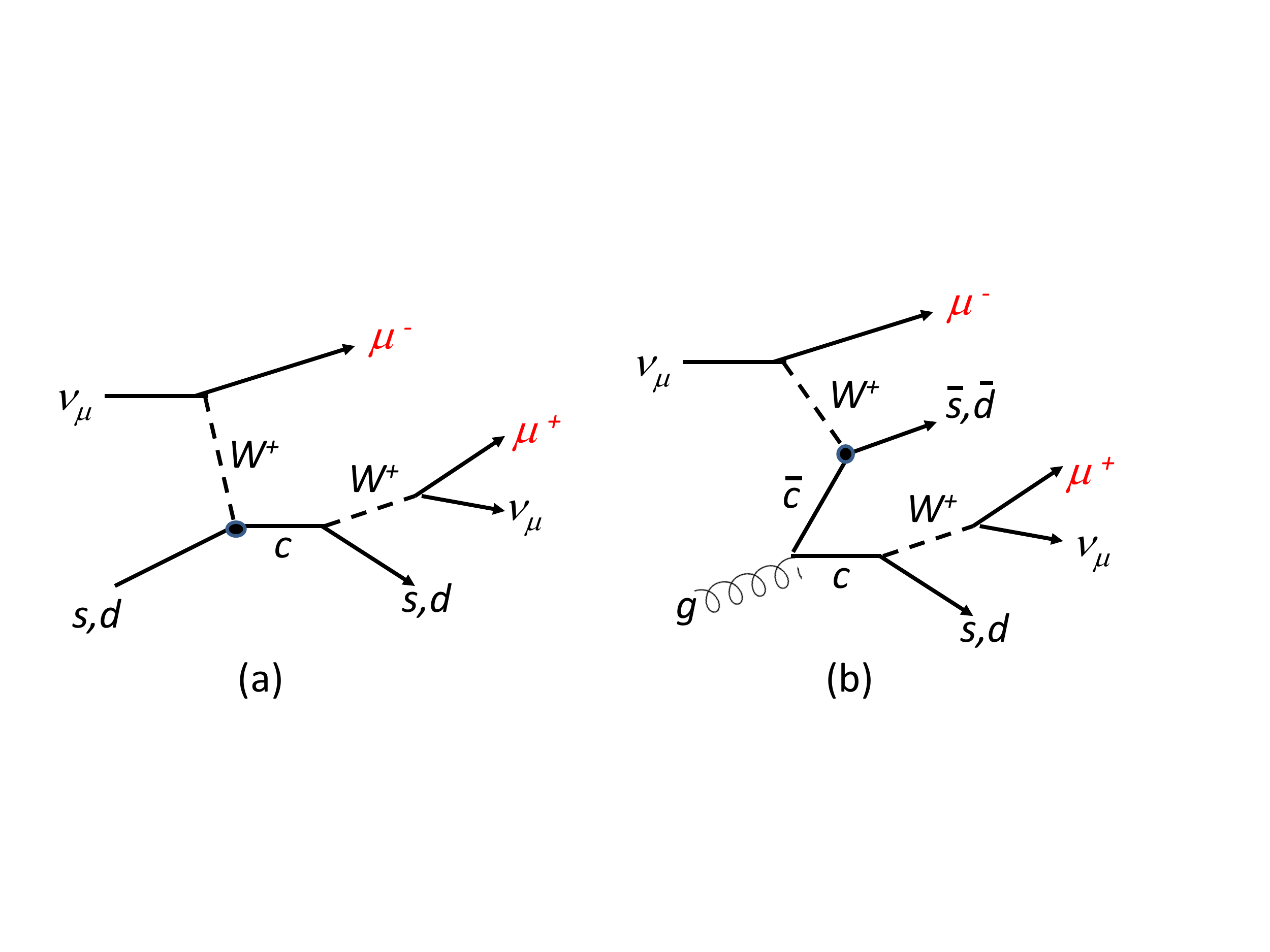}
\vspace{-3.2cm}
\caption{\sf Diagrams for dimuon production in $\nu_{\mu}N$ scattering. Only diagram (a) was considered in \cite{MSTW}, but here we include (b), although it gives a very small contribution.}
\label{fig:dimuon}
\end{center}
\end{figure}

Information on the $s$ and $\bar{s}$ quark distributions comes from dimuon 
production in $\nu_\mu N$ and ${\bar \nu}_\mu N$ scattering \cite{Dimuon},
where (up to Cabibbo mixing) an incoming muon (anti)neutrino scatters of a
(anti)strange quark to produce a charm quark, which is detected via the decay
of a charmed meson into a muon, see Fig. \ref{fig:dimuon}(a). 
These data were included in the MSTW2008 analysis, but here we 
make two changes to the analysis, one far more significant, in practice, 
than the other. 

\subsubsection{Improved treatment of the $D \to \mu$ branching ratio, $B_{\mu}$}
The comparison of theory predictions to the measured cross section  on dimuon
production requires knowledge of the branching fraction 
$B_\mu\equiv B(D\to \mu)$. In the previous analysis we used the 
fixed value
$B_\mu=0.099$ obtained by the NuTeV collaboration itself \cite{Mason}.
However, this requires a simultaneous fit of the dimuon data and the 
branching ratio, which can be dependent on assumptions made in the analysis. 
Indeed, in studies for this article we have noticed a significant dependence
on the parameterisation used for the input strange quark and the order
of perturbative QCD used. Hence, in the present analysis, we avoid 
using information on $B_\mu$ obtained from dimuon data.  Instead we use the 
value obtained from direct measurements \cite{Bolton}: 
$B_\mu = 0.092\pm 10\%$, where we feed the uncertainty into the PDF analysis.  
We note that this is somewhat lower than the number used in our previous 
analysis, though the two are easily consistent within the uncertainty of 
this value. We find that the fits prefer
\be
B_\mu=(0.085-0.091) \pm 15\%,
\ee
where the variation in the first number is the variation between 
the best value from different fits, and the uncertainty of $15\%$ is the 
uncertainty within any one fit due to the uncertainty on the data,
i.e. the variation that provides a significant deterioration in $\chi^2$ for 
dimuon data as determined by the dynamical tolerance procedure used to define
PDF uncertainties. 
Hence, the preferred value is always close to the central value in 
\cite{Bolton}. These  
lower branching ratios compared to the MSTW2008 analysis 
lead to a small increase in the normalisation of the strange quark. 
However, probably more 
importantly, the large uncertainty on the branching ratio allows for a much 
larger uncertainty on the strange quark than in our previous analysis. 
Indeed, this is one of the most significant differences between MMHT2014 and 
MSTW2008 PDFs. 

\subsubsection{Inclusion of the $g\to c\bar{c}$ initiated process with a displaced vertex}
We also correct the dimuon cross sections for a small missing contribution. 
In the previous analysis we calculated the dimuon cross section ignoring the
contribution where the charm quark is produced away from the interaction 
point of the quark with the $W$ boson, i.e. the contributions where
$g \to c \bar c$ then $(\bar c)c + W^{\pm} \to (\bar s)s$, as sketched in Fig. \ref{fig:dimuon}(b). Previously we had included only Fig. \ref{fig:dimuon}(a) and had (incorrectly) assumed that the absence of Fig. \ref{fig:dimuon}(b) was accounted for by the acceptance corrections. 
We now include this type of contribution, but it is usually of the order 
$5\%$ or less of the total dimuon cross section.
The correction to each of the structure functions, $F_2, F_L$ and $F_3$,
is proportionally larger than this, but if we look at 
the total dimuon cross section then it is proportional to 
$s +(1-y)^2 \bar c$ (or $\bar s +(1-y)^2 c$), where $y$ is the 
inelasticity $y=Q^2/(xs)$ and $c (\bar c)$ is the charm distribution coming
from the gluon splitting. However, $c (\bar c)$ only becomes significant 
compared to $s(\bar s)$ at higher $Q^2$ and low $x$, exactly where
$y$ is large and the charm contribution in the total cross section is 
suppressed. As such, this correction has a very small effect on the 
strange quark 
distributions  that are obtained, being of the same order as the change in nuclear 
corrections and much smaller than the changes due to the different treatment 
of the branching ratio $B_{\mu}$.

\subsection{Fit to NMC structure function data  \label{sec:2.7}}

In the MSTW2008 fit we used the NMC structure function data with the
$F_2(x,Q^2)$ values corrected for $R= F_L/(F_2-F_L)$ measured by 
the experiment, as originally recommended. However, it was pointed out in 
\cite{ABMNMC} that $R_{\rm NMC}$, the value of $R$ extracted from data
by the NMC collaboration \cite{Arneodo:1996qe}, was used more widely than 
was  really applicable. 
For example without changing the value over a range of $Q^2$, and that it was 
also often rather different from the prediction for $R$ obtained using the PDFs 
and perturbative QCD. In Section 5 of \cite{ThorneWattHiggs} we agreed with 
this, and showed the effect of using instead $R_{1990}$,  
a $Q^2$-dependent empirical parameterisation of SLAC data dating from 1990
\cite{SLAC1990} which agrees fairly well with the QCD predictions in the 
range where data are used. It was shown that the effect of this change on
our extracted PDFs and value of $\alpha_S(M_Z^2)$ was very small
(in contradiction to the claims in \cite{ABMNMC} but broadly in agreement with 
\cite{NNPDFNMC}), since the change in $F_2(x,Q^2)$ was only at most 
about the size of the uncertainty of a data point for a small fraction
of the data points, and negligible for many data points.    
In this analysis we use the same treatment as in \cite{ThorneWattHiggs},
i.e. the NMC structure data on $F_2(x,Q^2)$ with the 
$F_L(x,Q^2)$ correction very close 
to the theoretical $F_L(x,Q^2)$ value. This has very little effect, though 
the change in $F_2^d(x,Q^2)$ for $x<0.1$ does help the deuteron 
correction at low $x$ to be more like the theoretical expectation.

\section{Non-LHC data included since the MSTW2008 analysis   \label{sec:preLHC}}

Here we list the changes and additions to the non-LHC data sets used in the 
present analysis as compared to MSTW2008 \cite{MSTW}. 
All the data sets used in the MSTW2008 analysis are still included, 
unless the update is explicitly mentioned below. We continue to use the 
same cuts on structure function data, i.e. $Q^2=2~\GeV^2$ and $W^2=15~\GeV^2$. 
In \cite{MSTW} we imposed a stronger $W^2=25~\GeV^2$ cut on $F_3(x,Q^2)$ structure 
function data due to the expected larger contribution from higher-twist corrections
in $F_3(x,Q^2)$ than in $F_2(x,Q^2)$, see e.g. \cite{renormalon}. However, this still
leaves a possible contribution from quite small $x$ values for rather low $Q^2$. 
Hence we now impose a cut on $Q^2 = 5~\GeV^2$ for $F_3(x,Q^2)$.   

As an aside, we should comment on the very small $x$ domain. As usual 
we do not impose any cut at low $x$, although, at present, there are essentially no (non-LHC or LHC) data 
available probing the $x\lapproxeq 0.001$ domain\footnote{Exceptions are exclusive $J/\psi$ production 
\cite{LHCbJ} and low-mass Drell-Yan production \cite{LHCbDY} at high rapidity $y$ at the LHC, but here 
the data are sparse and, moreover, on the theory side, there are potentially large uncertainties,  
particularly in the former case where it is not the standard integrated PDFs which are being 
directly probed, and more work is needed for data from these processes to be useful 
\cite{ThorneLHCb,OMRDY1,OMRDY2,JMRT,IPSW}.}. The present analysis is based entirely
on fixed-order DGLAP evolution. So when we show plots, like Fig. \ref{fig:NNLOpdfs} going 
down to $x=10^{-4}$, and, later, show comparison plots going down to  
$x=10^{-5}$, we are going well beyond the available data, and also entering a domain which is potentially beyond the 
validity of a pure DGLAP framework. One possible source of contamination is large higher twist
corrections. However, even assuming these are small, in principle, the very small $x$ 
physics is influenced by the presence of large $\ln(1/x)$ terms in the 
perturbative expansion, which can be obtained from solutions of the  
BFKL equation (though this can include some higher-twist information as well). 
When data constraints are available at very small $x$, 
it is arguably the case that a unified fixed-order and resummation approach 
should be implemented. In \cite{ThorneWhite,CCSS,ABF} splitting functions 
are derived in this approach, with good agreement between groups. These 
suggest that the resummation effects lower the splitting functions for 
$x \sim 0.001-0.0001$ before a rise at $x< 10^{-5}$, and the likely effect is a 
slight slowing of evolution at low $Q^2$ and $x$.  Another related approach is to consider unified BFKL/DGLAP evolution which has been derived for the (integrated) gluon PDF in terms of the
gluon emission opening angle \cite{OMR}.

Having discussed the kinematic cuts that we apply, we are now ready to discuss
the fit obtained using only the non-LHC data sets. 
We study the inclusion of a variety of LHC data in the next section.
We note that in the fits, performed in this section, the coefficients of all
four Chebyshev polynomials for the $s_+$ distribution are set equal to
those for the light sea, as without LHC data there is insufficient 
constraining power in the data to fit these independently. This makes a
completely direct comparison between the full PDFs including LHC data in the 
analysis and the PDFs without LHC data impossible.

We replace the previously used HERA run I neutral and charged current 
data measured by the H1 
and ZEUS collaborations, by their combined data set \cite{H1+ZEUS} and use the 
full treatment of correlated errors.  We use a lower 
$Q^2$ cut of $2 ~\GeV^2$ and break the 
data down into five subsets; $\sigma^{{\rm NC},e^+p}$ at centre of mass 
energy $820$ GeV (78 points), $\sigma^{{\rm NC},e^+p}$ at centre of mass energy 
$920$ GeV (330 pts.), $\sigma^{{\rm NC},e^-p}$ at centre of mass energy $920$ GeV
(145 pts.), $\sigma^{{\rm CC},e^+p}$ at 
centre of mass energy $920$ GeV (34 pts.) and $\sigma^{{\rm NC},e^-p}$ at centre of 
mass energy $920$ GeV (34 pts.).   The fit to these data is very good at 
both NLO and NNLO; with a slightly better fit at NNLO, i.e.
$\chi^2/N_{\rm pts}=644.2/621$ at NNLO
compared to $666.0/621$ at NLO. Most of this 
improvement is in the $\sigma^{{\rm NC},e^+p}$ data which is 16 units better at 
NNLO. We do not include the separate H1 and ZEUS run II data yet, but wait 
for the combined data set, which as for run I we anticipate will produce 
improved constraints compared to the separate sets.

Similarly, we remove the previous measurements by ZEUS and H1 of 
$F_2^{c\bar c}(c,Q^2)$ and include instead 
the combined HERA data on $F_c(x,Q^2)$ 
\cite{H1+ZEUScharm} and use the full information on correlated uncertainties. 
Unlike the inclusive structure function data 
these data are fit better at NLO than NNLO, with  
$\chi^2/N_{\rm pts.} = 68.5/52$ at NLO but $\chi^2/N_{\rm pts.} = 78.5/52$ at NNLO
(this difference is less clear, and the values of $\chi^2$ are lower, if the 
additive definition of correlated uncertainties is used for this data set). 
As in the MSTW2008 analysis we use $m_c=1.4~\GeV$ in the pole mass scheme. 
Preliminary investigation implies that if $m_c$ is varied, a value 
$1.2-1.3~\GeV$ is preferred at both NLO and NNLO. 

\begin{figure} 
\begin{center}
\includegraphics[height=7cm]{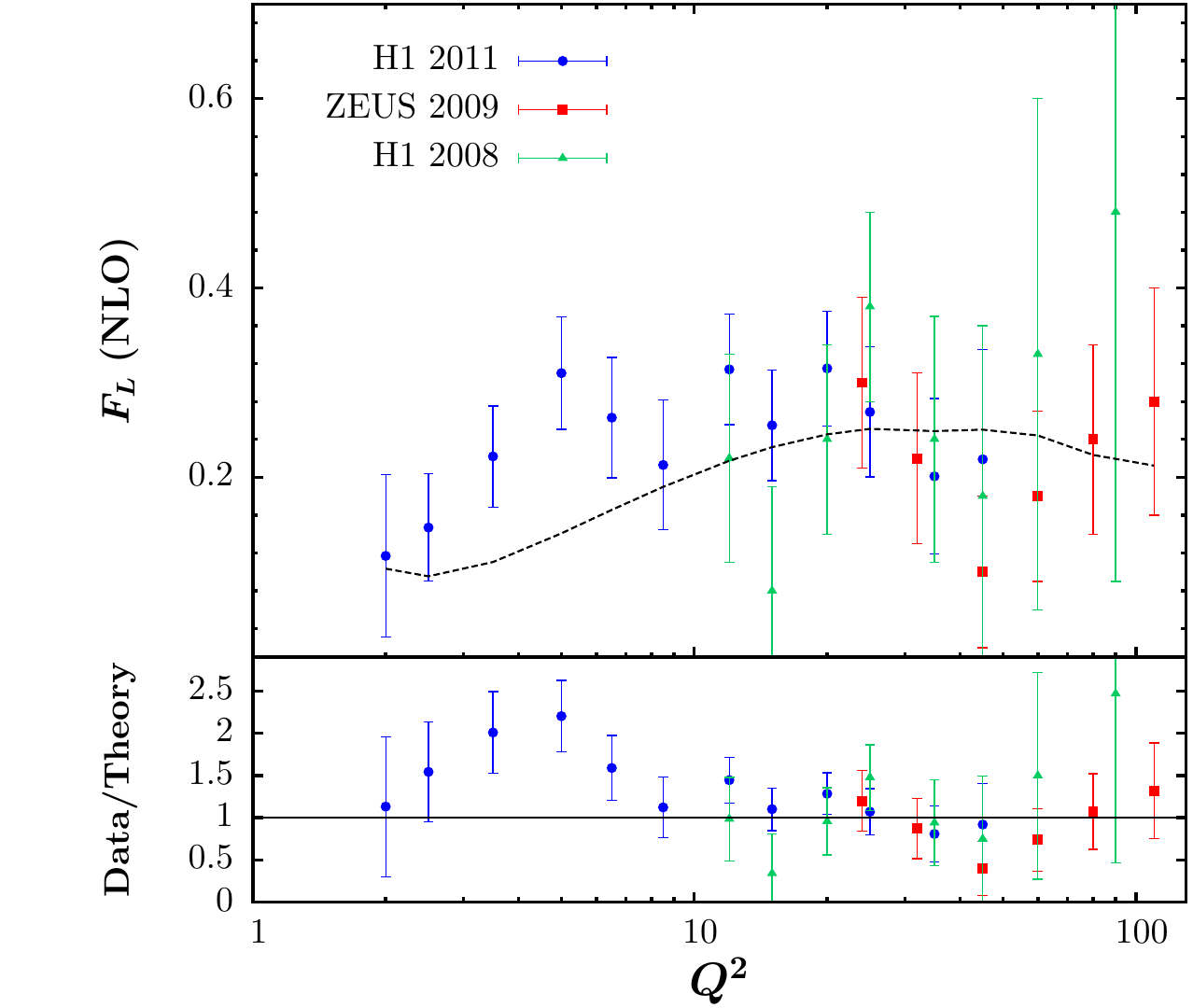}
\includegraphics[height=7cm]{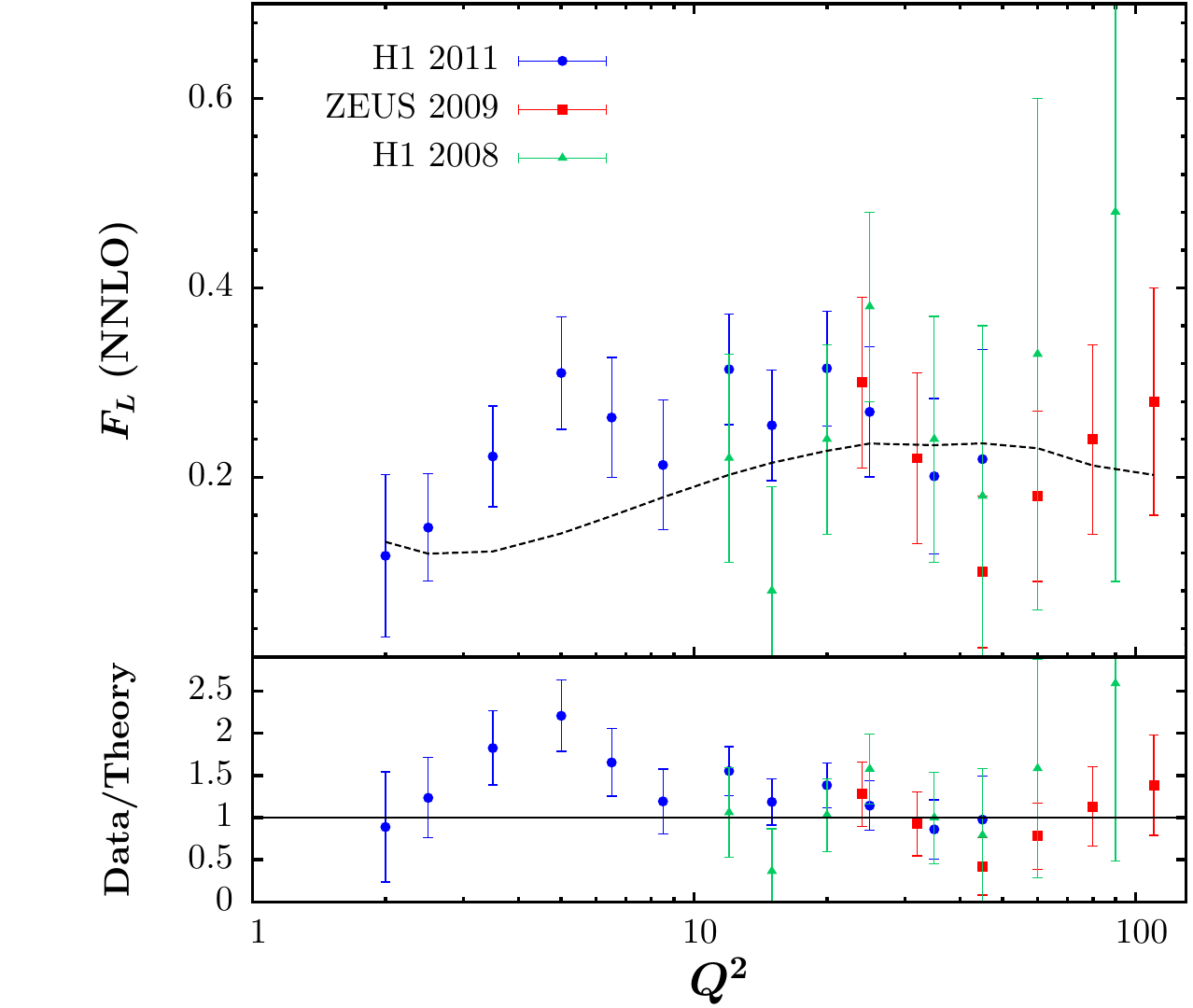}
\caption{\sf The fit quality for the HERA data on $F_L(x,Q^2)$
from \cite{H1FL,H1-FL,ZEUS-FL} at NLO (left) and NNLO (right).
The dotted curve, shown for illustration, is obtained from the prediction for the data in 
\cite{H1-FL} below $Q^2=45~\GeV^2$ and from the prediction for the
data in \cite{ZEUS-FL} above this. The ``Data/Theory'' comparison 
is obtained for the individual data points in each case.}
\label{fig:FL}
\end{center}
\end{figure}

We also include all of the HERA $F_L(x,Q^2)$ measurements published 
before the beginning of 2014 \cite{H1FL,H1-FL,ZEUS-FL}. The global fit undershoots 
some of the data a little at the lowest $Q^2$ values, slightly 
more so at NNLO than at NLO, as seen in Fig. \ref{fig:FL}, but the $\chi^2$ 
values are not much more than one per point. For the HERA $F_L(x,Q^2)$ data
we obtain $\chi^2/N_{\rm pts}= 29.8/26$ at NLO and $\chi^2/N_{\rm pts}= 
30.4/26$ at NNLO. 

\begin{figure} 
\begin{center}
\includegraphics[height=6.5cm]{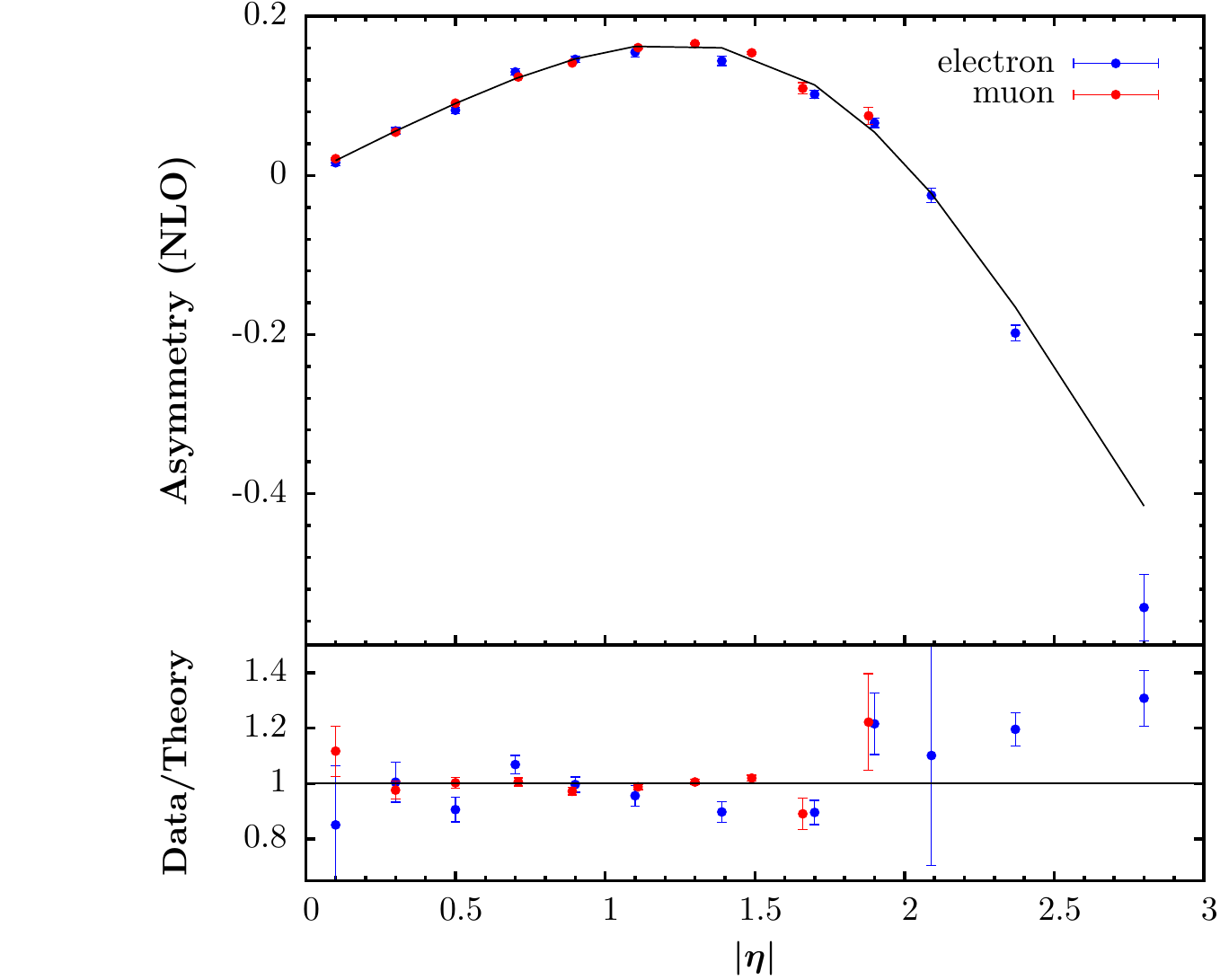}
\includegraphics[height=6.5cm]{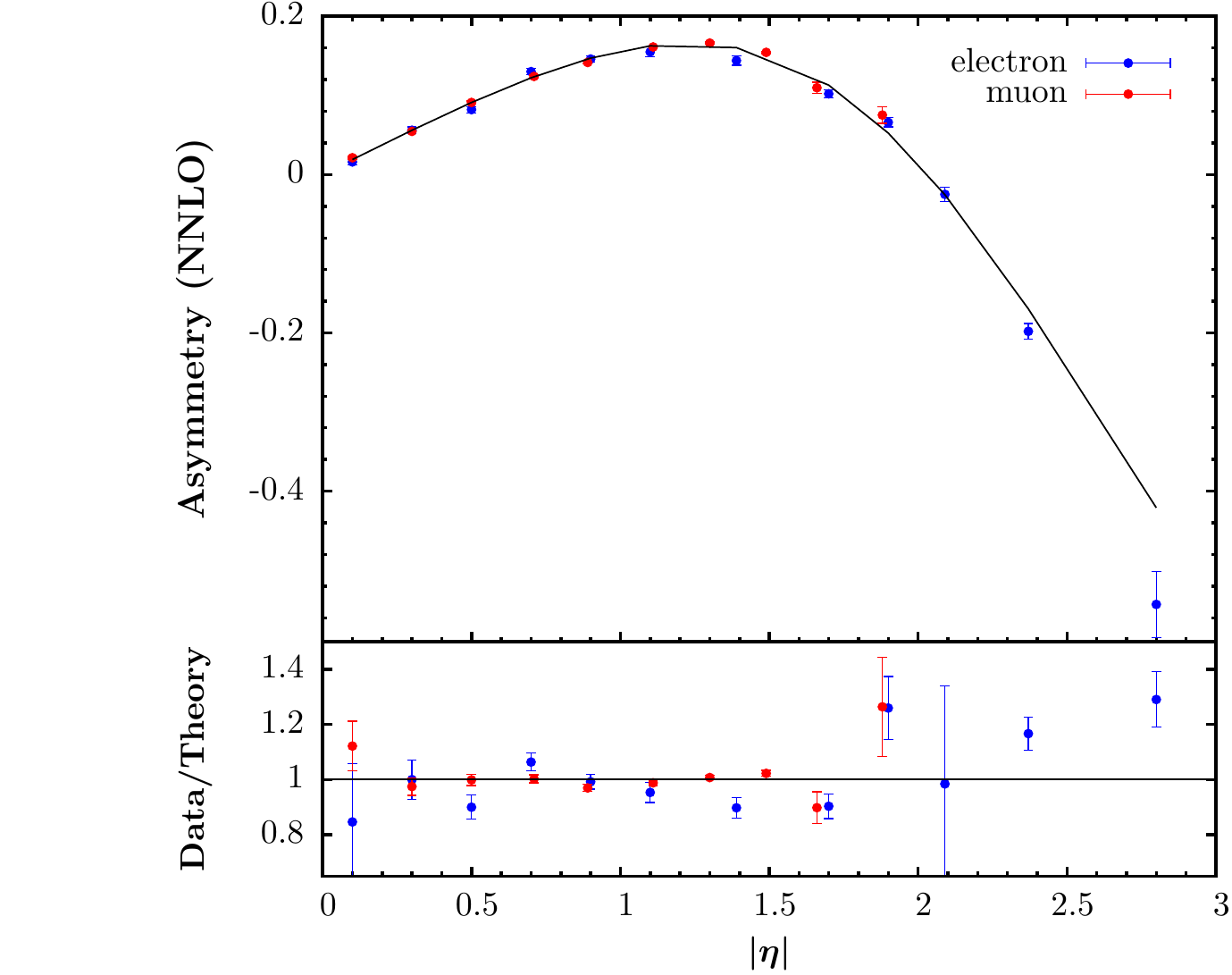}
\caption{\sf The fit quality for the two D0 lepton asymmetry data sets
\cite{D0-easym0.75,D0-muasym7.3} at NLO (left) and NNLO (right).}
\label{fig:D0asym}
\end{center}
\end{figure}

In the present analysis we include the CDF $W$ charge asymmetry data 
\cite{CDF-Wasym}, the D0 electron charge asymmetry data with $p_T>25$ GeV 
based on 0.75 ${\rm fb}^{-1}$ \cite{D0-easym0.75} and the new D0 muon charge 
asymmetry data with $p_T>25$ GeV based on 7.3 ${\rm fb}^{-1}$ 
\cite{D0-muasym7.3}. These replace the Tevatron asymmetry data used in the 
MSTW2008 analysis. Where the information on correlated uncertainties is available we use this in the conventional manner in calculating the $\chi^2$ values. The nominal fit quality for each of these
data sets appears quite poor with $\chi^2/N_{\rm pts}= 32.1/13, 30.5/12$ and
$20.3/10$ respectively at NLO and $\chi^2/N_{\rm pts}= 28.8/13, 28/12$ and
$19.8/10$ respectively at NNLO, but this seems to be mainly due to 
fluctuations in the data making a very good quality fit impossible (especially when 
fitting the data sets simultaneously), as seen in 
Fig.~\ref{fig:D0asym}. There is a tendency to overshoot the data at the very
highest rapidity, though this is a little less at NNLO than at NLO (we use
FEWZ \cite{FEWZ3} for the NLO and NNLO corrections).We do get an
approximately 2 sigma shift of data relative to theory corresponding to the 
systematic uncertainty due to electron identification for the fit to  
CDF $W$ charge asymmetry data, but no large shifts for the  new D0 muon charge 
asymmetry data. 

We also include the final measurements for the CDF $Z$ rapidity distribution 
\cite{CDF-Zrap}, since the final data changed slightly after the MSTW fit. 
We also now include the very small photon contribution in our calculation.
The effect of this second correction was discussed in Section 11.2 of 
\cite{MSTW}, although it was not used in the extraction of the MSTW2008 PDFs. 
The effect of both the final data set and the photon contribution is to 
improve the fits quality, $\chi^2/N_{\rm pts}= 36.9/28$ at NLO and $39.6/28$
at NNLO, compared to $49/29$ at NLO and $50/29$ at NNLO in \cite{MSTW},
while having essentially negligible impact on the PDFs.  

These changes to the theoretical procedures, and additions to the global data 
that are fitted, do not change the PDFs very much from those in \cite{MSTW}, 
except for the large change in ($u_V-d_V$) around $x \lapproxeq 0.01$ that 
was already found in \cite{MMSTWW}.  The small changes can be seen 
in Figs.~\ref{fig:NLO1}--\ref{fig:gsNNLO} where we show the central values of these PDFs fit
only to non-LHC data with the comparison of the MMHT2014 and MSTW2008 
PDFs. There is a moderate reduction in the uncertainty on the very small
$x$ gluon distribution due to the inclusion of the combined HERA data.
Without the inclusion of the error on the branching ratio in dimuon 
production there is also a small improvement in the uncertainty on 
light quarks, but this is lost when the branching ratio uncertainty is
included; as the increased uncertainty on the strange quarks also  leads to 
some increase in the uncertainty of the up and down quarks. As seen in Fig.~13 of 
\cite{MMSTWW} the increased parameterisation and improved deuteron 
corrections lead to an increase in the uncertainty in the up and down 
valence quarks, and this is far from compensated for by the inclusion of 
the new non-LHC data in this analysis.  
There is also only a small shift in the value of the QCD coupling extracted
in the best fit to data:
\bea
{\rm at ~~~NLO}~~~~~ \alpha_S(M_Z^2) &=& 0.1200 ~~~~{\rm from}~~~~0.1202  \\
{\rm at ~NNLO} ~~~~~\alpha_S(M_Z^2) &=& 0.1181 ~~~~{\rm from}~~~~0.1171
\eea

\section{The LHC data included in the present fit \label{sec:4}}

We now discuss the inclusion of the LHC data into the PDF fit. 
This includes a variety of data on $W$ and $Z$ production, 
also the completely new process for our PDF determination of top-quark 
pair production, and finally jet production. The addition of these LHC data 
sets to the data already discussed leads us to our final set of MMHT2014 
PDFs. We make these PDFs available at NLO and NNLO, but also at LO.
The full LO fit requires a much higher value of the strong 
coupling, $\alpha_S(M_Z^2)=0.135$, if the standard scale choices are made, i.e.
$\mu^2=Q^2$ in deep inelastic scattering, $\mu^2 =M^2$ in Drell-Yan 
production and $\mu^2=p_T^2$ in jet production, the same choices as 
made at NLO and NNLO. Even so the fit 
quality is much  worse at LO than at NLO and NNLO, both of which give 
a similar quality of description of the global data. 
We will present full details of the fit quality and the PDFs in the next 
section, but first we present the results of the fit to each of the 
different types of LHC data.

\subsection{$W$ and $Z$ data}

In order to include the LHC data on $W$ and $Z$ production in a variety of 
forms of differential distribution we use 
APPLGrid$-$MCFM \cite{APPLGrid, MCFM1,MCFM2} at NLO to produce 
grids which are interfaced to the fitting code, and at NNLO we use 
DYNNLO \cite{DYNNLO} and FEWZ \cite{FEWZ3} programmes to produce
precise $K$-factors (as a function of $\alpha_S$) to convert NLO to NNLO.
In the vast majority of cases the NLO to NNLO conversion is a very small
correction, especially for asymmetries and ratios.
 
 The quality of the description of the LHC $W$ and $Z$ data in the present  NLO and NNLO MMHT fits is shown in the last column of Table \ref{tab:LHCWZ}. For comparison, we also show the quality of the predictions of the MMHT fits and of the MMSTWW fits \cite{MMSTWW}, neither of which included these, or any other, LHC data.
 We discuss the description of the data sets listed in Table \ref{tab:LHCWZ} in turn.

\subsubsection{ATLAS $W$ and $Z$ data}  

\begin{figure} 
\begin{center}
\includegraphics[height=6cm]{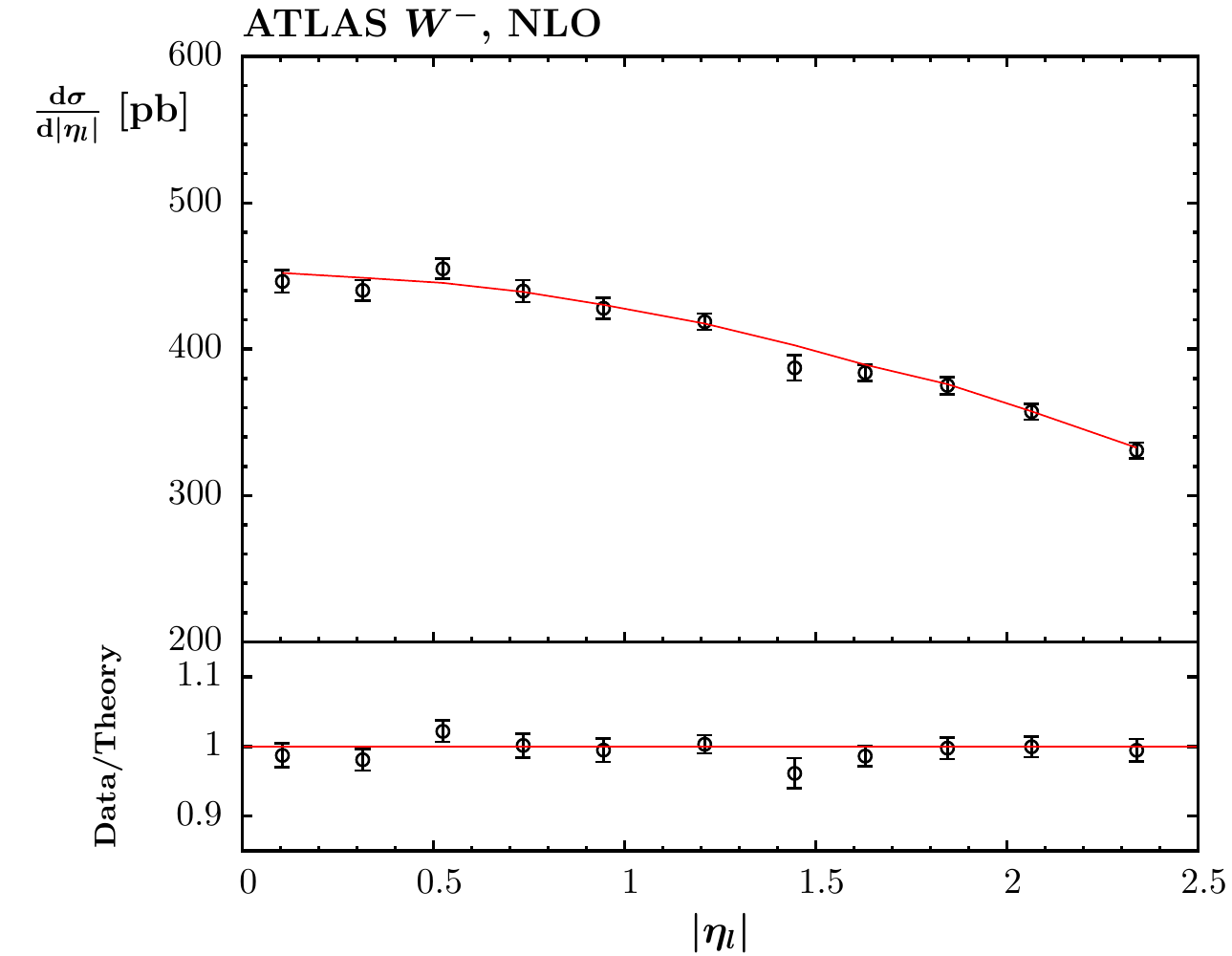}
\includegraphics[height=6cm]{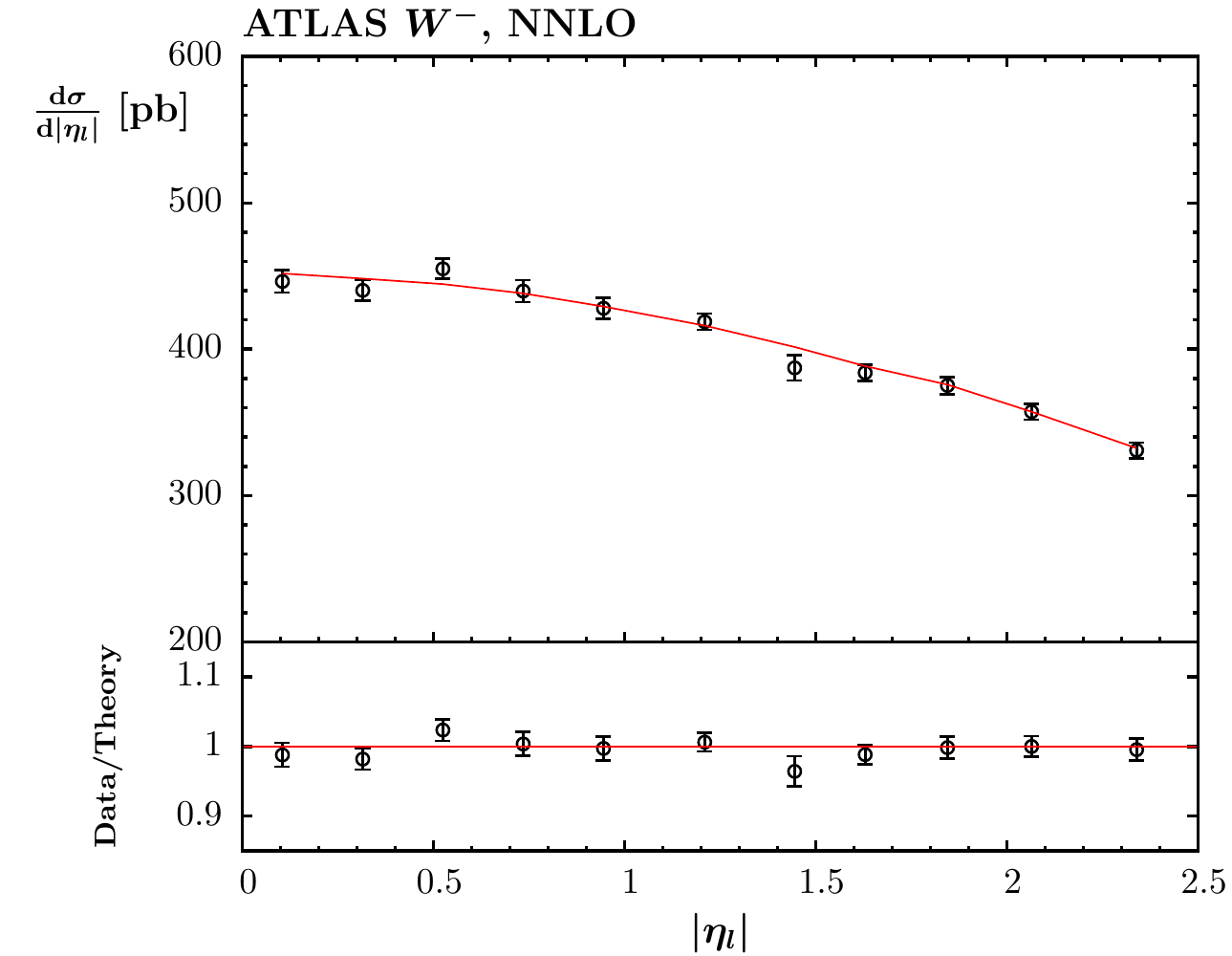}
\includegraphics[height=6cm]{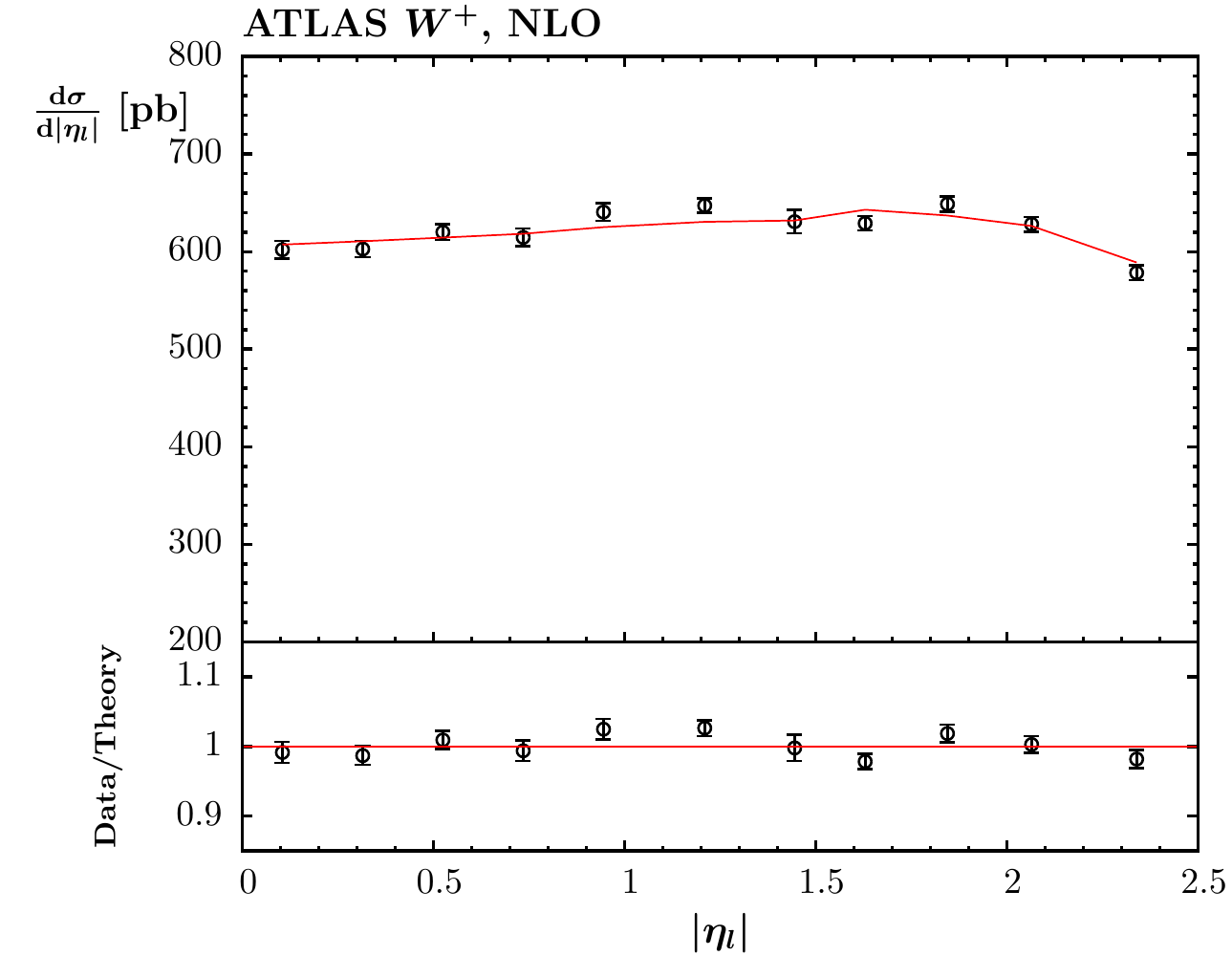}
\includegraphics[height=6cm]{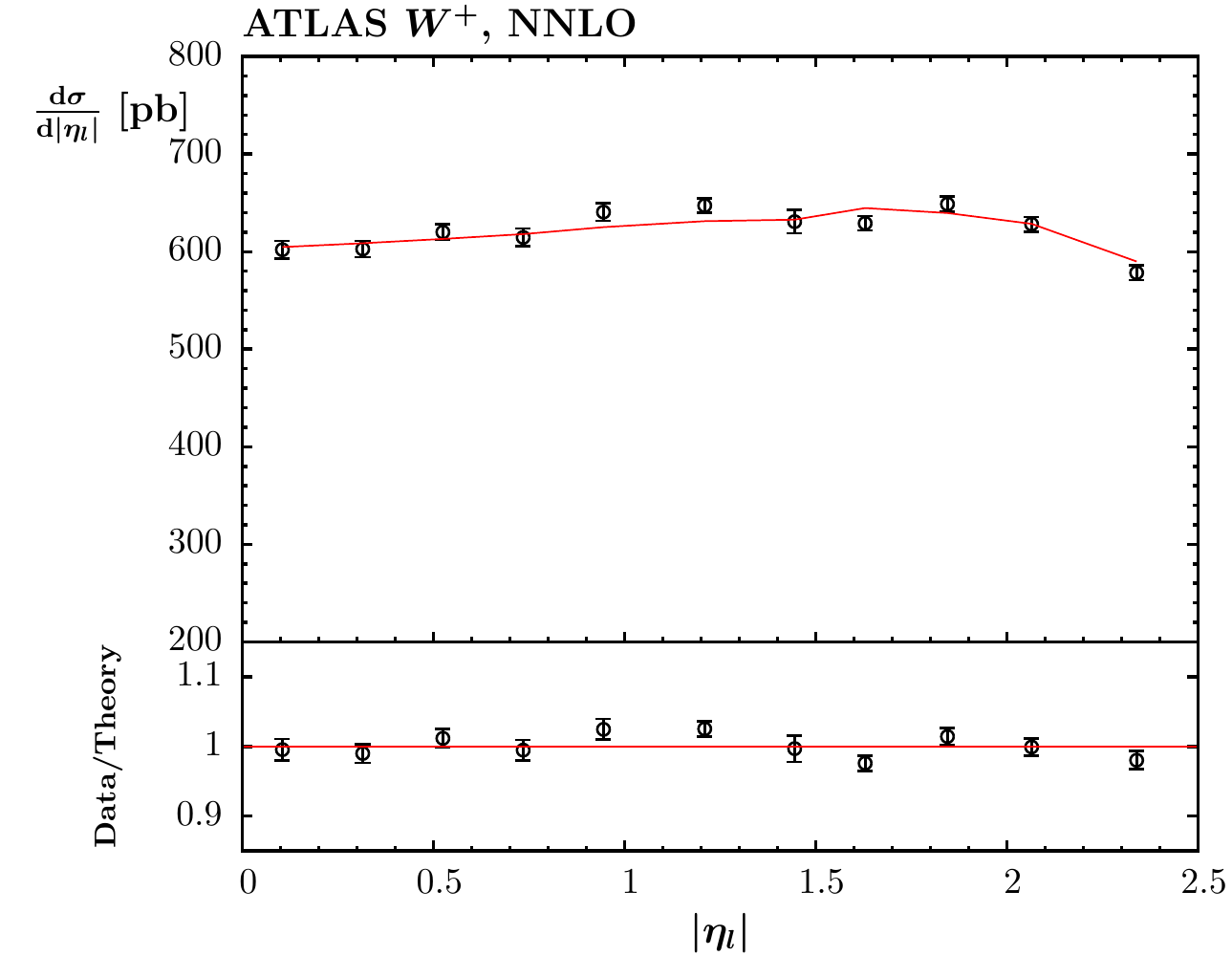}
\includegraphics[height=6cm]{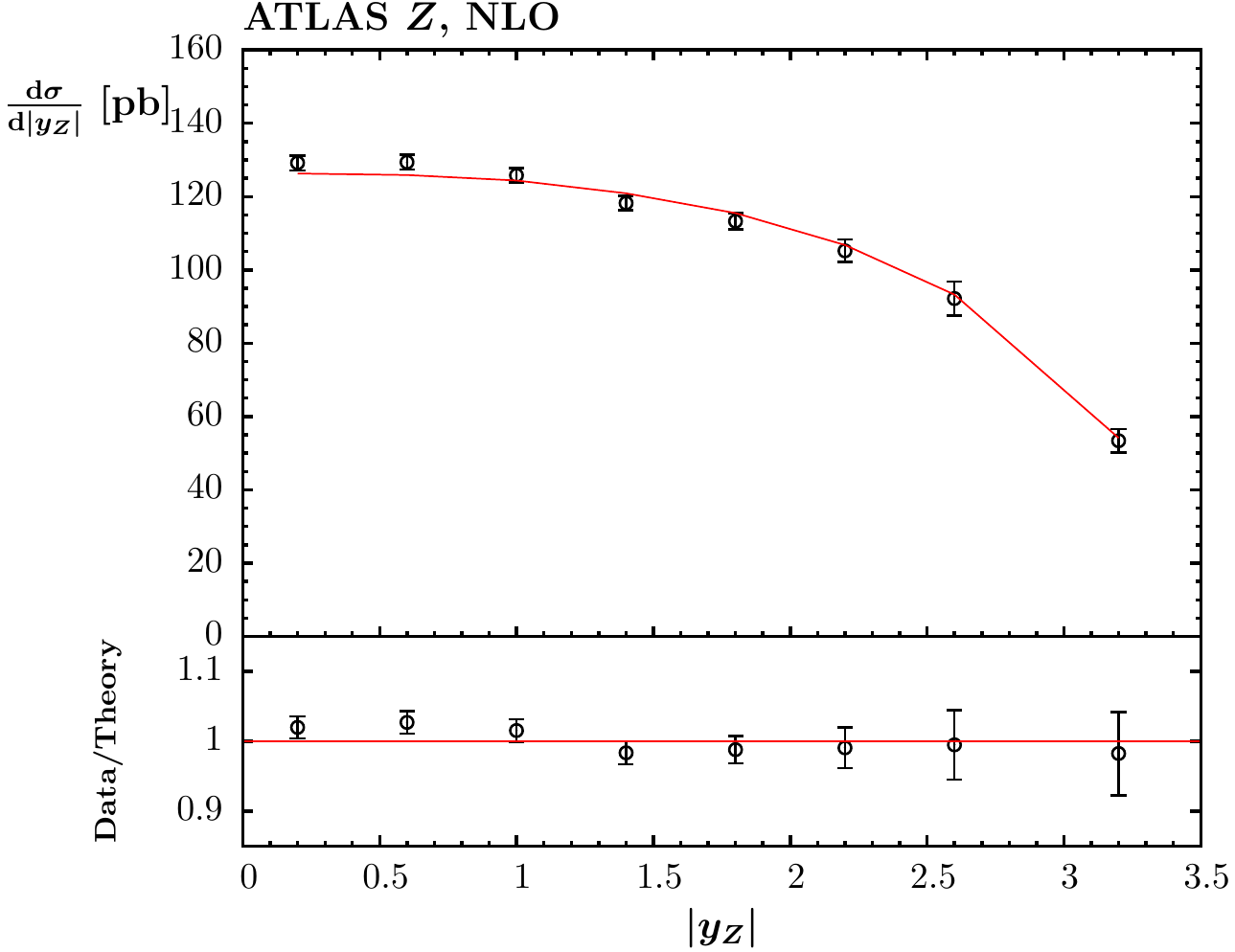}
\includegraphics[height=6cm]{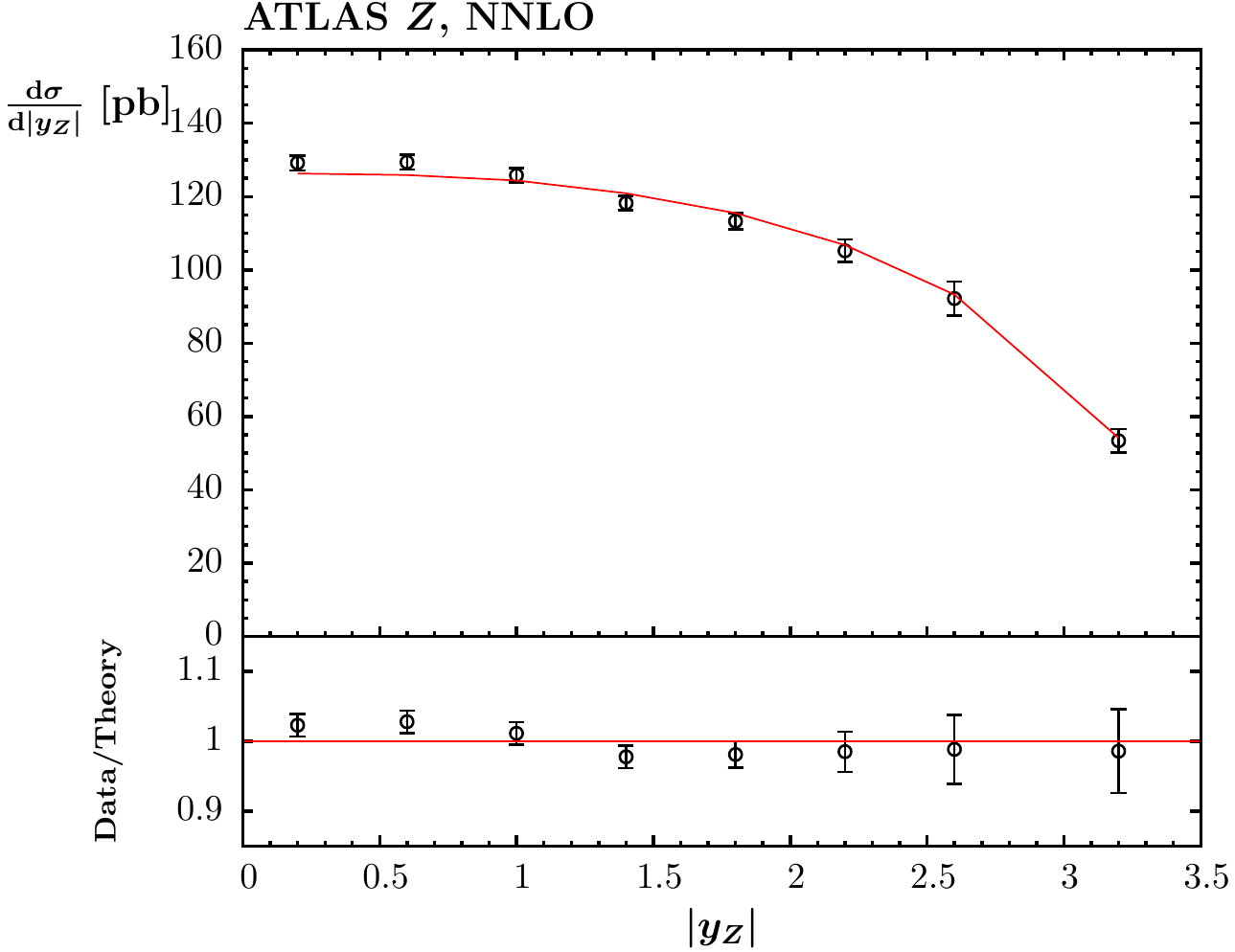}
\caption{\sf The fit quality of the ATLAS $W^-,W^+$ data sets for $d\sigma/d|\eta_l|$ (pb) versus $|\eta_l|$, and of the $Z$ data set for  $d\sigma/d|y_Z|$ versus $|y_Z|$ 
\cite{ATLAS-WZ}, obtained in the NLO (left) and NNLO (right) analyses. The points shown are 
when the shift of data relative to the theory due to correlated systematics is 
included. However, this shift is small 
compared to the uncorrelated error for the data, so the comparison before 
shifts is not shown.}
\label{fig:ATLASWZ}
\end{center}
\end{figure}

First we consider the description of  the ATLAS $W$ and $Z$ rapidity data \cite{ATLAS-WZ}.
These were poorly predicted by the MSTW2008 PDFs
(see e.g. \cite{Bench}), primarily due to the 
incorrect balance between $W^+$ and $W^-$ production at low rapidity, which 
is sensitive to the low-$x$ valence quark difference, and which shows up 
most clearly in the asymmetry between $W^+$ and $W^-$ production. 
This particular issue was automatically largely solved by the improved 
parameterisation
and deuteron corrections in the MMSTWW study \cite{MMSTWW}. Nevertheless, we see from Table \ref{tab:LHCWZ}, that the quality of
the description using the MMSTWW sets still has $\chi^2 \sim 1.6$ per point for the NLO fit, and 
$\chi^2 > 2$ per point in the NNLO fit. 
At NNLO it turns out that $u_V(x)-d_V(x)$ at small-$x$ is still not 
quite large enough to reproduce the observable charge asymmetry. However, at both NLO and 
NNLO the shape of the rapidity distribution (driven by the evolution of 
anti-quarks and hence ultimately by the gluon) is not quite ideal, and 
also a slightly larger fraction of strange quarks in the sea is preferred.
The inclusion of the non-LHC data, together with the changes in theoretical procedure mentioned in Section \ref{sec:theory}
(not included in \cite{MMSTWW}), already improves the fit quality, particularly
at NNLO, and after the inclusion of these ATLAS data, the $\chi^2$ improves 
to about 1.3 per point at both NLO and NNLO. This appears to be not quite as 
good as the best possible fits to these data, which seem to require an even
larger strange quark fraction in the sea; indeed, the same fraction  as the 
up and down sea \cite{ATLASstrange}, or even larger (in the `collider-only'
fit in \cite{NNPDF23}).  The fit quality is shown in Fig.~\ref{fig:ATLASWZ}.
One can see that there is a slight tendency to undershoot the $Z$ data at the 
lowest rapidity, which could be improved by a slight increase in the strange 
distribution for $x \sim 0.01$, as seen in \cite{ATLASstrange}, but also verified 
in our studies. 

\begin{table}  [t]
\begin{center}
\vspace{-.5cm}
\begin{tabular}{|l|c|c|c|c|}
\hline
 &  & MMSTWW     & MMHT2014 & MMHT2014 \\
~~~~~~data set         &   {$N_{\rm pts}$}          & Ref.\cite{MMSTWW}        & (no LHC)       &  (with LHC)       \\
\hline
\multicolumn{5}{|c|}{NLO}  \\
\hline
ATLAS {$W^+, W^-, Z$}               & 30 & 47 & 44 & 38 \\   
CMS {$W$} asymm {$p_T >35~\GeV$}    & 11 & 9  & 16 & 7  \\    
CMS asymm {$p_T >25~\GeV,30~\GeV$}  & 24 & 9  & 17 & 8  \\    
LHCb {$Z\to e^+e^-$}                & 9  & 13 & 13 & 13 \\   
LHCb {$W$} asymm {$p_T >20~\GeV$}   & 10 & 12 & 14 & 12 \\
CMS  {$Z\to e^+e^-$}                & 35 & 21 & 22 & 19 \\
ATLAS high-mass Drell-Yan           & 13 & 20 & 20 & 21 \\
CMS double diff. Drell-Yan          & 132& 385  & 396  & 372  \\   
\hline
\multicolumn{5}{|c|}{NNLO}  \\
\hline
ATLAS {$W^+, W^-, Z$}              & 30 & 72 & 53 & 39  \\ 
CMS {$W$} asymm {$p_T >35~\GeV$}   & 11 & 18 & 15 & 8 \\    
CMS asymm {$p_T >25~,30~\GeV$}     & 24 & 18 & 17 & 9 \\    
LHCb {$Z\to e^+e^-$}               & 9  & 23 & 22 & 21 \\   
LHCb {$W$} asymm {$p_T >20~\GeV$}  & 10 & 24 & 21 & 18 \\
CMS  {$Z\to e^+e^-$}               & 35 & 30 & 24 & 22 \\
ATLAS high-mass Drell-Yan          & 13 & 18 & 16 & 17 \\
CMS double diff. Drell-Yan         &132 & 159  & 151  & 150 \\ 
\hline
    \end{tabular}
\end{center}
\vspace{-.0cm}
\caption{\sf The quality of the description (as measured by the value of 
$\chi^2$) of the LHC $W,Z$ data before and after they are included in the 
global NLO and NNLO fits. 
We also show for comparison the $\chi^2$ values obtained in 
the CPdeut fit of the NLO MMSTWW analysis \cite{MMSTWW}, which did not 
include LHC data.}
\label{tab:LHCWZ}
\end{table}

\subsubsection{CMS asymmetry data}

\begin{figure} 
\begin{center}
\includegraphics[height=8cm]{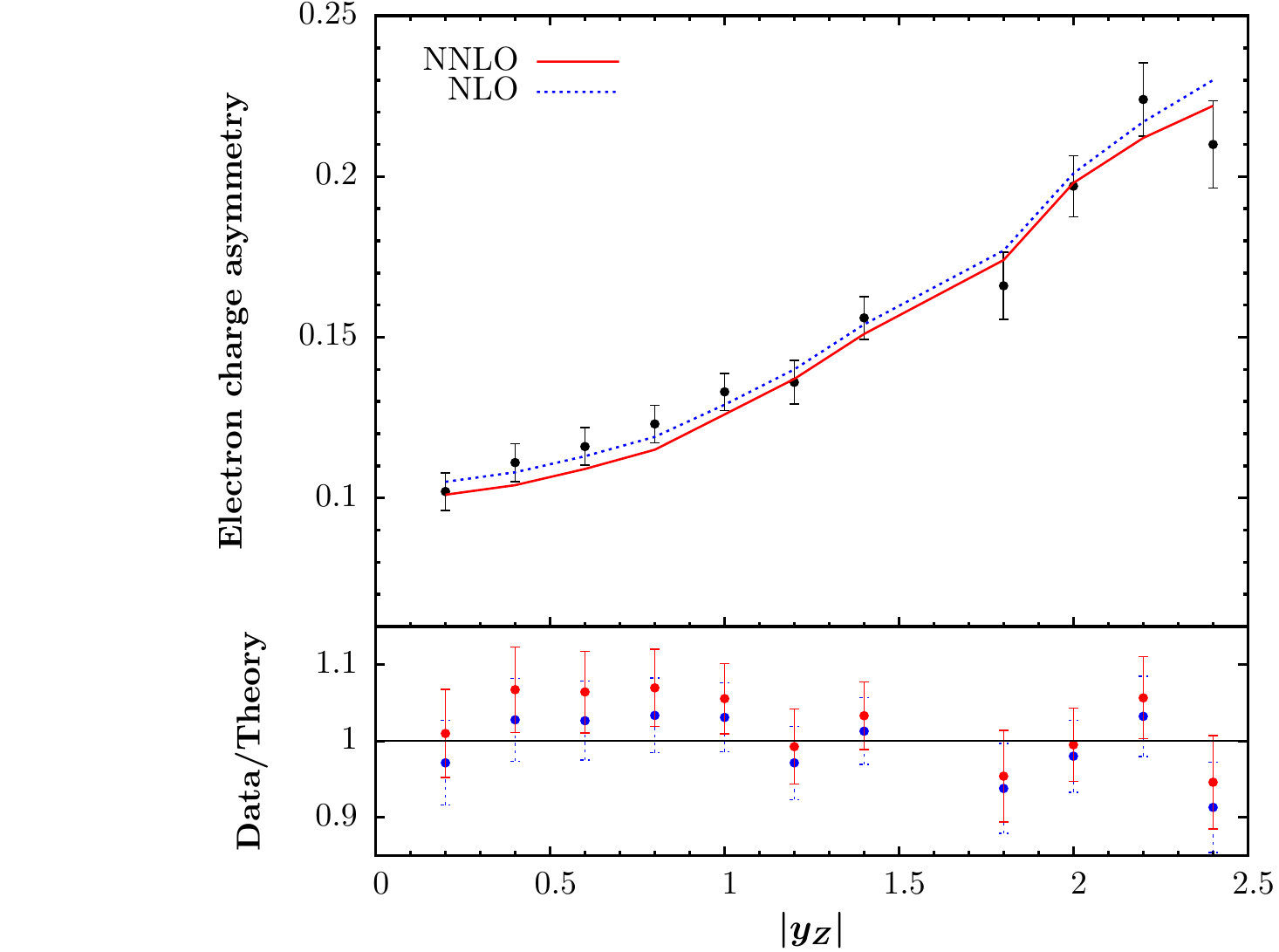}
\caption{\sf The fit quality for the CMS electron asymmetry data for
$p_T>35~\GeV$ in \cite{CMS-easym} at NLO and NNLO. Note that correlated 
uncertainties are made available in the form of a correlation matrix, so 
the shift of data relative to theory cannot be shown, and makes a comparison 
of data with PDF uncertainties less useful.}
\label{fig:CMSpT35}
\end{center}
\end{figure}

Next we discuss the description of the charge lepton asymmetries observed in the CMS data 
\cite{CMS-Wasym,CMS-easym}. These data were also not well described by MSTW2008 PDFs, but
as seen in Table \ref {tab:LHCWZ}, the prediction using the MMSTWW set at NLO
is very good. However, it is still not ideal when using the NNLO set. If we implement the changes
discussed above, in the present article, but before including the LHC data, the 
prediction for these
data deteriorates at NLO (due to $u_V(x)-d_V(x)$ becoming too large at 
$x\sim 0.01$) while it improves slightly at NNLO. When the LHC data are 
included, we see, from Table \ref {tab:LHCWZ}, that the fit quality becomes excellent. 
This is particularly the case 
at NLO, where the fit is about as good as possible, but the NNLO description 
is nearly as good. The fit quality is shown in Fig.~\ref{fig:CMSpT35}, 
and indeed the NLO fit is excellent, but at NNLO there is a slight tendency to
undershoot the low rapidity data, but this is exaggerated by the fact that 
only uncorrelated uncertainties are shown.

\subsubsection{LHCb $W$ and $Z$ data}
\begin{figure} [t]
\begin{center}
\includegraphics[height=6.3cm]{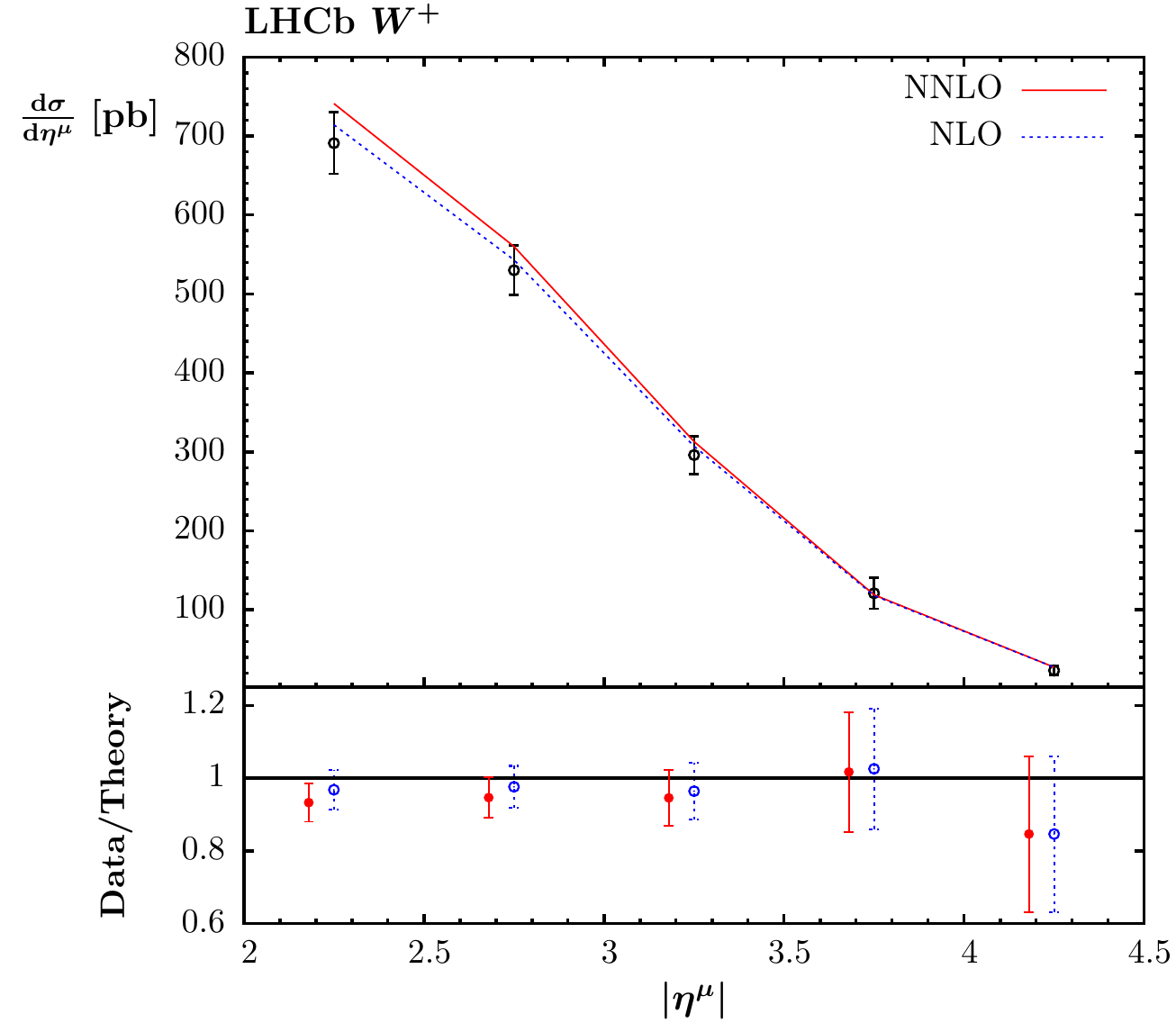}
\includegraphics[height=6.3cm]{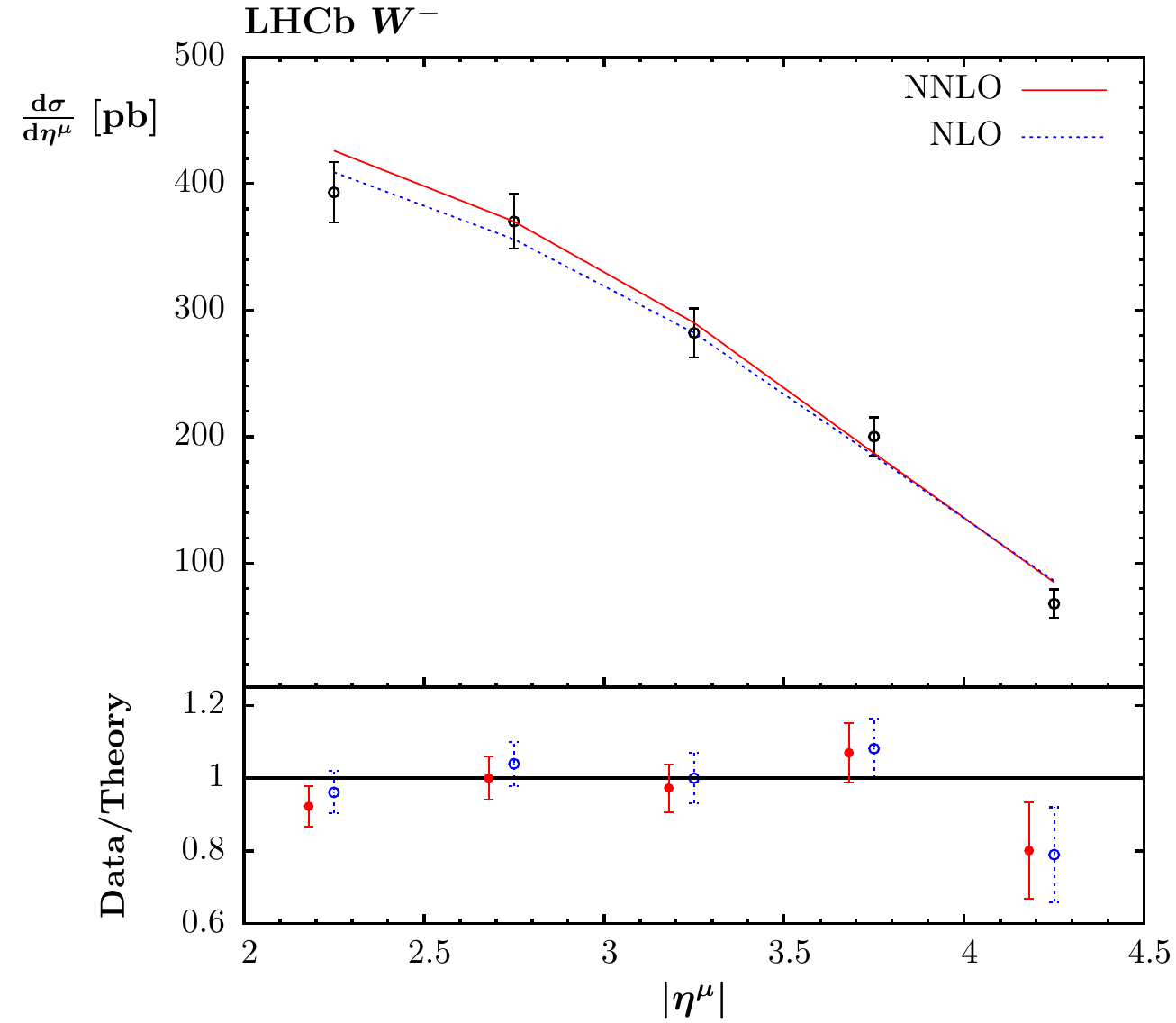}\\
\includegraphics[height=6.3cm]{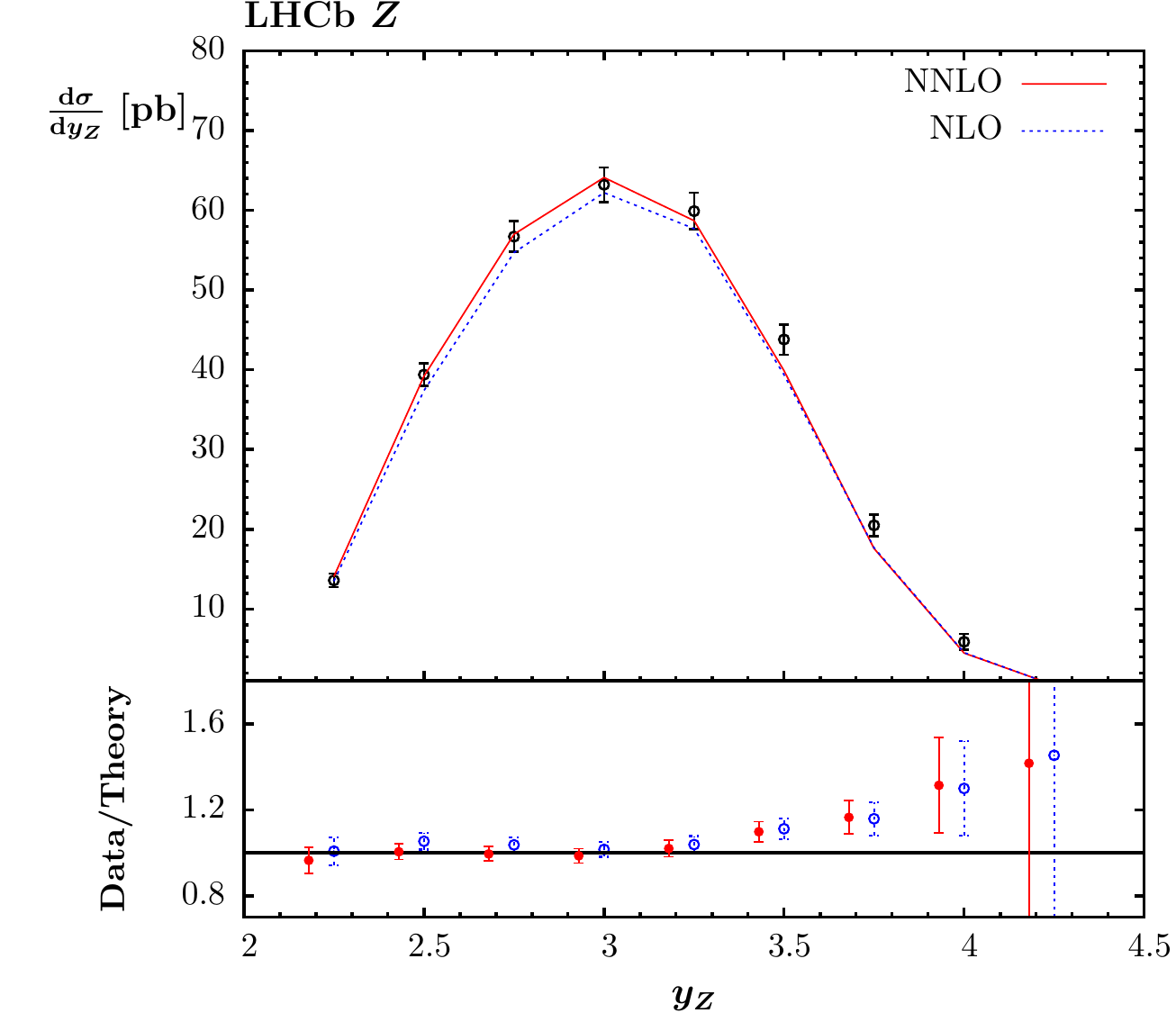}
\caption{\sf The fit quality for the LHCb  data for
$W$ and $Z$ production in \cite{LHCb-WZ} and \cite{LHCb-Zee} 
at NLO and NNLO. Note that correlated 
uncertainties are made available in the form of a correlation matrix, so 
the shift of data relative to theory cannot be shown. The plots show $d\sigma /d\eta^{\mu}$ versus $\eta^{\mu}$, and $d\sigma/dy_Z$ versus $y_Z$.}
\label{fig:LHCbwz}
\end{center}
\end{figure}

We also include the results for $W^\pm$ production \cite{LHCb-WZ} and for 
$Z \to e^+e^-$ \cite{LHCb-Zee} obtained by the LHCb experiment. These data 
are both predicted and fitted well at NLO. At NNLO the description is a 
little worse and is significantly under some of the data points for rapidity
$y \approx 3.5$ for the $Z\to e^+e^-$ data.  
However, this small discrepancy is not evident when we compare 
with the preliminary higher precision $Z\to \mu^+\mu^-$ data
\cite{LHCb-Zmumu}. The fit quality is shown in Fig.~\ref{fig:LHCbwz}.
The tendency to undershoot the high rapidity $Z$ data is clear, but this
is not an obvious feature of the comparison to the $W^{\pm}$ data. 
In principle, there are electroweak corrections,
including those where the photon distribution appears in the initial state, 
which are potentially significant. However, the electroweak
corrections are still somewhat smaller than the data uncertainty, so we 
use the pure QCD calculation in this article, though these data, and further measurements,
will be an essential feature of a future update of \cite{MRSTQED} which 
will appear shortly; see also \cite{MRy}.

\subsubsection{CMS $Z\to e^+e^-$ and ATLAS high-mass Drell-Yan data}

\begin{figure} 
\begin{center}
\includegraphics[height=5cm]{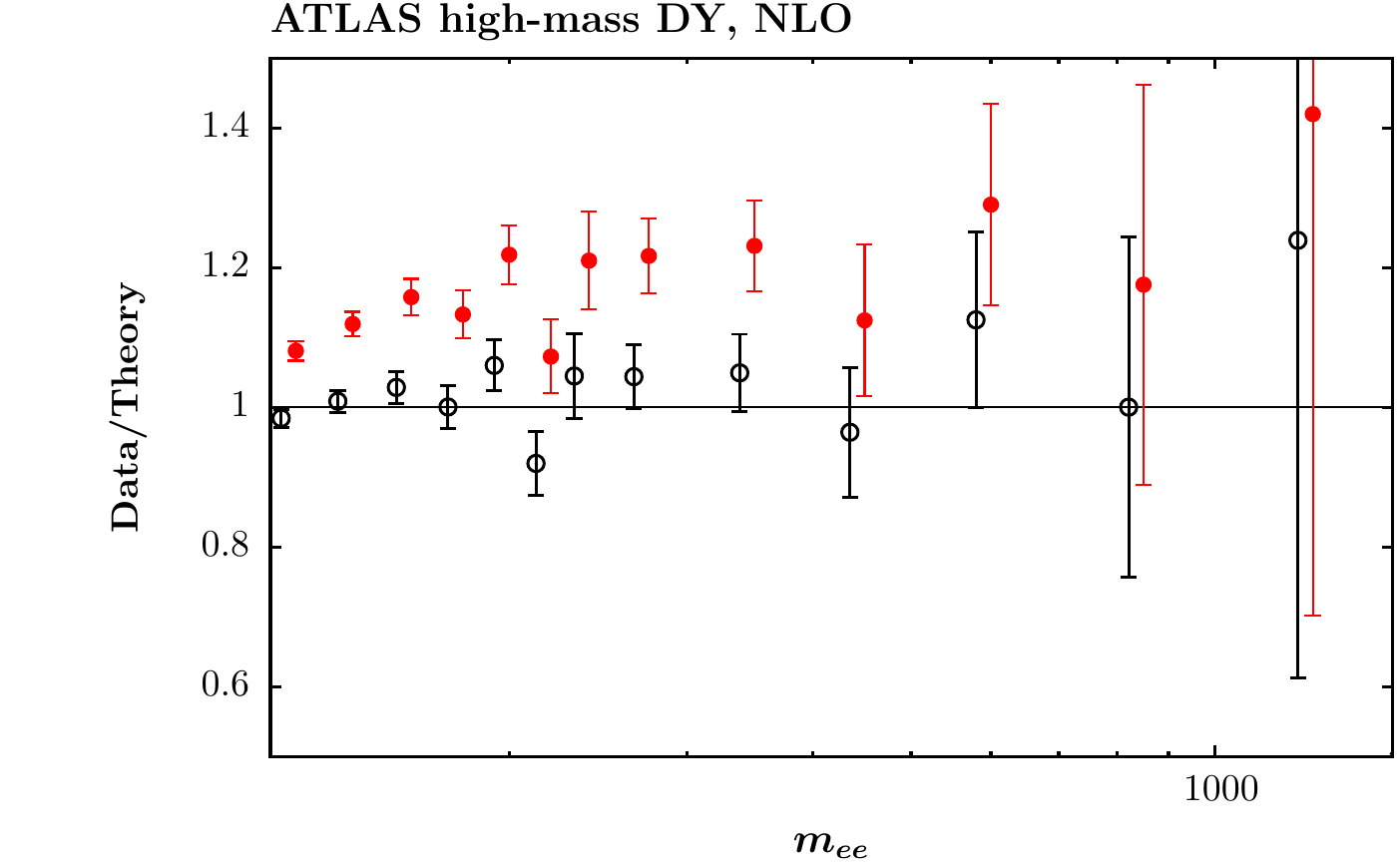}
\includegraphics[height=5cm]{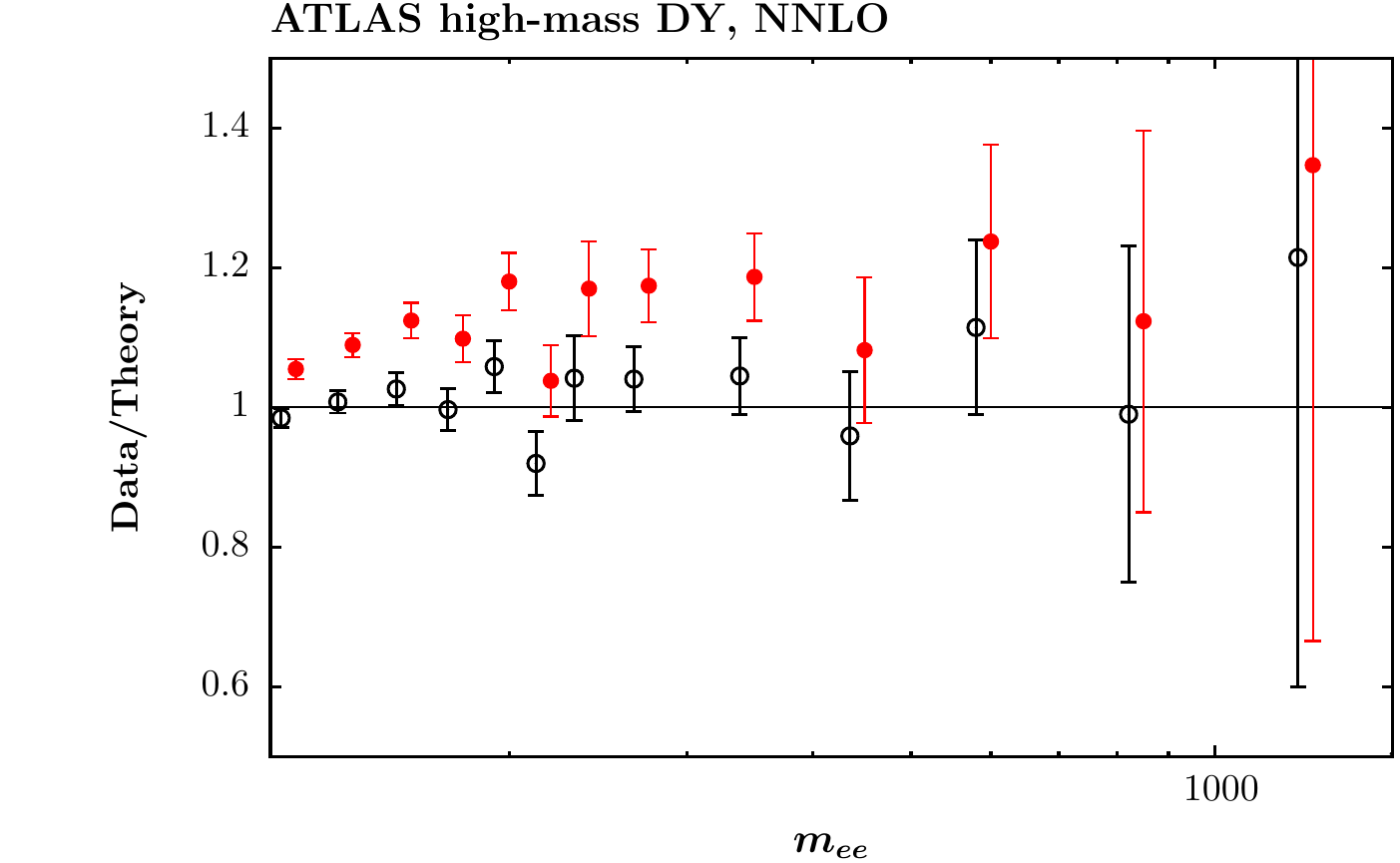}
\caption{\sf The fit quality for the ATLAS high-mass Drell-Yan  data set
\cite{ATLAShighmass} at NLO (left) and NNLO (right). The red points
represent the ratio of measured data to theory predictions, and the black 
points (clustering around Data/Theory=1) correspond to this ratio once
the best fit has been obtained by 
shifting theory predictions relative to data by using the correlated 
systematics.}
\label{fig:ATLAShm}
\end{center}
\end{figure}

In addition, we include in the fit the CMS data for $Z\to e^+e^-$ \cite{CMS-Zee}, 
and the ATLAS high-mass Drell-Yan data \cite{ATLAShighmass}. 
Both are well described, again slightly better at NLO than at NNLO.
The fit quality for the ATLAS high-mass Drell-Yan 
data is shown in Fig.~\ref{fig:ATLAShm}. The correlated uncertainties 
clearly play a big part in allowing the good quality fit, particularly at NLO.
However, these are presented in the form of correlation matrices so it is not possible
to illustrate shifts of data relative to theory. For these data sets the variation of
the theory predictions within the range of PDF uncertainties is smaller than the 
data uncertainties. 
As in the previous subsection, in principle there are electroweak corrections,
including those where the photon distribution appears in the initial state, which
is particularly relevant for this type of process, 
and are included in the analysis of \cite{ATLAShighmass},
which takes the photon PDF from \cite{MRSTQED}, and used as a very weak 
constraint on the photon PDF in \cite{NNPDFQED}. However, 
as in the last subsection these are still much smaller than the data uncertainty, though 
this may well not continue with future measurements.

\subsubsection{CMS double-differential Drell-Yan data}
\begin{figure} 
\begin{center}
\includegraphics[height=6.3cm]{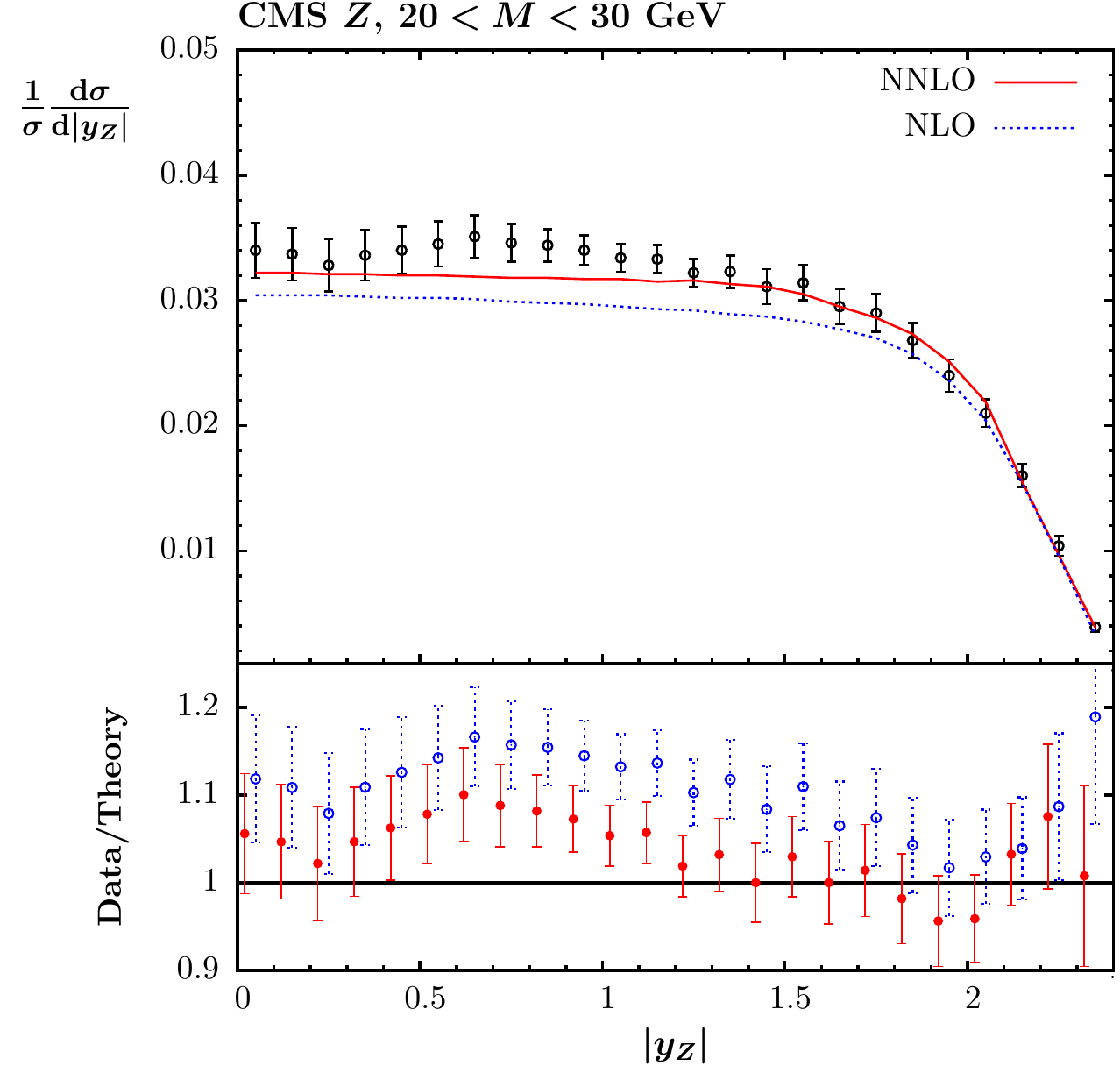}
\includegraphics[height=6.3cm]{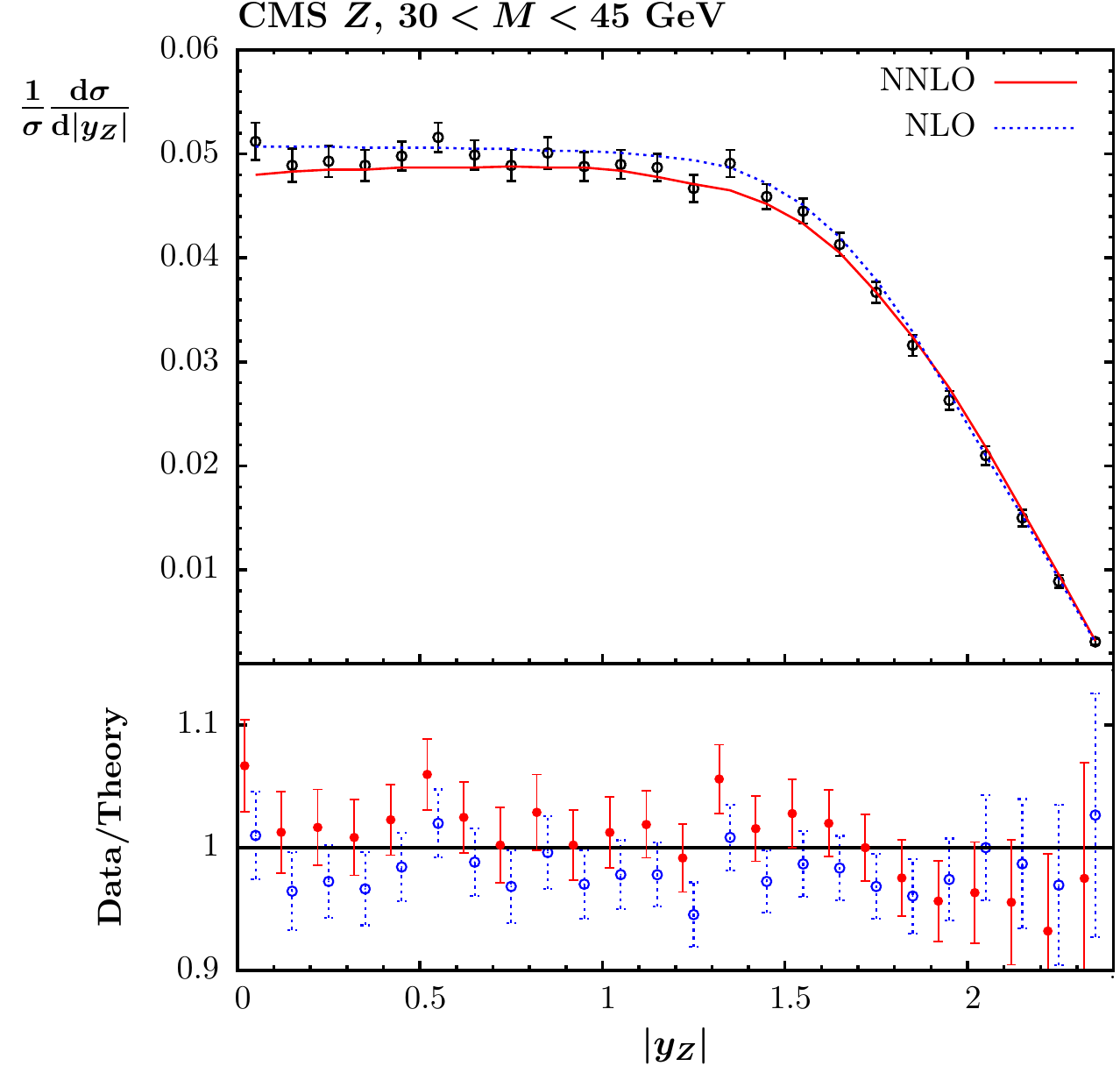}\\
\vspace{0.2cm}
\includegraphics[height=6.3cm]{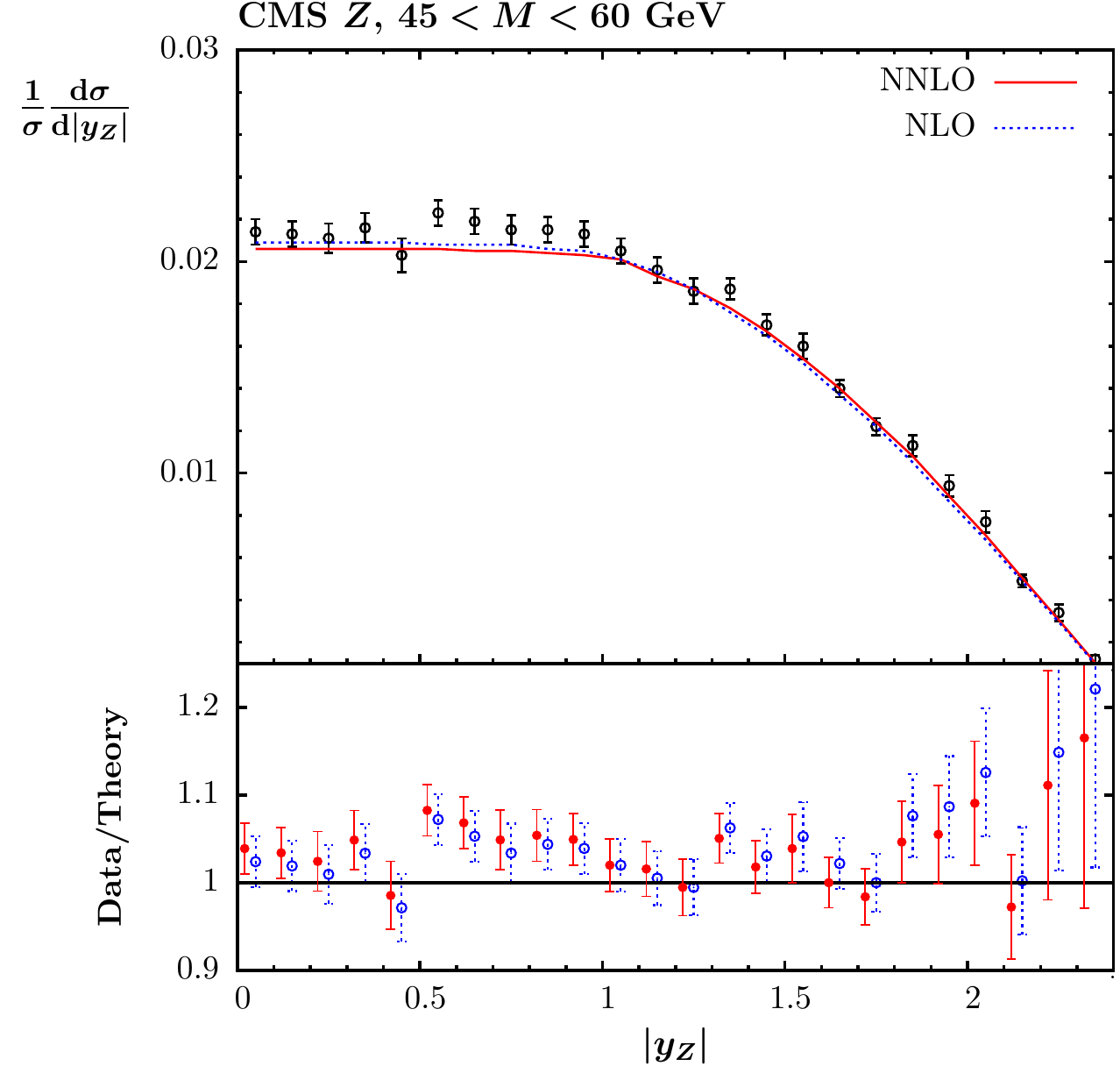}
\includegraphics[height=6.3cm]{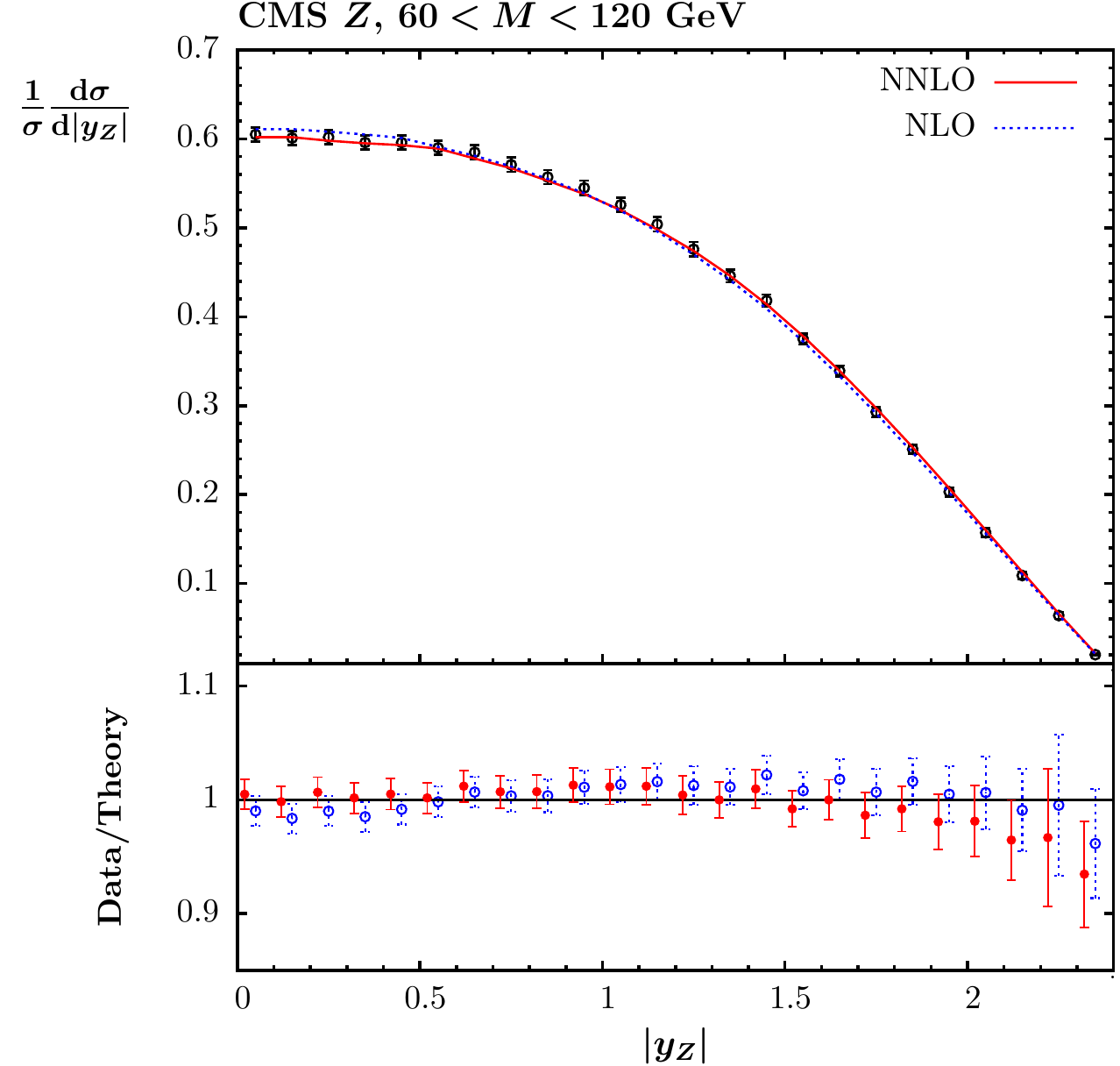}\\
\vspace{0.2cm}
\includegraphics[height=6.3cm]{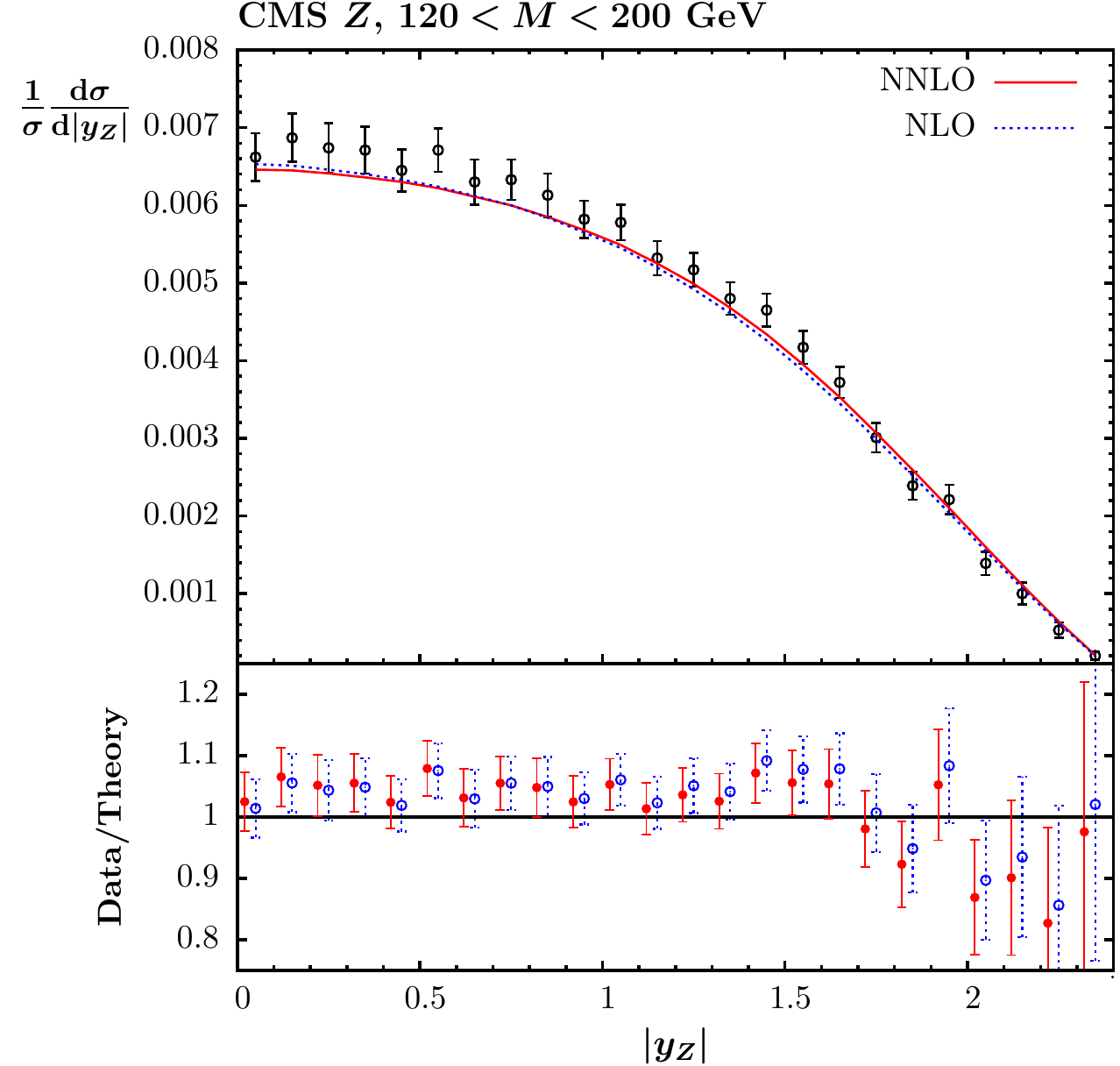}
\includegraphics[height=6.3cm]{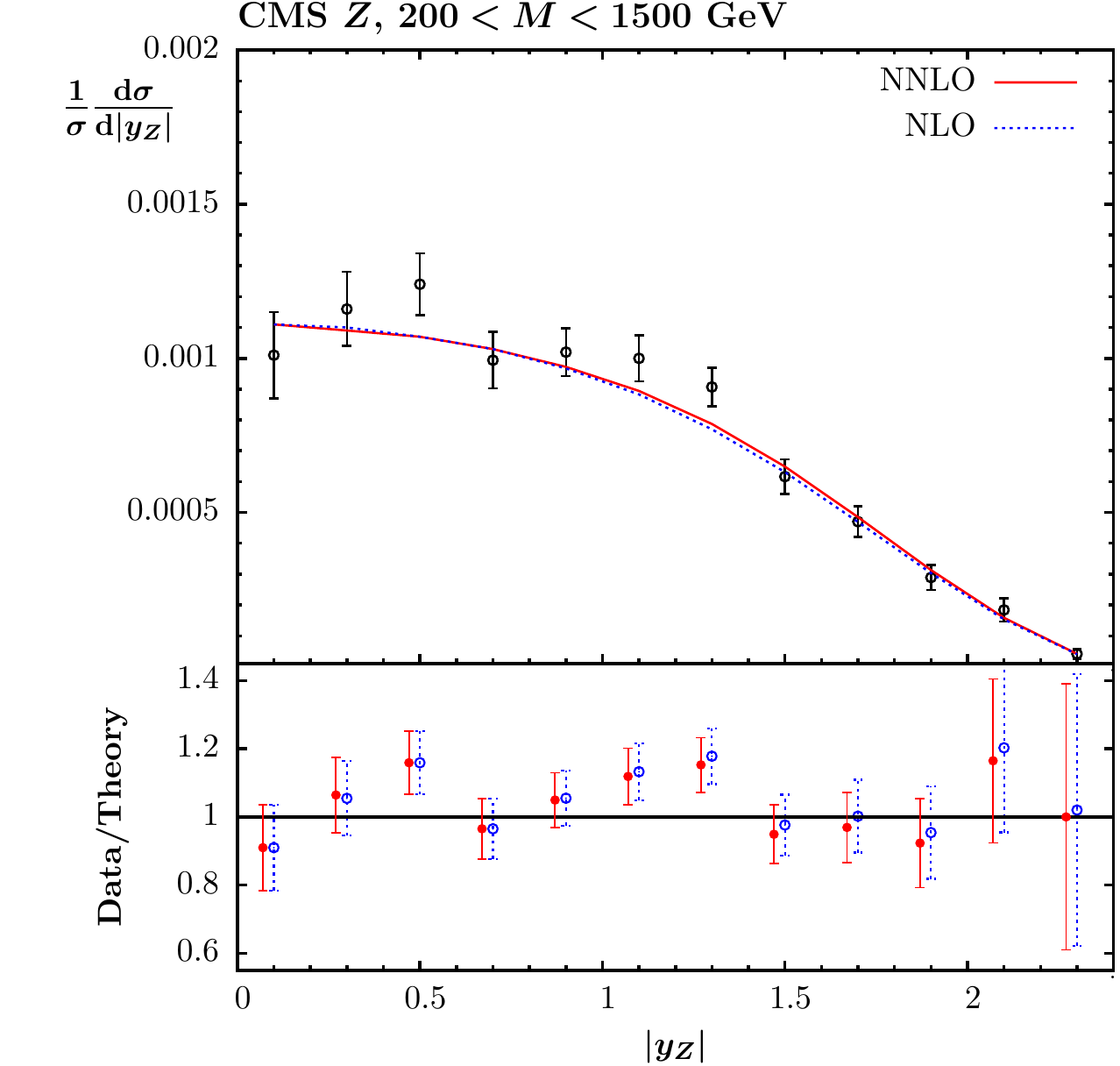}
\caption{\sf The fit quality for the CMS double differential Drell-Yan data 
for $(1/\sigma_Z)\cdot d\sigma/d|y_Z|$ versus $|y_Z|$, in \cite{CMS-ddDY}, 
for the lowest two mass bins ($20<M<30$ GeV and $30<M<45$ GeV) (top), 
the mass bins ($45<M<60$ GeV and $60<M<120$ GeV) (middle) and 
the mass bins ($120<M<200$ GeV and $200<M<1500$ GeV) (bottom), 
at NLO and NNLO. Note that correlated 
uncertainties are made available in the form of a correlation matrix, so 
the shift of data relative to theory cannot be shown.}
\label{fig:CMSdyDY20-2000}
\end{center}
\end{figure}

Finally, we include the CMS double-differential Drell-Yan data  
\cite{CMS-ddDY} extending down to relatively low masses, $M(\ell^+\ell^-) 
\sim 20-40$ GeV. (Again there is some sensitivity to electroweak corrections 
away from the $Z$-peak, but we do not include these corrections in the 
theoretical calculations.) The fit to these data is extremely poor at 
NLO, as shown in Table \ref {tab:LHCWZ}, and this is largely due to the 
comparison in the two lowest mass bins $20-30~\GeV$ and $30-45~\GeV$, 
see Fig. \ref{fig:CMSdyDY20-2000}. The data/theory comparison 
in the other mass bins is 
similar at NLO and NNLO, being very good in both cases.
The fit quality can only be improved marginally if this data set is 
given a very high weighting in the fit -- the PDFs are probed at similar 
values 
of  $x$ in adjacent mass bins, and if the normalisation is changed to improve 
the match to data in one mass bin it affects the quality in the nearby bins. 
The fit quality is hugely improved at NNLO, as shown in Fig. 
\ref{fig:CMSdyDY20-2000}. This might be taken as an indication that NNLO 
corrections are particularly important for low-mass Drell-Yan production. 
However, it is a little more complicated than this. The $p_T$ cut on 
one lepton in the final state is $14~\GeV$ (the other is $9~\GeV$),
meaning that at LO the minimum invariant mass is $28~\GeV$, and 
most of the lowest mass bin in the double differential cross section 
receives a contribution of zero from the LO calculation, and in this region 
the first non-zero results are at ${\cal O}(\alpha_S)$ when an extra particle 
is emitted. 
Hence, the $K$-factor going from LO to NLO is over 6 in the    
$20-30~\GeV$ region, and is still large $\sim 1.3$ when going from NLO to NNLO. 
The $K$-factors are much smaller in higher-mass bins. Hence, it is perhaps 
more correct to say that the NLO fit is poor because for the lowest mass it 
is effectively (nearly) a LO calculation, rather than because the NNLO
correction is intrinsically very important. A similar effect is noted in the 
low-mass single-differential measurement in \cite{ATLASlowmassDY}, where the 
prediction using MSTW2008 PDFs at NNLO is very good, but is poor at NLO at
low mass, and fits performed in this paper work well at NNLO, but not at NLO. 
  
\subsubsection{Procedure for LO fit to Drell-Yan data}
At LO we follow the procedure for fitting Drell-Yan (vector boson production)
data given in \cite{MSTW}. In this, and other previous studies, it has been found 
that it is not possible to obtain a good simultaneous fit of structure 
function and Drell-Yan data, since the quark (and antiquark) distributions are 
not compatible due to NLO corrections to coefficient functions 
being much larger for Drell-Yan production. This is because of 
a significant difference between the result in the space-like and   
time-like regimes; that is there is factor of $1 + (\alpha_S(M^2)/\pi) 
C_F\pi^2/2$ at NLO in the latter regime. Even for $Z$ production this is a factor
of $1.25$. Hence, as in \cite{MSTW} we include this common factor for 
all vector boson production in the LO fit.  Doing this enables a good fit 
to the low-energy fixed-target 
Drell-Yan data \cite{E866DY} (though it is less good for 
the asymmetry \cite{E866DYrat}). However, the general fit quality to 
rapidity-dependent data from the LHC and the Tevatron is generally poor
(with some exceptions, which are generally ratios, e.g. the 
D0 $Z$-rapidity data \cite{D0Zrap}, and the CMS lepton asymmetry data),
with neither the precise normalisation or the shape being correct. 
Nevertheless, the fit is distinctly better when including the 
correction factor than without it, while the normalisation is consistently
very poor. We do not include the CMS double-differential Drell-Yan data
at LO, since, as mentioned above, in the lowest mass bins the LO contribution
is an extremely poor approximation.

\subsection{Data on $t\bar{t}$ pair production}

We include in the fit the combined measurement of the 
D0 and CDF experiments for the $t\bar{t}$ production cross section as 
measured at the Tevatron \cite{Tevatron-top}
\be
\sigma(t\bar{t})=7.60\pm 0.41~{\rm pb~~~~~~with~~~}m_t=172.5~\GeV,
\ee
together with published $t\bar{t}$ cross section measurements from ATLAS and 
CMS at $\sqrt{s}=7$ TeV \cite{ATLAS-top7(1),ATLAS-top7(2),ATLAS-top7(3),
ATLAS-top7(4),ATLAS-top7(5),ATLAS-top7(6),CMS-top7(1),CMS-top7(2),CMS-top7(3),
CMS-top7(4),CMS-top7(5)} and at 8 TeV \cite{CMS-top8}\footnote{We note that the 
measurement at 8 TeV is actually published after the beginning of 2014 
(although submitted at the end of 2013), and hence officially does not 
satisfy our cut-off on the date for data included. However, this data point 
is extremely well fit at both NLO and NNLO, with the contribution to the 
$\chi^2$ much less than one unit, and has extremely little pull on the PDFs.
It is effectively included as a comparison rather than as a constraint.}.  
We use APPLGrid$-$MCFM at NLO and the code from \cite{topNNLO} for the 
NNLO corrections. 
We take $m_t=172.5$ GeV (defined in the pole scheme) with an 
error of 1 GeV, with the corresponding $\chi^2$ penalty applied. A variation of 
$1~\GeV$ in the mass is roughly equivalent to a $3\%$ change in the cross 
section. A number of the measurements of the cross section, including 
the most precise \cite{CMS-top7(2)}, use the same value of the mass as 
default. Some also parametrise the measured cross section as a function of 
$m_t$, and in these cases the cross section falls with increasing mass,
as for the theory prediction. However, the dependence is weaker, typically
$\sim 1\%$ per GeV or less, and so this variation is outweighed significantly
by the variation in the theory (though one can assume that the 
1 GeV uncertainty on the top mass used in the theory calculation is partially accounting for the 
variation of the cross section data as well, and the uncertainty on 
the top mass applied is consequently slightly less than 1 GeV in 
practice).      

\begin{figure} 
\begin{center}
\includegraphics[height=8cm]{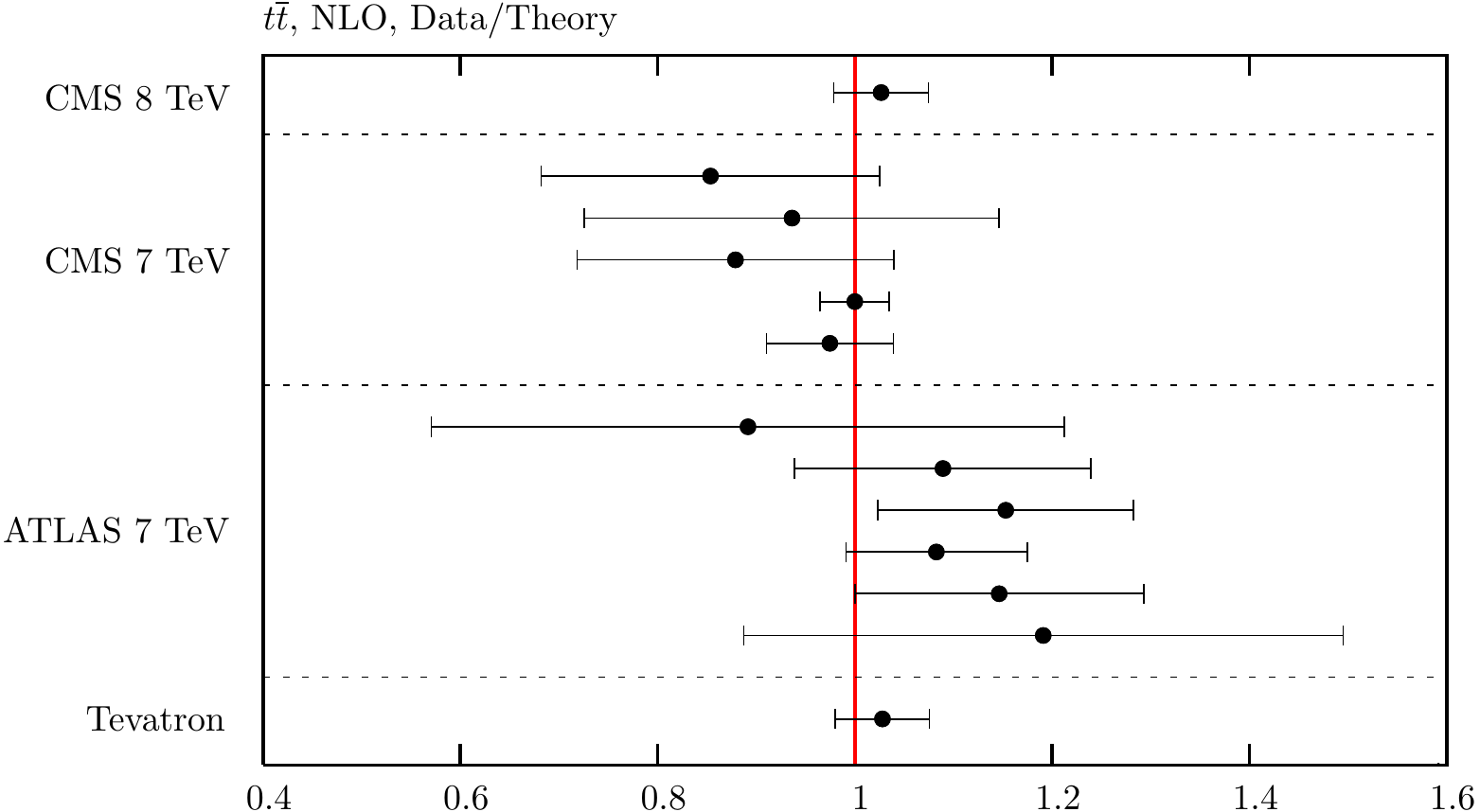}\\
\vspace{1cm}
\includegraphics[height=8cm]{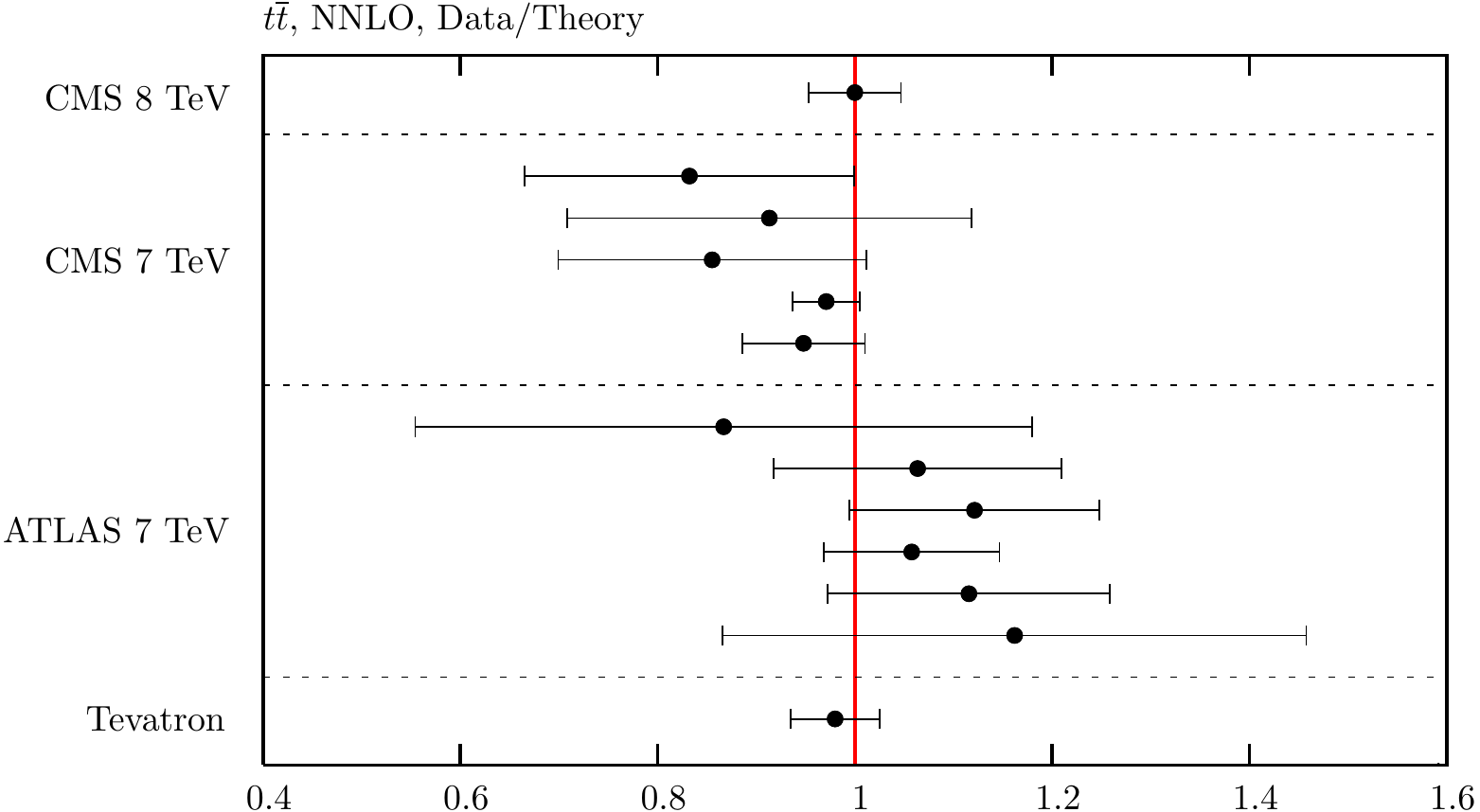}
\caption{\sf The fit quality of the cross section data for $t\bar t$ production  ($\sigma ({t \bar t})$) at NLO (top) and 
NNLO (bottom).}
\label{fig:top}
\end{center}
\end{figure}

The predictions and the fit are very good, as shown in 
Table \ref{tab:LHCtop}, and in Fig.~\ref{fig:top}, with a slightly lower mass 
$m_t = 171.7~\GeV$ preferred in the NLO fit, and a slightly higher value 
$m_t = 174.2~\GeV$ in the NNLO fit. Using the dynamical tolerance 
method both NLO and NNLO fits constrain the top mass to within about 
$0.7-0.8~\GeV$ of the best fit values, though the best value and 
uncertainties cannot be interpreted as independent determinations
as a preferred value and uncertainty for $m_t$ is input in the analysis. 
Nevertheless, it is encouraging that the preferred mass at NNLO
is consistent with the world average of  $173.34\pm 0.76~\GeV$
\cite{topmass}, whereas the NLO preferred value is a little low, 
highlighting the importance of the NNLO corrections, even though the fit 
quality is similar at both orders. There is a significant interplay between 
the gluon distribution, the top mass and the strong coupling constant. It is 
very clear that as the top quark mass increases the predicted cross section 
decreases, which can be compensated for in the cross-section by an increase
in both the gluon and in $\alpha_S(M_Z^2)$. This will be discussed more
in a forthcoming article which presents the variation of PDFs with $\alpha_S(M_Z^2)$
in detail and illustrates the constraint on the coupling. However, we note here that 
although the fit quality to the $t\bar{t}$ production cross section does depend
quite strongly on the values of $m_t$ and $\alpha_S(M_Z^2)$, the small size of the 
data set is such that the value of $\alpha_S(M_Z^2)$ for the best fit depends very 
little on variation of $m_t$, or even on the inclusion of the top data, i.e. of order
$0.0003$ at most.

\begin{table}
\begin{center}
\vspace{-.5cm}
\begin{tabular}{|l|c|c|c|c|}
\hline
 &  & MMSTWW     & MMHT2014 & MMHT2014 \\
~~~~~~data set         &   {$N_{\rm pts}$}          & Ref.\cite{MMSTWW}        & (no LHC)       &  (with LHC)       \\
\hline
\multicolumn{5}{|c|}{NLO}  \\
\hline
Tevatron, ATLAS, CMS ~~ {$\sigma ({t\bar t})$} & 13 & 8 & 10 & 7 \\
\hline
\multicolumn{5}{|c|}{NNLO}  \\
\hline
Tevatron, ATLAS, CMS ~~~{$\sigma ({t\bar t})$} & 13 & 8 & 11 & 8 \\\hline
    \end{tabular}
\end{center}
\vspace{-.0cm}
\caption{\sf The quality of the description (as measured by the value of 
$\chi^2$) of Tevatron and  LHC $t \bar t$ data before and after 
they are included in the 
global NLO and NNLO fits. 
We also show for comparison the $\chi^2$ values obtained in 
the CPdeut fit of the NLO MMSTWW analysis \cite{MMSTWW}, which did not 
include LHC data. Note that the subprocess $q\bar{q}\to t\bar{t}$ dominates at the Tevatron with $x_1,x_2 \sim 0.2$, while at the LHC $gg\to t\bar{t}$ gives the major contribution with $x_1,x_2\sim 0.05$.}
\label{tab:LHCtop}
\end{table}

The fit quality at LO is very poor, with $\chi^2/N_{\rm pts}=53/13$. This is 
because the LO calculation is too low and $m_t=163.5~\GeV$ is preferred, 
even though this incurs a very large $\chi^2$ penalty. 

\subsection{LHC data on jets}

\begin{figure} 
\begin{center}
\includegraphics[height=5cm]{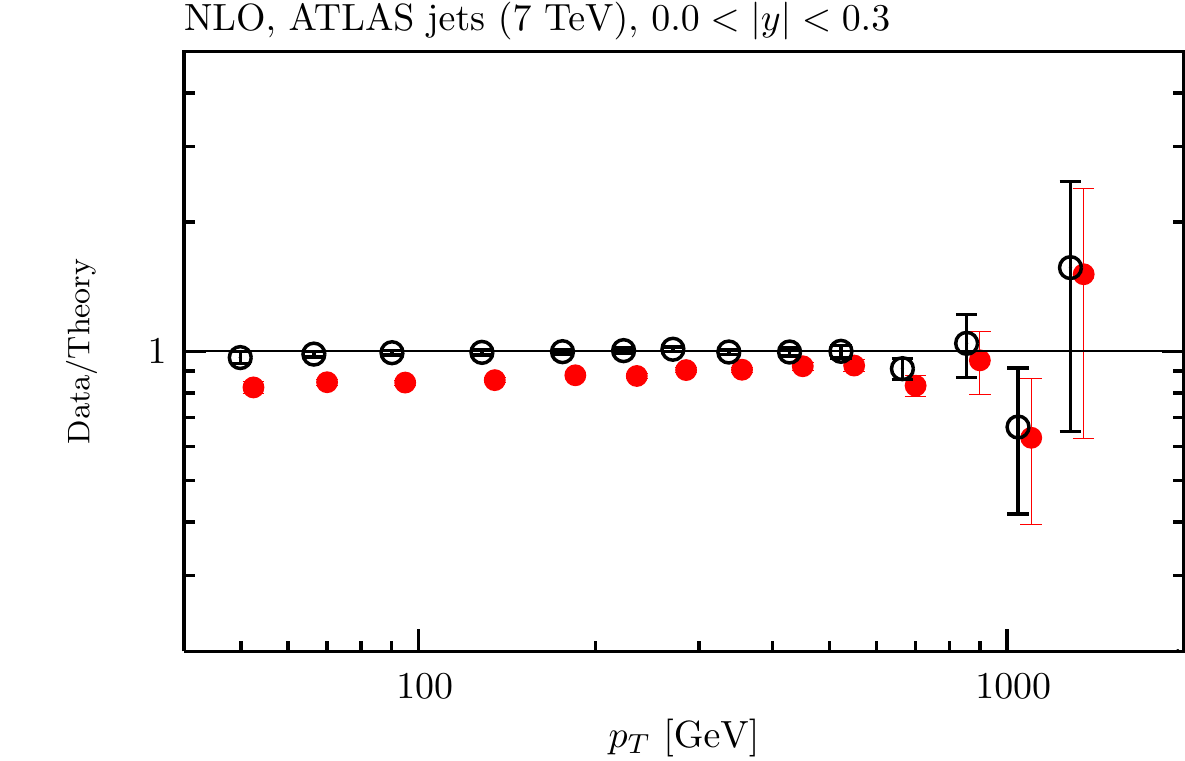}
\includegraphics[height=5cm]{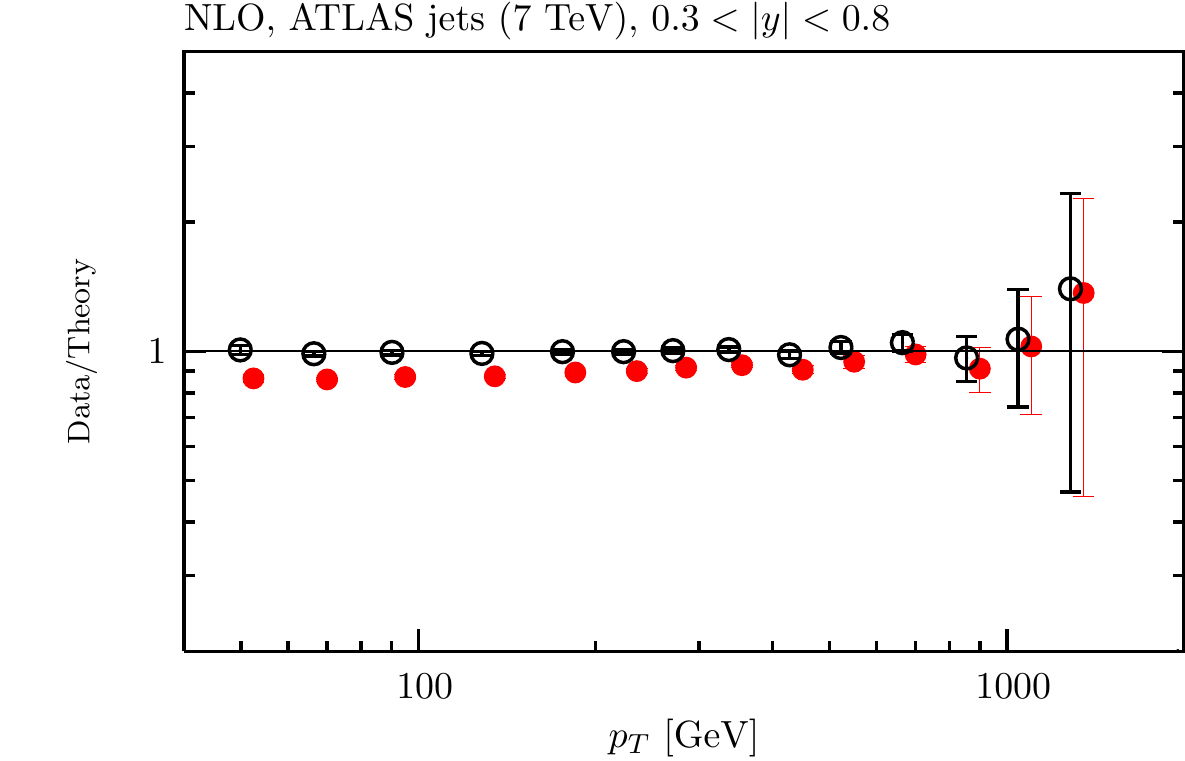}
\includegraphics[height=5cm]{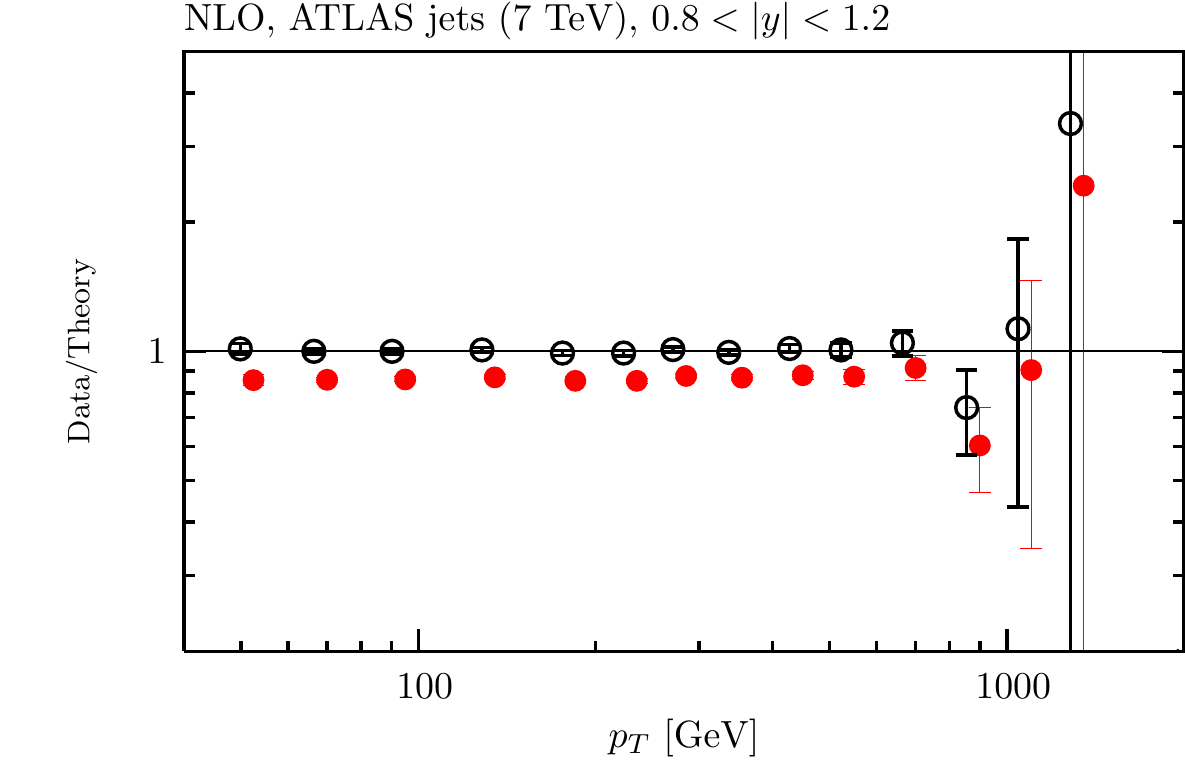}
\includegraphics[height=5cm]{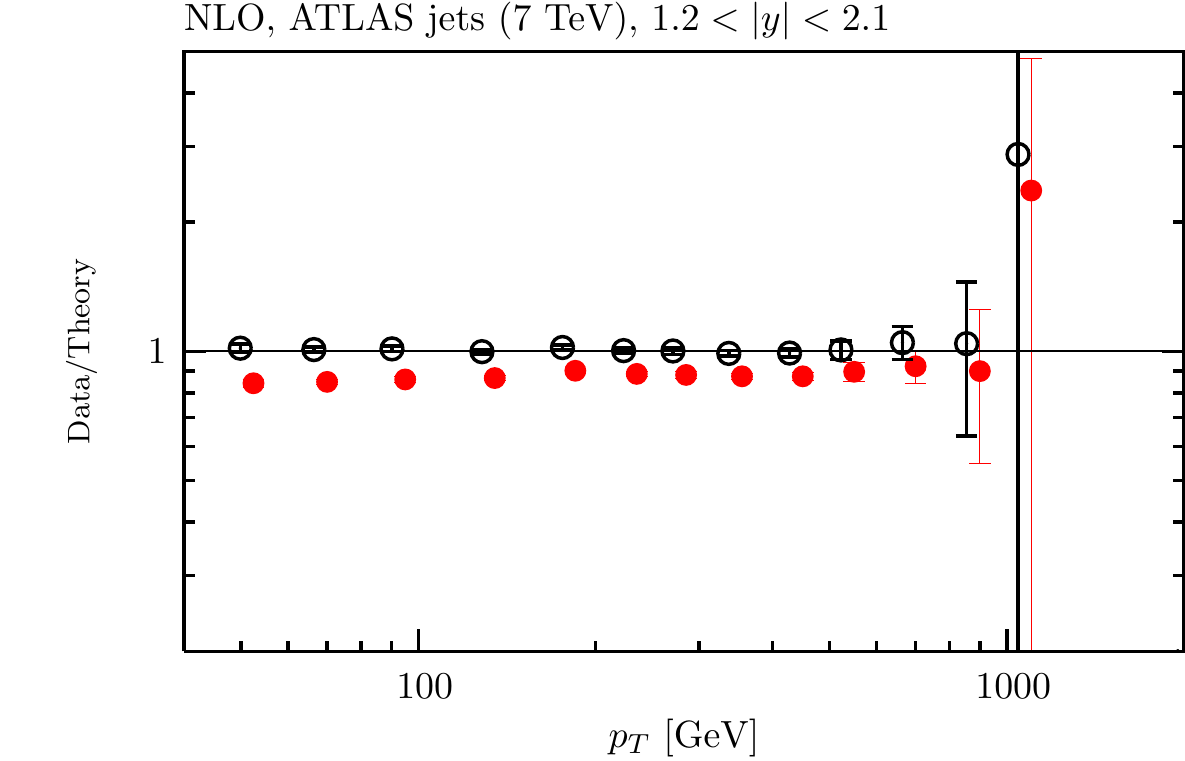}
\includegraphics[height=5cm]{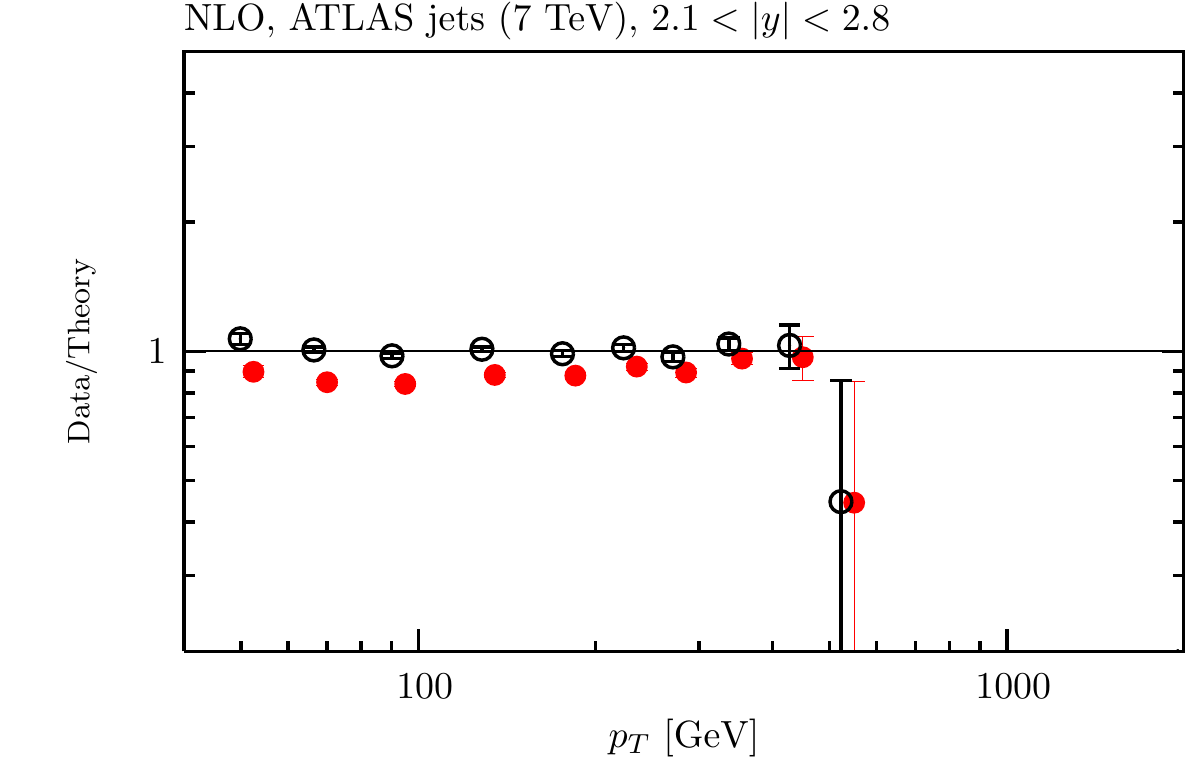}
\includegraphics[height=5cm]{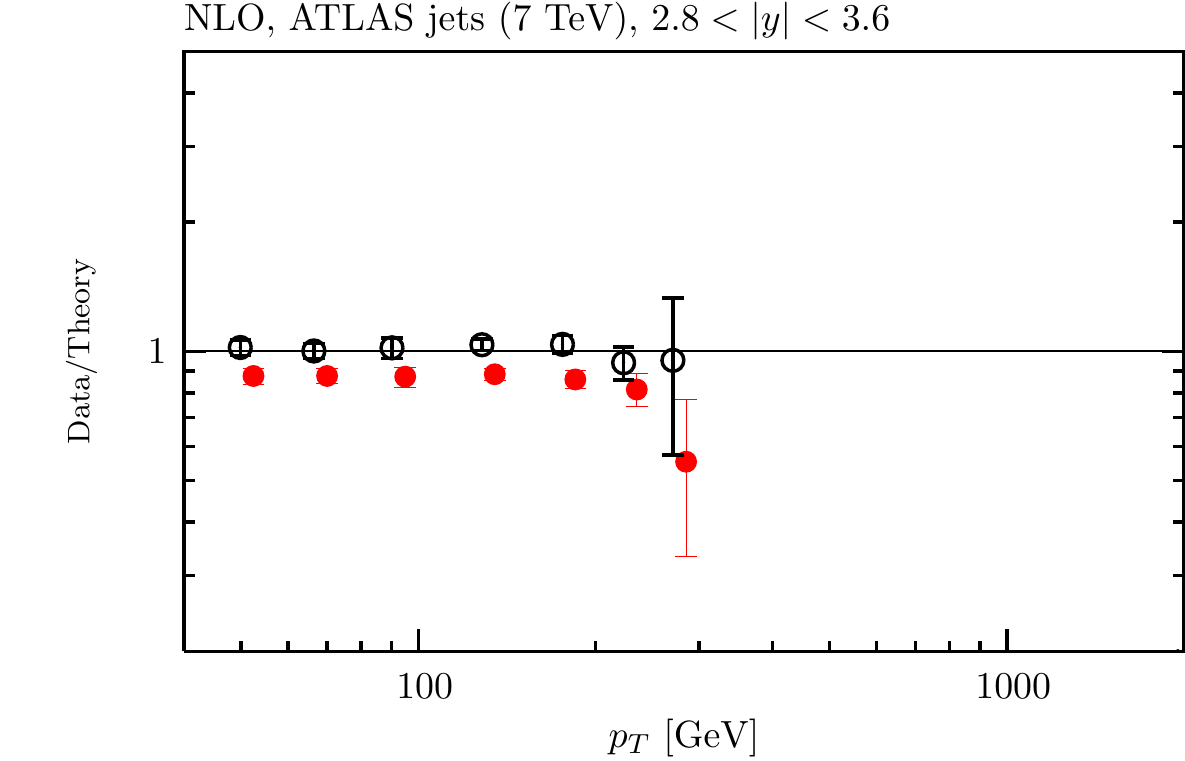}
\includegraphics[height=5cm]{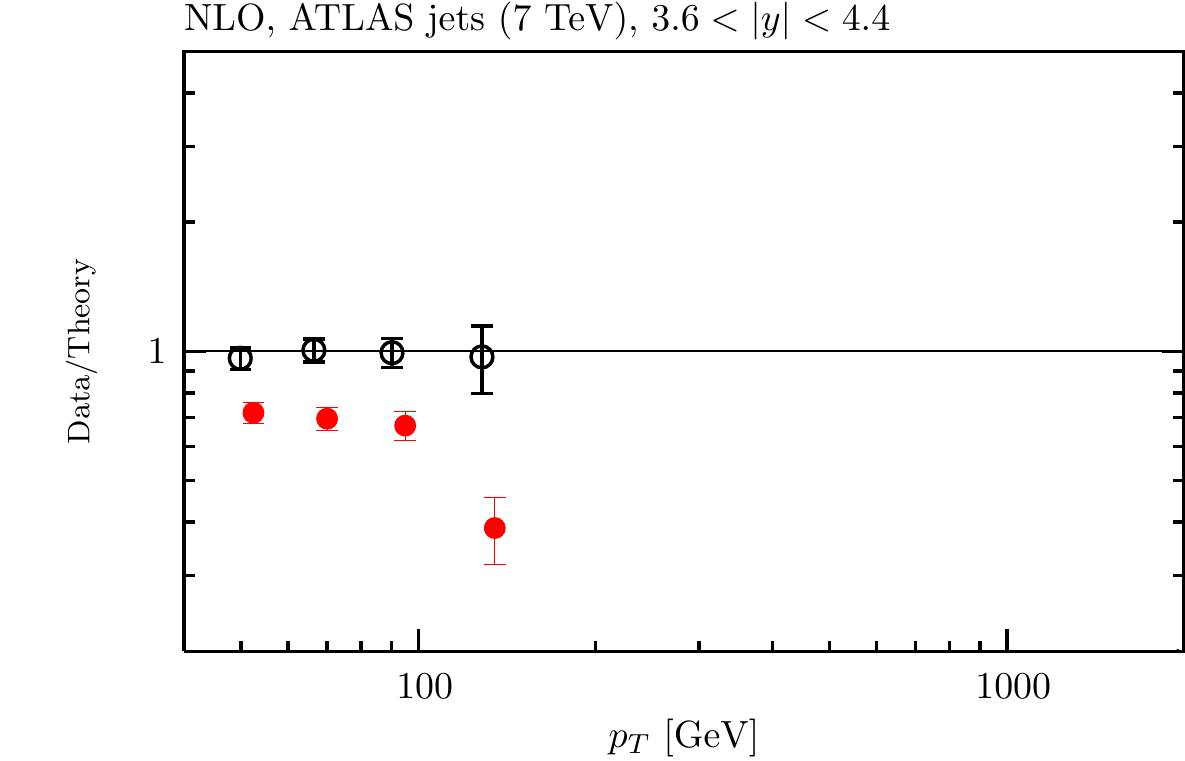}
\caption{\sf The fit quality for the ATLAS $7~\TeV$ jet data in various rapidity intervals 
\cite{ATLAS-jet7} at NLO. The red points
represent the ratio of measured data to theory predictions, and the black 
points (clustering around Data/Theory=1) correspond to this ratio once
the best fit has been obtained by 
shifting theory predictions relative to data by using the correlated 
systematics.}
\label{fig:ATLASjet7}
\end{center}
\end{figure}

\begin{figure} 
\begin{center}
\includegraphics[height=5cm]{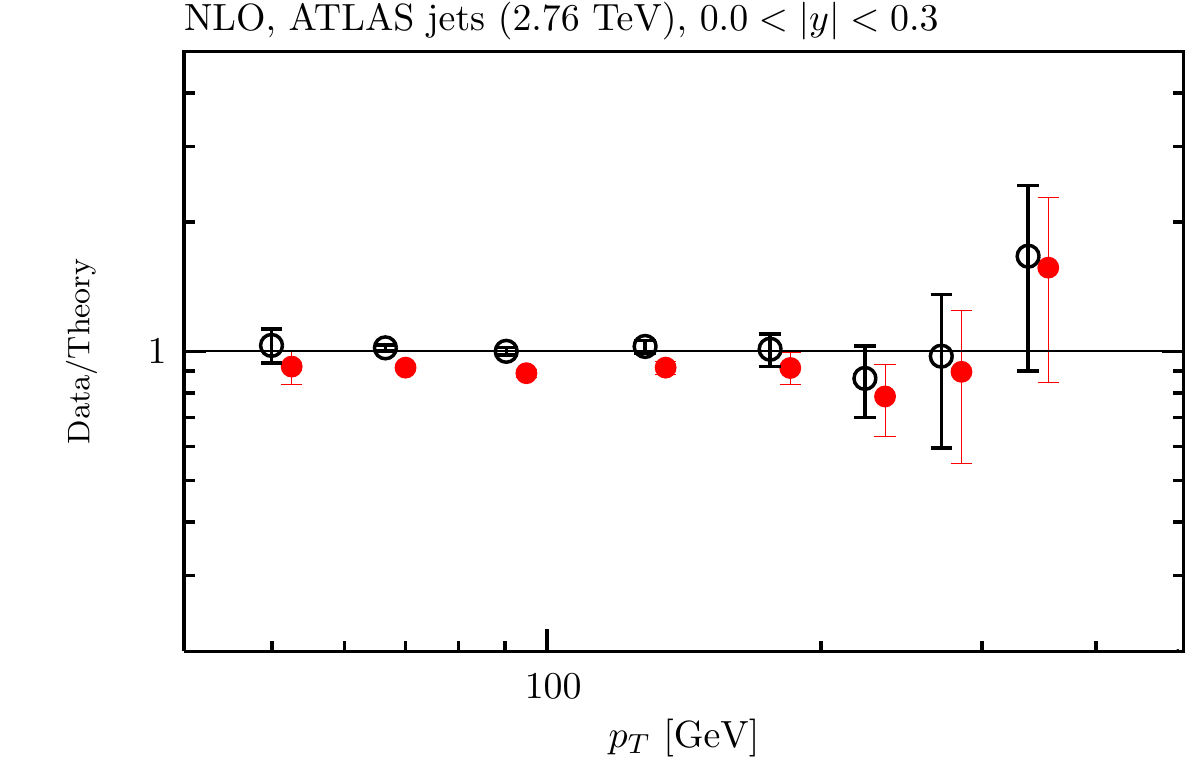}
\includegraphics[height=5cm]{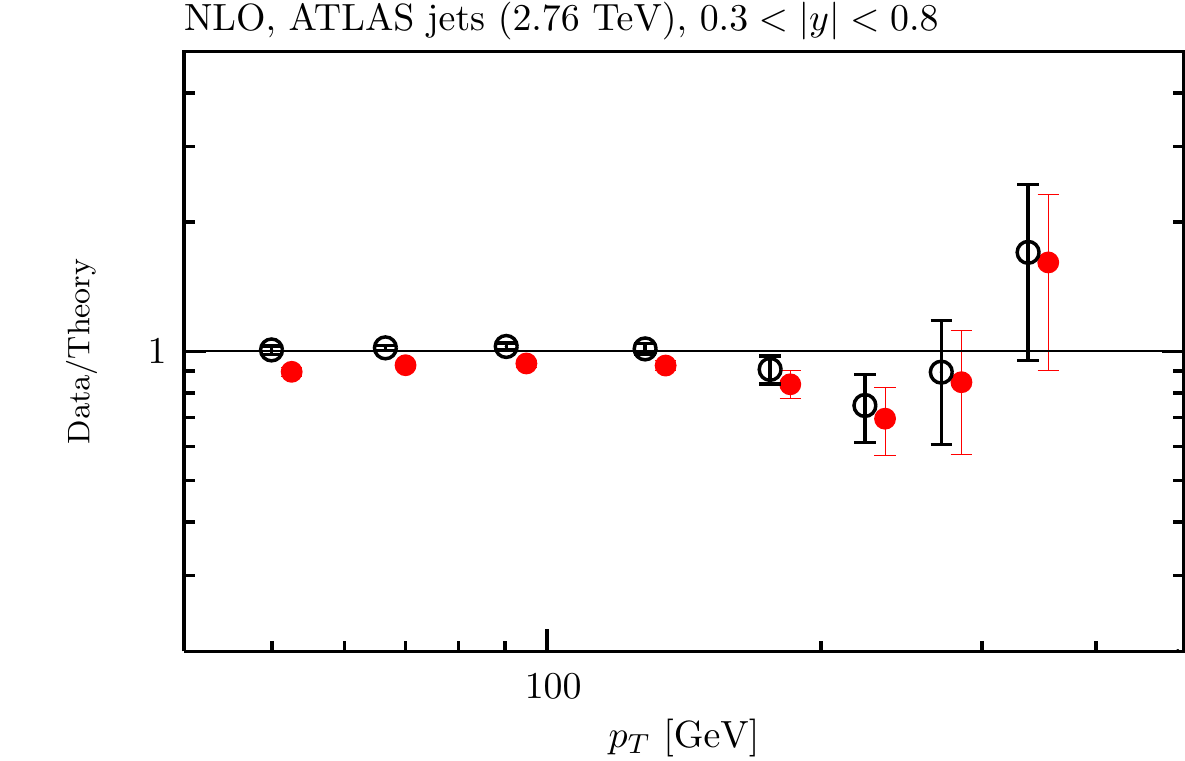}
\includegraphics[height=5cm]{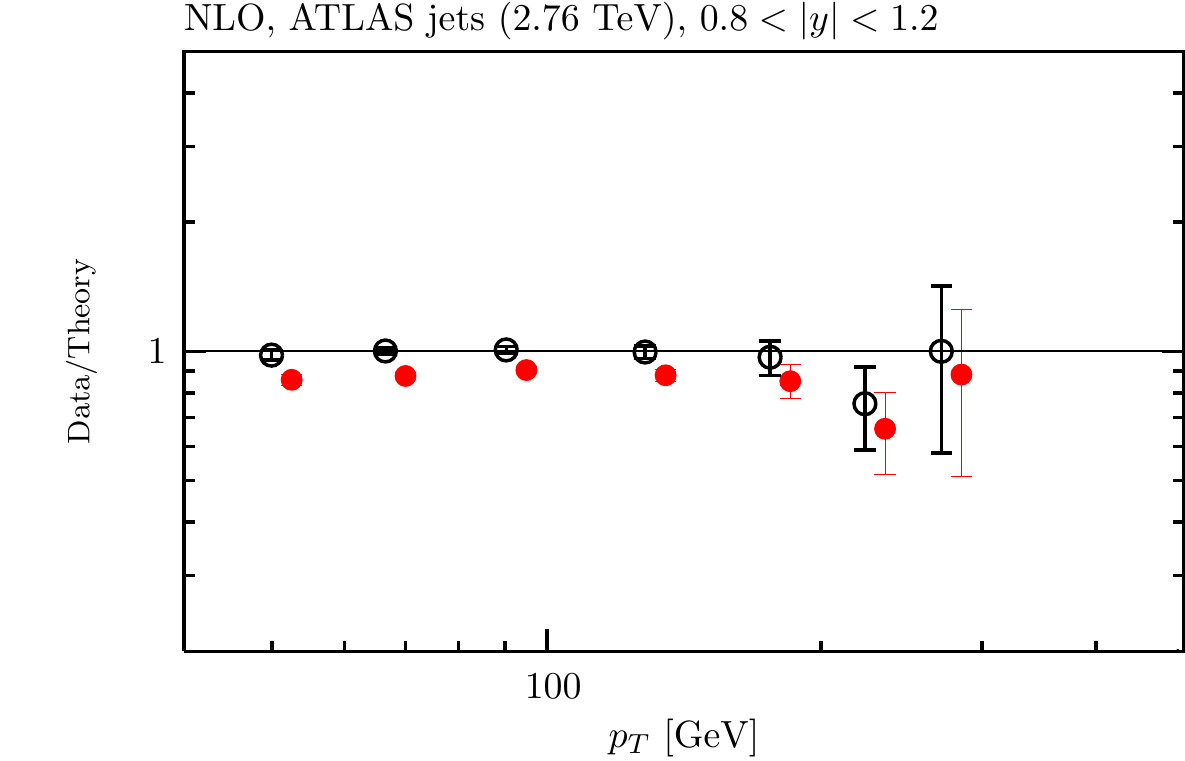}
\includegraphics[height=5cm]{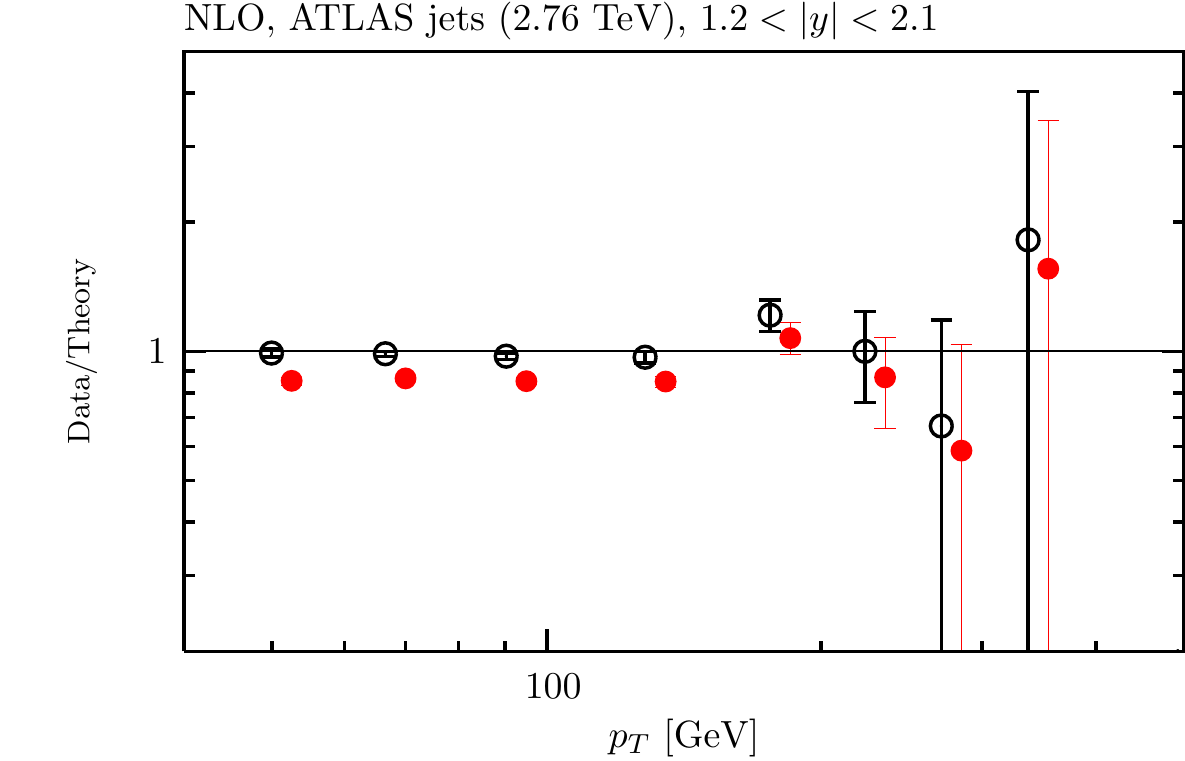}
\includegraphics[height=5cm]{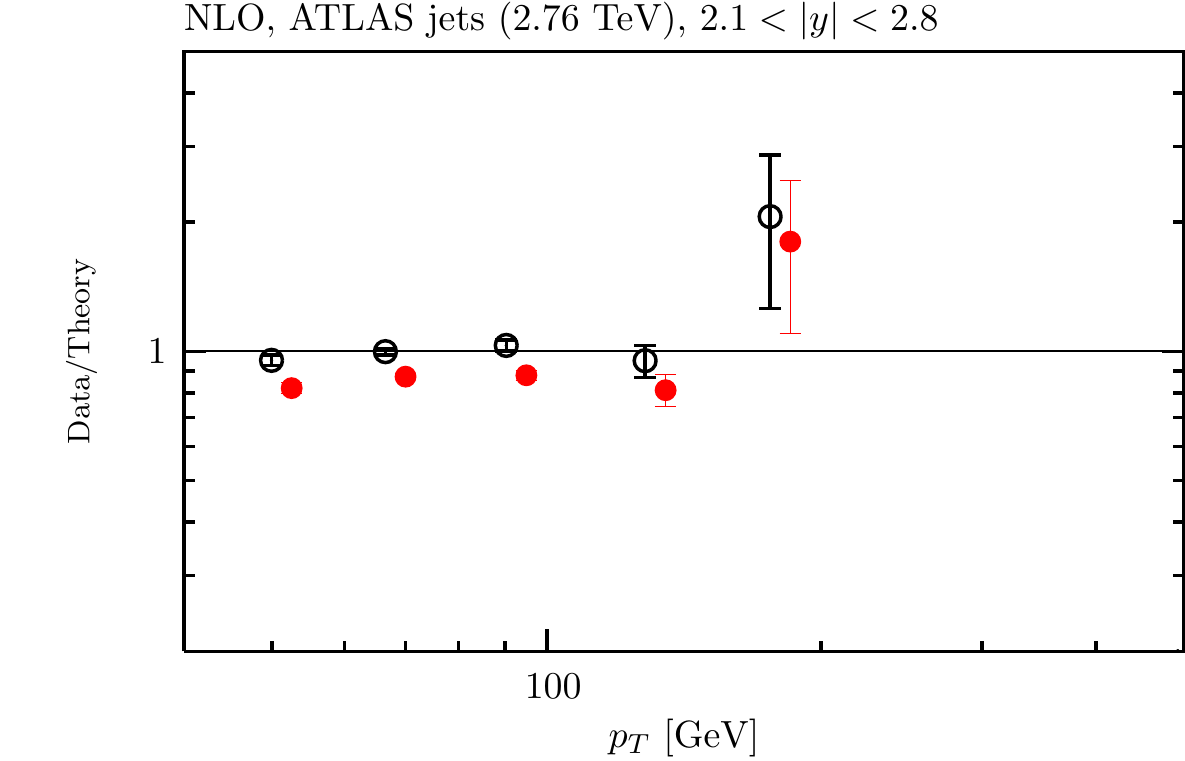}
\includegraphics[height=5cm]{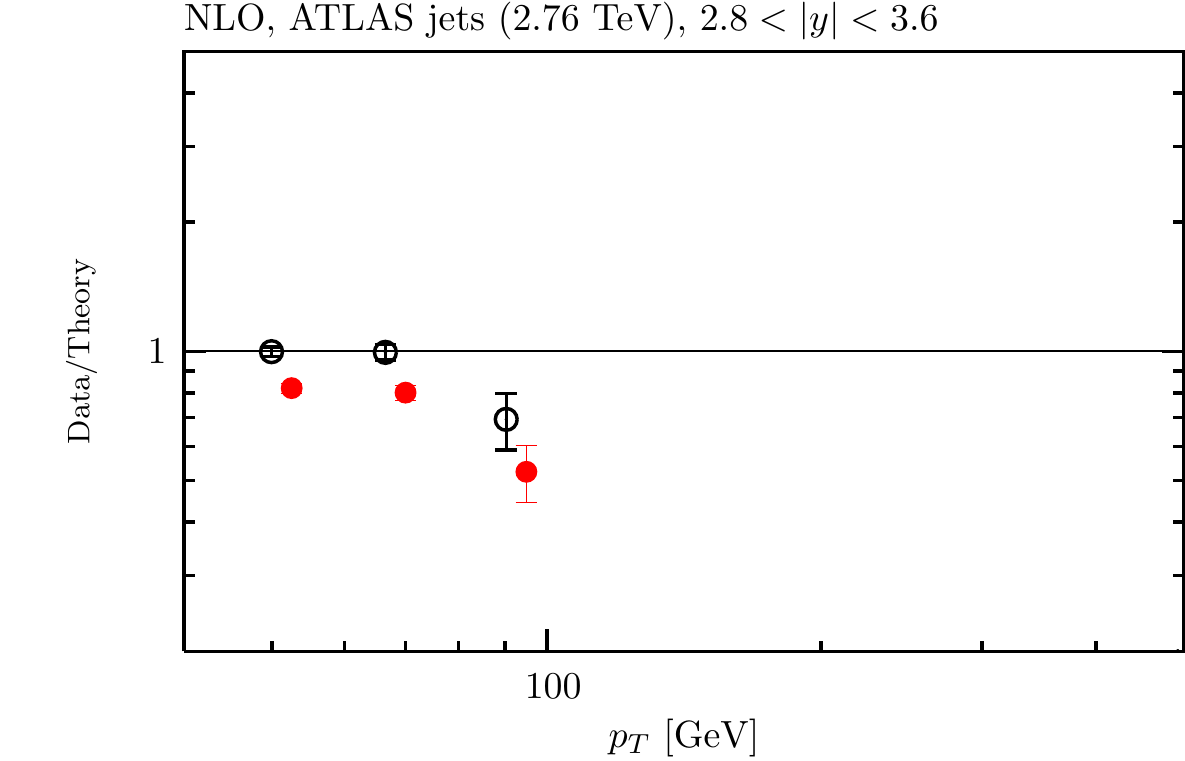}
\includegraphics[height=5cm]{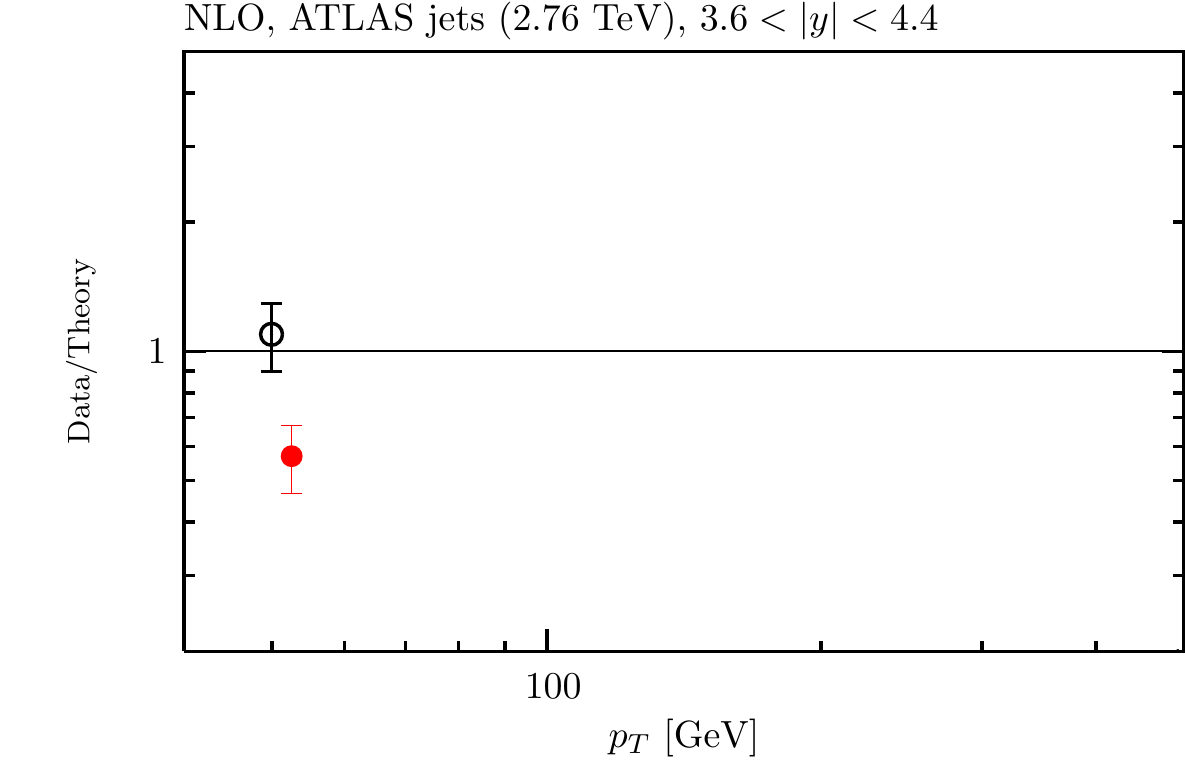}
\caption{\sf The fit quality for the ATLAS $2.76~\TeV$ jet data in various rapidity intervals  
\cite{ATLAS-jet2.76} at NLO. The red points
represent the ratio of measured data to theory predictions, and the black 
points (clustering around Data/Theory=1) correspond to this ratio once
the best fit has been obtained by 
shifting theory predictions relative to data by using the correlated 
systematics.}
\label{fig:ATLASjet276}
\end{center}
\end{figure}

\begin{figure} 
\begin{center}
\includegraphics[height=5cm]{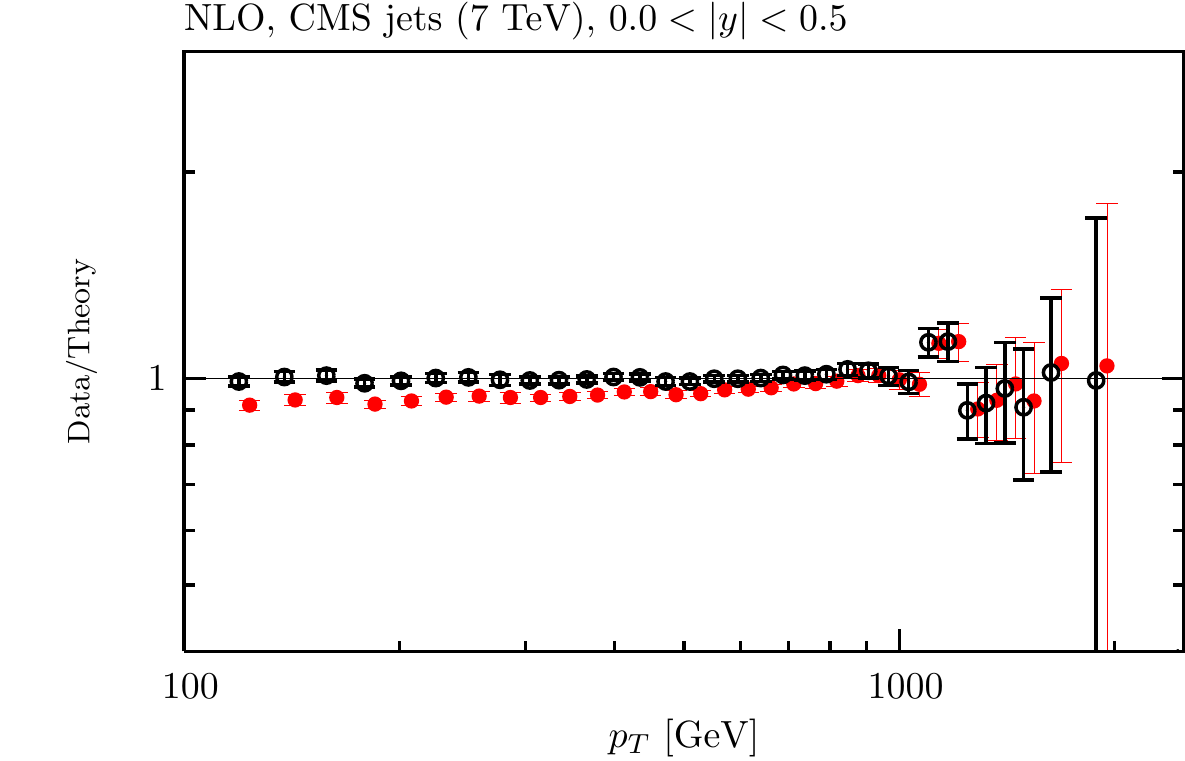}
\includegraphics[height=5cm]{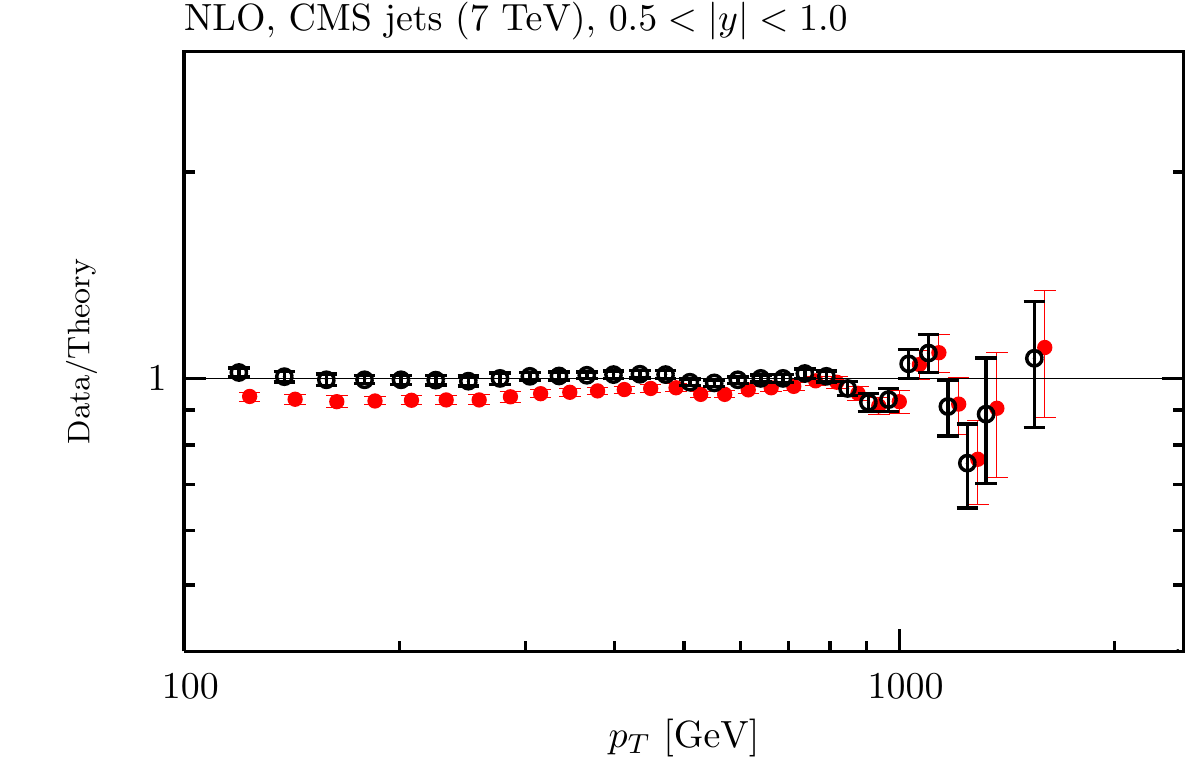}
\includegraphics[height=5cm]{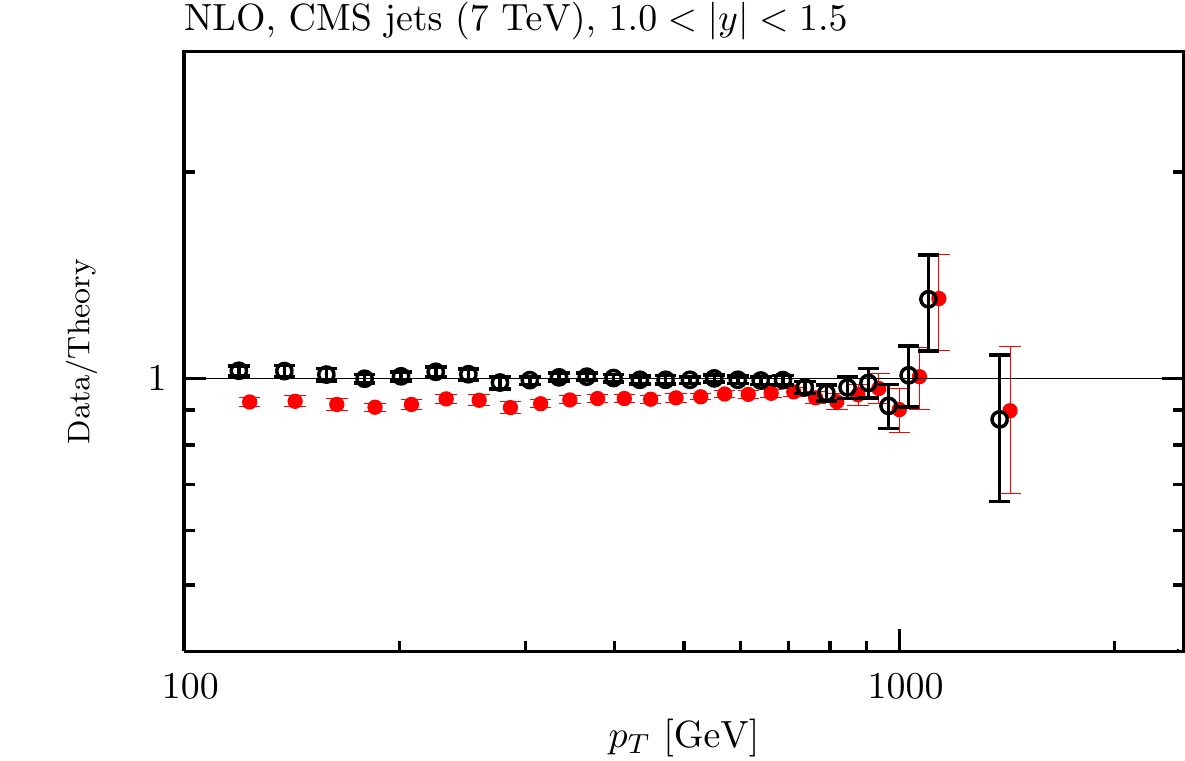}
\includegraphics[height=5cm]{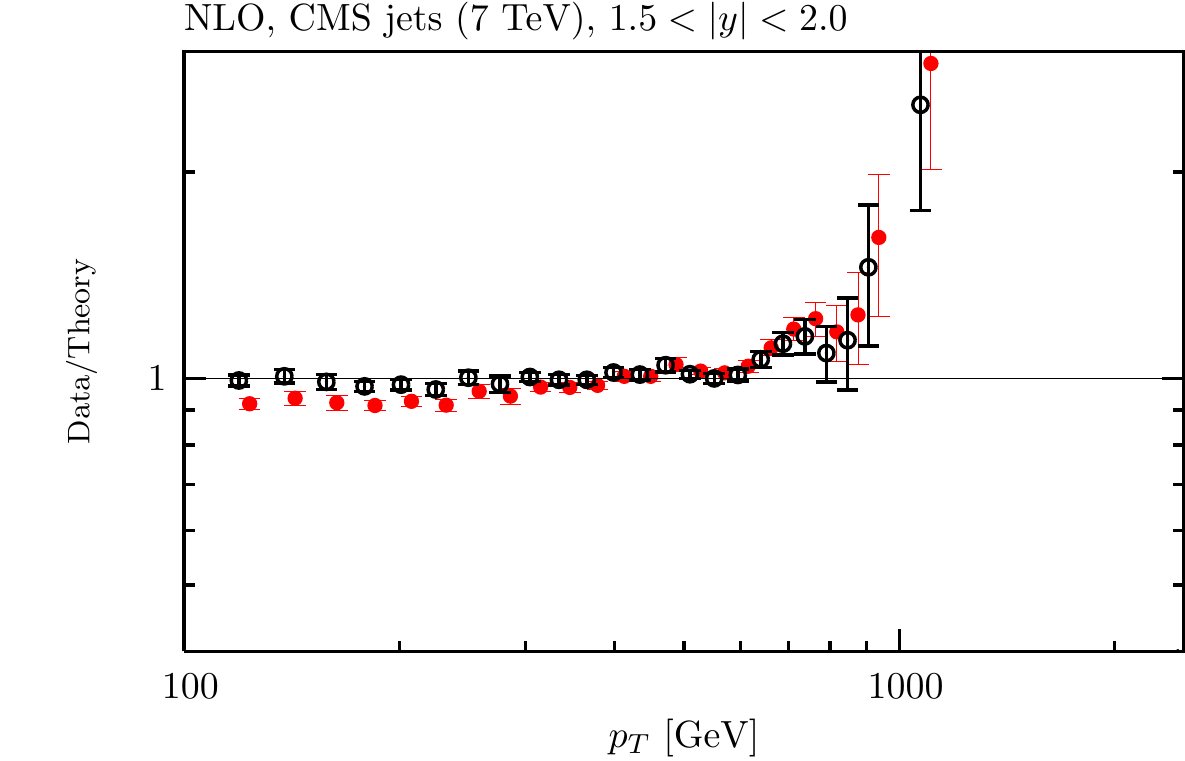}
\includegraphics[height=5cm]{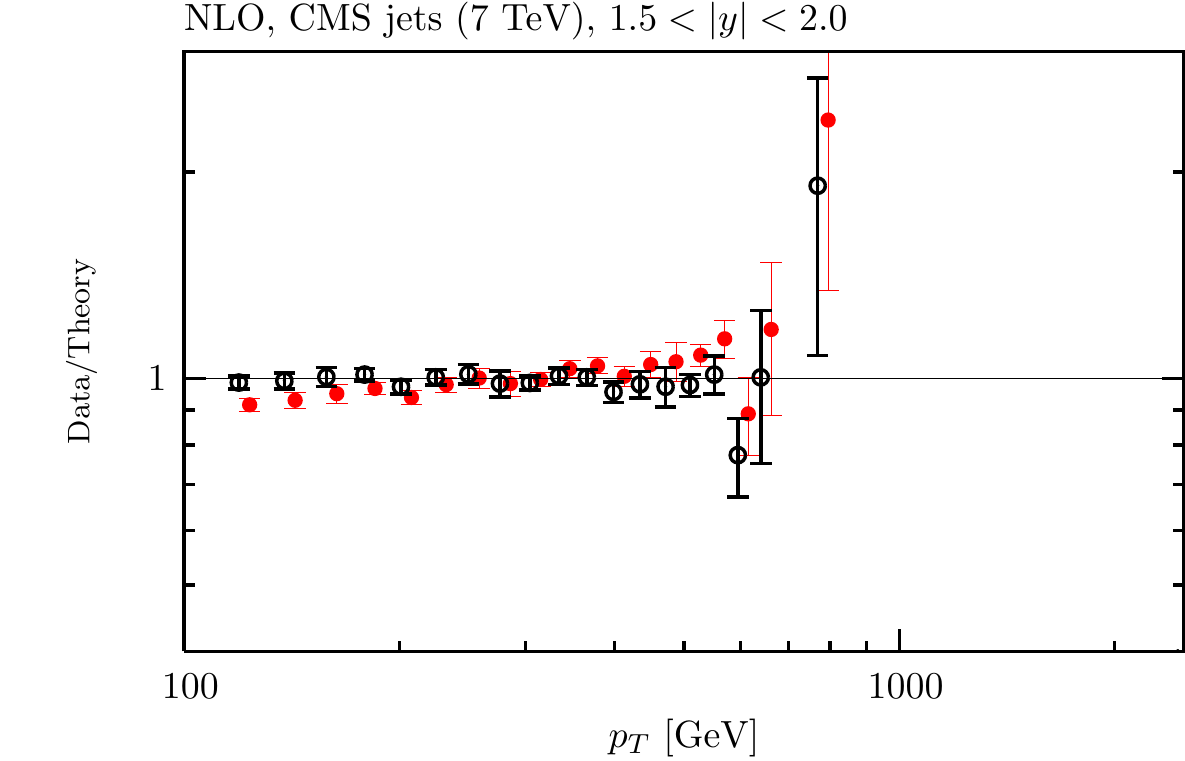}
\caption{\sf The fit quality for the CMS $7~\TeV$ jet data in various rapidity intervals 
\cite{CMS-jet7} at NLO. The red points
represent the ratio of measured data to theory predictions, and the black 
points (clustering around Data/Theory=1) correspond to this ratio once
the best fit has been obtained by 
shifting theory predictions relative to data by using the correlated 
systematics.}
\label{fig:CMSjet7}
\end{center}
\end{figure}

In the present global analysis at NLO we include the CMS inclusive jet data 
at $\sqrt{s}=7$ TeV with jet radius $R=0.7$ \cite{CMS-jet7}, together with the ATLAS data at 7 TeV 
\cite{ATLAS-jet7} and at 2.76 TeV with jet radius $R=0.4$ 
\cite{ATLAS-jet2.76}.  For the latter we 
use cuts proposed 
in the ATLAS study, which eliminate the two lowest $p_T$ points in 
each bin, due to the large sensitivity to hadronisation corrections in  these 
bins, and some of the highest $p_T$ points.\footnote{In the analysis of \cite{Wattjets}
we cut two more ATLAS points at the edge of rapidity bins due to very 
poor fits to these points. This was much more of an issue when using the 
additive definition for correlated uncertainties, and we have 
reinstated these points here. Indeed the whole fit quality for this data set is 
much better using the multiplicative definition.} We perform the calculations
within the fitting procedure using FastNLO \cite{fastnlo} version 
2 \cite{fastnlov2}, which uses 
NLOJet++ \cite{nlojet,nlojet2}, and APPLGrid. 
The jet data from the two 
experiments appear to be extremely compatible with each other. 
The data are both well-predicted and well fit, as shown in Table 
\ref{tab:LHCjet}. 
 Before these 
data are included in the fit we find $\chi^2 =107$ for 116 data points for 
ATLAS and $\chi^2=143$ for the 133 CMS jet data points at NLO, very similar to
the values of $\chi^2$ obtained from the earlier MMSTWW NLO PDF set. Including 
these jet data in the NLO fit leads to more improvement in the 
$\chi^2$ for CMS than for the ATLAS data, i.e. $143 \to 138$ as 
opposed to $107\to 106$. However, in both cases the possible improvement is 
rather small. We note that the treatment of the systematic uncertainties 
for the CMS jet data has been modified to take account of an increased 
understanding by the experiment since the original publication of the data 
\cite{CMS-jet7}. Initially the  the single pion related correlated
uncertainties were all correlated. However, in \cite{CMS-jeta} 
a decision was made to decorrelate single pion systematics, i.e. to 
split the single pion source into 5 separate parts. This lowers 
the $\chi^2$ obtained in the best fit significantly, from 
about 170 to about 135. However, it leads to no real change in PDFs extracted
in the global fit, though it allows a slightly higher value of 
$\alpha_S(M_Z^2)$. The fit quality for the LHC jet data is shown at NLO
in Figs.~\ref{fig:ATLASjet7}, \ref{fig:ATLASjet276} and 
\ref{fig:CMSjet7}. One can see that the correlated uncertainties play a 
significant role in enabling the good fit quality, with the shift of data
against theory being larger than the uncorrelated uncertainties. However,
for each of the three data sets the shape of the data/theory comparison 
is very good even before the correlated systematics are applied, with only
a small correction of order $10\%$ at most needed, this being
relatively independent of $p_T$, rapidity, or even data set.\footnote{It has 
very recently been brought to our attention that there is a change in the 
luminosity determination for the data in \cite{ATLAS-WZ,ATLAS-jet7}, and
the cross sections should be multiplied by a factor of 1.0187 and 
the uncertainty on the global normalisation (``Lumi") increases 
slightly from $3.4\%$ to $3.5\%$. This was too late to be included explicitly 
in our PDF determination. However, we note that this corrections results in 
the $\chi^2$ for the best fits at NLO and NNLO both reducing by about half 
a unit, and any changes in the PDFs are very much smaller than all 
uncertainties.}

\begin{table}
\begin{center}
\vspace{-.5cm}
\begin{tabular}{|l|c|c|c|c|}
\hline
 &  & MMSTWW     & MMHT2014 & MMHT2014 \\
~~~~~~data set         &   {$N_{pts}$}          & Ref.\cite{MMSTWW}        & (no LHC)       &  (with LHC)       \\
\hline
\multicolumn{5}{|c|}{NLO}  \\
\hline
ATLAS jets ({2.76~TeV+7~TeV})        & 116& 107 & 107 & 106 \\   
CMS jets ({7~TeV})                   & 133& 140 & 143 & 138 \\   
\hline
\multicolumn{5}{|c|}{NNLO small $K$-factor} \\
\hline
ATLAS jets ({2.76~TeV+7~TeV})        & 116& (107) & (123) & (122) 115\\   
CMS jets ({7~TeV})                   & 133& (142) & (137) & (138) 137\\   
\hline
\multicolumn{5}{|c|}{NNLO large $K$-factor} \\
\hline
ATLAS jets ({2.76~TeV+7~TeV})        & 116& (117) & (132) & (132) 126\\   
CMS jets ({7~TeV})                   & 133& (145) & (137) & (139) 139\\   
\hline
    \end{tabular}
\end{center}
\vspace{-.0cm}
\caption{\sf The quality of the description (as measured by the value of 
$\chi^2$) of the LHC inclusive jet data before and after 
they are included in the 
global NLO and NNLO fits. 
We also show for comparison the $\chi^2$ values obtained in 
the CPdeut fit of the NLO MMSTWW analysis \cite{MMSTWW}, which did not 
include LHC data. Also the LHC jet data are not included in the final NNLO MMHT global fit presented in this paper. However, the NNLO $\chi^2$ numbers and $K$ factors mentioned in the Table correspond to an exploratory approximate NNLO study described in Section \ref{sec:jetsNNLO}.}
\label{tab:LHCjet}
\end{table}

Of course, the full NNLO QCD calculation is not available for jet
cross sections, either in DIS or in hadron-hadron collisions.   
The NNLO calculation of jet production is ongoing, but not yet complete.  
It is an enormous project and much progress has been made, 
see \cite{GGGP1,GGGP2,GGGP3}, and it should hopefully be available soon.

Despite the absence of the full NNLO result, in the NNLO MSTW analysis the 
Tevatron jet data \cite{CDFjet,D0jet} 
were included in the fit using an approximation based on 
the knowledge of the threshold corrections \cite{KO}.
It was argued that although there was no guarantee that these give a very 
good approximation to the full NNLO corrections, in this case the NLO 
corrections themselves are of the same order as the systematic uncertainties 
on the data.  The threshold corrections are the only expected source of 
possible large NNLO corrections, so the fact that they provide a correction 
which is smooth in the $p_T$ of the jet and moderately small compared to 
systematic uncertainties in the data strongly implies that the full NNLO 
corrections would lead to  little change in the PDFs.  Since these jet data 
are the only good direct constraint on the high-$x$ gluon it was decided to 
include them in the NNLO fit judging that the impact of leaving them out 
would be far more detrimental than any inaccuracies in including them 
without knowing the full NNLO hard cross section.

In fact the threshold corrections to the Tevatron data gave about a 
10$\%$ positive correction, see for example Fig. 50 in \cite{Wattjets}.  
We also see from the same figure that the threshold 
corrections for the LHC data are similar to those at the Tevatron for the 
highest $x$ values at which jets are measured, but blow up at the low $x$ 
values probed, that is when they are far from threshold.
Recent detailed studies exploring the dependence of the threshold corrections 
on the jet radius $R$ values at NLO and NNLO  
show that the true corrections in the threshold region show a significant
dependence\footnote{The dependence on $R$ was not accounted for in \cite{KO}.} on $R$ at NLO 
\cite{thres1,thres2}, but that this is rather reduced at NNLO \cite{thres2}. 
However, the improved NNLO threshold calculations in \cite{thres2}
show that there are still 
problems at low and moderate values of jet $p_T$ . 

In the present global analysis, as a default at NNLO, we still include the 
Tevatron jet data in the fit. This seems reasonable, since they are always 
relatively near threshold, and the corrections do not obviously break down at 
the lowest $p_T$ values of the jet.\footnote{We realise that strictly 
speaking the D0 jet data are difficult to include in an NNLO fit since the
mid-point algorithm used becomes infrared unsafe at this order 
\cite{IRunsafe}. However, the whole ``NNLO'' jet treatment is approximate 
at present. We will revisit the question of whether to include these data in 
future fits when the full NNLO calculation is known. At this time presumably
there will also be more precise LHC jet data and the D0 jet data would play 
a diminishing role in the fit anyway.}  On the other hand, we omit the LHC jet 
data, since at the lowest $p_T$ measured the threshold corrections are not 
stable and, moreover, have large uncertainties at the highest rapidities 
observed. This is slightly more blunt, but quite similar in practice to the 
conclusion of \cite{NNPDFjet} which compares the degree of agreement between 
the approximate threshold calculation and the exact calculation for the 
$gg \to gg$ channel, where the later is known. It is found that the agreement 
is good for high values of $p_T$ (relative to centre of mass energy $\sqrt{s}$)
and relatively central rapidity. These regions of agreement are then deemed
to be the regions where the approximate NNLO is likely quite reliable. 
They correspond to most of the Tevatron data, except at high rapidity (where 
the systematic errors on data are large), much of the CMS jet data, but 
little of the ATLAS jet data. Hence, we feel confident including the Tevatron
jet data using approximate NNLO expressions, especially given that in 
\cite{Wattjets} we investigated the effect of rather dramatic modifications
of these corrections, finding only rather moderate changes in PDFs
and $\alpha_S(M_Z^2)$. We could arguably include (much of) the CMS jet
data, but for the moment err on the side of caution.

\subsubsection{Exploratory fits to LHC jet data at `NNLO'  \label{sec:jetsNNLO}}

\begin{figure} 
\begin{center}
\includegraphics[height=5cm]{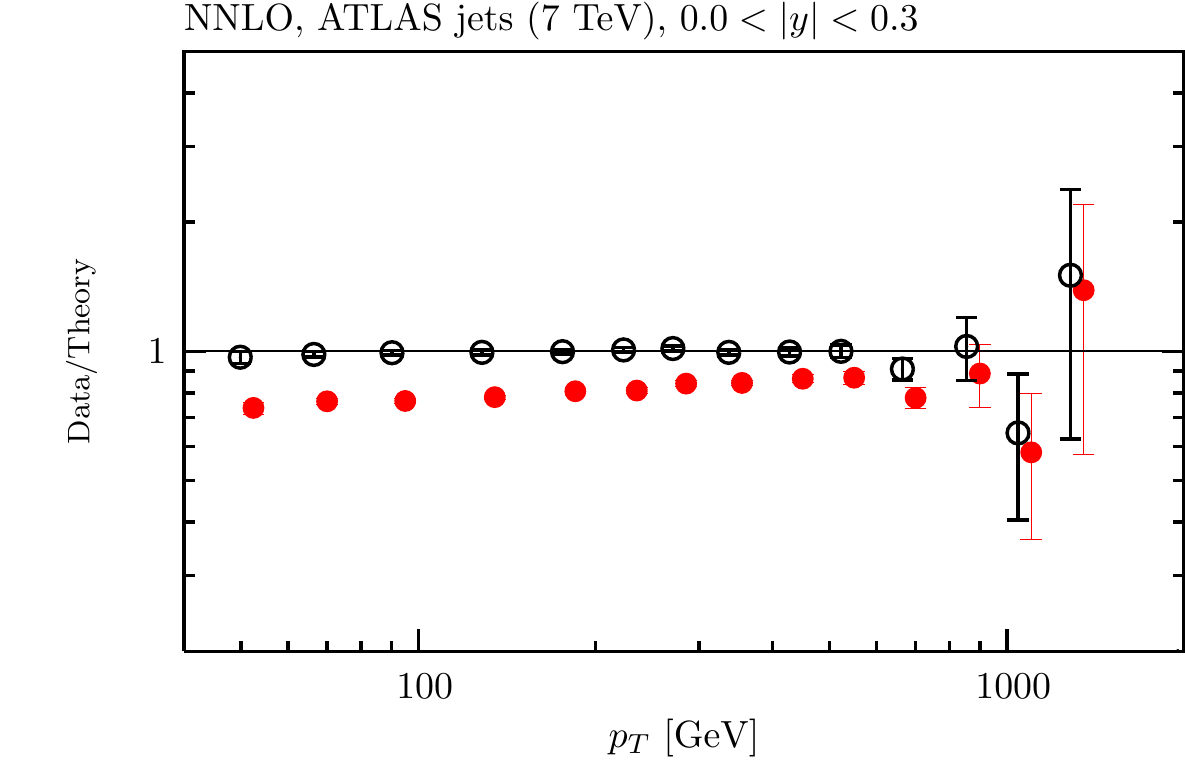}
\includegraphics[height=5cm]{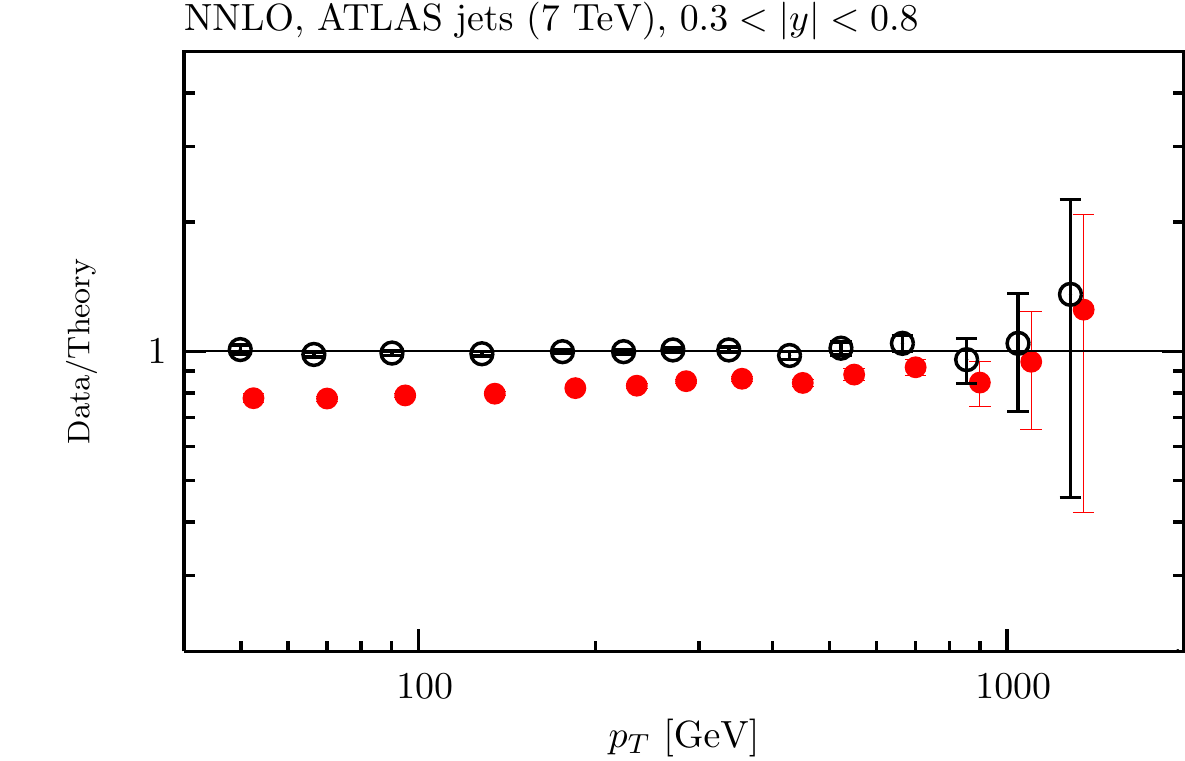}
\includegraphics[height=5cm]{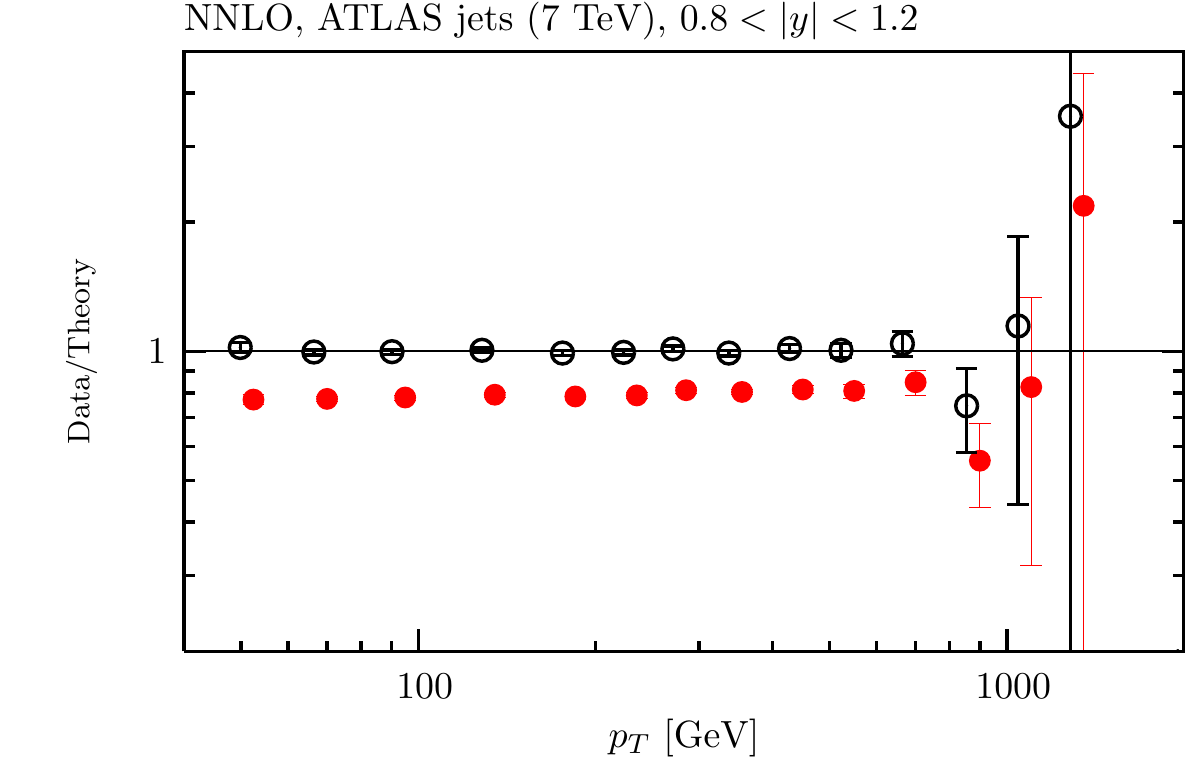}
\includegraphics[height=5cm]{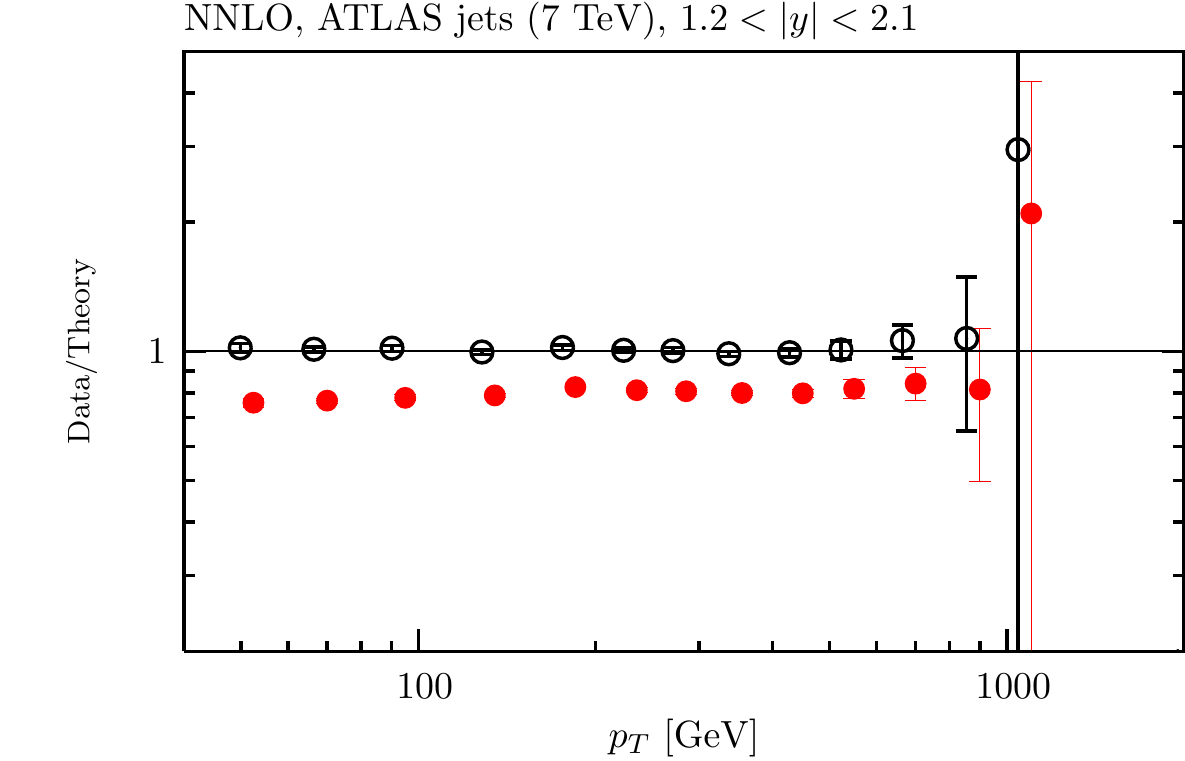}
\includegraphics[height=5cm]{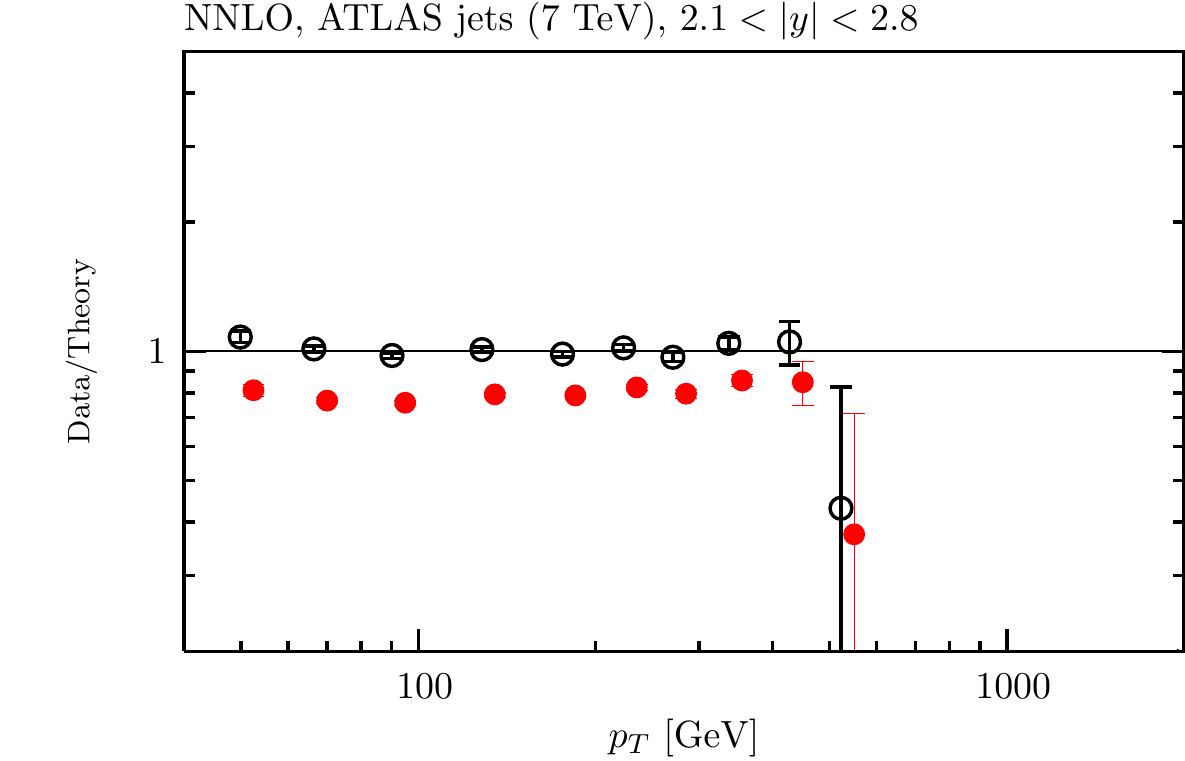}
\includegraphics[height=5cm]{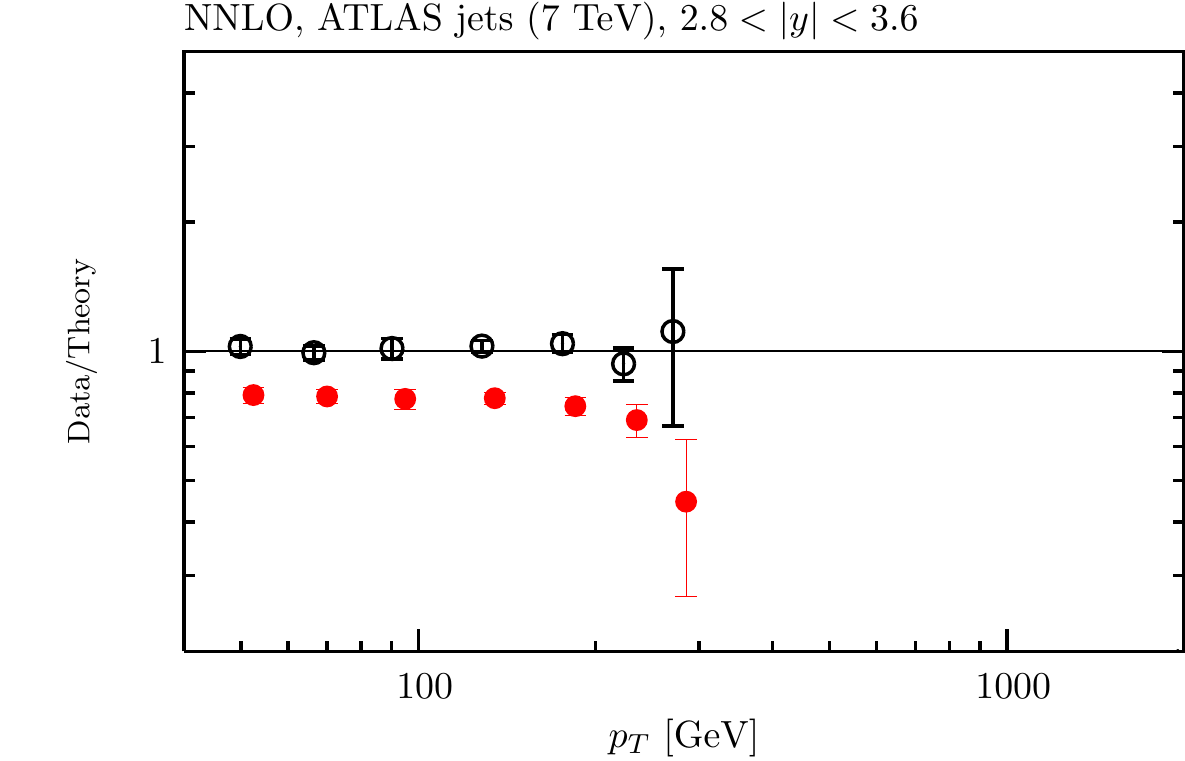}
\includegraphics[height=5cm]{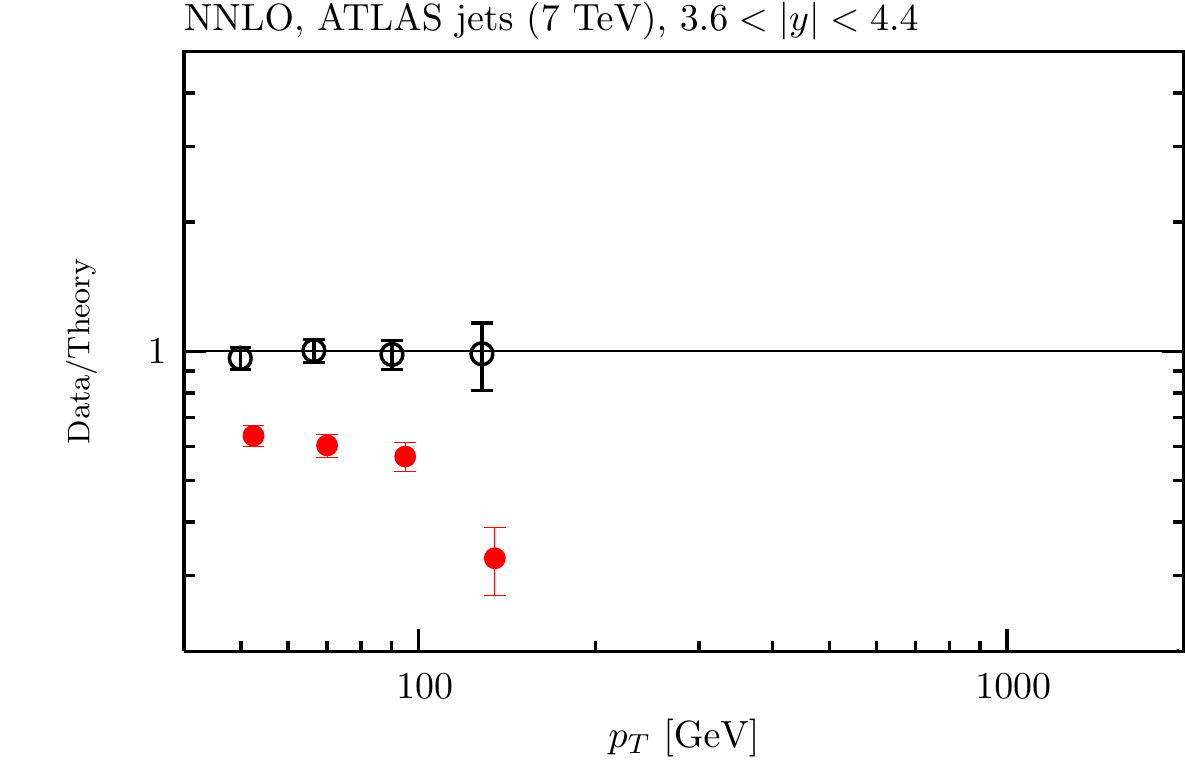}
\caption{\sf The fit quality for the ATLAS $7~\TeV$ jet data 
\cite{ATLAS-jet7} at NNLO, using the `larger' K-factor described in the text. 
The red points
represent the ratio of measured data to theory predictions, and the black 
points (clustering around Data/Theory=1) correspond to this ratio once
the best fit has been obtained by 
shifting theory predictions relative to data by using the correlated 
systematics.}
\label{fig:ATLASjet7NNLO}
\end{center}
\end{figure}

\begin{figure} 
\begin{center}
\includegraphics[height=5cm]{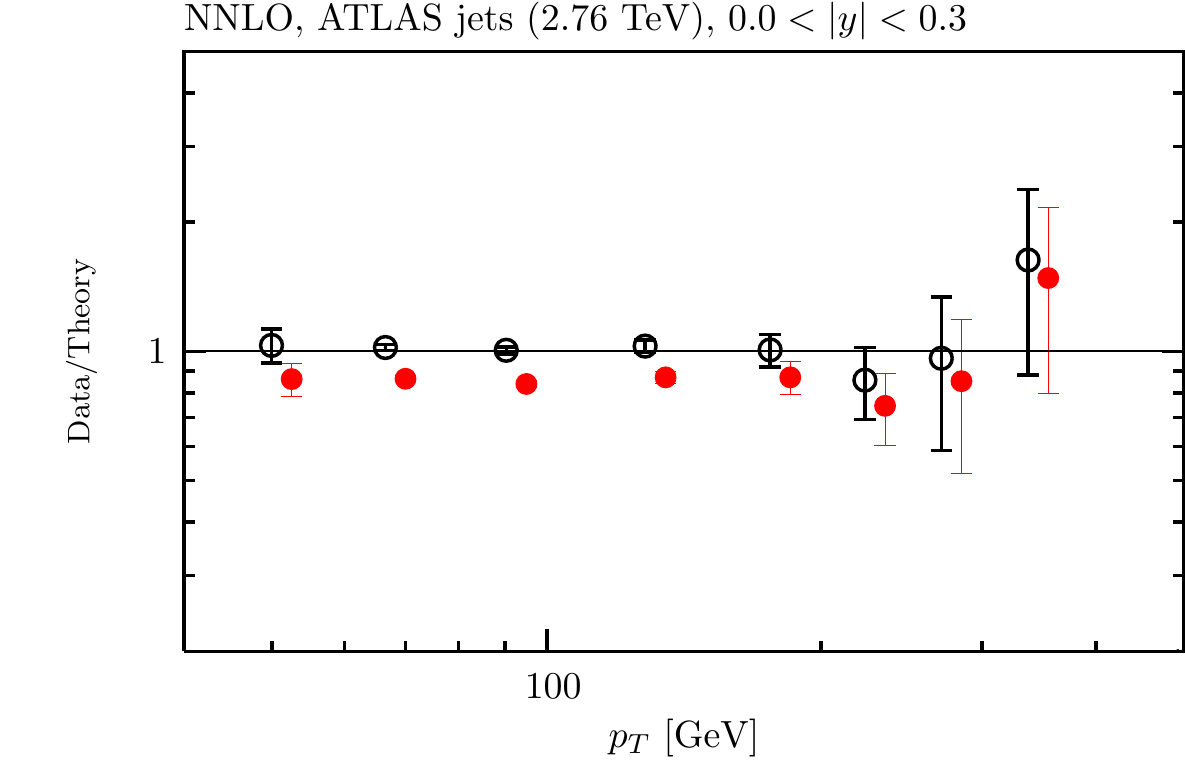}
\includegraphics[height=5cm]{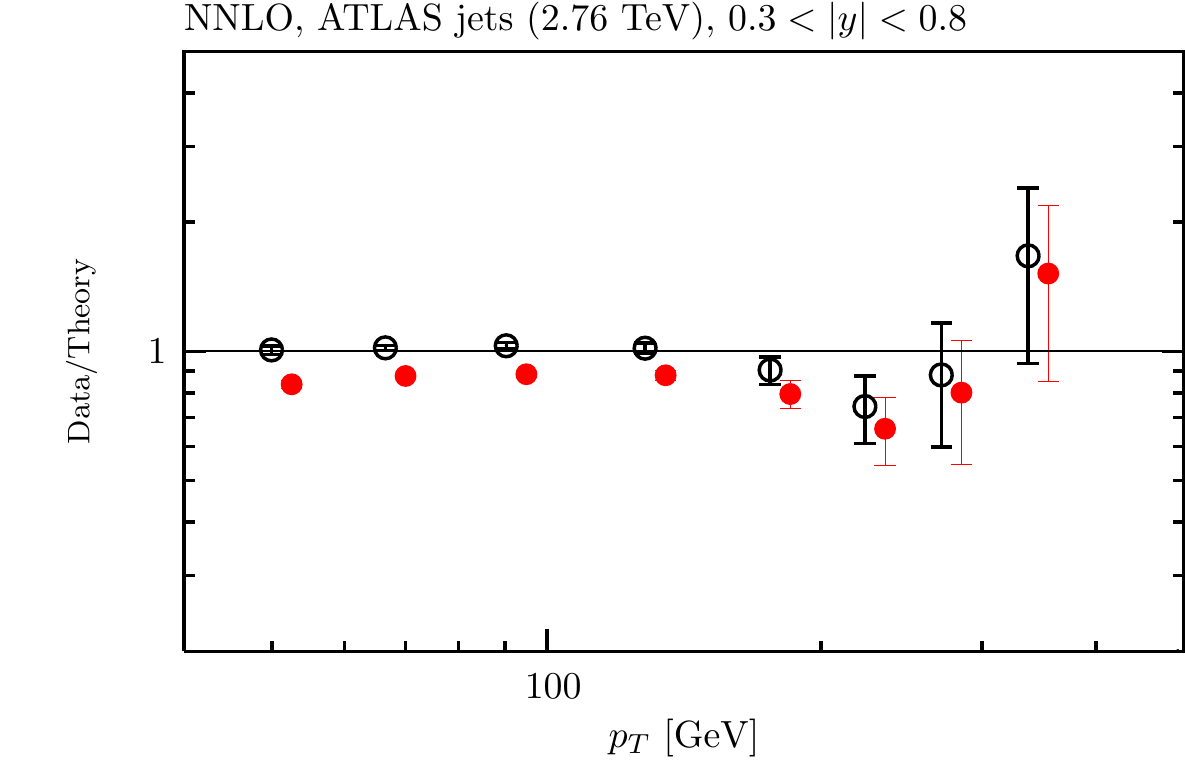}
\includegraphics[height=5cm]{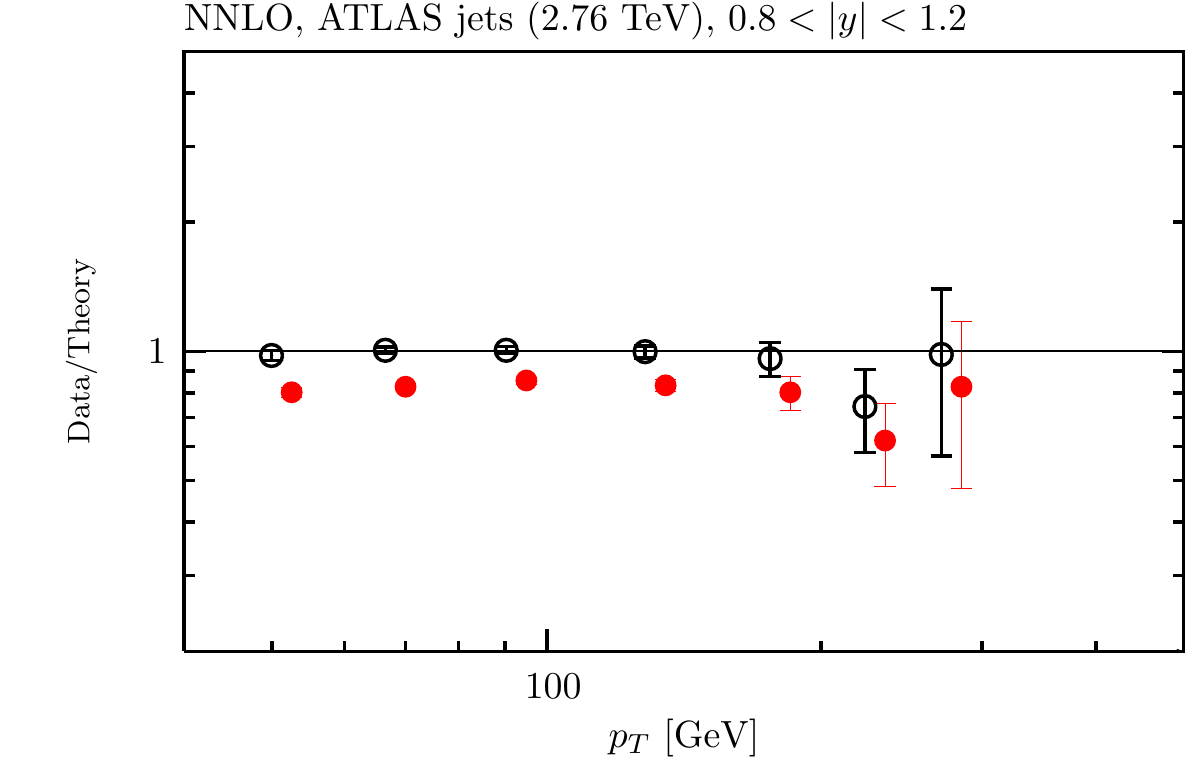}
\includegraphics[height=5cm]{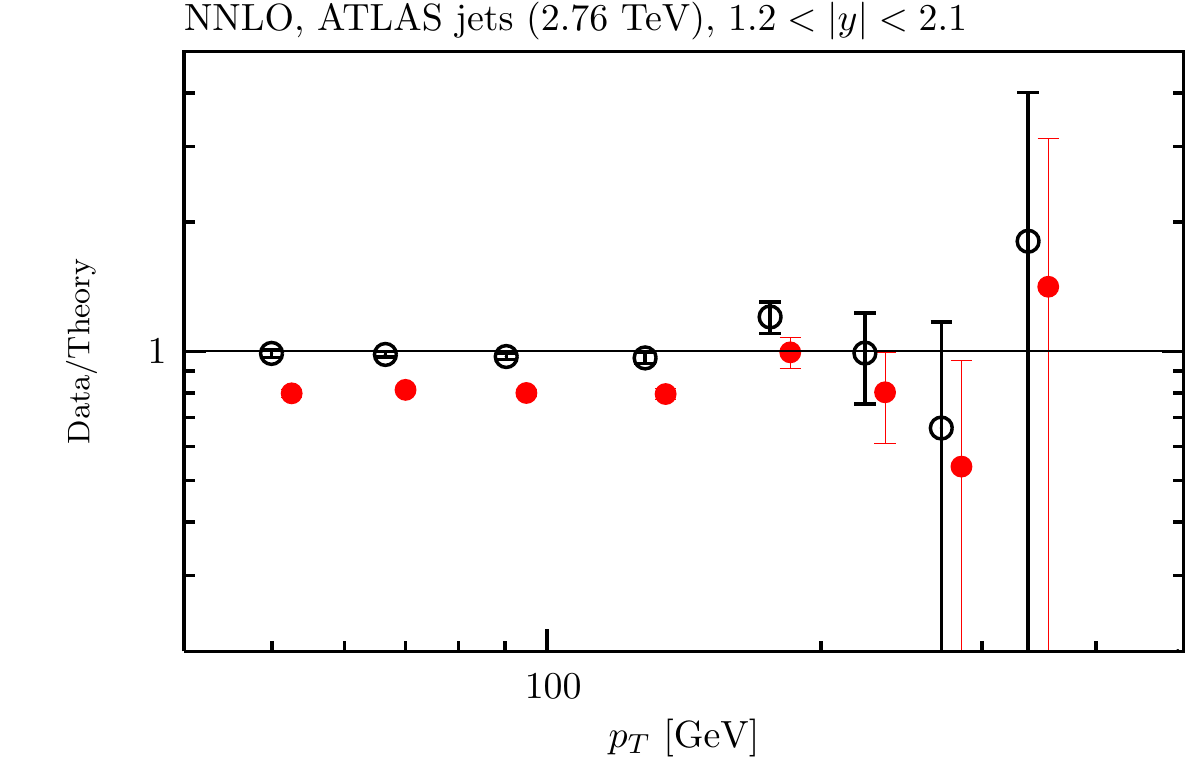}
\includegraphics[height=5cm]{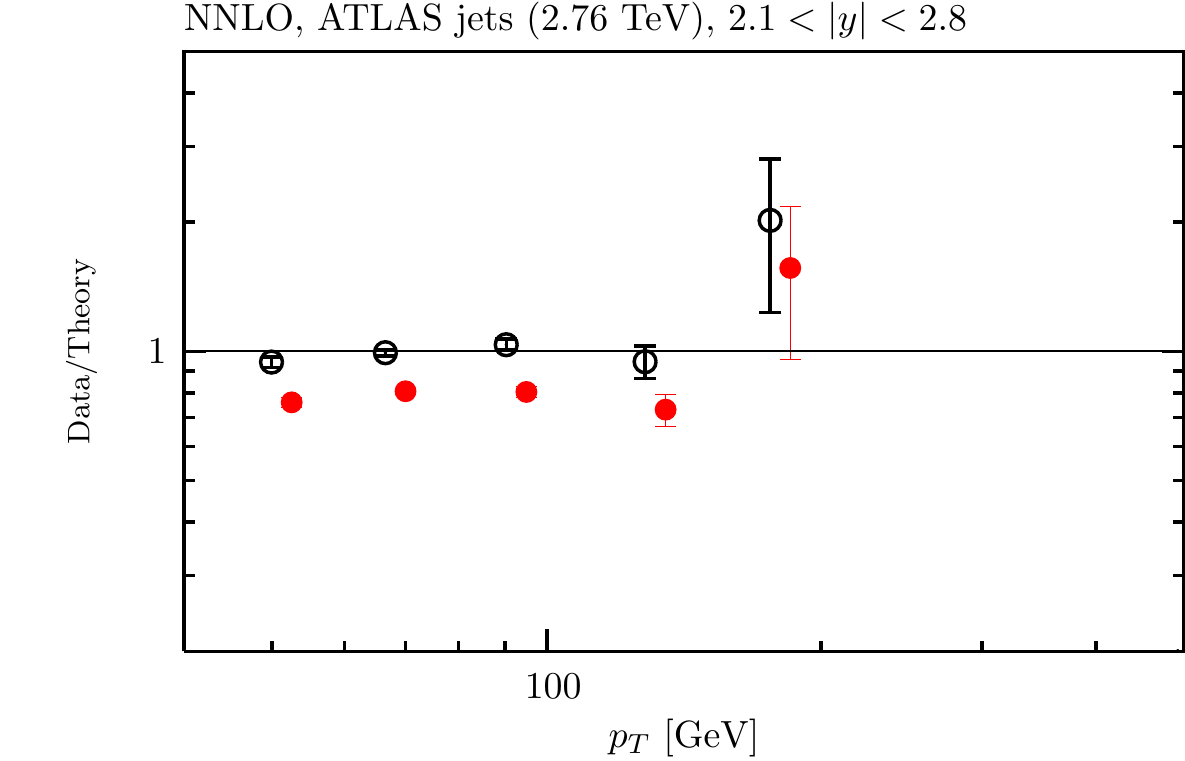}
\includegraphics[height=5cm]{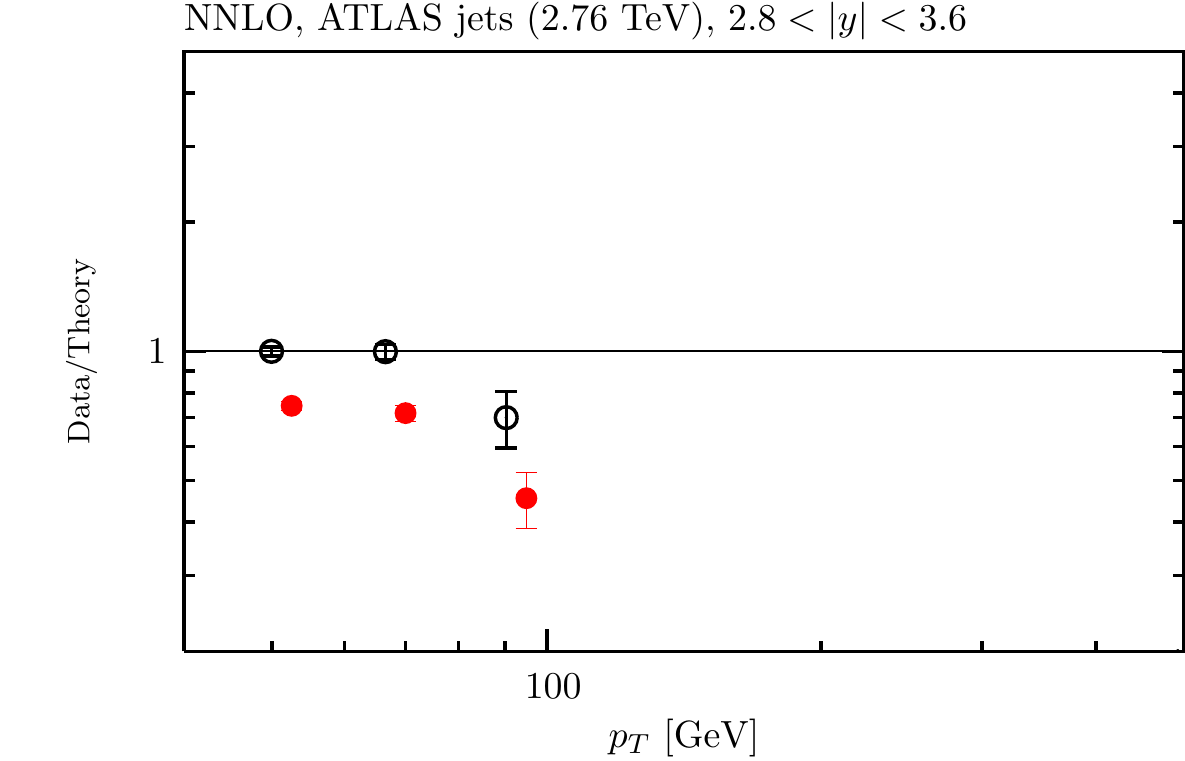}
\includegraphics[height=5cm]{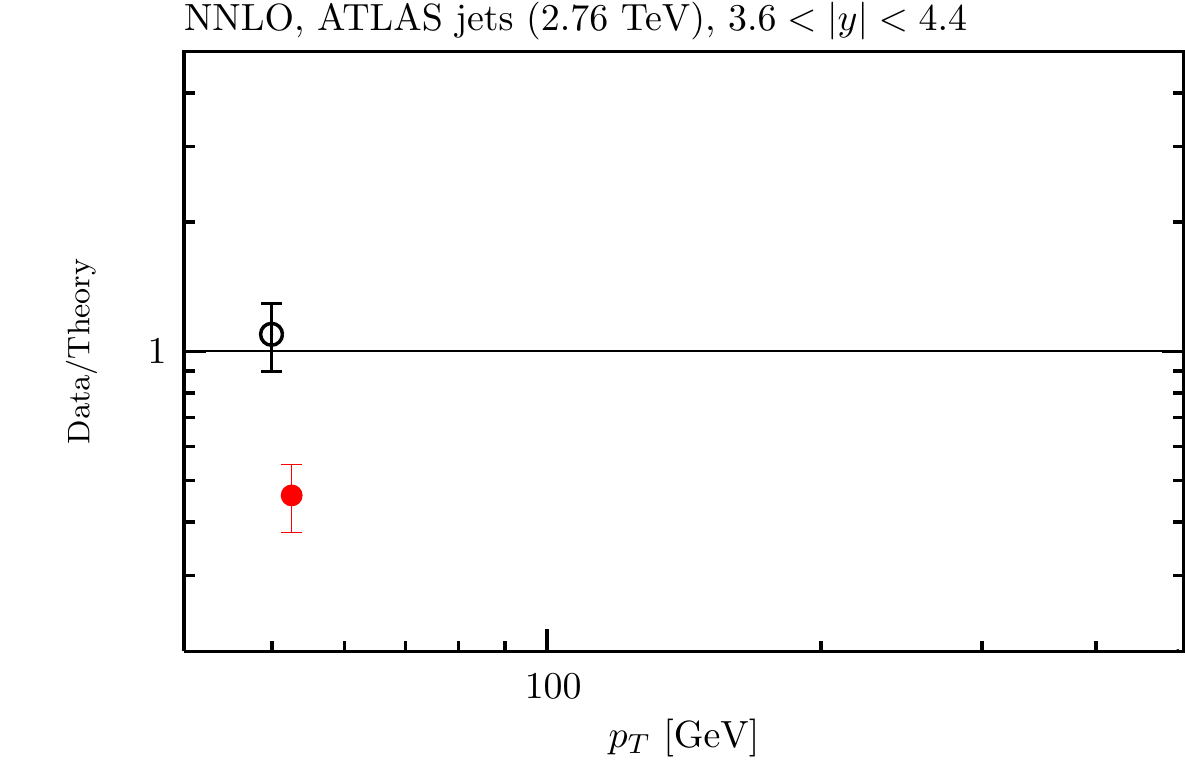}
\caption{\sf The fit quality for the ATLAS $2.76~\TeV$ jet data 
\cite{ATLAS-jet2.76} at NNLO, using the `larger' K-factor described in the text. The red points
represent the ratio of measured data to theory predictions, and the black 
points (clustering around Data/Theory=1) correspond to this ratio once
the best fit has been obtained by 
shifting theory predictions relative to data by using the correlated 
systematics.}
\label{fig:ATLASjet276NNLO}
\end{center}
\end{figure}

\begin{figure} 
\begin{center}
\includegraphics[height=5cm]{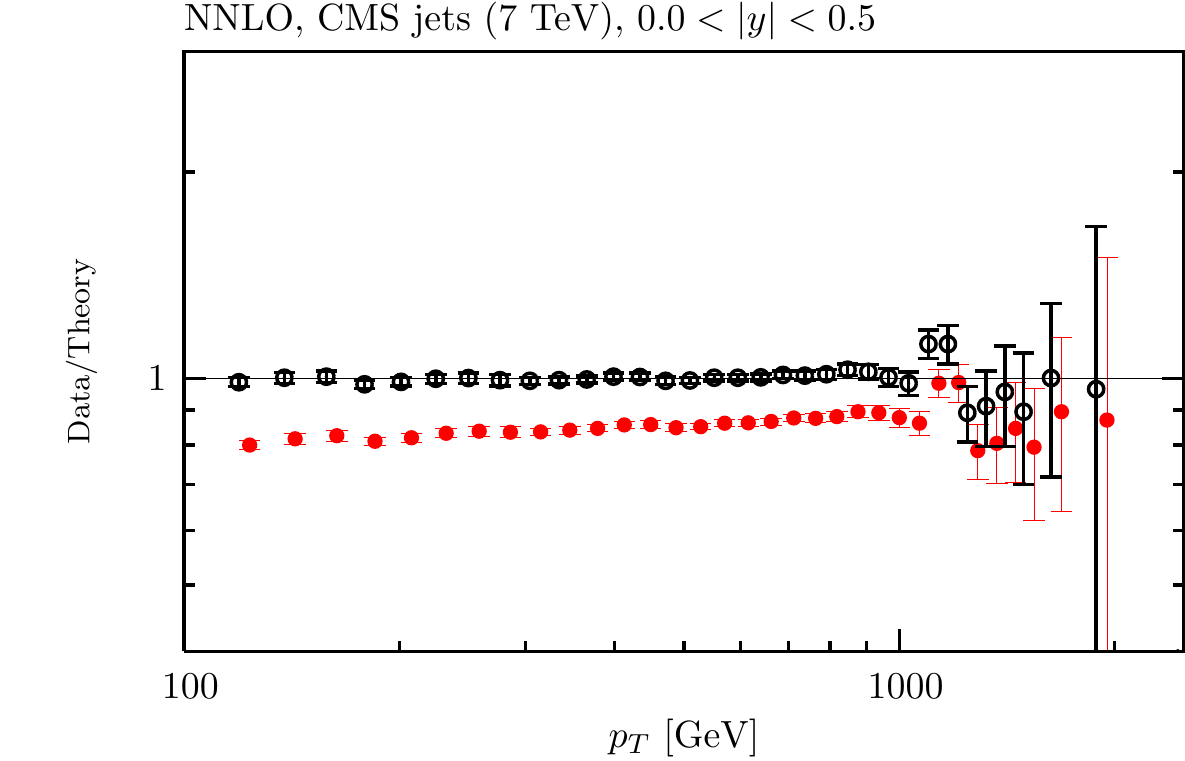}
\includegraphics[height=5cm]{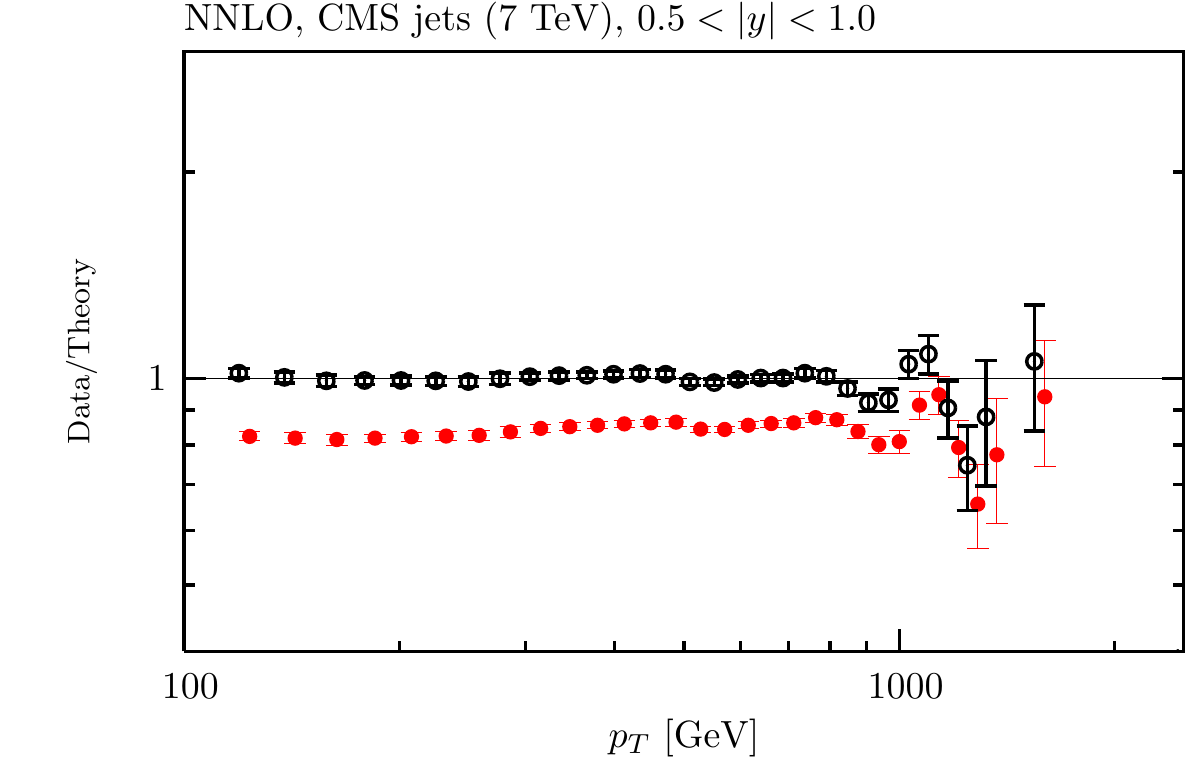}
\includegraphics[height=5cm]{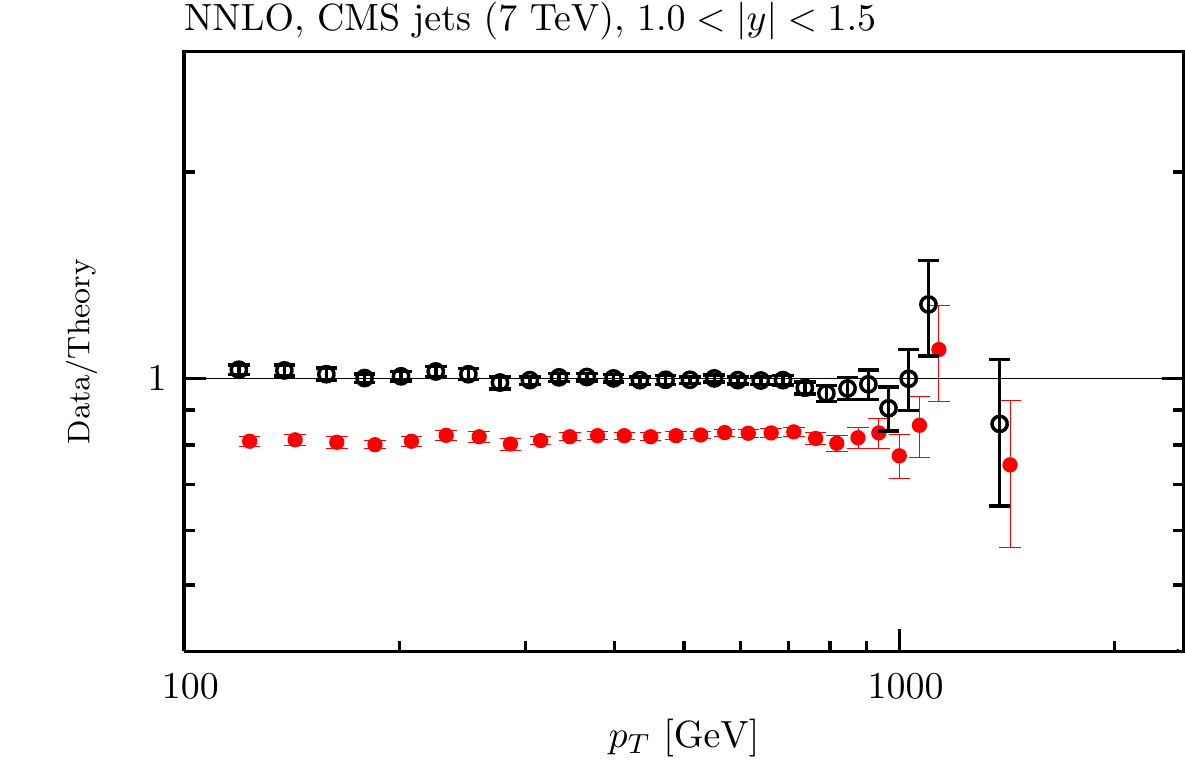}
\includegraphics[height=5cm]{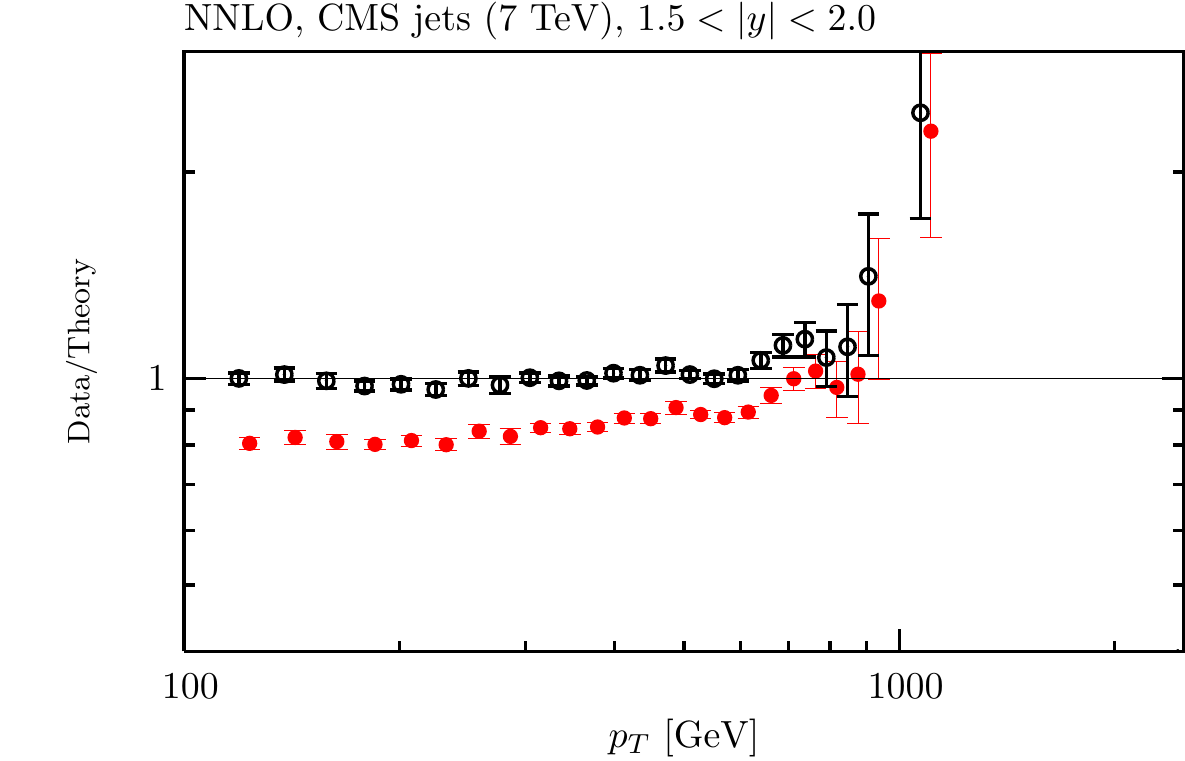}
\includegraphics[height=5cm]{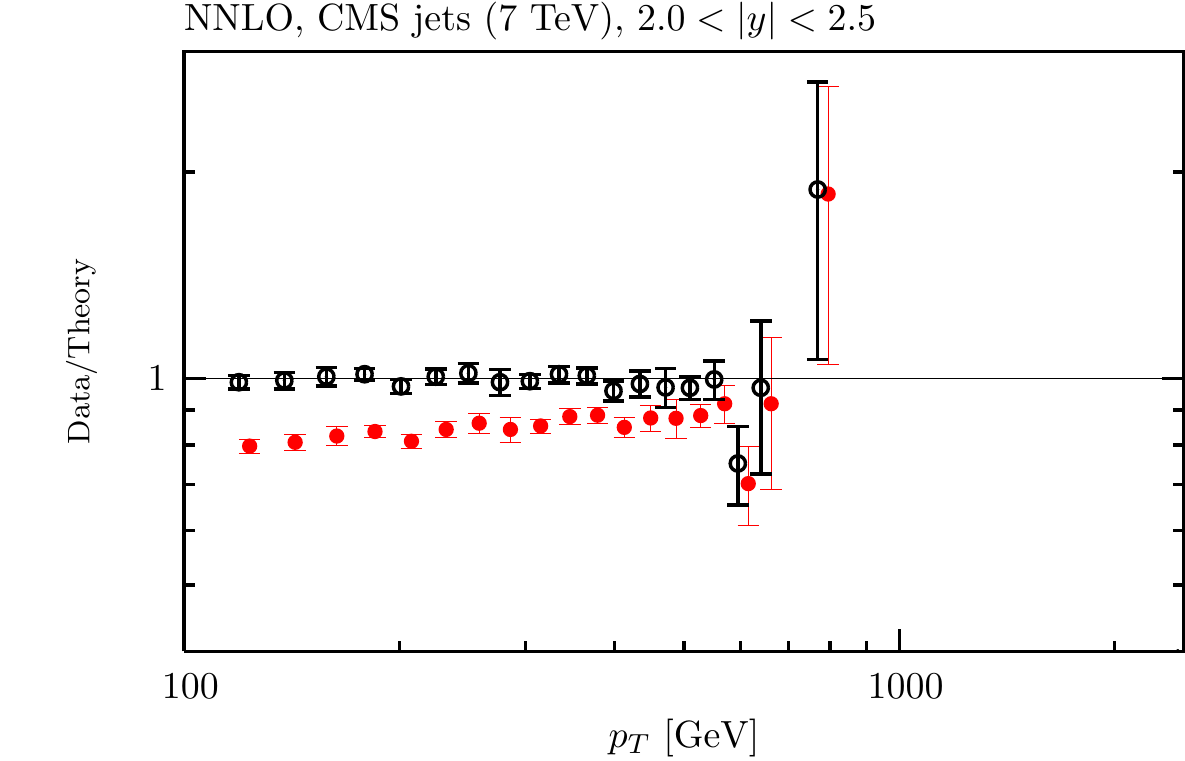}
\caption{\sf The fit quality for the CMS $7~\TeV$ jet data 
\cite{CMS-jet7} at NNLO, using the `larger' K-factor described in the text. The red points
represent the ratio of measured data to theory predictions, and the black 
points (clustering around Data/Theory=1) correspond to this ratio once
the best fit has been obtained by 
shifting theory predictions relative to data by using the correlated 
systematics.}
\label{fig:CMSjet7NNLO}
\end{center}
\end{figure}

Despite leaving the LHC jet data out of the PDF determination  at 
NNLO we have explored the effect of including very approximate NNLO 
corrections to the LHC data based on the threshold corrections and the 
known exact calculations so far available. To do this, we applied a  
$5-20\%$ positive correction, growing at the lower $p_T$ values, 
that is similar to the shape of the NNLO/NLO corrections in 
Figures 2 and 3 of \cite{GGGP2}. 
In detail we have used
\bea 
K_{\rm NNLO/NLO} &=& (1+ k(9.2 - 0.5\ln(p_T^2))/9.2), \qquad \qquad \qquad {\rm CMS},\\
K_{\rm NNLO/NLO} &=& (1+ k(8.0 - 0.5\ln(p_T^2))/8.0),~~~~~~ \qquad \qquad  {\rm ATLAS ~7~TeV},\\
K_{\rm NNLO/NLO} &=& (1+ k(8.0 - 0.5\ln((7/2.76)^2p_T^2))/8.0), \qquad  {\rm ATLAS ~2.76~TeV}.
\eea
We tried two alternatives, a `smaller' and `larger' $K$-factor,
i.e. $k=0.2$ and $k=0.4$,
with corrections of about 10$\%$ and 20$\%$ at $p_T=100$ GeV, independent 
of rapidity. The quality of the comparison to the data is shown 
in Table \ref{tab:LHCjet} using both the smaller and 
larger $K$-factors. The numbers in brackets 
represent predictions rather than a new fit. Clearly for both MMSTWW and MMHT
PDFs the quality of the prediction for the CMS data is similar to 
that for predictions, and the best fit, at NLO, using either choice of 
$K$-factor. For the ATLAS data the prediction using MMSTWW PDFs is also
similar to the best NLO results with the smaller $K$-factor, but deteriorates
a little with the larger $K$-factor. The predictions using MMHT are slightly
worse, and again there is more deterioration with increasing $K$-factor. 
The greater deterioration for ATLAS data seems to be due to the fact that 
while the fit to data is not changed much by $K$-factors of $10-20\%$ 
at NNLO, the ATLAS data are sensitive to the relative change of the theoretical
calculation between the two energies, which is rather difficult to 
approximate/guess accurately. Even so, in this case the comparison to data 
is still quite good, even with the larger $K$-factors. 
The fit quality for the LHC jet data is shown at NNLO, using the 
larger $K$-factor,
in Figs.~\ref{fig:ATLASjet7NNLO}, \ref{fig:ATLASjet276NNLO} and 
\ref{fig:CMSjet7NNLO}. One can see that the shape of data relative to 
theory remains very good, but the discrepancy before correlated uncertainties 
are applied is now larger in magnitude. This seems to cause little problem
for the fit quality for CMS data, but the fact that the relative size of the 
mismatch between ``raw'' theory and data is different for the two energies
for the ATLAS measurement leads to some limited deterioration in the fit 
quality.  

We have also tried the experiment of including the CMS and ATLAS jet data 
into the MMHT2014 fit with each of the $K$-factors. The quality is then 
shown by the unbracketed numbers in the right-hand column of Table 
\ref{tab:LHCjet}. The fit quality to the jet data improves slightly, mainly 
for ATLAS data, though it is still slightly worse than for the NLO fit. The 
PDFs and $\alpha_S(M_Z^2)$ change extremely little when the LHC jet data are 
included in the NNLO fit (discussed a little more later), and the fit quality
to the other data increases by at worst a couple of units in 
$\chi^2$. \footnote{We note, however, that the stability of the fit quality to 
CMS jet 
data with inclusion of NNLO $K$-factors was less apparent before the 
improved treatment of systematics advocated in   
\cite{CMS-jeta} was incorporated, and a fit with the data included did tend to 
lower $\alpha_S(M_Z^2)$ slightly.}

\subsubsection{Jet data in the LO fit}

In the LO fit, where the cross section is calculated 
at order ${\cal O}(\alpha_S^2)$, 
the jet data are all included. The fit quality to both 
LHC and Tevatron data is worse than at NLO, but only with an increase in 
$\chi^2$ of $10-20\%$, except for ATLAS data where we obtain 
$\chi^2/N_{\rm pts}=162/116$. The fit does normalise the Tevatron data downwards
quite significantly, but this is not so apparent for the LHC data, partially
due to the much smaller normalisation uncertainties at the LHC.

\section{Results for the global analysis \label{sec:5}}

The previous section shows the quality of the description of the LHC data 
before and after they are included in both the NLO and the NNLO global fit.  
In this section we discuss the overall fit quality and the resulting parton 
distributions functions. We also compare the results with the MSTW 2008 PDFs.

\begin{table}
  \centering
{\footnotesize
  \begin{tabular}{l|c|c|c}
    \hline \hline
    Data set & LO & NLO & NNLO \\ \hline
    BCDMS $\mu p$ $F_2$ \cite{Benvenuti:1989rh}         & 162 / 153 & 176 / 163 & 173 / 163 \\ 
    BCDMS $\mu d$ $F_2$ \cite{Benvenuti:1989fm}         & 140 / 142 & 143 / 151 & 143 / 151 \\ 
    NMC $\mu p$ $F_2$ \cite{Arneodo:1996qe}             & 141 / 115 & 132 / 123 & 123 / 123 \\ 
    NMC $\mu d$ $F_2$ \cite{Arneodo:1996qe}             & 134 / 115 & 115 / 123 & 108 / 123 \\ 
    NMC $\mu n/\mu p$ \cite{Arneodo:1996kd}             & 122 / 137 & 131 / 148 & 127 / 148 \\ 
    E665 $\mu p$ $F_2$ \cite{Adams:1996gu}              & 59 / 53   & 60 / 53   & 65 / 53 \\ 
    E665 $\mu d$ $F_2$ \cite{Adams:1996gu}              & 52 / 53   & 52 / 53   & 60 / 53 \\ 
    SLAC $ep$ $F_2$ \cite{Whitlow:1991uw,SLAC1990}      & 21 / 18   & 31 / 37   & 31 / 37 \\ 
    SLAC $ed$ $F_2$ \cite{Whitlow:1991uw,SLAC1990}      & 13 / 18   & 30 / 38   & 26 / 38 \\ 
    NMC/BCDMS/SLAC/HERA $F_L$ \cite{Arneodo:1996qe,Benvenuti:1989rh,SLAC1990,H1FL,H1-FL,ZEUS-FL} & 113 / 53 & 68 / 57 & 63 / 57 \\ \hline 
    E866/NuSea $pp$ DY \cite{E866DY}                    & 229 / 184 & 221 / 184 & 227 / 184 \\ 
    E866/NuSea $pd/pp$ DY \cite{E866DYrat}              & 29 / 15   & 11 / 15   & 11 / 15   \\ \hline 
    NuTeV $\nu N$ $F_2$ \cite{NuTeV}            & 35 / 49   & 39 / 53   & 38 / 53 \\ 
    CHORUS $\nu N$ $F_2$ \cite{CHORUS}          & 25 / 37   & 26 / 42   & 28 / 42 \\ 
    NuTeV $\nu N$ $xF_3$ \cite{NuTeV}           & 49 / 42   & 37 / 42   & 31 / 42 \\ 
    CHORUS $\nu N$ $xF_3$ \cite{CHORUS}         & 35 / 28   & 22 / 28   & 19 / 28 \\ 
    CCFR $\nu N\to \mu\mu X$ \cite{Dimuon}              & 65 / 86   & 71 / 86   & 76 / 86 \\ 
    NuTeV $\nu N\to \mu\mu X$ \cite{Dimuon}             & 53 / 40   & 38 / 40   & 43 / 40 \\ \hline 
    HERA $e^+p$ NC  820~GeV\cite{H1+ZEUS}               & 125 / 78  & 93 / 78   & 89 / 78 \\ 
    HERA $e^+p$ NC 920~GeV\cite{H1+ZEUS}                & 479 /330  & 402 /330  & 373/ 330 \\ 
    HERA $e^-p$ NC 920~GeV \cite{H1+ZEUS}               & 158/ 145  & 129/ 145  & 125 /145 \\ 
    HERA $e^+p$ CC \cite{H1+ZEUS}                       & 41 / 34   & 34 / 34   & 32 / 34 \\ 
    HERA $e^-p$ CC \cite{H1+ZEUS}                       & 29 / 34   & 23 / 34   & 21 / 34 \\ 
    HERA $ep$ $F_2^{\rm charm}$ \cite{H1+ZEUScharm}      & 105 /52   & 72 / 52   & 82 / 52 \\ 
    H1 99--00 $e^+p$ incl.~jets \cite{Aktas:2007aa}     & 77 / 24   & 14 / 24   & --- \\ 
    ZEUS incl.~jets \cite{Chekanov:2006xr,Chekanov:2002be}& 140/60  & 45 / 60   & ---  \\ 
    D{\O} II $p\bar{p}$ incl.~jets \cite{D0jet}         & 125 / 110 & 116 / 110 & 119 / 110 \\ 
    CDF II $p\bar{p}$ incl.~jets \cite{CDFjet}          & 78 / 76   & 63 / 76   & 59 / 76 \\ 
    CDF II $W$ asym. \cite{CDF-Wasym}                   & 55 / 13   & 32 / 13   & 30 / 13 \\ 
    D{\O} II $W\to \nu e$ asym. \cite{D0-easym0.75}       & 47 / 12   & 28 / 12   & 27 / 12 \\ 
    D{\O} II $W\to \nu \mu$ asym. \cite{D0-muasym7.3}     & 16 / 10   & 19 / 10   & 21 / 10 \\ 
    D{\O} II $Z$ rap. \cite{D0Zrap}                     & 34 / 28   & 16 / 28   & 16 / 28 \\ 
    CDF II $Z$ rap. \cite{CDF-Zrap}                      & 95 / 28   & 36 / 28   & 40 / 28 \\ \hline 
\hline
    ATLAS {$W^+, W^-, Z$}  \cite{ATLAS-WZ}              & 94/30     & 38/30     & 39/30  \\  
    CMS {$W$} asymm {$p_T >35~\GeV$} \cite{CMS-easym}   & 10/11     & 7/11      & 9/11   \\  
    CMS asymm {$p_T >25~\GeV,30~\GeV$}\cite{CMS-Wasym}  & 7/24      & 8/24      & 10/24   \\  
    LHCb {$Z\to e^+e^-$}\cite{LHCb-Zee}                 & 76/9      & 13/9      & 20/9   \\  
    LHCb {$W$} asymm {$p_T >20~\GeV$}\cite{LHCb-WZ}     & 27/10     & 12/10     & 16/10  \\  
    CMS  {$Z\to e^+e^-$} \cite{CMS-Zee}                 & 46/35     & 19/35     & 22/35  \\  
    ATLAS high-mass Drell-Yan \cite{ATLAShighmass}      & 42/13     & 21/13     & 17/13  \\  
    CMS double diff. Drell-Yan \cite{CMS-ddDY}          & ---       & 372/132   & 149/132 \\ 
    Tevatron, ATLAS, CMS ~~ {$\sigma_{t\bar t}$} \cite{Tevatron-top}--\cite{ATLAS-top7(6)} & 53/13 & 7/13 & 8/13  \\ 
    ATLAS jets ({2.76~TeV+7~TeV})\cite{ATLAS-jet2.76,ATLAS-jet7}& 162/116& 106/116 & ---\\ 
    CMS jets ({7~TeV}) \cite{CMS-jet7}                  & 150/133& 138/133 & ---  \\ 
    All data sets & \textbf{3706 / 2763} & \textbf{3267 / 2996} & \textbf{2717 / 2663} \\
    \hline \hline
  \end{tabular}
}
\vspace{-0.2cm}
\caption{\sf The values of $\chi^2 / N_{\rm pts.}$ for the data sets included in the global fit.  For the NuTeV $\nu N\to \mu\mu X$ data, the number of degrees of freedom is quoted instead of $N_{\rm pts.}$ since smearing effects mean nearby points are highly correlated.  The details of corrections to data, kinematic cuts applied and definitions of $\chi^2$ are contained in the text.}
\label{tab:chisquared}
\end{table}

The parameterisation  of the input PDFs is as discussed in Section \ref{sec:inputPDF}, and 
we now treat the coefficients of the first two Chebyshev polynomials for the 
$s_{+}$ distribution as free, unlike the case before inclusion of LHC data. 
At LO we make some changes to the parameterisation to stop the PDFs behaving 
peculiarly in regions where they are not directly constrained -- there is a
tendency for a large negative contribution in a very limited region of $x$
which would provide a negative contribution to the momentum sum rule, and
for $s_{+}$ to become extremely large at very small $x$. 
Hence, we only allow the first Chebyshev polynomial for $s_{+}$ to be free at LO
and parametrise the gluon with four free Chebyshev polynomials, but no second
term. This means that both $s_{+}$ and the gluon have one fewer free parameter at 
LO than at NLO or NNLO.

\subsection{The values of the QCD coupling, $\alpha_S(M^2_Z)$}
  \begin{figure} [t]
\begin{center}
\vspace*{-1.0cm}
\includegraphics[height=9cm]{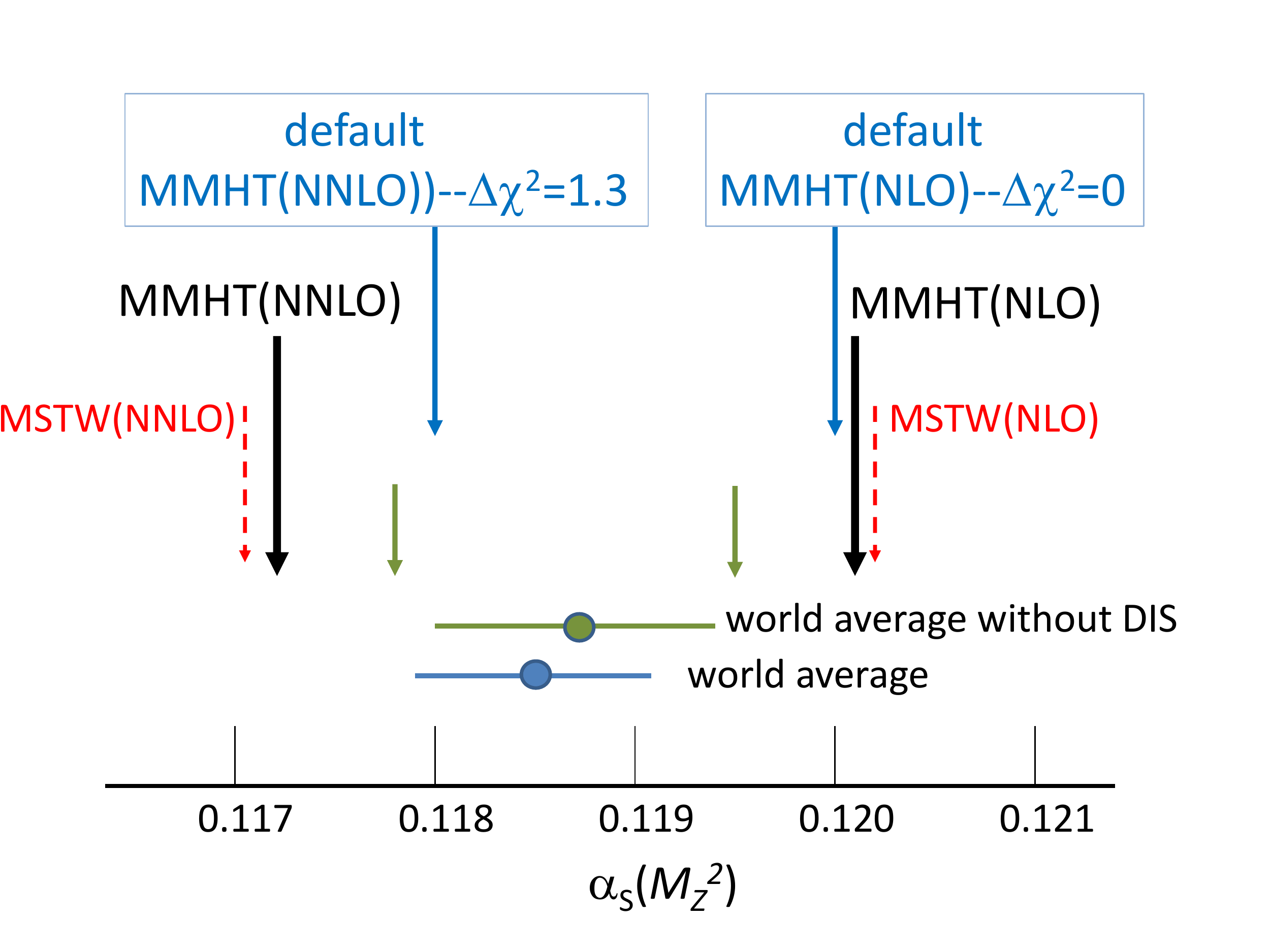} 
\caption{\sf The dark arrows indicate the optimal values of $\alpha_S(M_Z^2)$ 
found in NLO and NNLO fits of the present analysis (MMHT2014). The dashed 
arrows are the values found in the MSTW2008 analysis \cite{MSTW}. These are 
compared to the world average value, which was obtained assuming, for 
simplicity, that the NLO and NNLO values are the same -- which, in principle, 
is not the case.  The short arrows indicate the NLO and NNLO values obtained 
from the present global analyses if the world average value (obtained 
without including DIS data) were to be included in the fit. However, the default 
values $\alpha_{S,{\rm NLO}}=0.120$ and $\alpha_{S,{\rm NNLO}}=0.118$ are used for the final MMHT2014 PDF sets presented here; the values of $\Delta \chi^2$ are the changes in $\chi^2_{\rm global}$ in going from the optimal to the default fit. }
\label{fig:alpha}
\end{center}
\end{figure}
  
At both NLO and 
at NNLO the value of $\alpha_S(M_Z^2)$ is allowed to vary as a free parameter in the fit. At NLO
the best value of the QCD coupling is found to be
\be
\alpha_{S,{\rm NLO}}(M^2_Z)=0.1201.
\ee 
This is extremely similar to the value of $0.1202$ found in \cite{MSTW}.
At NNLO
the best value of the QCD coupling is found to be
\be
\alpha_{S,{\rm NNLO}}(M^2_Z)=0.1172,
\ee  
again very similar to that of $0.1171$ in \cite{MSTW} -- to be precise only 0.00015 larger. The 
difference between the NLO and NNLO values has decreased slightly. At LO it is difficult to 
define an absolute best fit, but the preferred value of $\alpha_S(M_Z^2)$ 
is certainly in the vicinity of 0.135, so we fix it at this value. 

It is a matter of considerable debate as to whether one should attempt to extract
the value of $\alpha_S(M_Z^2)$ from PDF fits or simply use it as in input 
with the value taken from elsewhere -- for example, simply to use the world average value \cite{PDG2014}. 
We believe that useful information on the coupling can be obtained from PDF fits, and 
as our extracted values of $\alpha_S(M_Z^2)$ at NLO and NNLO are quite close to 
the world average of $\alpha_S(M_Z^2)=0.1185\pm 0.0006$ we regard these as our best 
fits. We will discuss the variation with $\alpha_S(M_Z^2)$ and the uncertainty in  
a PDF fit determination in a future publication. However, we elaborate slightly here. 

As well as leaving $\alpha_S(M^2_Z)$ as a completely independent parameter, we also
include the world average value (without the inclusion of DIS data to avoid double 
counting) of $\alpha_S(M_Z^2)=0.1187\pm 0.0007$ as a data point in our fit. 
This changes the preferred values to 
\be
\alpha_{S,{\rm NLO}}(M^2_Z)=0.1195 \quad {\rm and} \quad \alpha_{S,{\rm NNLO}}(M^2_Z)=0.1178.
\ee 
Each of these is about one standard deviation away from the world average, 
so our PDF fit is entirely consistent with the independent determinations of 
the coupling. Moreover, the quality of the fit to the data other than the single 
point on $\alpha_S(M_Z^2)$ increases by about 1.5 units at NLO and just 
over one unit at NNLO when the coupling value is added as a data point.
It is ideal to present PDF sets at common, and hence round values of 
$\alpha_S(M_Z^2)$ in order to compare with, and combine with, other PDF sets, for 
example as in \cite{PDF4LHC1, PDF4LHC2, bench1, bench2}. At NLO we hence
choose $\alpha_S(M_Z^2)=0.120$ as the default value, which is essentially identical 
to the value for the best PDF fit when the coupling is free, and still very similar
when the world average is included as a constraint. At NNLO, when 
$\alpha_S(M_Z^2)=0.118$ is chosen, the fit quality is still 
only 1.3 units in $\chi^2$ higher than that when the coupling is free. This value is extremely close to the 
value determined when the world average is included as a data point. Hence, we choose
to use $\alpha_S(M_Z^2)=0.118$ as the default for our NNLO PDFs, a value which is very consistent with the world average. The summary of this discussion 
is shown above in Fig. \ref{fig:alpha}. At NLO we also make a set available with 
$\alpha_S(M_Z^2)=0.118$, but in this case the $\chi^2$ increases by 17.5 units from 
the best fit value.

\subsection{The fit quality}  
   
\begin{table}
  \centering
{\footnotesize
  \begin{tabular}{c|lll}
    \hline \hline
    Parameter        & LO         & NLO        & NNLO \\\hline
    $\alpha_S(M_Z^2)$& $0.135$    & $0.120$    & $0.118$ \\ \hline
    $A_u$            & $1.3358$   & $4.2723$   & $3.8539$ \\
    $\delta_u$         & $0.34430$  & $0.74687$  & $0.70900$ \\
    $\eta_u$         & $2.2318$   & $2.7421$   & $2.8773$ \\
    $a_{u,1}$        & $-0.26767$ & $0.26349$  & $0.80527$ \\
    $a_{u,2}$        & $-0.51620$ & $-0.00256$ & $-0.19419$ \\
    $a_{u,3}$        & $0.47167$  & $0.25858$  & $0.27225$ \\
    $a_{u,4}$        & $-0.12224$ & $0.05000$  & $-0.01211$ \\ \hline
    $A_d$            & $3.6009$   & $3.3002$   & $7.5602$ \\
    $\delta_d$         & $0.25049$  & $0.90012$  & $1.1147$ \\
    $\eta_d-\eta_u$  & $2.3847$   & $-0.58802$ & $-0.25180$ \\
    $a_{d,1}$        & $-1.3817$  & $1.2898$   & $1.2663$ \\
    $a_{d,2}$        & $0.49690$  & $0.60385$  & $0.78475$ \\
    $a_{d,3}$        & $-0.040740$& $0.33590$  & $0.32372$ \\
    $a_{d,4}$        & $-0.03926$ & $0.26150$  & $0.25099$ \\ \hline
    $A_S$            & $18.597$   & $31.329$   & $43.726$ \\
    $\delta_S$       & $-0.09018$ & $-0.13358$ & $-0.03946$ \\
    $\eta_S$         & $10.922$   & $11.945$   & $12.776$ \\
    $a_{S,1}$        & $-1.5611$  & $-1.6020$  & $-1.5979$ \\
    $a_{S,2}$        & $0.85903$  & $0.86538$  & $0.87445$ \\ 
    $a_{S,3}$        & $-0.30427$ & $-0.29923$ & $-0.30196$ \\
    $a_{S,4}$        & $0.07061$  & $0.06022$  & $0.006227$ \\ \hline
    $\int_0^1\!\dif{x}\;\Delta(x,Q_0^2)$ & $0.15782$ & $0.09531$ & $0.081983$ \\
    $A_\Delta$       & $0.29972$  & $7.1043$   & $25.408$ \\
    $\delta_\Delta$    & $0.60594$  & $1.7116$   & $2.1602$ \\
    $\gamma_\Delta$  & $13.029$   & $10.659$   & $8.1584$ \\
    $\epsilon_\Delta$  & $46.611$   & $-33.341$  & $-36.418$ \\ \hline
    $A_g$            & $17.217$   & $0.88746$  & $0.53411$ \\
    $\delta_g$       & $-0.33293$ & $-0.45853$ & $-0.56889$ \\
    $\eta_g$         & $5.3687$   & $2.8636$   & $1.3022$ \\
    $a_{g,1}$        & $-1.664$   & $-0.36317$ & $0.56995$ \\
    $a_{g,2}$        & $0.99169$  & $0.20961$  & $0.37592$ \\
    $a_{g,3}$        & $-0.42245$ & -----      & ----- \\
    $a_{g,4}$        & $0.10176$  & -----      & ----- \\
    $A_{g^\prime}$   & ---        & $-1.0187$  & $-0.09827$ \\
    $\delta_{g^\prime}$& ---      & $-0.42510$ & $-0.57405$ \\
    $\eta_{g^\prime}$& ---        & $32.614$   & $22.417$ \\ \hline
    $A_+$            & $2.2447$   & $4.6779$   & $8.2868$ \\
    $\eta_+$         & $14.055$   & $11.588$   & $13.752$ \\
    $a_{+,1}$        & $-1.5090$  & $-1.5910$  & $-1.5958$ \\
    $a_{+,2}$        & -----      & $0.86501$  & $0.88792$ \\ \hline
    $A_-$            & $-0.53737$ & $-0.01614$ & $-0.011373$ \\
    $\eta_-$         & $14.402$   & $7.1599$   & $6.4376$ \\
    $\delta_-$       & $0.91595$  & $-0.26403$ & $-0.26403$ \\
    $x_0$            & $0.056131$ & $0.026495$ & $0.028993$ \\ \hline
    \hline \hline
  \end{tabular}
}
  \caption{\sf The optimal values of the input PDF parameters (as defined in Section \ref{sec:inputPDF}) at $Q_0^2 = 1$ GeV$^2$ determined from the global 
analyses.  $A_u$, $A_d$, $A_g$ and $x_0$ are determined from sum rules and are not fitted parameters.  Similarly, 
$A_\Delta$ is determined from $\int_0^1\!\dif{x}\;\Delta(x,Q_0^2)$.}
  \label{tab:parameters}
\end{table}

The quality of the best fit is shown at LO, NLO and NNLO in Table 
\ref{tab:chisquared}. Note that at NNLO the values are for the absolute best fit with 
$\alpha_S(M_Z^2)=0.1172$, though the values are generally extremely similar when 
$\alpha_S(M_Z^2)=0.118$ and the total is $2718.6$ rather than $2717.3$. It has already 
been noted that both at NLO and NNLO (with the exception of the CMS double-differential data at NLO)
the fit quality is excellent. In most cases there is little improvement in the quality of the fit
from the inclusion of the LHC data (the ATLAS $W,Z$ and CMS asymmetry data being minor exceptions). 
It is clear that the inclusion of the LHC data has not spoilt the fit to any of the non-LHC data
in any way at all. The fit quality is very similar to that in \cite{MMSTWW} for the data sets that are
common to both fits, with some small differences being attributable to the changes in procedure 
applied in this study, as outlined in, for example, Sections 2.6 and 2.7. The fit quality for non-LHC data 
is within a handful of  chisquared units of the fit when only non-LHC data were included. In fact, in some
cases the two extra free parameters in the total strange distribution in the fit including 
LHC data leads to an improvement in non-LHC data, despite the extra constraint from new data. 
For example, at NNLO $\chi^2/N_{\rm pts}=637.7/621$ for the HERA combined structure
function data in the full fit compared to $\chi^2/N_{\rm pts}=644.2/621$ in the non-LHC fit 
(at NLO the non-LHC fit gives $666.0/621$ compared to $678.8/621$ in the full fit).   
At NNLO the main deterioration, about 6 units, is in NuTeV structure function data, which
is in some tension with ATLAS $W,Z$ data. This is not an issue at NLO. 

Overall the quality of the NNLO fit is 247 units in $\chi^2$ lower when counted for the 
data which are included in both fits, though this is reduced to only 25 units when the CMS 
double differential Drell-Yan data are removed from the comparison. Some of the data sets
within the global fit have a lower $\chi^2$ at NLO than at NNLO. 
It would be surprising if the total $\chi^2$ were lower at NLO, but this is not impossible: 
even though one would expect NNLO
to be closer to the ``ideal'' theory prediction fluctuations in data could allow an apparently
better fit quality to a worse prediction. On the other hand, given that NLO and NNLO are in general not very 
different predictions for most quantities it is quite possible that the shape of the PDFs obtained by 
the best fit at NNLO results in a best fit where the improvement in fit quality to some data sets
is partially compensated by a slight deterioration in the fit to some other data sets. 
As already noted with the LHC data, the LO fit is sometimes very poor, in particular for the HERA
jet data where NLO corrections are large.

\subsection{Central PDF sets and Uncertainties} 

The parameters for the central PDF sets at LO, NLO and NNLO are shown in 
Table \ref{tab:parameters}. In order to describe the uncertainties 
on the PDFs we apply
the same procedure as in \cite{MSTW} (originally presented in 
\cite{Hessian}), i.e. we use the Hessian approach
with a dynamical tolerance, and hence obtain a set of PDF eigenvector
sets each corresponding to $68\%$ confidence level uncertainty and being 
orthogonal to each other.

\subsubsection{Procedure to determine PDF uncertainties}

In more detail, if we have input 
parameters $\{a_i^0\}=\{a_1^0,\ldots,a_n^0\}$.  then we write
\begin{equation} \label{eq:hessian}
  \Delta\chi^2_{\rm global} \equiv \chi^2_{\rm global} - \chi_{\rm min}^2 = \sum_{i,j=1}^n H_{ij}(a_i-a_i^0)(a_j-a_j^0),
\end{equation}
where the Hessian matrix $H$ has components
\begin{equation}
  H_{ij} = \left.\frac{1}{2}\frac{\partial^2\,\chi^2_{\rm global}}{\partial a_i\partial a_j}\right|_{\rm min}.
\end{equation}
The uncertainty on a quantity $F(\{a_i\})$ is then obtained from 
standard linear error propagation:
\begin{equation} \label{eq:heserror}
  \Delta F = T \sqrt{\sum_{i,j=1}^n\frac{\partial F}{\partial a_i}C_{ij}\frac{\partial F}{\partial a_j}},
\end{equation}
where $C\equiv H^{-1}$ is the covariance matrix, and $T = \sqrt{\Delta\chi^2_{\rm global}}$ is the ``tolerance'' for the required confidence interval,
usually defined to be $T=1$ for $68\%$ confidence level. 

It is very useful to diagonalise the covariance (or Hessian) matrix 
\cite{Hessian}, and work in terms of the eigenvectors.  The covariance matrix 
has a set of normalised {\it orthonormal} eigenvectors $v_k$ defined by
\begin{equation} \label{eq:eigeq}
  \sum_{j=1}^n C_{ij} v_{jk} = \lambda_k v_{ik},
\end{equation}
where $\lambda_k$ is the $k^{\rm th}$ eigenvalue and $v_{ik}$ is the $i^{\rm th}$ 
component of the $k^{\rm th}$ orthonormal eigenvector ($k = 1,\ldots,n$).  
The parameter displacements from the global minimum can be expanded 
in terms of rescaled eigenvectors $e_{ik}\equiv \sqrt{\lambda_k}v_{ik}$:
\begin{equation} \label{eq:component}
  \Delta a_i\equiv a_i - a_i^0 = \sum_k e_{ik} z_k,
\end{equation}
i.e. the $z_k$ are the coefficients when we express a change in parameters away from 
their best fit values in terms of the rescaled eigenvectors, and a change 
in parameters corresponding to $\Delta \chi^2_{\rm global}=1$ corresponds to $z_k=1$.
This results in the simplification
\begin{equation} \label{eq:hessiandiag}
  \chi^2_{\rm global} = \chi^2_{\rm min} + \sum_k z_k^2.
\end{equation}
Eigenvector PDF sets $S_k^\pm$ can then be produced with parameters given by
\begin{equation}
  a_i(S_k^\pm) = a_i^0 \pm t\,e_{ik},
\end{equation}
with $t$ adjusted to give the desired $T = \sqrt{\Delta\chi^2_{\rm global}}$.
In the limit that  Eq.~(\ref{eq:hessian}) is exact, i.e. there are no 
significant corrections to quadratic behaviour, $t\equiv T$. 
We limit our number of eigenvectors so that this is true to a reasonable 
approximation. This results in the PDF eigenvector sets being 
obtained by fixing some of the parameters at their best-fit values, otherwise
the large degree of correlation between some parameters would lead to
significant violations in $t\approx T$. 

\begin{table}
\begin{center}
\begin{tabular}{| c | c | c |l| c | c |l|}
\hline
eigen-- & $+$  & &most constraining & $-$  & & most constraining\\
vector  & $t$ &  $T$ & data set & $t$ & $T$ & data set \\
\hline
1  & 4.00 & 3.97 & HERA $e^+p$ NC 920~GeV      & 4.30  & 4.66 & HERA $e^+p$ NC  820~GeV \\
2  & 2.50 & 2.84 & HERA $e^+p$ NC 920~GeV      & 1.80  & 1.53 & NMC $\mu d$ $F_2$\\
3  & 3.80 & 4.00 & NMC.....HERA $F_L$   & 3.70  & 3.69 & NMC $\mu d$ $F_2$\\
4  & 4.05 & 4.00 & D{\O} II $W\to \nu e$ asym.   & 5.00  & 5.11 & D{\O} II $W\to \nu \mu$ asym. \\
5  & 3.40 & 3.35 & D{\O} II $W\to \nu \mu$ asym. & 4.20  & 4.45 & NuTeV $\nu N\to \mu\mu X$ \\
6  & 1.85 & 1.88 & NuTeV $\nu N\to \mu\mu X$   & 3.70  & 3.71 & D{\O} II $W\to \nu \mu$ asym. \\
7  & 1.55 & 1.67 & E866/NuSea $pd/pp$ DY       & 2.15  & 2.03 & E866/NuSea $pd/pp$ DY \\
8  & 2.75 & 2.64 & D{\O} II $W\to \nu \mu$ asym. & 1.90  & 2.01 & E866/NuSea $pd/pp$ DY\\
9  & 3.40 & 3.46 & E866/NuSea $pd/pp$ DY       & 3.80  & 3.78 & BCDMS $\mu p$ $F_2$ \\
10 & 3.15 & 3.47 & NuTeV $\nu N\to \mu\mu X$   & 2.40  & 2.13 & NuTeV $\nu N$ $F_2$ \\
11 & 3.80 & 3.86 & CDF II $W$ asym.            & 4.00  & 3.96 & E866/NuSea $pd/pp$ DY \\
12 & 3.70 & 3.53 & SLAC $ed$ $F_2$             & 3.60  & 3.81 & BCDMS $\mu p$ $F_2$\\
13 & 4.30 & 5.47 & HERA $e^+p$ NC  820~GeV     & 5.30  & 4.33 & NMC $\mu d$ $F_2$ \\
14 & 3.30 & 3.36 & D{\O} II $W\to \nu e$ asym.   & 2.80  & 3.42 & CMS {$W$} asym. {$p_T >35~\GeV$} \\
15 & 2.90 & 3.08 & NuTeV $\nu N$ $xF_3$        & 3.30  & 3.12 & E866/NuSea $pp$ DY \\
16 & 3.65 & 3.70 & CDF II $p\bar{p}$ incl.~jets& 2.65  & 2.64 & NuTeV $\nu N$ $xF_3$ \\
17 & 1.80 & 1.85 & E866/NuSea $pd/pp$ DY       & 2.40  & 2.16 & E866/NuSea $pd/pp$ DY \\
18 & 1.15 & 1.42 & CMS asym. {$p_T >25,30~\GeV$}  & 2.60  & 3.19 & BCDMS $\mu p$ $F_2$ \\
19 & 2.60 & 2.86 & CMS asym. {$p_T >25,30~\GeV$}  & 2.10  & 3.35 & D{\O} II $p\bar{p}$ incl.~jets \\
20 & 1.60 & 1.72 & CCFR $\nu N\to \mu\mu X$    & 1.55  & 1.45 & NuTeV $\nu N\to \mu\mu X$\\ 
21 & 2.80 & 3.45 & NuTeV $\nu N\to \mu\mu X$   & 3.30  & 3.47 & ATLAS {$W^+, W^-, Z$}  \\
22 & 4.70 & 6.48 & NuTeV $\nu N$ $xF_2$        & 4.00  & 3.67 & NuTeV $\nu N$ $xF_3$ \\
23 & 1.90 & 1.96 & NuTeV $\nu N\to \mu\mu X$   & 4.85  & 3.50 & CCFR $\nu N\to \mu\mu X$ \\
24 & 2.35 & 3.13 & HERA $e^+p$ NC 920~GeV      & 3.75  & 4.27 & HERA $e^+p$ NC 920~GeV \\
25 & 2.50 & 2.63 & E866/NuSea $pd/pp$ DY       & 1.30  & 2.15 & E866/NuSea $pd/pp$ DY \\
\hline
\end{tabular}
\end{center}
\caption{\sf Table of expected $\sqrt{\Delta \chi^2}=t$ and true $\sqrt{\Delta \chi^2}=T$ 
values for $68\%$ confidence level uncertainty for
each eigenvector and the most constraining data sets for the MMHT2014 NLO fits.}
\label{chi2eigNLO}
\end{table}

\begin{table}
  \centering
{\footnotesize
  \begin{tabular}{c|c|c|c|c|c|c|c}
    \hline \hline
    evector & $g$ & $u_v$ & $d_v$ & $S({\rm ea})$ & $\bar{d}-\bar{u}$ & $s+\bar{s}$ & $s-\bar{s}$ \\ \hline
    1  &  -- &  -- &  -- &0 0.3 0&  -- &  -- &  -- \\
    2  &  -- &  -- &  -- &0 0.4 0&  -- &  -- &  -- \\
    3  &4 0 0&  -- &  -- &  -- &  -- &  -- &  -- \\
    4  &2 0 0&0 0 2&  -- &  -- &  -- &  -- &  -- \\
    5  &1 0 0&  -- &  -- &1 0 0&  -- &  -- &1 0 0\\
    6  &  -- &  -- &  -- &  -- &  -- &  -- &2 1 2\\
    7  &  -- &  -- &  -- &  -- &0 2 2&  -- &  -- \\
    8  &  -- &  -- &0 0 2&  -- &0 1 2&  -- &  -- \\
    9  &  -- &1 2 3&  -- &  -- &0 1 2&  -- &  -- \\
    10 &  -- &  -- &  -- &2 1 0&  -- &2 3 1&  -- \\
    11 &  -- &0 1 2&2 3 4&  -- &0 1 1&  -- &  -- \\
    12 &  -- &4 3 5&1 2 2&0 1 0&  -- &  -- &  -- \\
    13 &8 5 2&1 1 1&0 0 1&1 1 0&  -- &  -- &  -- \\
    14 &  -- &  -- &2 3 7&  -- &  -- &  -- &  -- \\
    15 &1 2 2&1 1 2&2 1 2&0 0 1&1 1 0&  -- &  -- \\
    16 &0 1 5&1 2 2&0 1 2&0 3 3&1 2 0&  -- &  -- \\
    17 &  -- &  -- &  -- &0 0 1&2 3 4&  -- &  -- \\
    18 &  -- &4 4 0&0 1 0&  -- &  -- &  -- &  -- \\
    19 &  -- &  -- &2 3 2&  -- &  -- &  -- &  -- \\
    20 &  -- &  -- &  -- &0 0 1&1 0 0&0 0 6&1 0 0\\
    21 &0 0 1&1 2 0&2 1 2&4 4 4&0 1 0&5 6 6&4 3 3\\
    22 &1 2 0&1 0 1&2 2 2&4 2 4&0 0 1&2 1 2&1 0 0\\
    23 &  -- &0 1 0&0 0 1&1 0 3&1 0 0&1 2 2&2 8 10\\
    24 &0 5 6&  -- &0 1 1&0 1 0&0 0 1&  -- &  -- \\
    25 &  -- &  -- &  -- &  -- &7 4 9&  -- &  -- \\
    \hline \hline
  \end{tabular}
}
\caption{\sf The three numbers in each entry are the fractional 
contribution to the total uncertainty  
for the $g,u_v,\ldots$ input distributions in the small $x$ ($x<0.01$), 
medium $x$ ($0.01<x<0.1$) and large $x$ ($x>0.1$) regions, respectively, 
arising from eigenvector $k$ in the NLO global fit.  Each number has been 
multiplied by ten; for example, 4 denotes 0.4. For a precise value of $x$ the sum of 
each column should be 10. However, the entries shown are the 
maximum fraction in each interval of $x$, so often do not satisfy this 
condition. In general we do not show contributions below $5\%$, but for the 
first two eigenvectors at NLO no uncertainty contribution is this large, so we
show the largest contributions.}
\label{tab:fractionsNLO}
\end{table}

As in \cite{MSTW} we 
do not determine the size of the eigenvectors using the standard
$\Delta \chi^2=1$ or $T=1$ rule. Rather, we allow $T \ne 1$ to account, primarily, 
for the  
tensions in fitting the {\it different} data sets within fixed order perturbative 
QCD. Neither do we use a fixed value of $T$. Instead we use the ``dynamical 
tolerance'' procedure devised in \cite{MSTW}.
In brief, we define the 68\%  confidence level region for each data set $n$ (comprising $N$ data points) by the condition that
\begin{equation} \label{eq:68percentCL}
  \chi_n^2 < \left(\frac{\chi_{n,0}^2}{\xi_{50}}\right)\xi_{68},
\end{equation}
where $\xi_{68}$ is the 68th percentile of the $\chi^2$-distribution with 
$N$ degrees of freedom, and $\xi_{50}\simeq N$ is the most probable value.  
For each eigenvector (in each of the two directions) we then determine the 
values of $t$ and $T$ for which the $\chi_n^2$ for 
each data set $n$ are minimised, together with $68\%$ confidence level 
limits defined by values at which Eq.~(\ref{eq:68percentCL}) ceases
to be satisfied. For a perfect data set we would only need the value 
of $\xi_{68}$, but for a number of data sets $\chi_{n,0}^2$ is 
not very close to $\xi_{50}$ ($\xi_{50}\sim n_{\rm pts}$), being potentially both higher and lower, as 
seen in Table \ref{tab:chisquared}. For more details of the 
``dynamical tolerance'' procedure see Section 6.2 of \cite{MSTW}.

\subsubsection{Uncertainties of the MMHT2014 PDFs}

The increase in the parameterisation flexibility in the present MMHT analysis leads to an increase in the number of parameters left free in the determination of the PDF uncertainties, as compared to the MSTW2008 analysis.  Indeed, we now have 25 eigenvector pairs, 
rather than the 20 in \cite{MSTW} or even the 23 in \cite{MMSTWW}. 
The 25 parameters\footnote{The expressions for the input PDFs in terms of the parameters are given in Section \ref{sec:inputPDF}.} left free for the determination of the eigenvectors 
consist of: $\eta, \delta, a_2$ and $a_3$ for each of the valence quarks,
$A, \eta, \delta, a_2$ and $a_3$ for the light sea; 
$\int_0^1\!\dif{x}\;\Delta(x,Q_0^2), \eta$ and $\gamma$ for 
$\bar d - \bar u$;  
$\eta, \delta, \eta_-$ and $\delta_-$ for the gluon 
(or $\eta, \delta, a_2$ and $a_3$ at LO); $A,\eta$ and $a_2$ for 
$s_+$ (or $A,\eta$ and $a_1$ at LO); and $A$ and $\eta$ for $s_-$.     
During the determination of the eigenvectors all deuteron parameters, 
free coefficients for nuclear corrections and all parameters associated 
with correlated uncertainties, including normalisations, are allowed 
to vary (some with appropriate $\chi^2$ penalty). 

The most constraining data set for each eigenvector direction, and also 
the values of $t$ and $T$ are shown for the NLO fit in Table \ref{chi2eigNLO}.
The fractional contribution to the total uncertainty of each PDF is then 
also shown in summary in Table \ref{tab:fractionsNLO}. The same information 
is shown for the NNLO fit in Tables \ref{chi2eigNNLO} and \ref{tab:fractionsNNLO}.
One can see that for the vast majority of cases there is good agreement 
between $t$ and $T$ at both NLO and NNLO. Hence, within the region of 
$68\%$ uncertainty confidence levels for the PDFs, the $\chi^2$ distribution
is quite accurately a quadratic function of the parameters. There is, however,
a reasonable degree of asymmetry between the $t$ and $T$ values in the two 
directions for a single eigenvector, and it is nearly always the case that
it is a different data set which is the main constraint in the two directions. 
In fact, the data set which has the most rapid deterioration in fit quality
in one direction is often improving in fit quality until quite a 
high value of $t$ along the other direction. This is an indication of the 
tension between data sets, with nearly all eigenvectors having some data
sets which pull in opposite directions. The values of $t$ and $T$ for the
$68\%$ confidence levels are on average about $t\approx T \approx 3$, i.e. 
$\Delta \chi^2_{\rm global} \approx 10$, though $T^2$ does vary between about
1 unit and at most $T^2\approx 40$. 

We comment briefly on the manner in which the values of $t$ and $T$ arise for 
some illustrative cases. For a number of eigenvectors there is one 
data set which is overwhelmingly most constraining. Examples are eigenvectors 
17 and 25 at NLO and 7 and 25 at NNLO. A number of these are 
where the constraint is from the E866/NuSea Drell-Yan ratio data, since this 
is one of the few data sets sensitive to the $\bar d -\bar u$ difference. 
In these cases the tolerance tends to be low. For the cases where the 
tolerance is high there are some definite examples where this is due to 
tension between two data sets. One of the clearest and most interesting examples
is eigenvector 13 at NLO. In this case the fit to HERA $e^+p$ NC  820~GeV improves 
in one direction and deteriorates in the other, while the fit to NMC structure 
function data for $x < 0.1$ deteriorates in one direction and improves in the other. 
In this case the NMC data are at low $Q^2$ and the HERA data at higher $Q^2$ and the fit
does not match either perfectly simultaneously. The effect is smaller at NNLO though 
is evident in eigenvector 3. Other cases where $t$ is high and data sets are in very 
significant tension are eigenvector 4 at NLO, where D{\O} electron and muon asymmetry compete
and eigenvector 20 at NLO where CCFR and NuTeV dimuon data prefer a different  
high-$x$ strange quark. This complete tension is less evident in NNLO eigenvectors. 
However, there are some cases where one data set has deteriorating fit quality in one 
direction and improving quality in the other, while another data set deteriorates quickly
in one direction, but varies only slowly in the other. Examples of this are eigenvectors 1 
and 23 at NLO and eigenvector 1 at NNLO. Often the variation of $\chi^2$ of all data sets is
fairly slow except for one data set in one direction and a different data set in another 
direction. Examples of this are eigenvector 22 at NLO and eigenvectors 10, 22 and 24 at NNLO.
A final type of cases is similar, but where one data set deteriorates in both directions but 
one other deteriorates slightly more quickly in one direction but very slowly in the other. 
Examples are eigenvector 4 at NNLO, where BCDMS data deteriorates in both directions but SLAC 
only in one direction and 
eigenvector 21 at NNLO, where ATLAS $W,Z$ data deteriorates in both directions, but HERA data only 
in one direction.

We do not show the details of the
eigenvectors at LO since we regard this as a much more approximate fit.
However, we note that at LO the good agreement between $t$ and $T$ breaks 
down much more significantly, particularly for eigenvectors with the highest 
few eigenvalues. This is a feature of even more tension between data sets 
in the LO fit, and indeed, in the NLO and NNLO fit we would regard these 
eigenvectors as unstable, and discount them. However, we wish to obtain 
a conservative uncertainty on the PDFs at LO, so keep the same number of
eigenvectors as at NLO and NNLO.    
  
We see that there is some similarity between the eigenvectors for the
NLO and NNLO PDFs, with some, e.g. 1, 5, 7, 19, 20, being 
constrained by the same 
data set and corresponding to the same type of PDF uncertainty. In some cases 
the order of the eigenvectors (determined by size of eigenvalue) is simply
modified slightly by the changes between the NLO and NNLO fit e.g. 
3 at NLO and 2 at NNLO, 23 at NLO and 24 at NNLO. However, despite
the fact that the data fit at NNLO is very similar to that at NLO, and the 
parameterisation of the input PDFs is identical, the changes in the details of 
the NLO and NNLO fit are sufficient to remove any very clear mapping between 
the eigenvectors in the two cases, and some are completely different. We present
the details of the eigenvectors at NLO here for the best-fit value of 
$\alpha_S(M_Z^2) =0.120$. However, we also make available a NLO PDF set    
with $\alpha_S(M_Z^2) =0.118$ with both a central value and a full set of
eigenvectors (though the fit quality is 17 units worse for this value of 
$\alpha_S(M_Z^2)$). It is perhaps comforting to note that there is a practically
identical mapping between the NLO eigenvectors for the two values of 
$\alpha_S(M_Z^2)$, with the main features of PDF 
uncertainties being the same, without any modification of the order of the 
eigenvectors. The precise values of $t$ and $T$ are modified a little, and in a 
couple of cases the most constraining sets changed (always for one which 
was almost the most constraining set at the other coupling value). The 
uncertainties (defined by changes in $\chi^2$ relative to the best-fit values 
in each case) are very similar.

\subsubsection{Data sets which most constrain the MMHT2014 PDFs  \label{sec:5.3.3}}
\begin{table}
\begin{center}
\begin{tabular}{| c | c | c |l| c | c |l|}
\hline
eigen-- & $+$  & &most constraining & $-$  & & most constraining\\
vector  & $t$ &  $T$ & data set & $t$ & $T$ & data set \\
\hline
1  & 3.50 & 3.41 & HERA $e^+p$ NC 920~GeV      & 4.50  & 4.78 & HERA $e^+p$ NC  820~GeV \\
2  & 3.95 & 3.92 & NMC.....HERA $F_L$   & 3.95  & 4.03 & HERA $e^+p$ NC 920~GeV\\
3  & 3.85 & 4.10 & HERA $e^+p$ NC 920~GeV      & 1.55  & 1.37 & NMC $\mu d$ $F_2$\\
4  & 5.00 & 5.07 & BCDMS $\mu p$ $F_2$         & 5.00  & 4.99 & SLAC $ed$ $F_2$ \\
5  & 2.50 & 2.48 & D{\O} II $W\to \nu \mu$ asym. & 2.40  & 2.46 & NuTeV $\nu N\to \mu\mu X$ \\
6  & 5.30 & 5.47 & CCFR $\nu N\to \mu\mu X$    & 2.30  & 2.31 & NuTeV $\nu N\to \mu\mu X$ \\
7  & 1.40 & 1.46 & E866/NuSea $pd/pp$ DY       & 1.70  & 1.64 & E866/NuSea $pd/pp$ DY \\
8  & 2.50 & 2.60 & D{\O} II $W\to \nu \mu$ asym. & 2.70  & 2.61 & D{\O} II $W\to \nu e$ asym.\\
9  & 5.70 & 6.00 & HERA ep $F_2^{\rm charm}$   & 3.20  & 3.04 & CCFR $\nu N\to \mu\mu X$  \\
10 & 3.40 & 3.13 & E866/NuSea $pd/pp$ DY       & 4.60  & 4.67 & CDF II $W$ asym. \\
11 & 4.30 & 4.41 & E866/NuSea $pd/pp$ DY       & 3.00  & 2.92 & NuTeV $\nu N\to \mu\mu X$ \\
12 & 4.85 & 5.25 & HERA ep $F_2^{\rm charm}$   & 4.70  & 4.44 & BCDMS $\mu p$ $F_2$\\
13 & 1.85 & 2.14 & CMS asym. {$p_T >25,30~\GeV$}& 4.70  & 4.34 & NuTeV $\nu N$ $xF_3$ \\
14 & 2.85 & 3.01 & BCDMS $\mu d$ $F_2$         & 2.55  & 2.79 & CMS {$W$} asym. {$p_T >35~\GeV$} \\
15 & 1.20 & 0.95 & Tevatron, ATLAS, CMS ~$\sigma_{t\bar t}$& 3.30  & 3.72 & CDF II $p\bar{p}$ incl.~jets \\
16 & 1.75 & 2.01 & CMS asym. {$p_T >25,30~\GeV$}& 3.55  & 3.43 & BCDMS $\mu p$ $F_2$ \\
17 & 1.75 & 1.90 & CMS asym. {$p_T >25,30~\GeV$}& 3.30  & 3.12 & E866/NuSea $pd/pp$ DY \\
18 & 3.10 & 3.11 & BCDMS $\mu p$ $F_2$         & 1.40  & 1.87 & CMS asym. {$p_T >25,30~\GeV$}\\
19 & 1.80 & 1.84 & CMS asym. {$p_T >25,30~\GeV$}  & 2.55  & 3.26 & D{\O} II $p\bar{p}$ incl.~jets \\
20 & 2.00 & 2.20 & CCFR $\nu N\to \mu\mu X$    & 1.50  & 1.51 & NuTeV $\nu N\to \mu\mu X$\\ 
21 & 3.00 & 3.03 & ATLAS {$W^+, W^-, Z$}       & 4.70  & 5.49 & HERA $e^+p$ NC 920~GeV  \\
22 & 1.20 & 1.60 & E866/NuSea $pd/pp$ DY       & 6.90  & 5.31 & NMC $\mu n/\mu p$ \\
23 & 2.20 & 2.86 & HERA $e^+p$ NC 920~GeV      & 1.85  & 3.73 & HERA $e^+p$ NC 920~GeV \\
24 & 4.30 & 3.38 & CCFR $\nu N\to \mu\mu X$    & 1.75  & 1.86 & NuTeV $\nu N\to \mu\mu X$ \\
25 & 1.90 & 3.39 & HERA $e^+p$ NC 920~GeV      & 1.60  & 2.78 & HERA $e^+p$ NC 920~GeV \\
\hline
\end{tabular}
\end{center}
\caption{\sf Table of expected $\sqrt{\Delta \chi^2}=t$ and true $\sqrt{\Delta \chi^2}=T$ 
values for $68\%$ confidence level uncertainty for
each eigenvector and the most constraining data sets for the MMHT2014 NNLO fits.}
\label{chi2eigNNLO}
\end{table}

It is very clear from Tables \ref{chi2eigNLO} and \ref{chi2eigNNLO} that a wide
variety of different data types are responsible for constraining the PDFs.
At NLO 6 of the 50 eigenvector directions are constrained by HERA structure 
function data, 13 by fixed-target data structure function data, 
and 4 by the newest LHC data. Three 
of the LHC driven constraints are on the valence quarks and come from 
lepton asymmetry data. One is a constraint on the strange quark from 
the ATLAS $W$ and $Z$ data. There are still 9 constraints from Tevatron data, again
mainly on the details of the light quark decomposition. The CCFR and NuTeV dimuon
data \cite{Dimuon} constrain 8 eigenvector directions because they still 
provide by far the dominant constraint on the strange and antistrange quarks, 
which have 5 free parameters in the eigenvector determination. Similarly, the 
E866 Drell-Yan total cross section asymmetry data constrain 10 eigenvector
directions mainly because the asymmetry data are still by far the best 
constraint on $\bar d - \bar u$ which has 3 free parameters. 

At NNLO the picture
is quite similar, but now HERA data constrain 11 eigenvector directions. 
Fixed-target data are similar to NLO with 10, but the Tevatron reduces to 
6. The LHC data now constrain 8 eigenvector directions. As at NLO, this is 
dominantly lepton asymmetry data constraining valence quarks (winning out over 
Tevatron data compared to NLO in a couple of cases) but also ATLAS $W,Z$
data constrain the sea and strange sea in one eigenvector direction and 
$\sigma ({t \bar t})$ provide a constraint on the high-$x$ gluon. The dimuon 
and E866 Drell-Yan data provide similar constraints to NLO with 9 and 6 
respectively, though in the latter case it is always the asymmetry data which
contribute.

\begin{table} [t]
  \centering
{\footnotesize
  \begin{tabular}{c|c|c|c|c|c|c|c}
    \hline \hline
    evector & $g$ & $u_v$ & $d_v$ & $S({\rm ea})$ & $\bar{d}-\bar{u}$ & $s+\bar{s}$ & $s-\bar{s}$ \\ \hline
    1  &1 0 0&  -- &  -- &  -- &1 0 0&  -- &  -- \\
    2  &4 0 0&  -- &  -- &  -- &  -- &  -- &  -- \\
    3  &  -- &  -- &  -- &0 1 0&  -- &  -- &  -- \\
    4  &1 0 0&0 0 2&  -- &  -- &1 0 0&  -- &  -- \\
    5  &  -- &  -- &  -- &  -- &1 0 0&  -- &1 0 1\\
    6  &1 1 0&0 0 1&0 0 1&1 1 0& --  &  -- &2 1 2\\
    7  &  -- &  -- &  -- &  -- &1 2 2&  -- &  -- \\
    8  &  -- &  -- &0 0 3&  -- &  -- &  -- &1 1 1\\
    9  &2 2 0&1 1 1&0 0 1&0 1 1&  -- &1 2 1&1 0 1\\
    10 &  -- &1 1 2&0 1 1&1 1 1&0 3 3&1 2 1&  -- \\
    11 &  -- &  -- &1 1 2&1 1 1&0 1 1&1 2 2&1 1 1\\
    12 &4 3 2&0 1 3&1 2 2&0 3 1&1 1 1&  -- &  -- \\
    13 &1 1 1&5 4 4&1 1 1&0 1 0&1 0 0&  -- &  -- \\
    14 &  -- &  -- &2 2 6&  -- &  -- &  -- &  -- \\
    15 &1 2 4&1 1 1&1 1 1&  -- &1 0 0&  -- &  -- \\
    16 &0 0 2&2 2 1&0 1 1&0 2 2&1 1 1&  -- &  -- \\
    17 &  -- &2 1 0&  -- &  -- &2 3 4&  -- &  -- \\
    18 &0 0 1&3 3 1&0 1 1&0 0 10&0 0 10&  -- &  -- \\
    19 &  -- &  -- &5 4 2&  -- &  -- &  -- &  -- \\
    20 &  -- &  -- &  -- &0 0 1&  -- &0 0 5&1 0 1\\
    21 &0 0 2&1 2 1&2 2 2&3 3 5&0 0 2&4 6 6&3 3 3\\
    22 &  -- &0 1 1&0 0 1&0 0 1&8 6 9&  -- &  -- \\
    23 &1 2 5&  -- &  -- &1 1 1&  -- &1 2 0&  -- \\
    24 &0 0 1&  -- &0 0 1&0 0 1&1 0 0&0 1 1&2 10 10\\
    25 &1 2 2&  -- &  -- &1 0 0&1 0 0&  -- &  -- \\
    \hline \hline
  \end{tabular}
}
\caption{\sf The three numbers in each entry are the fractional 
contribution to the total uncertainty  
for the $g,u_v,\ldots$ input distributions in the small $x$ ($x<0.01$), 
medium $x$ ($0.01<x<0.1$) and large $x$ ($x>0.1$) regions, respectively, 
arising from eigenvector $k$ in the NNLO global fit.}
\label{tab:fractionsNNLO}
\end{table}

We do not make $90\%$ confidence-level eigenvectors directly available,
as was done in \cite{MSTW}, but simply advocate expansion of the 
$68\%$ confidence level uncertainties by the standard factor of 1.645.
This is true to a reasonably good approximation. There was not a very obvious 
demand for explicit $90\%$ confidence-level eigenvectors in the last release,
and some cases where the availability of two different sets of eigenvectors
led to mistakes and confusion.

\subsubsection{Availability of MMHT2014 PDFs   \label{sec:access}}
\begin{figure} 
\begin{center}
\includegraphics[height=7.5cm]{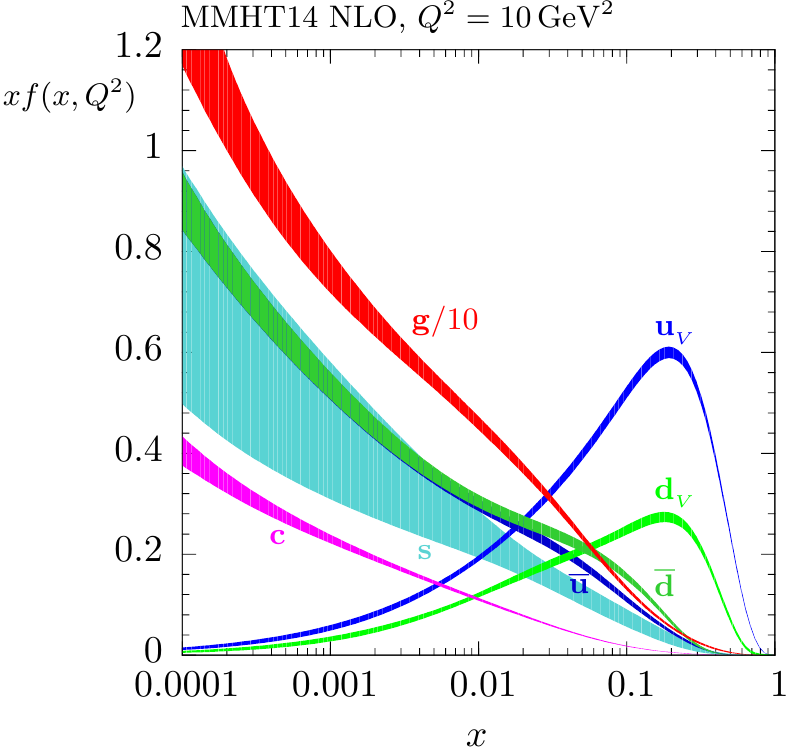}
\includegraphics[height=7.5cm]{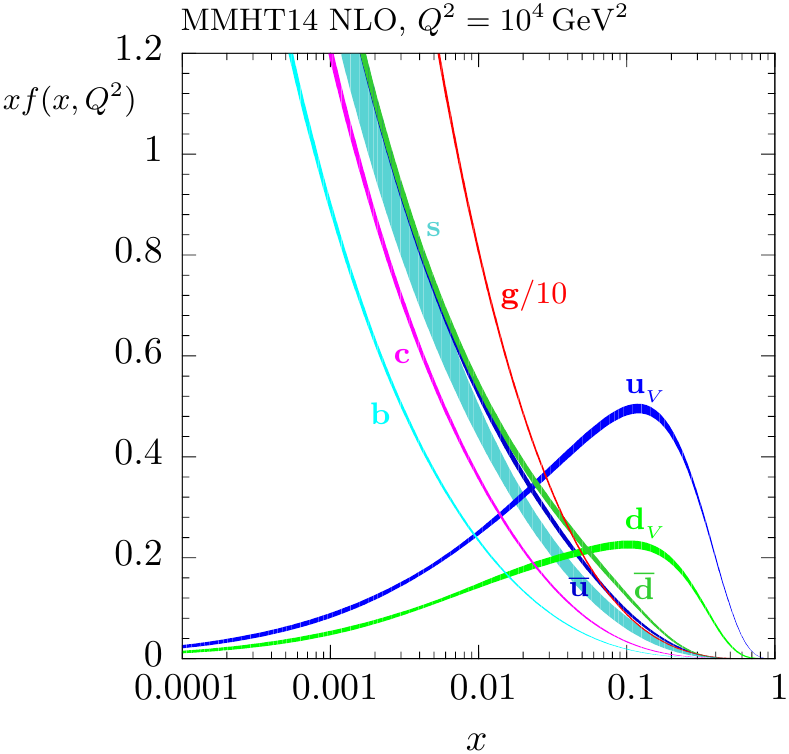}
\caption{\sf MMHT2014 NLO PDFs at $Q^2=10 ~\GeV^2$ and $Q^2=10^4 \GeV^2$, with associated 68$\%$ confidence-level uncertainty bands. The corresponding plot of NNLO PDFs was shown in Fig. \ref{fig:NNLOpdfs}.}
\label{fig:NLOpdfs}
\end{center}
\end{figure}

Recall that the NNLO set of PDFs that we present correspond to the default value of $\alpha_S(M_Z^2)=0.118$. These NNLO PDFs at scales of $Q^2=10~\GeV^2$ and $10^4~\GeV^2$ were shown in Fig. \ref{fig:NNLOpdfs}. The corresponding NLO PDFs with a default value $\alpha_S(M_Z^2)=0.120$
are shown in Fig. \ref{fig:NLOpdfs}. 
As $Q^2$ increases we expect the uncertainties on the PDFs to decrease, particularly at very 
small $x$.  This is well illustrated in the plots by comparing the PDFs at $Q^2=10\GeV^2$ with 
those at $Q^2=10^4 ~\GeV^2$.  
We also make available a second set of NLO PDFs with $\alpha_S(M_Z^2)=0.118$.  In addition, we provide a LO set of PDFs, which have $\alpha_S(M_Z^2)=0.135$, though these give a poorer description of the global data, see Table \ref{tab:chisquared}.

These four sets of PDFs are available as programme-callable functions from 
\cite{UCLsite}, and from the LHAPDF library \cite{LHAPDF}. A new HepForge \cite{hepforgesite} project site is also expected.

Although we leave a full study of the relationship between the PDFs and the strong coupling
constant $\alpha_S$ to a follow-up publication we also make available PDF sets with changes of 
$\alpha_S(M_Z^2)$ of 0.001 relative to the PDF eigenvector sets, i.e. at $\alpha_S(M_Z^2)=0.117$
and $0.119$ at both NLO and NNLO, and also at $\alpha_S(M_Z^2)=0.121$ at NLO. We also make sets available at $\alpha_S(M_Z^2)=0.134$ and 0.136 at LO. This is in order to enable the $\alpha_S$ variation 
in the vicinity of the default PDFs to be examined and for the uncertainty to be calculated 
if the simple procedure of addition of $\alpha_S(M_Z^2)$ errors in quadrature is applied.\footnote{See \cite{CTEQalphas}, where it is shown this is equivalent to treating 
$\alpha_S(M_Z^2)$ as an extra parameter in the eigenvector approach in the limit that 
the Hessian formalism is working perfectly.}

\subsection{Comparison of MMHT2014 with MSTW2008 PDFs}

\begin{figure} [t]
\begin{center}
\includegraphics[height=5cm]{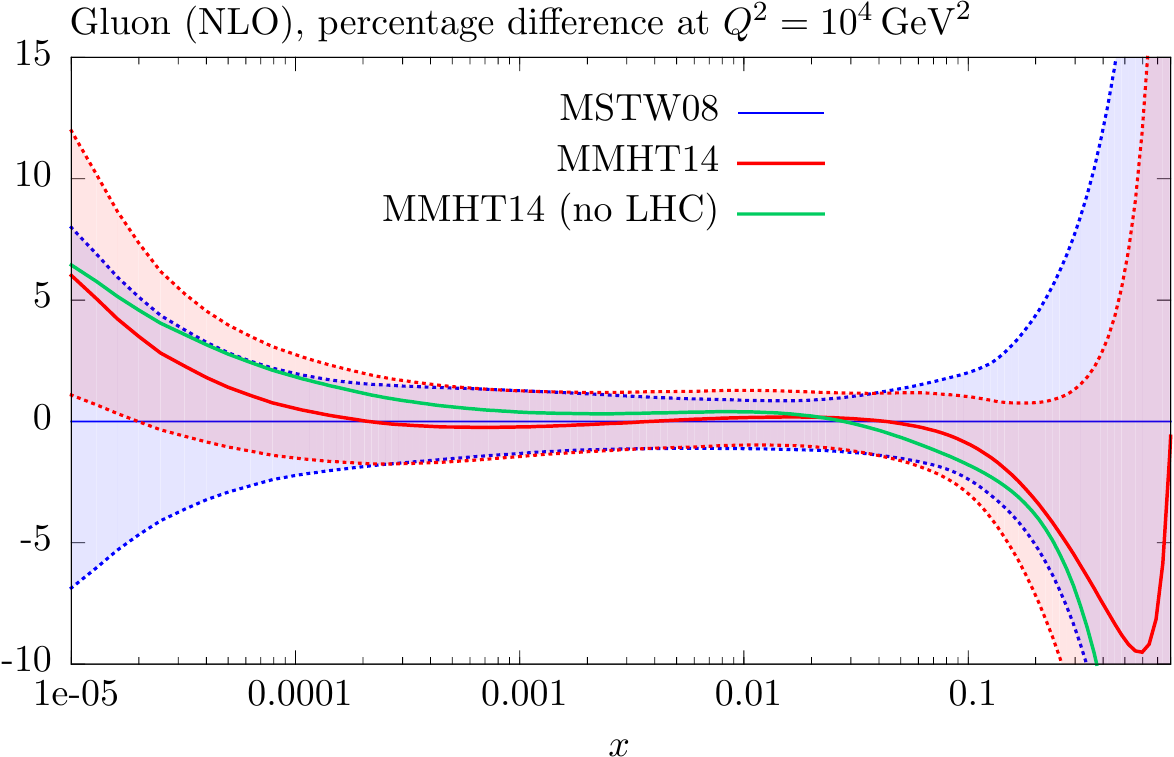}
\includegraphics[height=5cm]{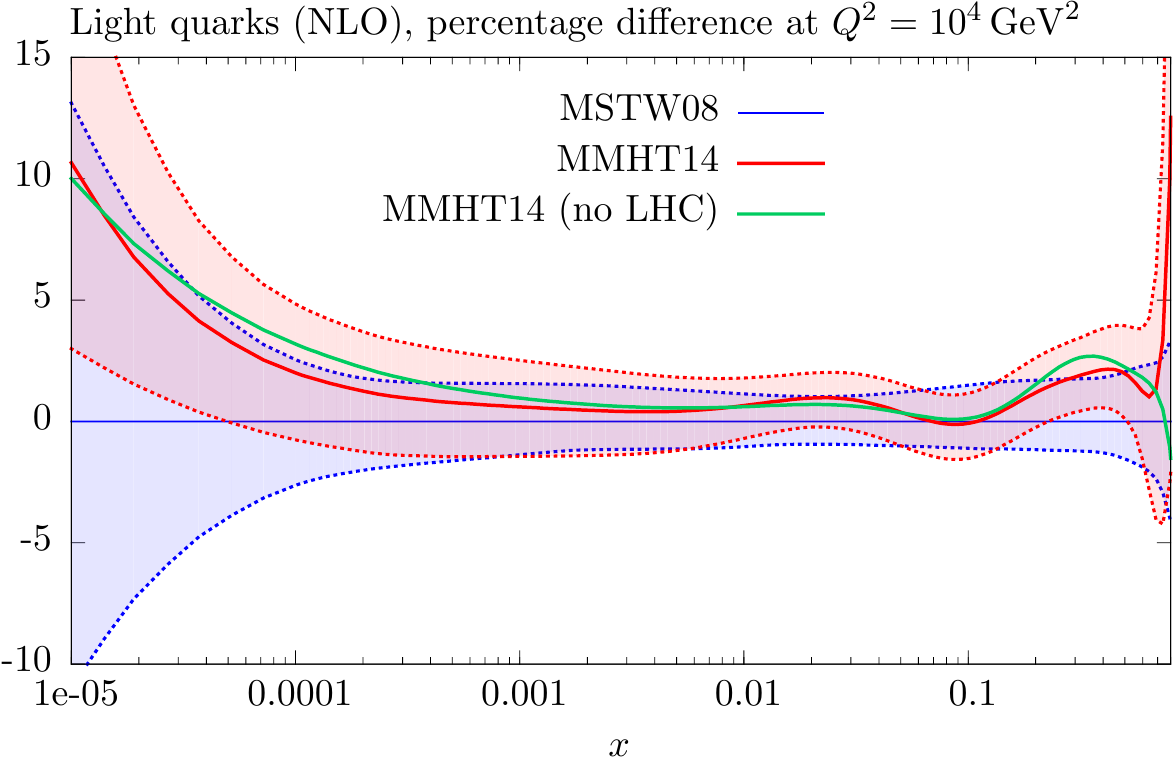}
\includegraphics[height=5cm]{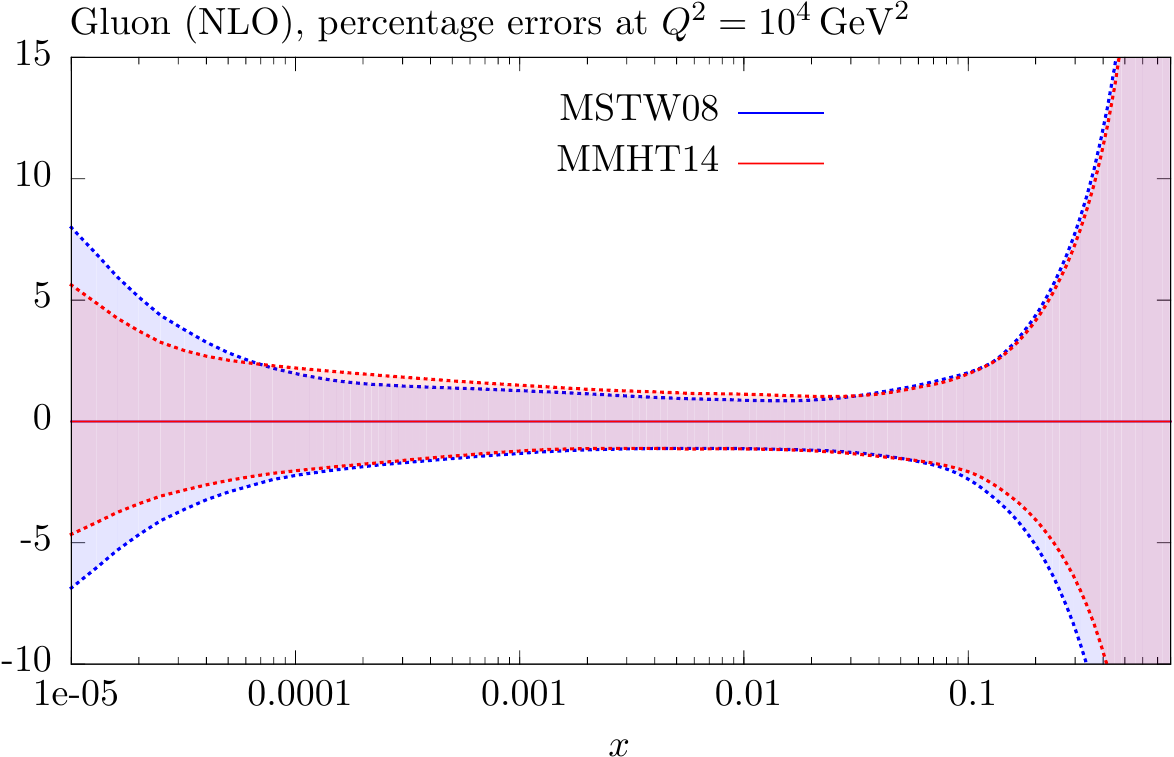}
\includegraphics[height=5cm]{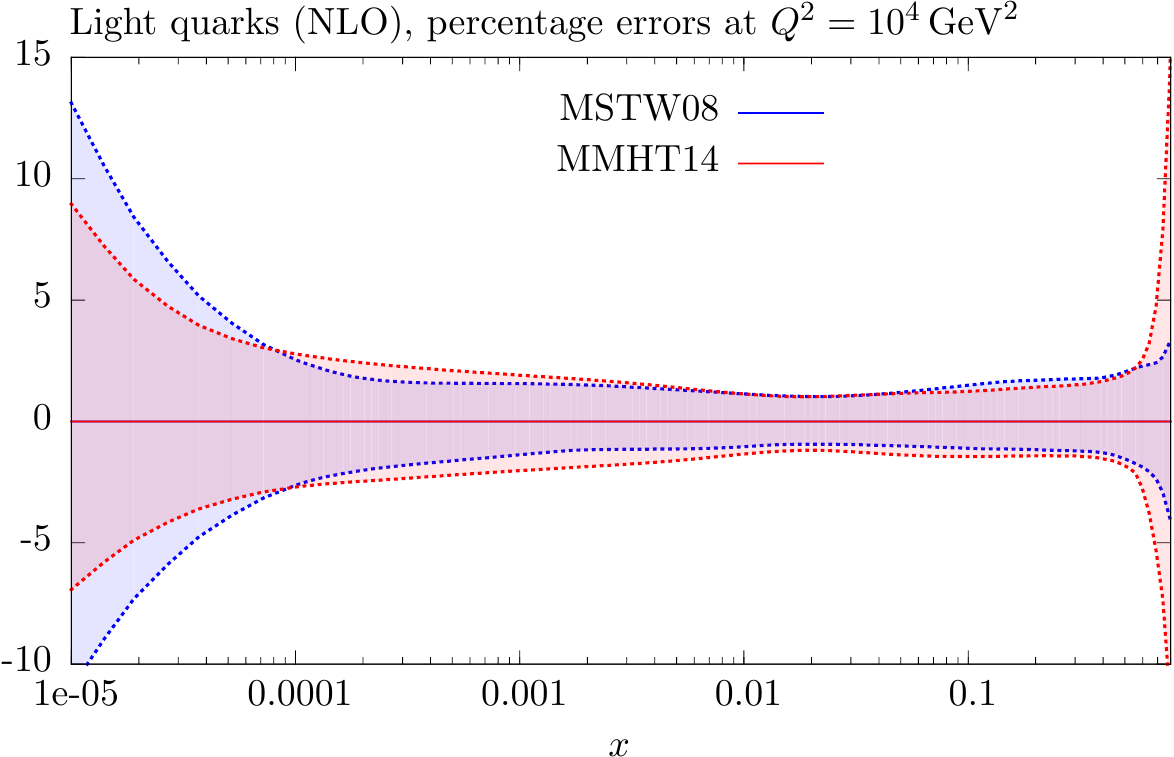}
\caption{\sf The change, in the $g$ and light quark PDFs at NLO for $Q=10^4~\GeV^2$, in going from the MSTW values to those in the present global NLO fit, which includes the LHC data. Also shown are comparisons of the percentage errors in the two analyses. }
\label{fig:NLO1}
\end{center}
\end{figure}

\begin{figure} [t]
\begin{center}
\includegraphics[height=5cm]{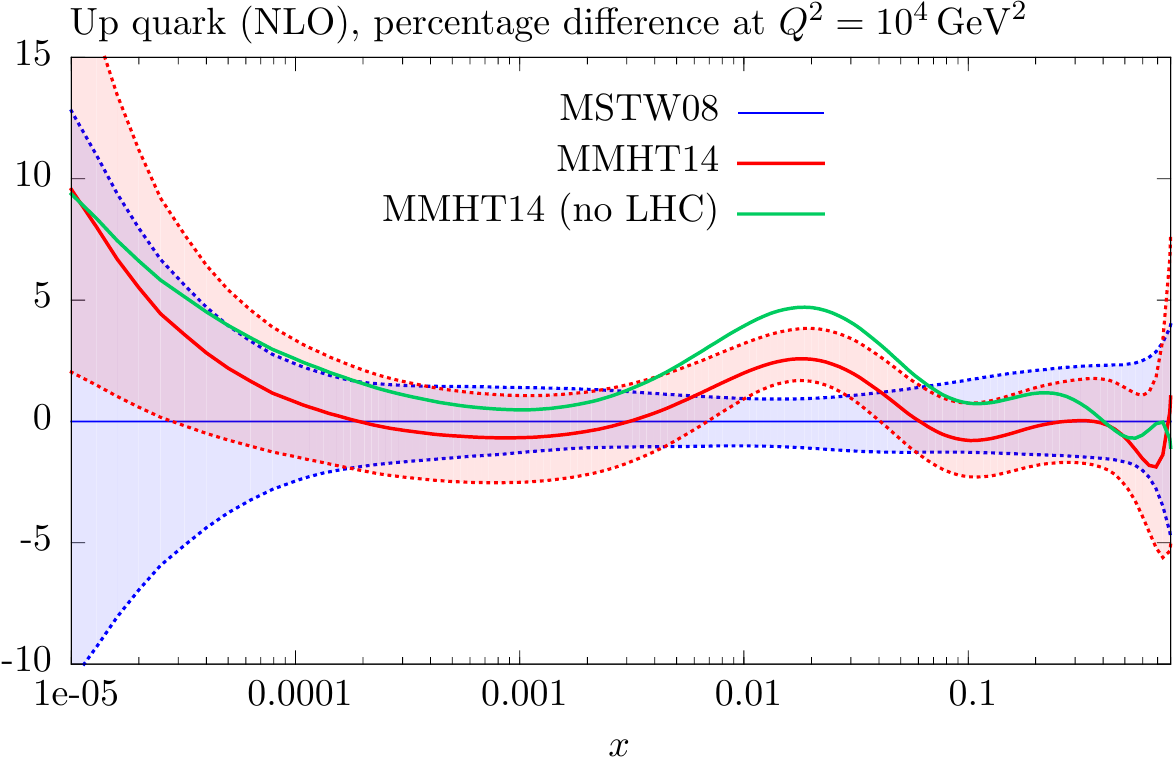}
\includegraphics[height=5cm]{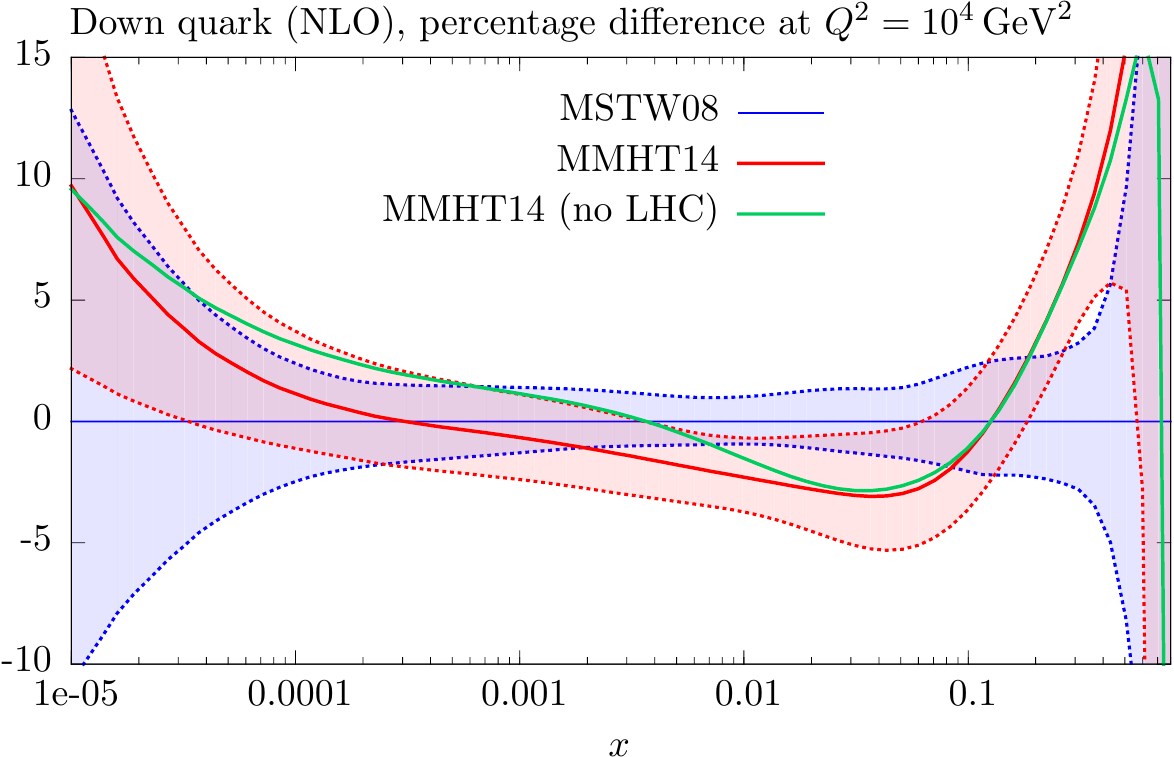}
\includegraphics[height=5cm]{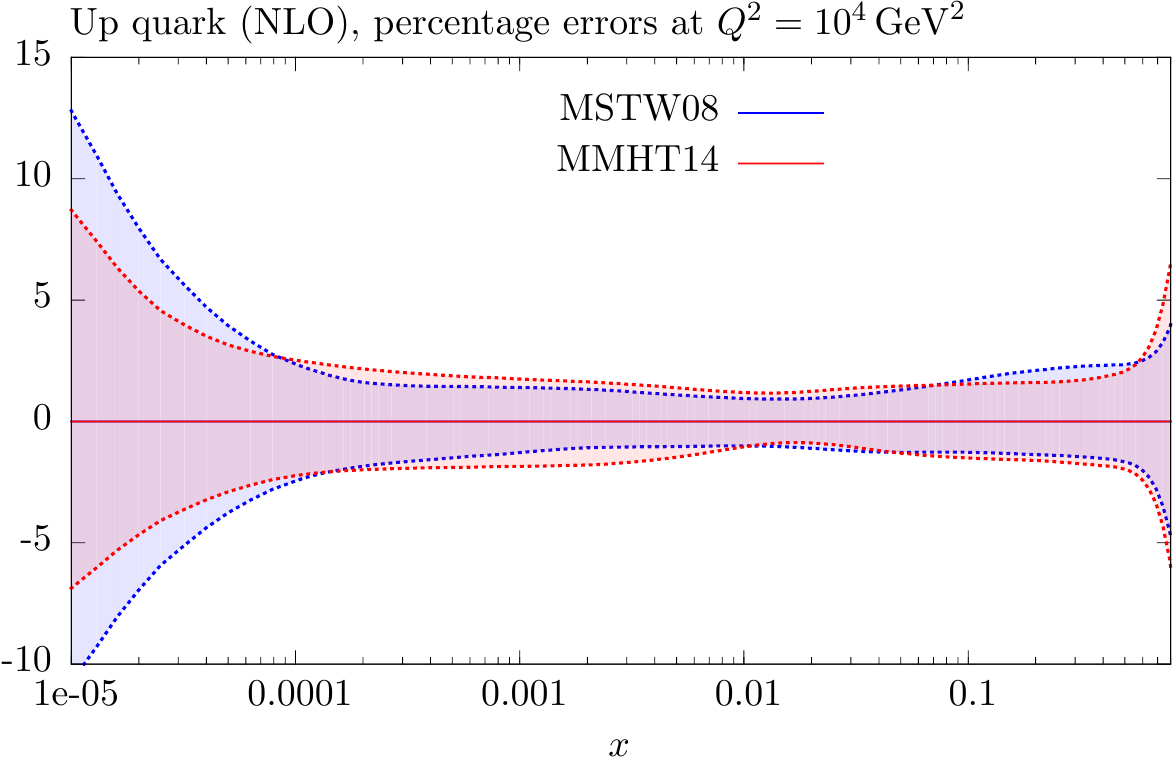}
\includegraphics[height=5cm]{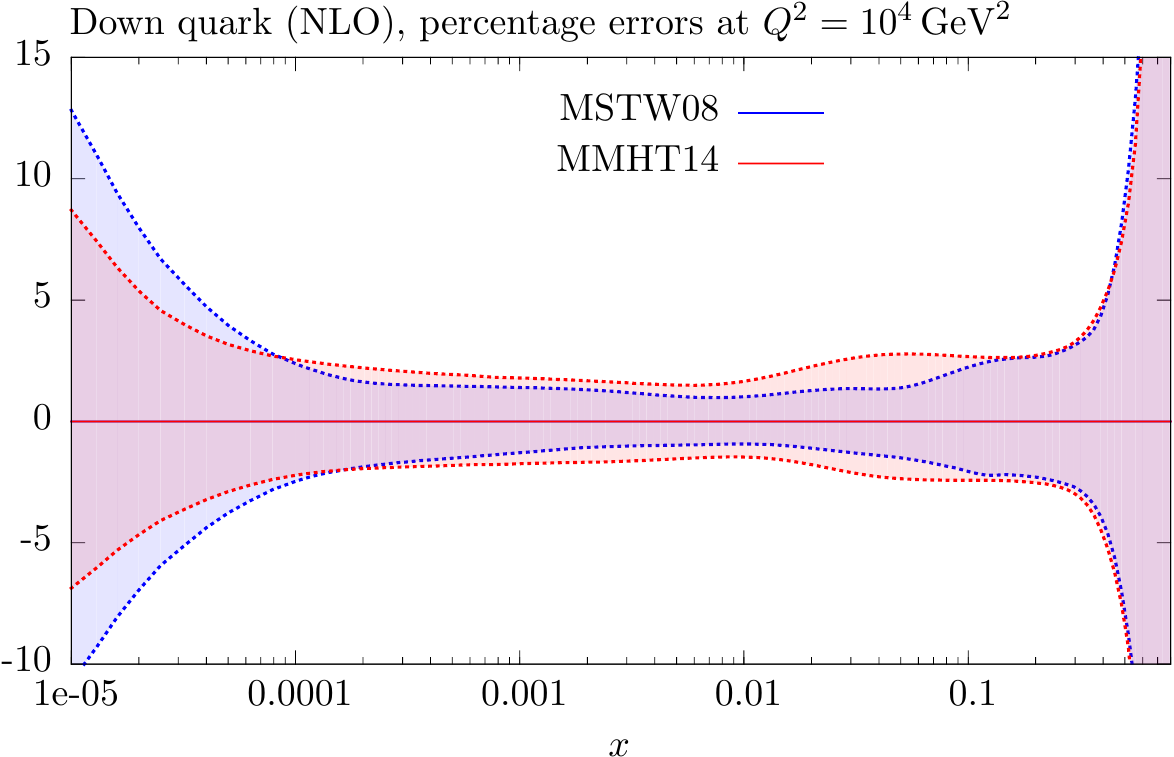}
\caption{\sf The change, in the $u$ and $d$ PDFs at NLO for $Q=10^4~\GeV^2$, in going from the MSTW values to those in the present global NLO fit, which includes the LHC data. Also shown are comparisons of the percentage errors in the two analyses. }
\label{fig:NLO2}
\end{center}
\end{figure}

We now show the change in both the central values and the uncertainties of 
the NLO PDFs at $Q^2=10^4~\GeV^2$ 
in going from the NLO MSTW analysis. The ratio of the 
MMHT2014 PDFs, along with uncertainties, to the MSTW2008 PDFs is shown in 
Figs. \ref{fig:NLO1}, \ref{fig:NLO2} and \ref{fig:NLO3}. We also show
the central value of the MMHT2014 fit before LHC data are added in the 
top plot in each case. In the lower plots we simply compare the uncertainties 
of the MMHT2014 PDFs and the MSTW2008 PDFs.

\subsubsection{Gluon and light quark}
 
In Fig. \ref{fig:NLO1} we compare the  gluon and total light quark
distributions. In this and subsequent plots we show uncertainty bands for the full
MMHT2014 and MSTW2008 PDFs, but only show the central value of the MMHT2014 PDFs obtained
without LHC data. This is because it is interesting to see the (usually quite small) direct 
effect on the best PDFs from LHC data, but we note that the parameterisation for the strange
quark is more limited when LHC data are not included as without LHC Drell Yan type data there
is insufficient constraint on the details of the shape of the strange quark. This means it is 
not possible to properly reflect the change in strange quark uncertainty in MMHT2014
PDFs before and after LHC data is added, which is actually the 
dominant change in PDF uncertainties between MSTW2008 and MMHT2014 PDFs, and which feeds into the 
total light quark uncertainty. Really, it is only the addition of the LHC data which allow us to 
present an uncertainty on the strange PDFs with full confidence. We do note, however, that 
the gluon uncertainty is essentially unchanged by the addition of LHC data except to a 
very minor improvement at high-$x$ at NLO.       

The change in the central value of the gluon 
is almost the same with and without LHC data. It is slightly softer at 
high $x$ and a little larger at the smallest $x$ values shown, but within
uncertainties, particularly when the LHC data are included. This slight 
change in shape is due to the inclusion of the combined HERA data,
as indicated in \cite{MSTWDIS}. However,
the slight softening at high $x$ is also exhibited when the default heavy flavour 
scheme is replaced by the optimal scheme in \cite{Thorne} and when
LHC jet data are included in \cite{Wattjets}. Hence, it seems that a variety
of new effects all prefer this slight change in shape, but even the combination
of all of them only results in a small change. The gluon and light quark uncertainty 
decreases a little at lowest $x$, due to the combined HERA data, and
the gluon uncertainty decreases very slightly at $x > 0.1$ due to inclusion of LHC jet data. 
The light sea is a little larger at the smallest $x$, driven by the same 
shape change in the gluon distribution and the evolution.
We note that there are few data for $x < 10^{-4}$, but there is some, which acts 
to constrain the small-$x$ sea. There is less direct constraint 
on the gluon at very small $x$ and $Q^2$, though still some from $dF_2(x,Q^2)/d\ln\, Q^2$ and 
$F_L(x,Q^2)$ and the uncertainty is very large. However, at much higher $Q^2$ most of the 
gluon and light sea at $x=10^{-5}$ is determined by evolution from higher $x$, and even 
a very large uncertainty at input is largely washed out by this. 

The changes in detailed shape at high $x$ are mainly due to individual 
quark flavour contributions and will be discussed below. The uncertainty is 
reduced for $x< 0.0001$, mirroring the same effect in the gluon. The 
increase in uncertainty at very high $x$ is due to the improved parameterisation 
flexibility. The slight increase in uncertainty over a wide range of $x$ is 
due to the large uncertainty introduced into the branching 
ratio, $B_{\mu}$, for charmed mesons decaying to muons (as discussed in Section \ref{sec:2.6}), which increases the strange quark 
uncertainty and hence that of the entire light sea.

\subsubsection{Up and down quark}
In Fig. \ref{fig:NLO2} we compare the up and down  quark
distributions. The very small $x$ increase has already been explained, and is 
common to all quarks. The increase around $x=0.01$ compared to MSTW2008
was already apparent in \cite{MMSTWW}, and is due to the improved parameterisation
(and to some extent improved deuteron corrections) and the increase is mainly
in the up valence distribution. The increase is very compatible with fitting 
ATLAS and CMS data on $W^{\pm}$ production at low rapidity, but is not actually
driven by this at all. In fact, we see that the increase is actually 
significantly larger before the inclusion of LHC data. The down quark has 
changed shape quite clearly. The decrease for $x\sim 0.05$ and increase at 
high $x$ was again already apparent in \cite{MMSTWW} and is due to improved
deuterium corrections and parameterisation. The fine details are modified by 
the inclusion of LHC data, but the main features are present in the fit without 
LHC data. The change in the uncertainties is similar to that for the total 
light sea, though the flexibility in the improved deuteron corrections 
does contribute to the increase in uncertainty of the down distribution.

\begin{figure} [t]
\begin{center}
\includegraphics[height=5cm]{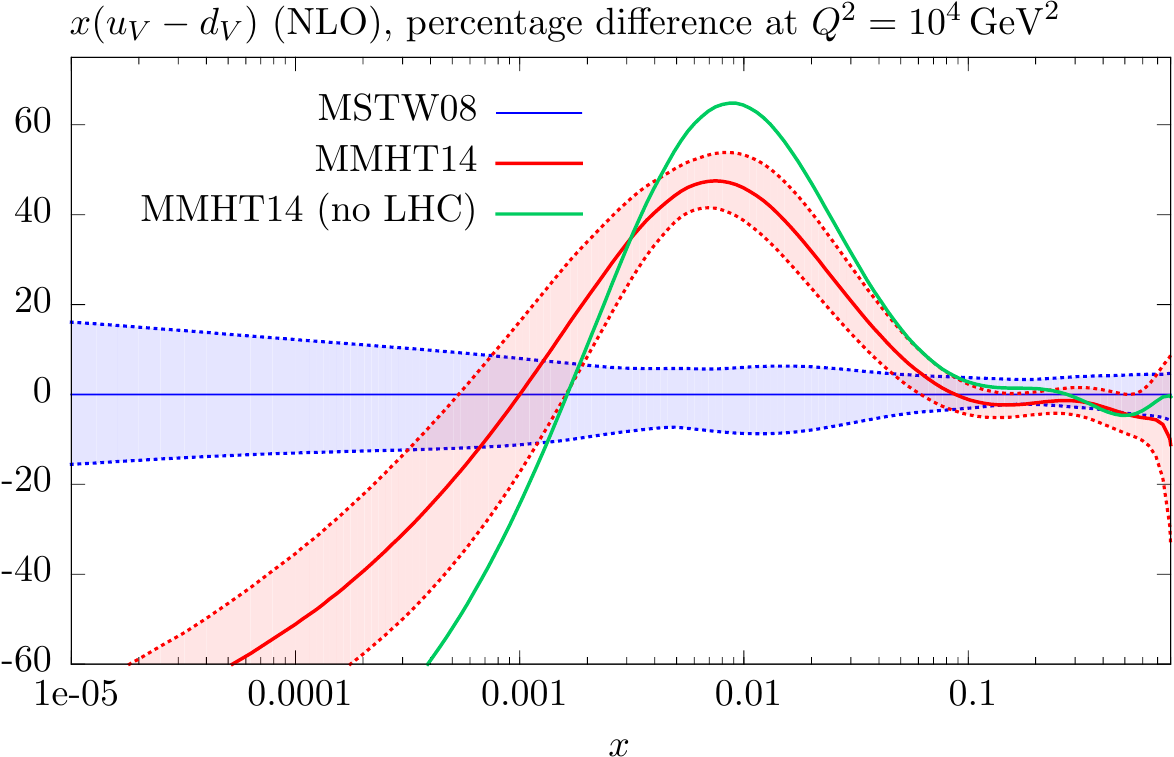}
\includegraphics[height=5cm]{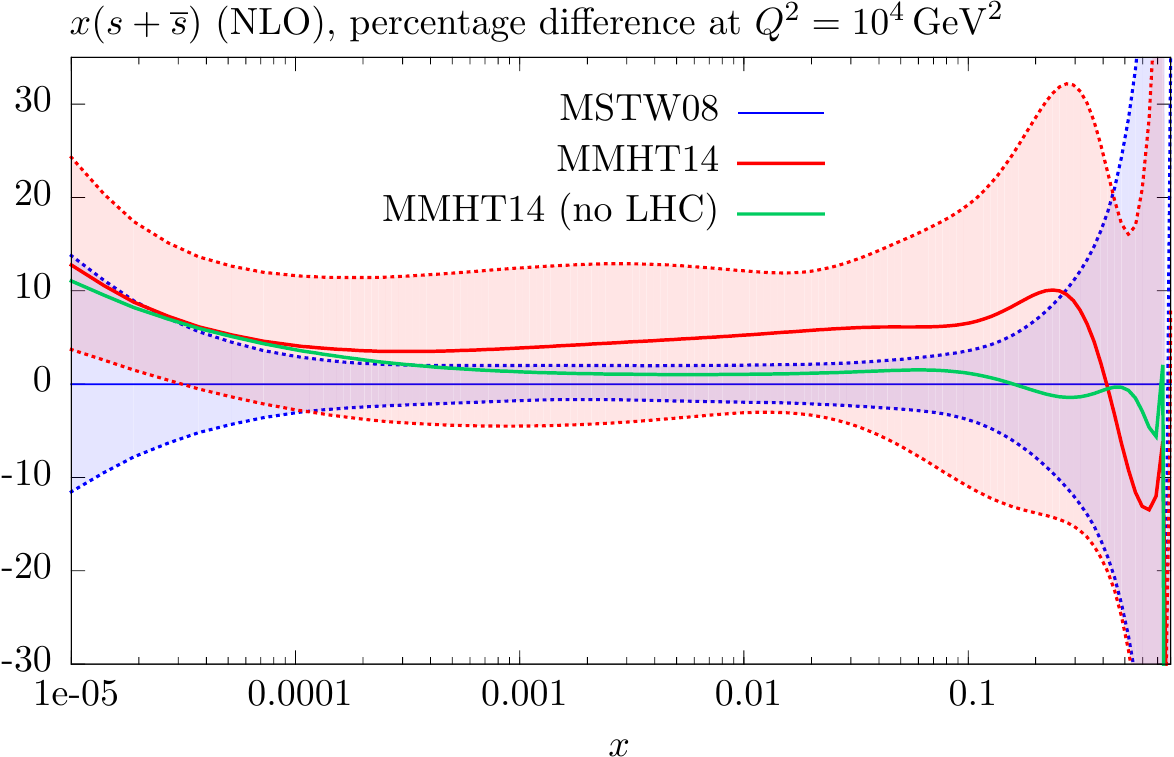}
\includegraphics[height=5cm]{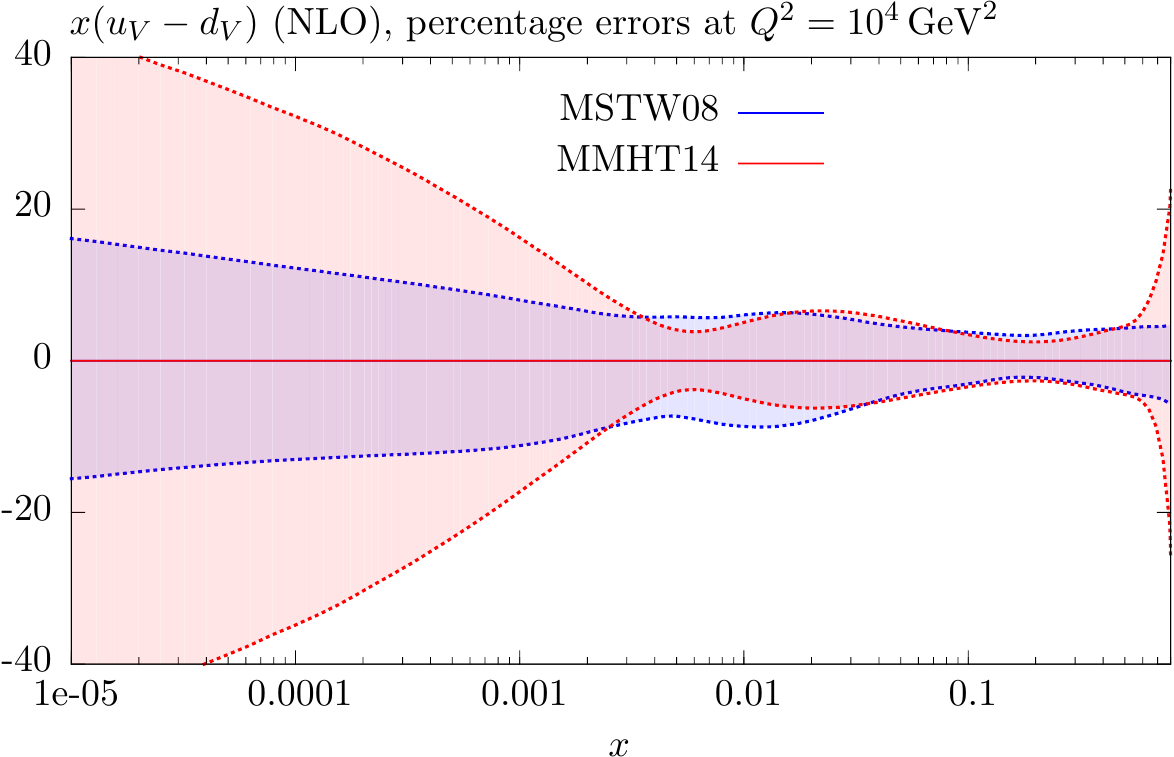}
\includegraphics[height=5cm]{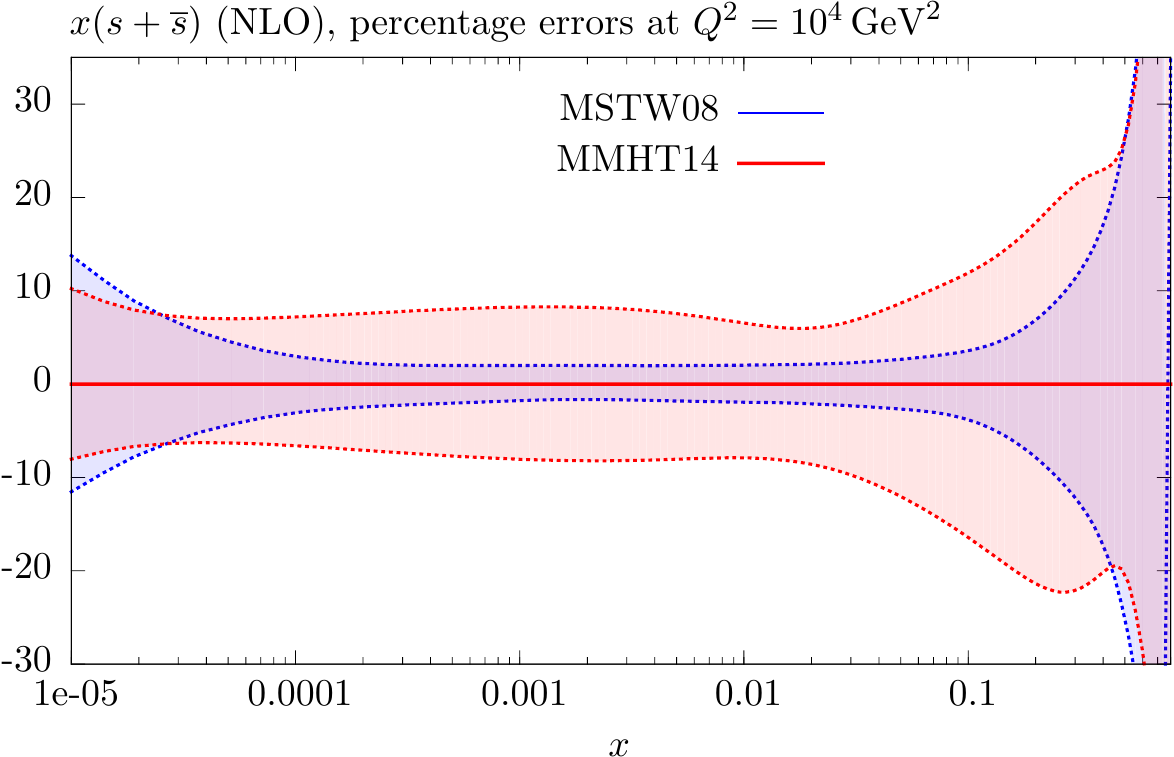}
\caption{\sf The change, in the $(u_V-d_V)$ and $(s+\bar{s})$  PDFs at NLO for $Q=10^4~\GeV^2$, in going from the MSTW values to those in the present global NLO fit, which includes the LHC data. Also shown are comparisons of the percentage errors in the two analyses. }
\label{fig:NLO3}
\end{center}
\end{figure}

\subsubsection{$u_V-d_V$ and $s+\bar s$ distributions}

In Fig. \ref{fig:NLO3} we compare the $u_V(x,Q^2)-d_V(x,Q^2)$ and 
$s(x,Q^2) + \bar s(x,Q^2)$ distributions. The very dramatic change in the former
was already seen in \cite{MMSTWW}. 
In fact Ref. \cite{MMSTWW} was able to give a reasonable description of the observed lepton charge asymmetry at the LHC, whereas MSTW2008 gave a poor prediction. This is really the only blemish of the MSTW2008 \cite{MSTW} predictions. The change in $u_V-d_V$ for $x\lapproxeq 0.03$ is very evident in the Figure.
This change is not driven by the LHC data, but rather by the improved flexibility of the MMHT (and MMSTWW \cite{MMSTWW}) parametrisations (and improved deuteron corrections). 
Indeed, as seen with the up quark, the change, from the MSTW2008 partons, is larger before the inclusion 
of LHC data. The uncertainty in  $u_V(x,Q^2)-d_V(x,Q^2)$ increases very 
significantly at small $x$ due to the increased flexibility of the MMHT 
parameterisation. However, there is a decrease near $x=0.01$ due to the 
constraint added by the LHC asymmetry data, which is the only real change compared to
the MMSTWW distribution. 

There is a very significant increase in the uncertainty in the $s+\bar{s}$ 
distribution (at all but the lowest $x$ where the distribution is governed 
mainly by evolution from the gluon), due mainly to the freedom allowed for 
the branching fraction $B_{\mu}$, see Section \ref{sec:2.6}, though there is 
also one more free parameter for this PDF in the eigenvector determination.  
The central value of the total strange distribution is very similar to MSTW2008
before LHC data are included, with only the common slight increase at lowest $x$.
This is despite the correction of the theoretical calculation of dimuon production
and a change in nuclear corrections, showing the small impact of these two
effects (though they do actually tend to pull in opposite directions).
There is a few percent increase when the LHC data are included, mainly
driven by the ATLAS $W,Z$ data. The central value is outside the 
uncertainty band of the MSTW2008 distribution. However, the MSTW2008 distribution
is included comfortably within the error band of the MMHT2014 distribution.

\subsubsection{$\bar d - \bar u$ and $s-\bar s$ distributions}

\begin{figure} 
\begin{center}
\includegraphics[height=5cm]{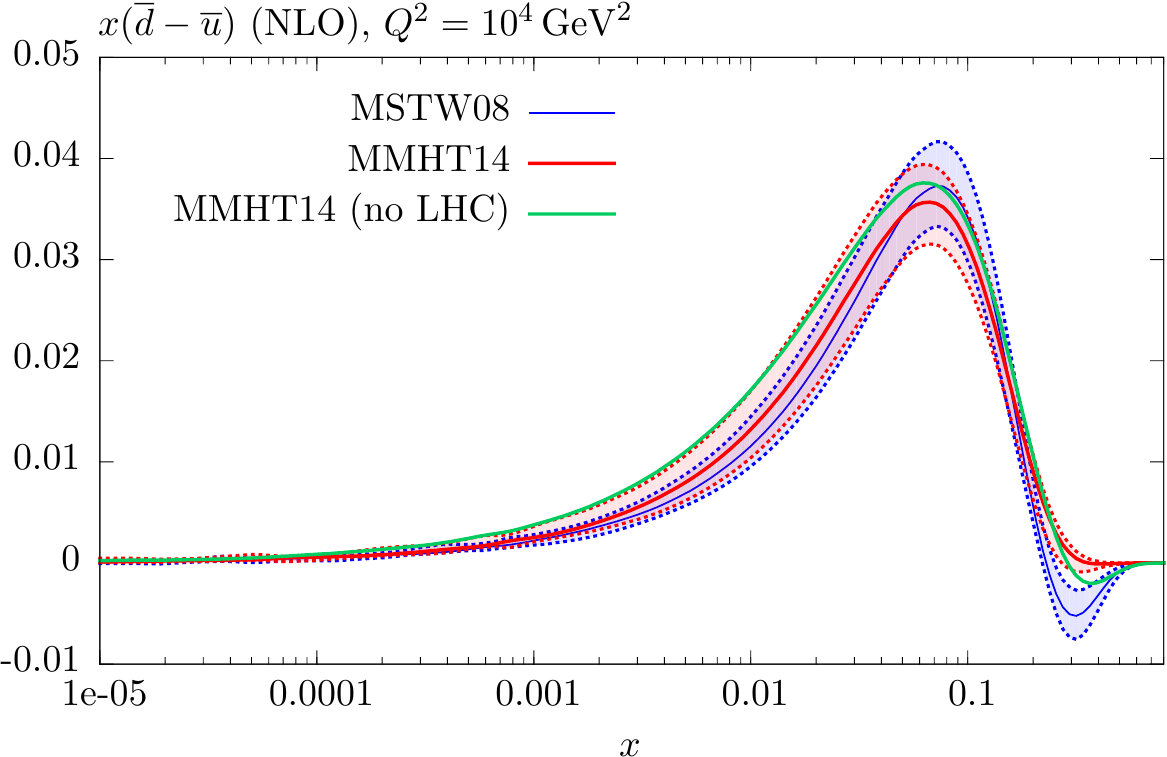}
\includegraphics[height=5cm]{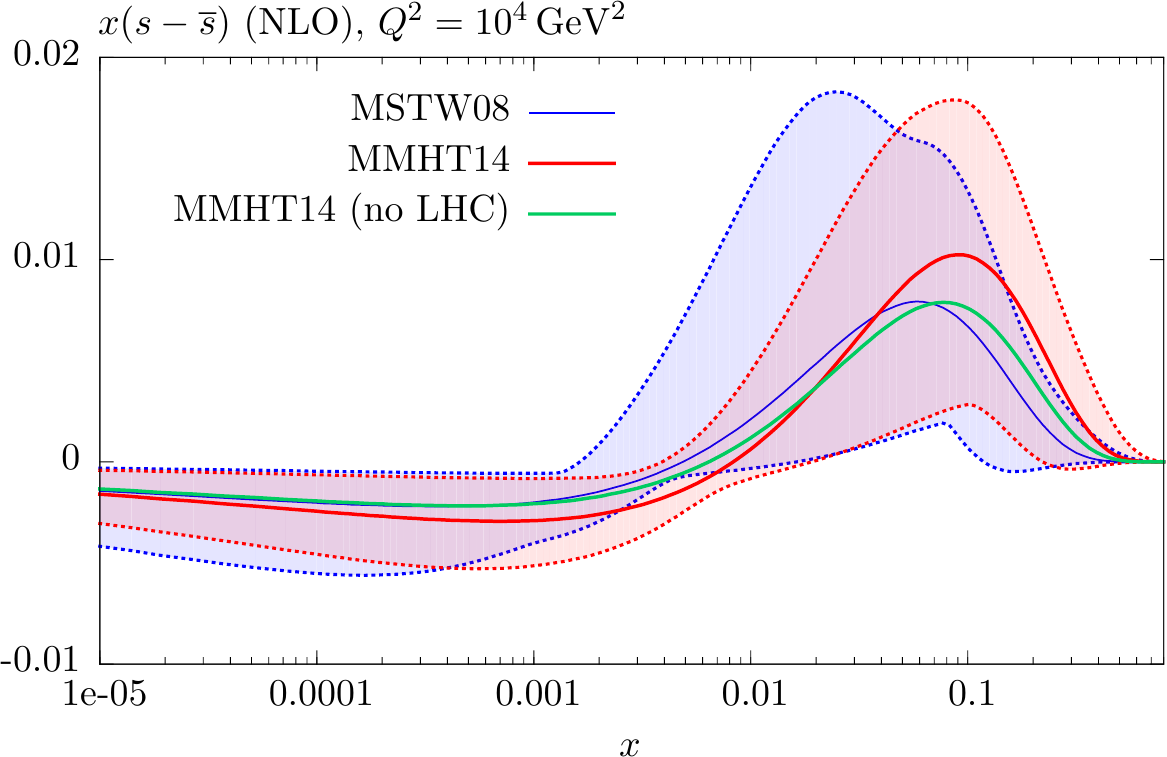}
\caption{\sf The change, in the $(\bar d  -\bar u)$ and $(s-\bar{s})$  PDFs at 
NLO for $Q=10^4~\GeV^2$, in going from the MSTW values to those in the 
present global NLO fit, which includes the LHC data. }
\label{fig:NLO4}
\end{center}
\end{figure}

In Fig.~\ref{fig:NLO4} we show the comparison of 
$\bar d(x,Q^2)-\bar u(x,Q^2)$ and  $s(x,Q^2) - \bar s(x,Q^2)$. In this case 
showing the percentage uncertainties is not useful, due to the fact that both distributions pass through zero. One can see that there is 
no very significant change in either the central values or uncertainties. 
There is a fairly distinct tendency for $\bar d(x,Q^2)-\bar u(x,Q^2)$ to be
negative for $x\sim 0.3$ in the MSTW2008 set, which may be a sign of the 
overall more restricted parameterisation in this case, but other than this the 
MSTW2008 and MMHT2014 $\bar d(x,Q^2)-\bar u(x,Q^2)$ distributions are 
very consistent. This is unsurprising as the dominant constraint is still
the E866/NuSea Drell-Yan ratio data \cite{E866DYrat}. The MMHT2014 
$s(x,Q^2) - \bar s(x,Q^2)$ distribution has a tendency to peak 
at slightly higher $x$, but the MSTW2008 and MMHT2014 distributions are 
very consistent and have similar size uncertainties. The main constraint is
still overwhelmingly the CCFR and NuTeV $\nu N\to \mu\mu X$ data \cite{Dimuon}, and
the change in the treatment of the branching ratio has little effect on the asymmetry.
There is some small constraint from $W$ asymmetry data, and the new data from the LHC
provides some pull, and contributes to the MMHT2014 uncertainty being a little smaller 
for $x<0.05$. This constraint will improve in the future.  

\subsubsection{Comparison with MSTW2008 at NNLO}

\begin{figure}[t]
\begin{center}
\includegraphics[height=5cm]{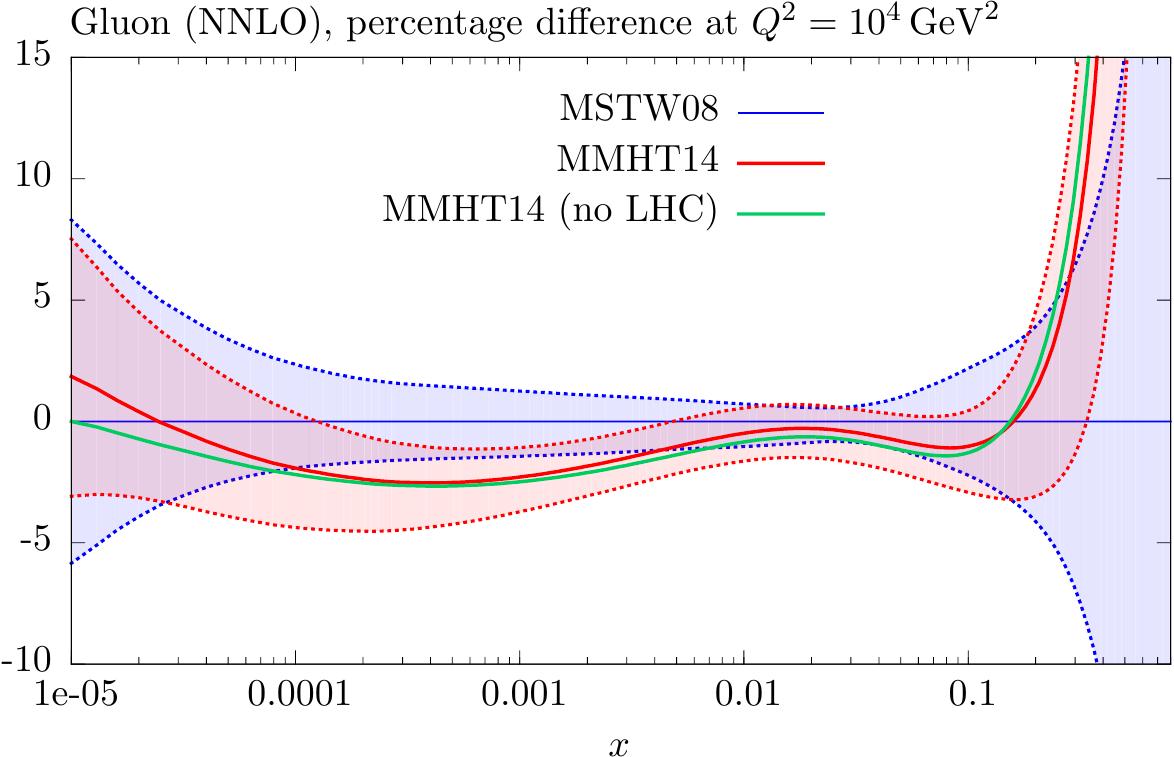}
\includegraphics[height=5cm]{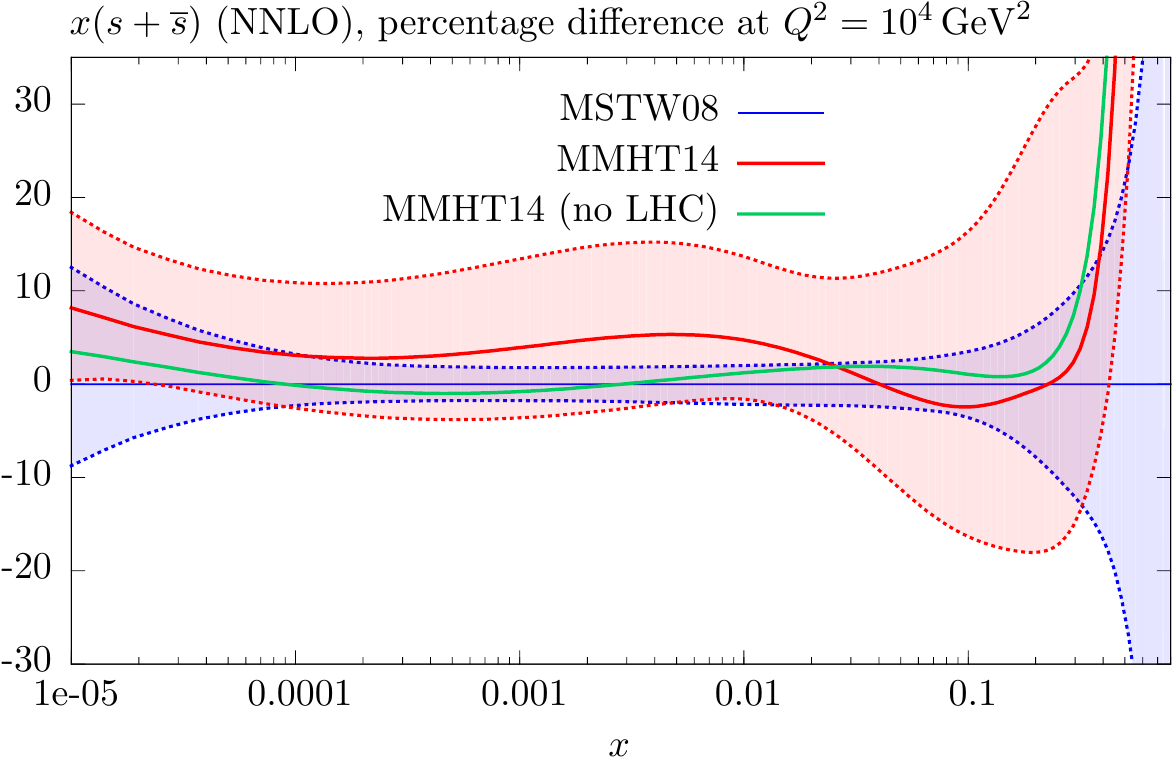}
\includegraphics[height=5cm]{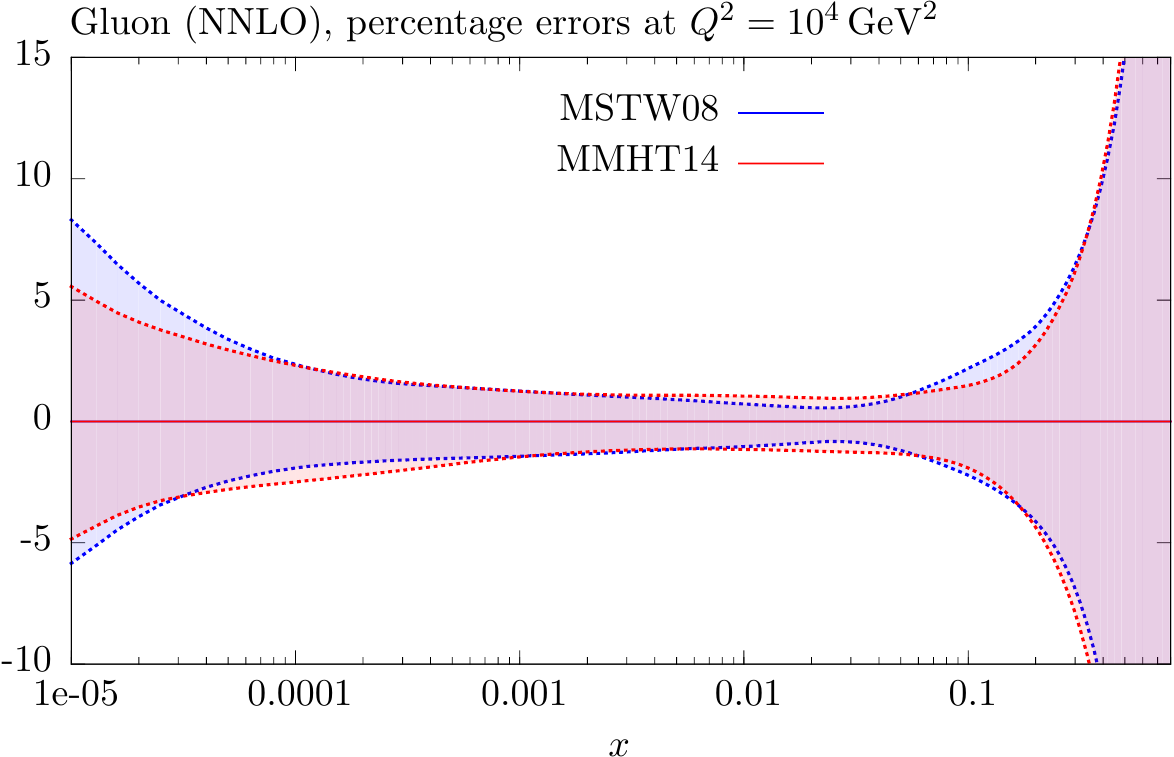}
\includegraphics[height=5cm]{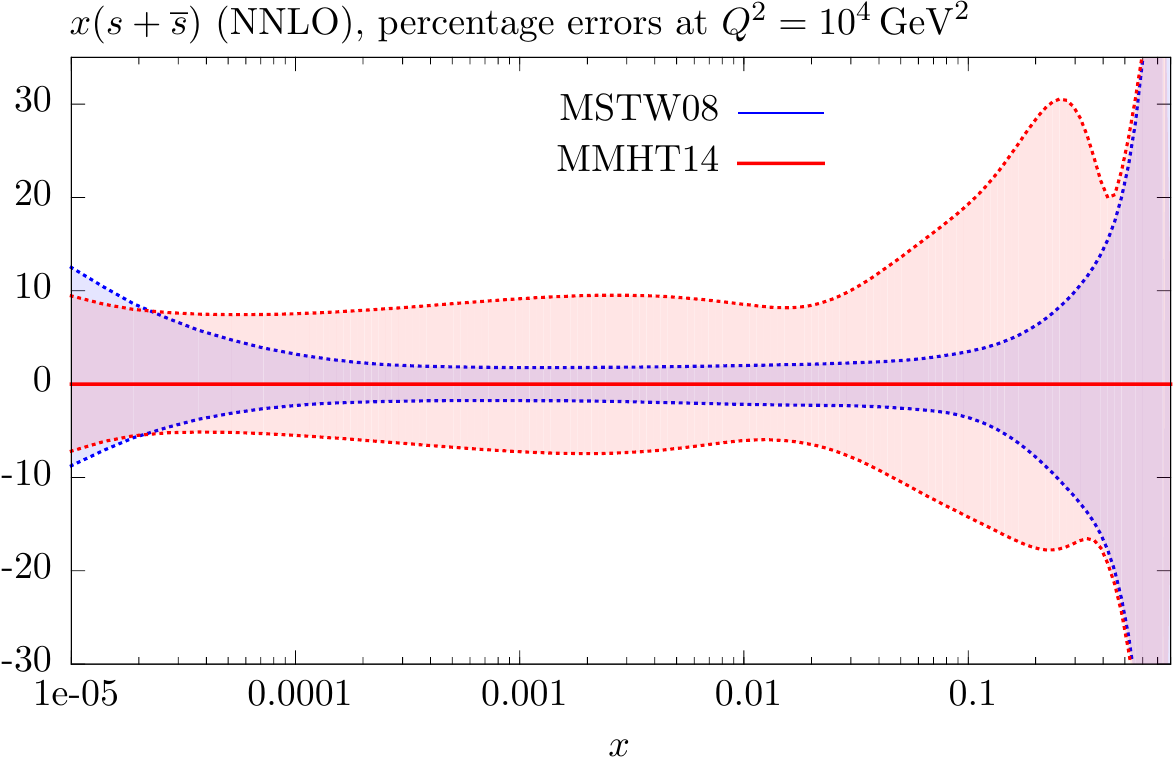}
\caption{\sf The change, in the $g$ and $s+\bar{s}$ PDFs at NNLO for $Q=10^4~\GeV^2$, in going from the MSTW values to those in the present global NNLO fit, which includes the LHC data. Also shown are comparisons of the percentage errors in the two analyses. }
\label{fig:gsNNLO}
\end{center}
\end{figure}

The changes in the NNLO PDFs going from MSTW2008 to MMHT2014 are very similar 
to those at NLO. However, the $g$ and $s+\bar{s}$ changes are shown in 
Fig. \ref{fig:gsNNLO}. The gluon has now become a little harder at high $x$ and 
a bit smaller between $x=0.0001$ and $x=0.01$.
The slight decrease in the NNLO gluon between 
$x=0.0001$ and $x=0.01$ (which, via evolution, shows up to some extent in the 
sea quarks) is driven largely by the fit to the combined HERA data,
while the increase at very high $x$ is related to the use of multiplicative 
uncertainties for the Tevatron jet data, and by the momentum sum rule. 
The change in the MMHT2014 $s +\bar s$ distribution is 
similar to that at NLO, except that there is a slight decrease near $x=0.1$
as opposed to an increase at all $x$. This is due to a slightly larger 
correction to the dimuon cross section in this region at NNLO than at NLO, but 
also, this seems to be the preferred shape to fit the ATLAS $W,Z$ data at NNLO.

\begin{figure} 
\begin{center}
\includegraphics[height=5cm]{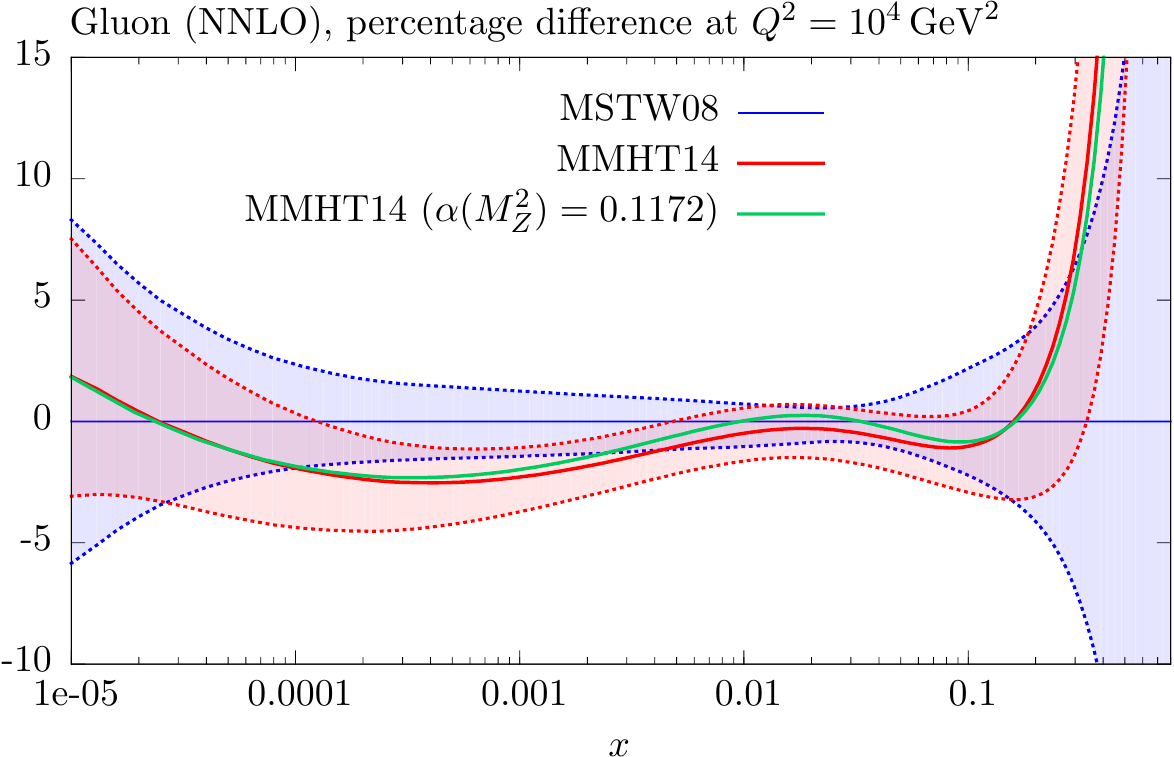}
\includegraphics[height=5cm]{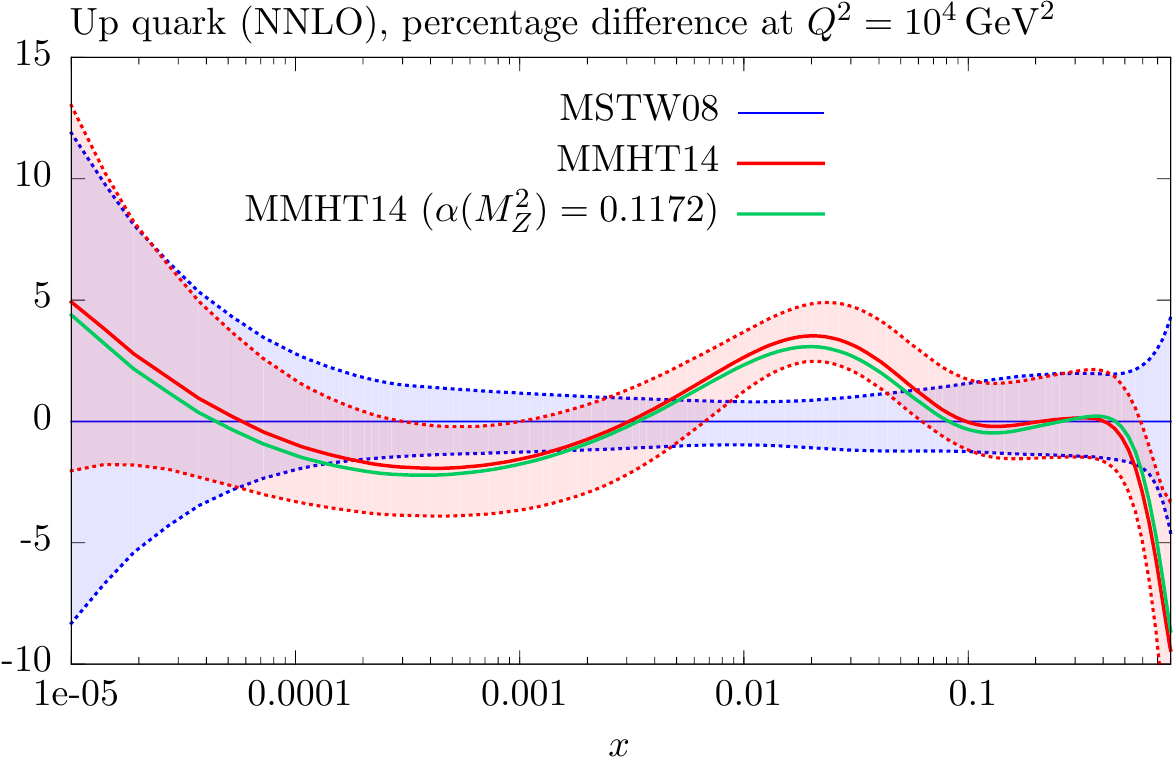}
\caption{\sf The change in the $g$ and $u$ PDFs at NNLO for $Q=10^4~\GeV^2$, in going from 
the MSTW values to those in the present global NNLO fit (with default $\alpha_S(M_Z^2)=0.118)$, which includes the LHC data. Also shown is the NNLO fit for the optimal value $\alpha_S(M_Z^2)=0.1172$.}
\label{fig:NNLO1172}
\end{center}
\end{figure}

Part of the change in the gluon distribution is due to the fact 
that the MMHT2014 PDFs were defined at $\alpha_S(M_Z^2)=0.118$ while the MSTW2008
PDFs are defined at $\alpha_S(M_Z^2)=0.1171$. 
Recall that the gluon increases at very high $x$
and decreases at lower $x$ with an increase in $\alpha_S(M_Z^2)$, as seen in 
Fig.~11(f) of \cite{MSTWalpha}. However, this is responsible for only a 
relatively minor part of the total difference between the MMHT2014 and MSTW2008 
NNLO gluon distributions. The gluon distribution for the MMHT optimal fit value of 
$\alpha_S(M_Z^2)=0.1172$ is shown in Fig.~\ref{fig:NNLO1172}. As one sees the 
gluon for $\alpha_S(M_Z^2)=0.1172$ is much closer to the MMHT2014 gluon 
(for the default $\alpha_S(M_Z^2)=0.118$) than to the MSTW2008 gluon, and is always 
well within the uncertainty band. For the up and down quark distributions
the difference between the results for the default value  
$\alpha_S(M_Z^2)=0.118$ and the optimal  $\alpha_S(M_Z^2)=0.1172$
at $Q^2=10^4~\GeV^2$ agree to within $0.5\%$ for all $0.0001<x<0.6$, as 
one can also see in Fig.~\ref{fig:NNLO1172}. We also see, by comparing
to Fig.~\ref{fig:NLO2}, that the change in 
the up quark distribution in going from MSTW2008 to MMHT2014 is indeed very 
similar at NNLO to that at NLO.    

Just as at NLO, the only real impact on the quark uncertainties due to the LHC data is a slight
improvement in the flavour decomposition near $x=0.01$. However, the fact that LHC jet data
is absent at NNLO means the very slight reduction in uncertainty in the high-$x$ gluon due
to the inclusion of LHC data is absent at NNLO. 

\begin{figure} [t]
\begin{center}
\includegraphics[height=8cm]{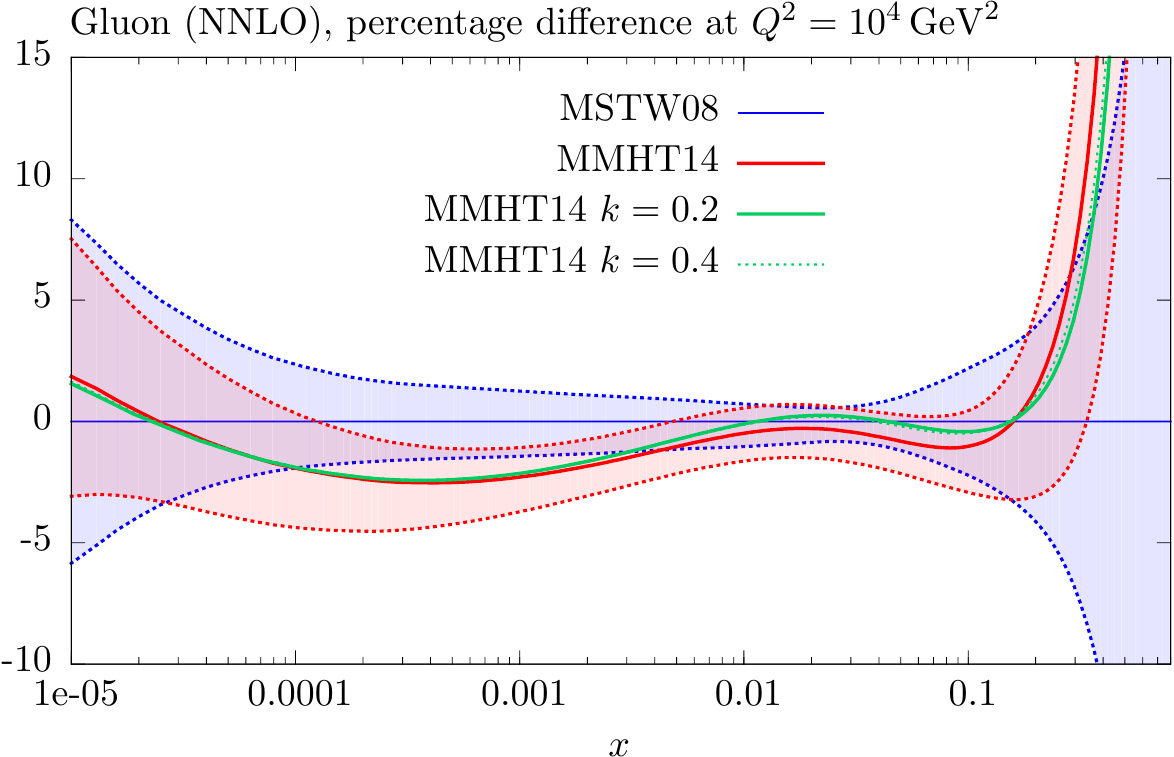}
\caption{\sf The change in the $g$ PDF at NNLO for $Q=10^4~\GeV^2$, in going from 
the MSTW values to those in the present global NNLO fit, which includes the LHC data. Also shown are the central values of the change in the $g$ PDF in the NNLO fits where LHC jet data are included 
with both the larger and smaller approximate $K$-factors; these two curves are almost indistinguishable from each other.}
\label{fig:gNNLOk}
\end{center}
\end{figure}

We also show the effect of including the LHC jet data in the NNLO fit
with the use of both the smaller and larger $K$-factors described in 
Section 4.3.1. In both fits the preferred value of $\alpha_S(M_Z^2)$
is close to $0.1172$. The resulting gluon distribution in each case is shown 
in Fig.~\ref{fig:gNNLOk}. One can see that the change in the gluon is very 
small (indeed it is very similar to that in the $\alpha_S(M_Z^2)=0.1172$ fit, 
as can be seen by comparing with Fig.~\ref{fig:NNLO1172}) and fairly 
insensitive to the overall size of the $K$-factor. As was seen in 
Fig.~\ref{fig:CMSjet7NNLO}, 
a relatively smooth and moderately sized correction 
to theory can be largely accommodated by a larger shift of data compared 
to theory using correlated systematics, with little, if any extra penalty.
As noted in Section 4, however, this is not as easy to do with jet data 
taken at two different energy scales, and will also not be 
as successful with reduced correlated systematic uncertainties. 
 
\subsubsection{Comparison between NLO and NNLO}

\begin{figure} [t]
\begin{center}
\includegraphics[height=5cm]{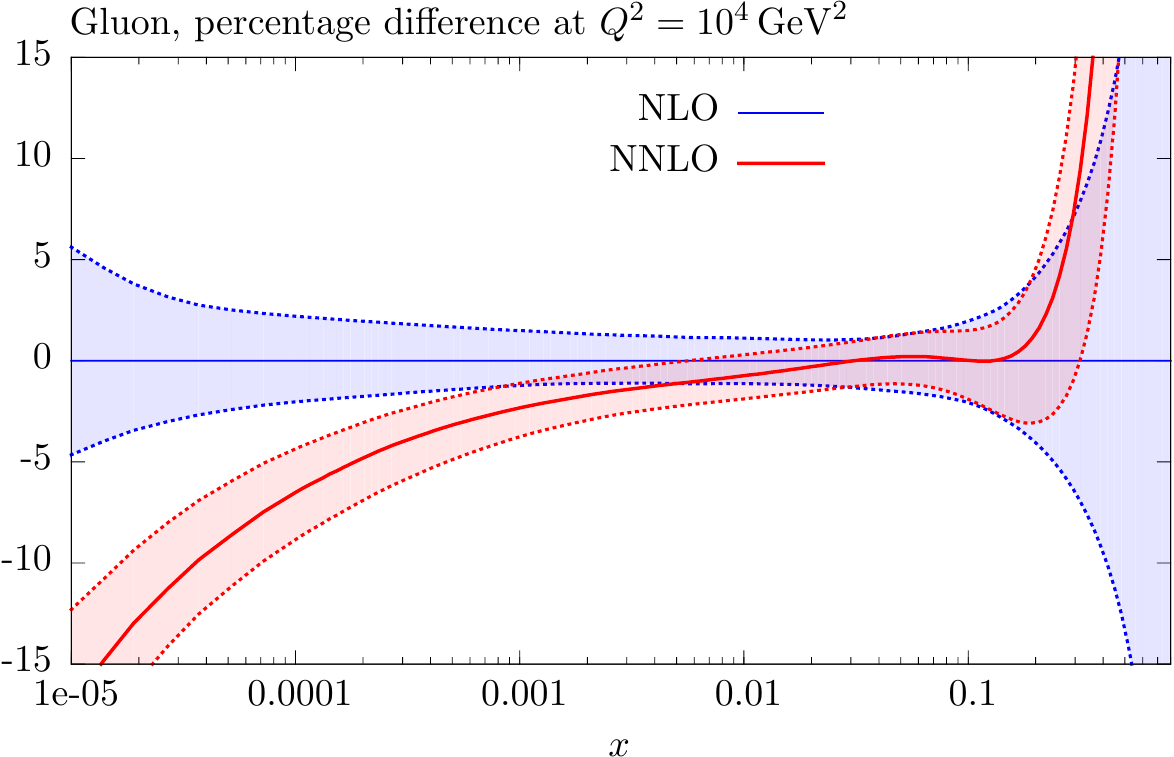}
\includegraphics[height=5cm]{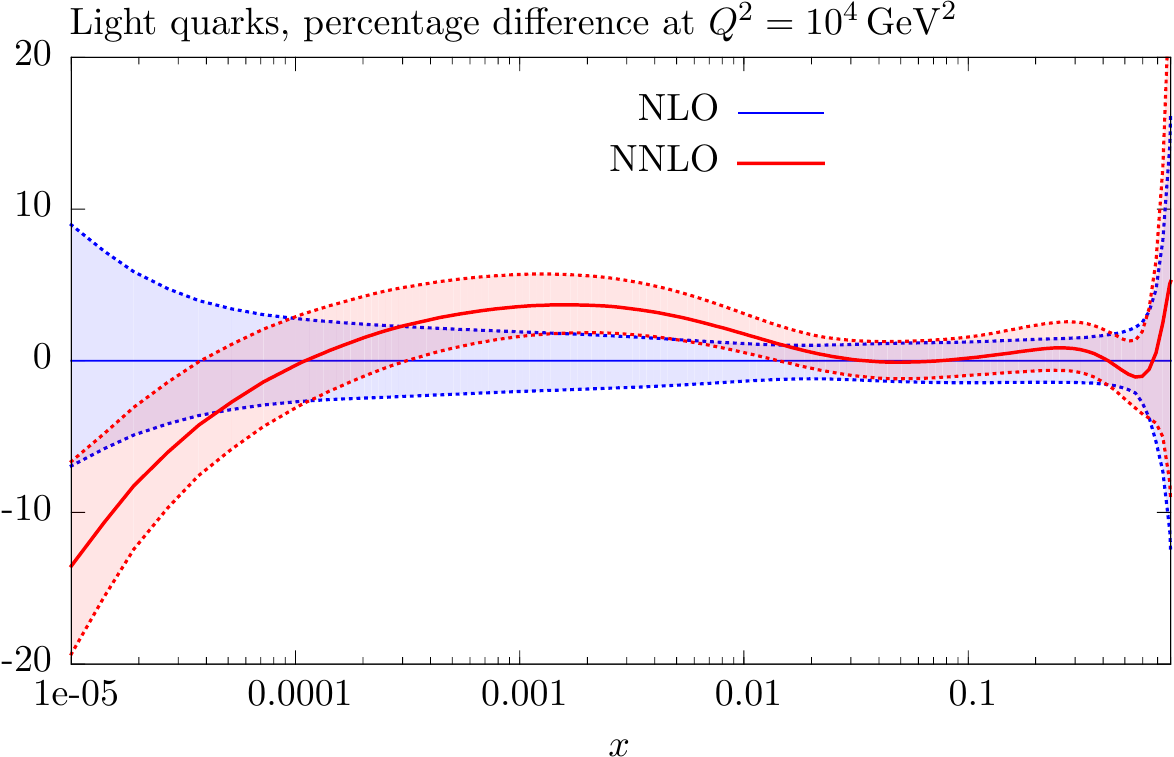}
\includegraphics[height=5cm]{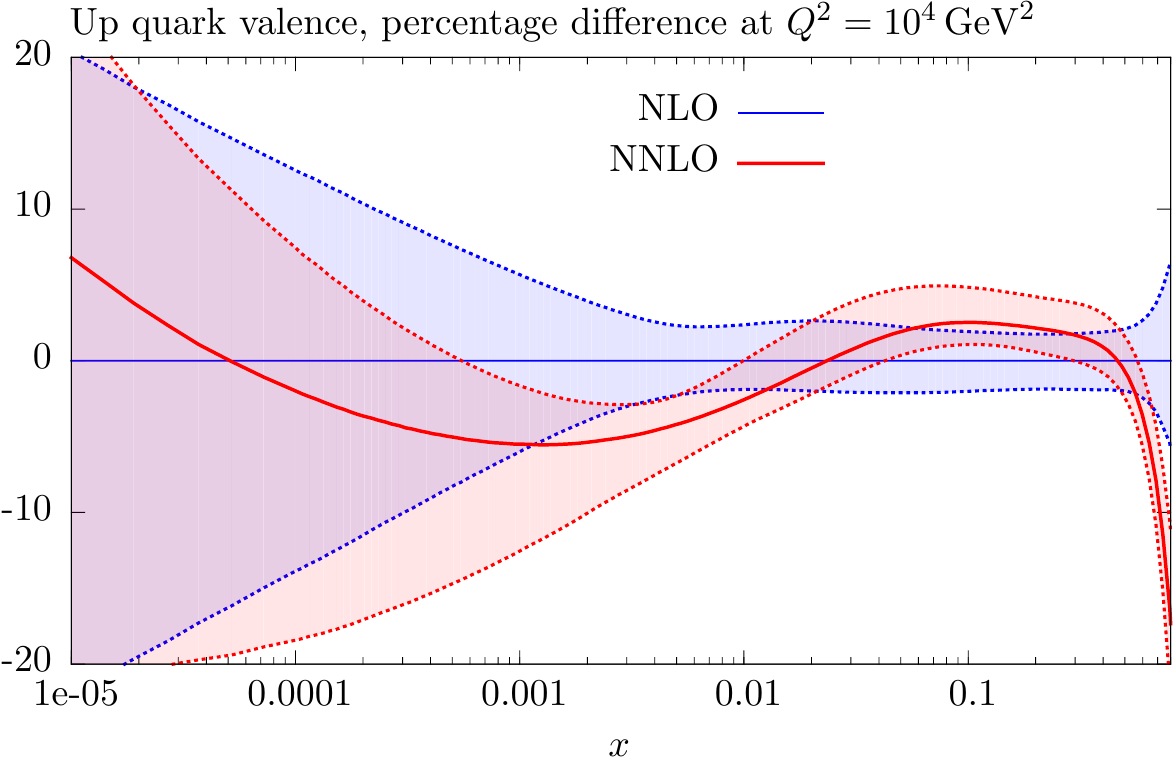}
\includegraphics[height=5cm]{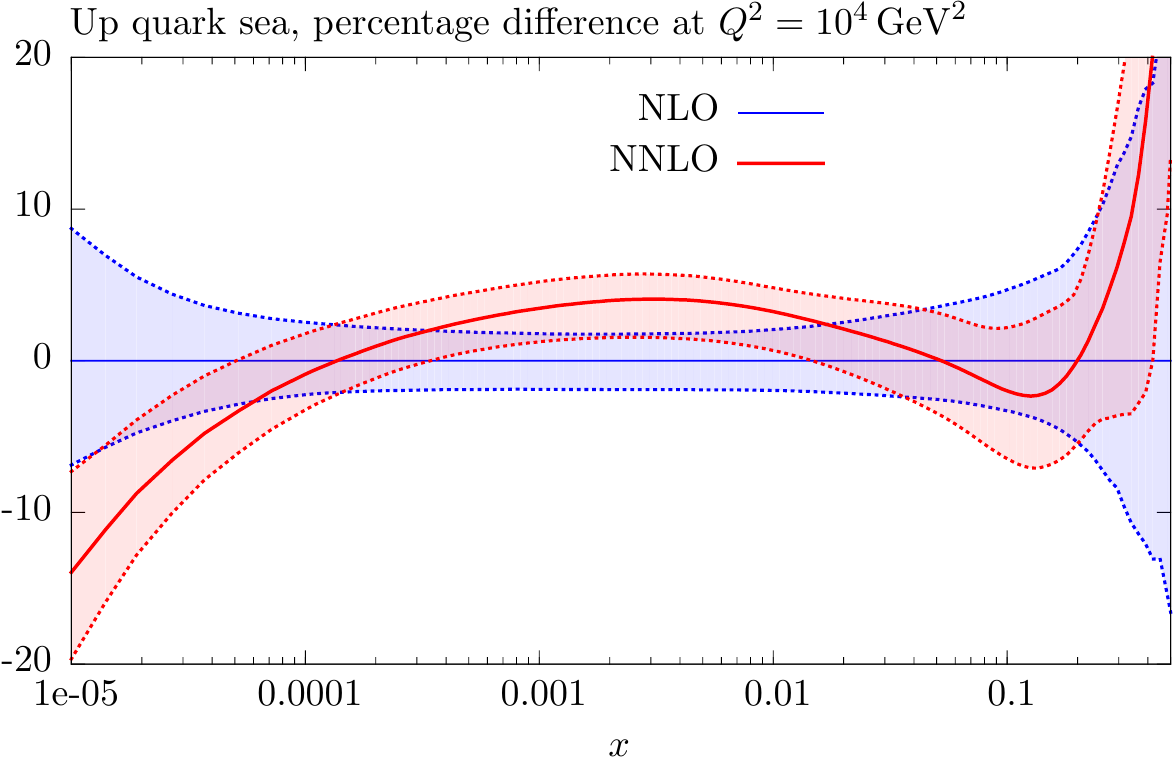}
\caption{\sf The comparison between the NLO and NNLO $g$, light quark, $u_V$ and $\bar u$ PDFs for $Q=10^4~\GeV^2$.}
\label{fig:NLONNLO}
\end{center}
\end{figure}

The comparison between some of the NLO and NNLO PDFs is shown in Fig.~\ref{fig:NLONNLO}. One can see that the NNLO
gluon is a little higher at highest $x$ and becomes smaller at the lowest $x$ values. The latter effect may be 
understood as being due to the slower evolution of the gluon at very small $x$ at NNLO as a consequence of
the correction to the splitting function. This is mirrored in the very small-$x$ behaviour of the light quarks
and the up sea quark, where the evolution is driven by the gluon. The change in shape of $u_V$ between NLO and 
NNLO is a consequence of the NNLO non-singlet coefficient function which is positive at very large $x$, leading 
to fewer quarks, and then becomes negative near $x=0.1$, leading to more valence quarks. The effect at high
$x$ is less clear in the $d_V$ distribution due to the freedom for the deuteron correction to be different at NNLO than at NLO. 
The sea quark is larger at NNLO for all $x<0.1$ until the lowest values. This is due to a negative NNLO 
structure function coefficient function in this region, which means the fit to data requires more sea quarks.
The shape is common to all light sea quarks, not just $\bar u$.  
This is also evident in the change in the light quark distribution. The heavy quarks are generated almost
entirely by evolution from the gluon, so their shape change is extremely similar to that of the gluon.  
The uncertainties at NLO and NNLO are very similar to each other, depending primarily on the uncertainties in
the data.

\section{Predictions and Benchmarks \label{sec:6}}

In Tables \ref{tab:sigmaNLO}  and \ref{tab:sigmaNNLO} we show the predictions 
for various benchmark processes at the LHC for the 
MSTW PDFs \cite{MSTW} and the MMHT sets of PDFs, also showing the results 
before LHC data are included in the fit for comparison (though the 
uncertainties are not calculated in this case). 
We calculate the total cross sections for $Z\to l^+l^-$, $W\to \l \nu$, 
Higgs production via gluon-gluon fusion and $t \bar t$ production. 
For $W, Z$ and Higgs 
production we use the same approach to calculation as used\footnote{We use the code provided by W.J. Stirling, based on the calculation in \cite{WZNNLO}, 
\cite{HiggsNNLO1} and \cite{HiggsNNLO2}.} in \cite{MSTW},
and improved in \cite{bench1}. For the $Z\to l^+l^-$ branching
ratio we use 0.033658 and for the $W\to l \nu$ we take 0.1080~\cite{PDG2014}.
We use LO electroweak perturbation theory, with the $qqW$ and $qqZ$ 
couplings defined by 
\begin{equation}
  g_W^2 =  G_F M_W^2 / \sqrt{2}, \qquad g_Z^2 = G_F M_Z^2 \sqrt{2}, 
\end{equation}
and other electroweak parameters are as in \cite{MSTW}. We take the Higgs mass to be 
$m_H=125~\GeV$,  and the top pole mass $m_t=172.5~\GeV$. For the $t \bar t$
cross section we use the calculation and code in \cite{topNNLO}. 
In all cases we use the particle mass as the renormalisation and 
factorisation scale. 
The main purpose of the presentation is to investigate how both the central
values and the uncertainties of the predictions have changed in going from 
MSTW2008 PDFs to MMHT2014 PDFs, so we provide results for the Tevatron and 
LHC with centre of mass energies 7~TeV and 14 ~TeV. This gives quite a spread
of energies whereas relative effects at 8~TeV and 13~TeV would be very similar
to those at 7~TeV and 14 ~TeV. We do not intend to present definite 
predictions or compare in detail to other PDF sets as both these results 
are frequently provided in the literature with very specific choices of 
codes, scales and parameters which may differ from those used here.

\begin{table}
\begin{center}
\vspace{-1.0cm}
\begin{tabular}{|l|c|c|c|}
\hline
& MSTW08 NLO& MMHT14 NLO no LHC& MMHT14 NLO    \\
\hline
$\!\! W\,\, {\rm Tevatron}\,\,(1.96~\TeV)$   & $2.659^{+0.057}_{-0.045}$    & $2.685$  & $2.645^{+0.058}_{-0.049}$    \\   
$\!\! Z \,\,{\rm Tevatron}\,\,(1.96~\TeV)$   & $0.2426^{+0.0054}_{-0.0043}$ & $0.2486$ & $0.2442^{+0.0049}_{-0.0043}$ \\    
$\!\! W^+ \,\,{\rm LHC}\,\, (7~\TeV)$        & $5.960^{+0.129}_{-0.097}$    & $6.107$  & $5.974^{+0.092}_{-0.086}$   \\    
$\!\! W^- \,\,{\rm LHC}\,\, (7~\TeV)$        & $4.192^{+0.092}_{-0.071}$    & $4.181$  & $4.163^{+0.069}_{-0.061}$   \\    
$\!\! Z \,\,{\rm LHC}\,\, (7~\TeV)$          & $0.931^{+0.020}_{-0.014}$    & $0.941$  & $0.932^{+0.013}_{-0.013}$   \\    
$\!\! W^+ \,\,{\rm LHC}\,\, (14~\TeV)$       & $12.07^{+0.24}_{-0.21}$      & $12.43$  & $12.17^{+0.20}_{-0.18}$   \\    
$\!\! W^- \,\,{\rm LHC}\,\, (14~\TeV)$       & $9.107^{+0.19}_{-0.16}$      & $9.16$   & $9.10^{+0.15}_{-0.14}$    \\    
$\!\! Z \,\,{\rm LHC}\,\, (14~\TeV)$         & $2.001^{+0.040}_{-0.032}$    & $2.035$  & $2.016^{+0.031}_{-0.033}$    \\
\hline    
$\!\!{\rm Higgs} \,\,{\rm Tevatron}$         & $0.658^{+0.021}_{-0.027}$    & $0.636$  & $0.644^{+0.021}_{-0.022}$    \\
$\!\!{\rm Higgs} \,\,{\rm LHC}\,\,(7~\TeV)$  & $11.39^{+0.16}_{-0.19}$      & $11.26$  & $11.28^{+0.21}_{-0.20}$    \\
$\!\!{\rm Higgs} \,\,{\rm LHC}\,\,(14~\TeV)$ & $37.93^{+0.42}_{-0.60}$      & $37.67$  & $37.63^{+0.67}_{-0.59}$    \\ 
\hline    
$\!\! t\bar t \,\,{\rm Tevatron}$            & $6.85^{+0.19}_{-0.13}$       & $6.89$   & $6.82^{+0.18}_{-0.17}$    \\
$\!\! t\bar t\,\,{\rm LHC}\,\,(7~\TeV)$      & $162.0^{+4.3}_{-5.4}$        & $157.0$  & $158.6^{+4.5}_{-4.5}$        \\
$\!\! t\bar t\,\,{\rm LHC}\,\,(14~\TeV)$     & $903.8^{+16}_{-17}$          & $886.7$  & $891.9^{+18}_{-18}$    \\ 

\hline
    \end{tabular}
\end{center}
\caption{\sf The values of various cross sections (in nb) obtained with the NLO 
MSTW 2008 parton sets \cite{MSTW} and the NLO MMHT 2014 sets.  
We show the values before and after the LHC data are included in the present 
fits, but not the uncertainty in the former case.The uncertainties are PDF uncertainties only.}
\label{tab:sigmaNLO}   
\end{table}

For the NLO PDFs one can see that there are no shifts in $W$ or $Z$ cross sections 
as large as the uncertainties when going from the MSTW2008 predictions to those of 
MMHT2014. The NLO values of the cross section for $Z$ production at the Tevatron and of $W^+$ production at the LHC do change by slightly more than one standard deviation on the non-LHC MMHT2014 fit, but the
inclusion of LHC data brings these cross sections back towards the MSTW2008 predictions. 
The uncertainties are generally slightly smaller when using the MMHT2014 PDFs, but this is 
a fairly minor effect.  For Higgs production via gluon-gluon fusion at NLO the changes are
all within one standard deviation, with a slight decrease in the MMHT2014 sets due to the 
slightly smaller high-$x$ gluon distribution. The uncertainties are slightly 
decreased with the new PDFs at low energy, but increase a little at higher energy. 
For $t \bar t$ production there is a slight decrease in the predicted cross section 
for the MMHT2014 set at the LHC, and as with Higgs production this is more of an effect
before LHC data are included. As with Higgs production this is due mainly
to the smaller gluon at high-$x$, with $\sigma_{\bar t t}$ probing higher $x$ than Higgs production.

\begin{table}
\begin{center}
\vspace{-1.0cm}
\begin{tabular}{|l|c|c|c|}
\hline
& MSTW08 NNLO& MMHT14 NNLO no LHC& MMHT14 NNLO    \\
\hline
$\!\! W\,\, {\rm Tevatron}\,\,(1.96~\TeV)$   & $2.746^{+0.049}_{-0.042}$    & $2.803$  & $2.782^{+0.056}_{-0.056}$    \\   
$\!\! Z \,\,{\rm Tevatron}\,\,(1.96~\TeV)$   & $0.2507^{+0.0048}_{-0.0041}$ & $0.2574$ & $0.2559^{+0.0052}_{-0.0046}$ \\    
$\!\! W^+ \,\,{\rm LHC}\,\, (7~\TeV)$        & $6.159^{+0.111}_{-0.099}$    & $6.214$  & $6.197^{+0.103}_{-0.092}$   \\    
$\!\! W^- \,\,{\rm LHC}\,\, (7~\TeV)$        & $4.310^{+0.078}_{-0.069}$    & $4.355$  & $4.306^{+0.067}_{-0.076}$   \\    
$\!\! Z \,\,{\rm LHC}\,\, (7~\TeV)$          & $0.9586^{+0.020}_{-0.014}$   & $0.9695$ & $0.9638^{+0.014}_{-0.013}$   \\    
$\!\! W^+ \,\,{\rm LHC}\,\, (14~\TeV)$       & $12.39^{+0.22}_{-0.21}$      & $12.49$  & $12.48^{+0.22}_{-0.18}$   \\    
$\!\! W^- \,\,{\rm LHC}\,\, (14~\TeV)$       & $9.33^{+0.16}_{-0.16}$       & $9.39$   & $9.32^{+0.15}_{-0.14}$    \\    
$\!\! Z \,\,{\rm LHC}\,\, (14~\TeV)$         & $2.051^{+0.035}_{-0.033}$    & $2.069$  & $2.065^{+0.035}_{-0.030}$    \\    
\hline    
$\!\! {\rm Higgs} \,\,{\rm Tevatron}$        & $0.853^{+0.028}_{-0.029}$    & $0.877$  & $0.874^{+0.024}_{-0.030}$    \\
$\!\!{\rm Higgs} \,\,{\rm LHC}\,\,(7~\TeV)$  & $14.40^{+0.17}_{-0.23}$      & $14.54$  & $14.56^{+0.21}_{-0.29}$    \\
$\!\!{\rm Higgs} \,\,{\rm LHC}\,\,(14~\TeV)$ & $47.50^{+0.47}_{-0.74}$      & $47.61$  & $47.69^{+0.63}_{-0.88}$    \\ 
\hline    
$\!\! t\bar t \,\,{\rm Tevatron}$            & $7.19^{+0.17}_{-0.12}$       & $7.54$   & $7.51^{+0.21}_{-0.20}$    \\
$\!\! t\bar t\,\,{\rm LHC}\,\,(7~\TeV)$      & $171.1^{+4.7}_{-4.8}$        & $176.5$  & $175.9^{+3.9}_{-5.5}$        \\
$\!\! t\bar t\,\,{\rm LHC}\,\,(14~\TeV)$     & $953.3^{+16}_{-18}$          & $969.0$  & $969.9^{+16}_{-20}$    \\ 

\hline
    \end{tabular}
\end{center}
\caption{\sf The values of various cross sections (in nb) obtained with the NNLO 
MSTW 2008 parton sets \cite{MSTW} and the NNLO MMHT 2014 sets.  
We show the values before and after the LHC data are included in the present 
fits, but not the uncertainty in the former case.The uncertainties are PDF uncertainties only.}
\label{tab:sigmaNNLO}   
\end{table}

The trend is the same for the predictions for $W$ and $Z$ cross sections at NNLO. 
There is generally a slight increase from the use of the MMHT2014 sets, but, with
the marginal exception of $Z$ production at the Tevatron, this change is always within
one standard deviation for the full MMHT2014 PDFs. It is sometimes slightly more 
than this 
when using the non-LHC data MMHT2014 sets, and again the inclusion of LHC data brings 
MMHT2014 closer to MSTW2008. For the Higgs cross sections via gluon-gluon fusion there is 
consistently a very small increase. This is because even though the gluon distribution
decreases in the most relevant $x$ region, i.e. $x \approx 0.06$ for 
$\sqrt{s}=1.96~\TeV$ 
and i.e. $x \approx 0.009$ for $\sqrt{s}=14~\TeV$, the coupling constant has increased, 
and this slightly overcompensates the smaller gluon. If the predictions are made using 
the absolutely best fit PDFs with $\alpha_S(M_Z^2)=0.1172$ the Higgs predictions 
decrease compared to MSTW2008, but again by much less than the uncertainty. As at NLO
the MMHT2014 uncertainties have reduced a little at the highest energies but increased 
at higher energies.  For $t \bar t$ production there is an increase in 
the cross section for the MMHT PDFs of about $4-5\%$ at the Tevatron and $2-3\%$ at the 
LHC, 
with again the effect being slightly larger before LHC data are included. This is 
partially
due to the larger coupling in the MMHT sets, with the change being reduced to about 
$3\%$ at the Tevatron and $1-2\%$ if the MMHT2014 absolute best fit set with 
$\alpha_S(M_Z^2)=0.1172$ is used. The remainder of the effect is due to the enhancement
of the very high$-x$ gluon at NNLO in MMHT2014. The change is in some cases more than 
one standard deviation from the best MSTW prediction, but only when compared to just the
PDF uncertainties. If predictions with common $\alpha_S(M_Z^2)$ are compared, 
or PDF + $\alpha_S(M_Z^2)$ uncertainties taken into account the changes are at most about
one standard deviation.

\section{Other constraining data: dijet, $W+c$, differential $t \bar t$  \label{sec:7}}

As well as improvements in the type of data we currently include in 
the PDF analysis there are currently a variety of new forms 
of LHC data  being released , which will also provide new, sometimes complementary, constraints 
on PDFs. Some of the most clear examples of these are dijet data
\cite{ATLAS-jet7, CMS-jet7,ATLASdijet}, top quark differential 
distributions \cite{ATLASdifftop1, CMSdifftop} 
and $W^-+c$ (and $W^++\bar c$) 
production \cite{ATLASWc,CMSWc}. The first two 
should help constrain the high-$x$ gluon and the last is a direct constraint 
on the strange quark distribution. None of these have been included 
in our current analysis, either because suitably accurate data satisfying 
our cut-off on the publication date, was not available or because there is some 
limitation in the theoretical precision, or both. Nevertheless, we briefly 
comment on the comparison with each set of data.

\begin{figure} 
\begin{center}
\includegraphics[height=6.5cm]{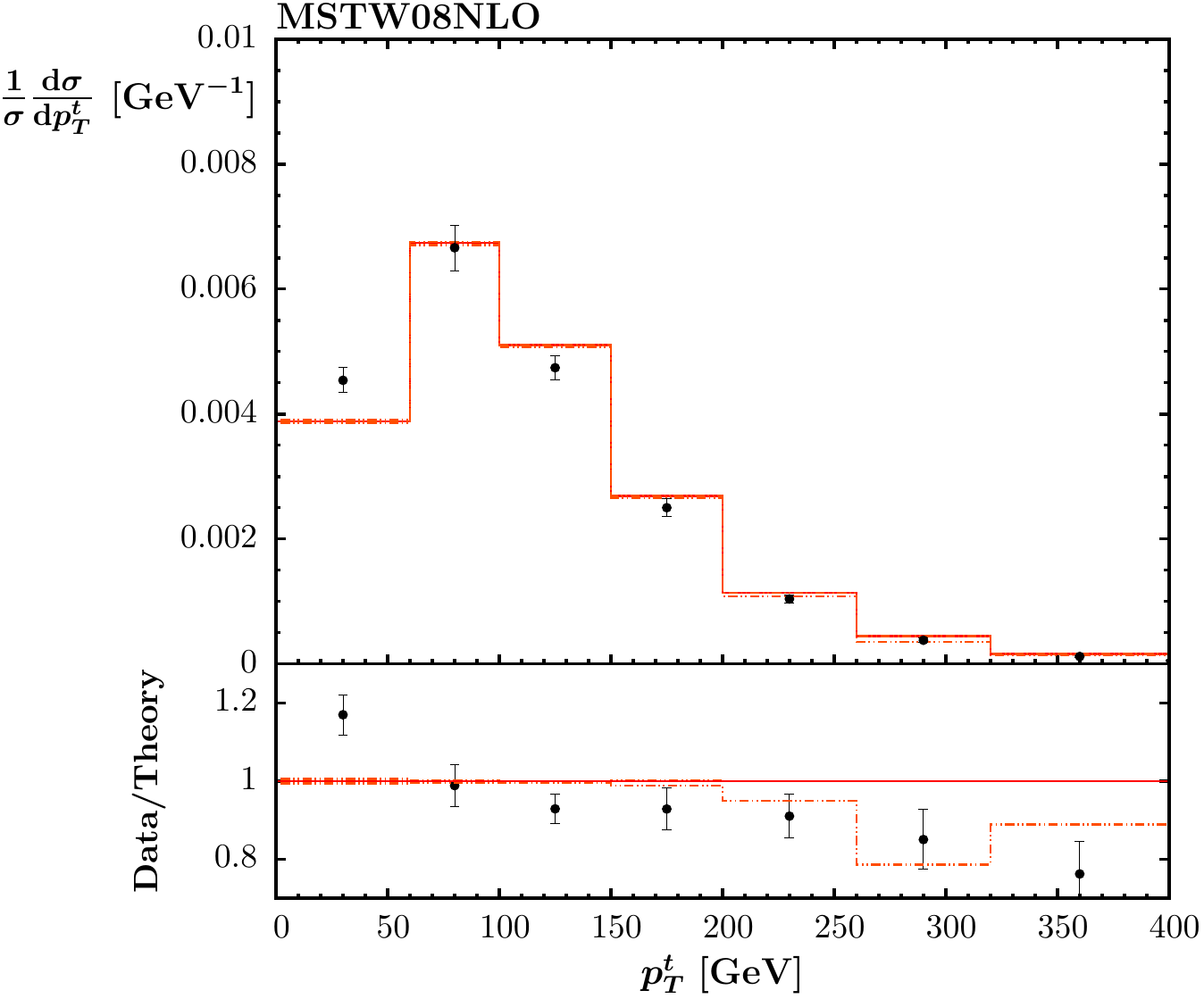}
\includegraphics[height=6.5cm]{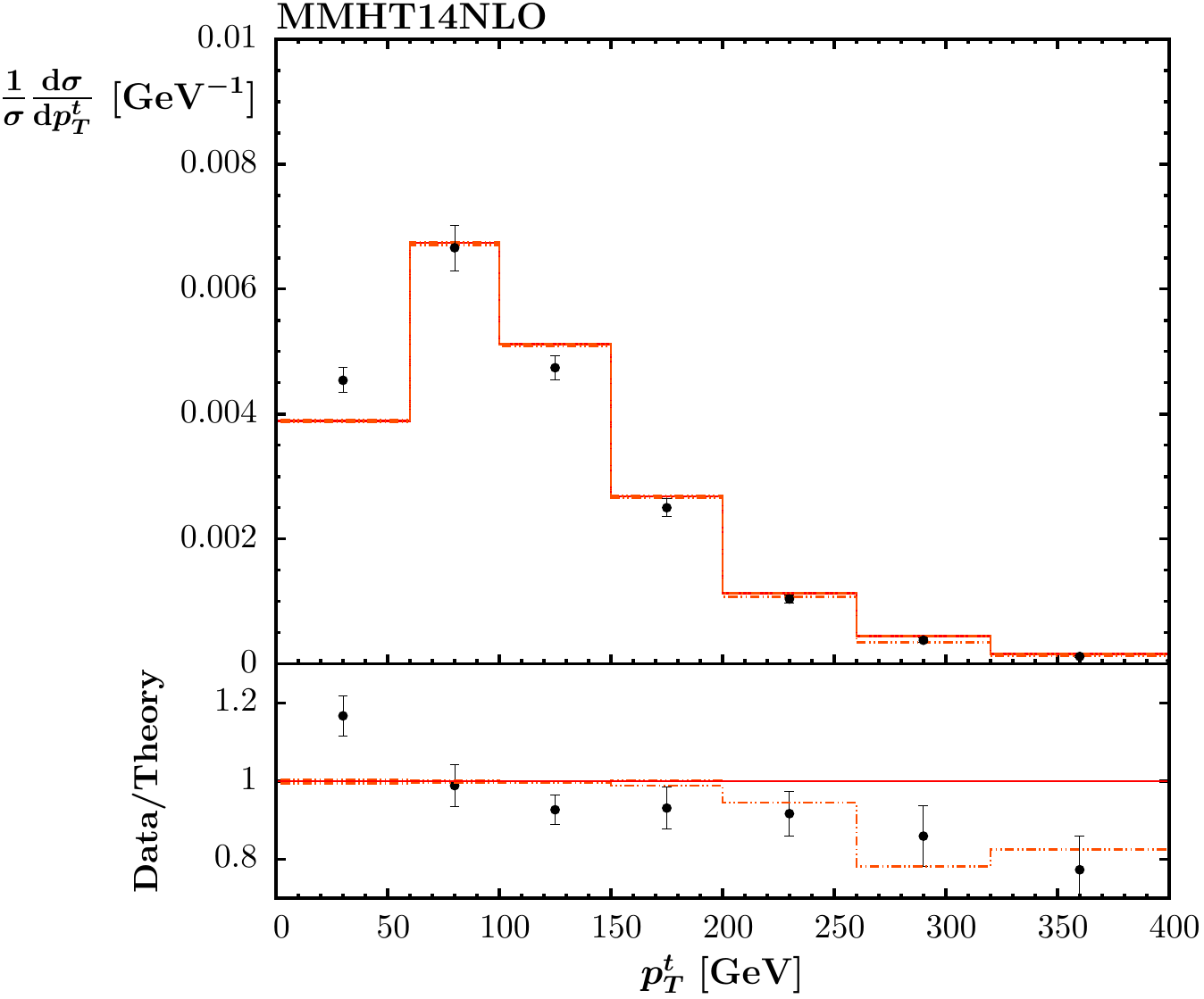}\\
\includegraphics[height=6.5cm]{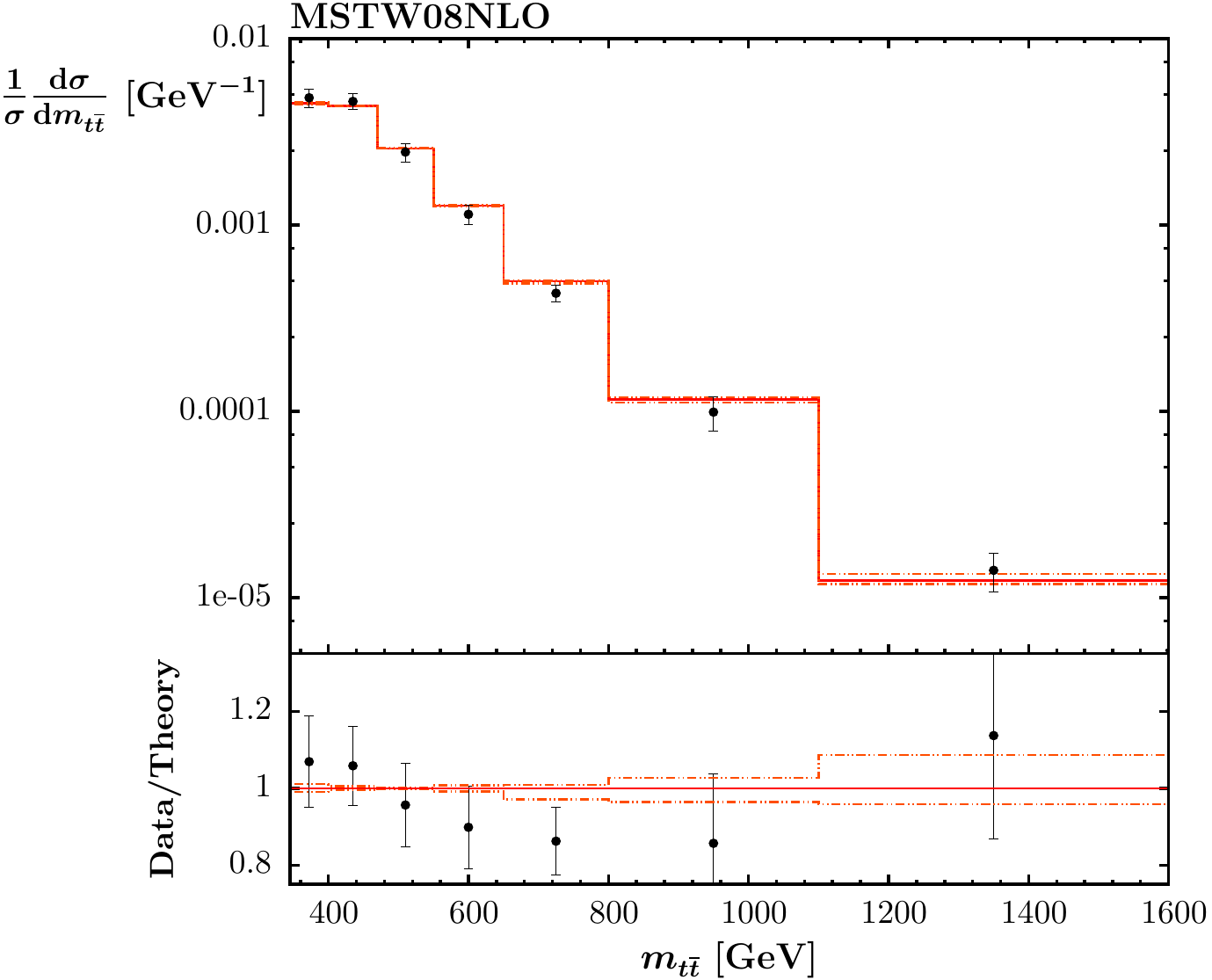}
\includegraphics[height=6.5cm]{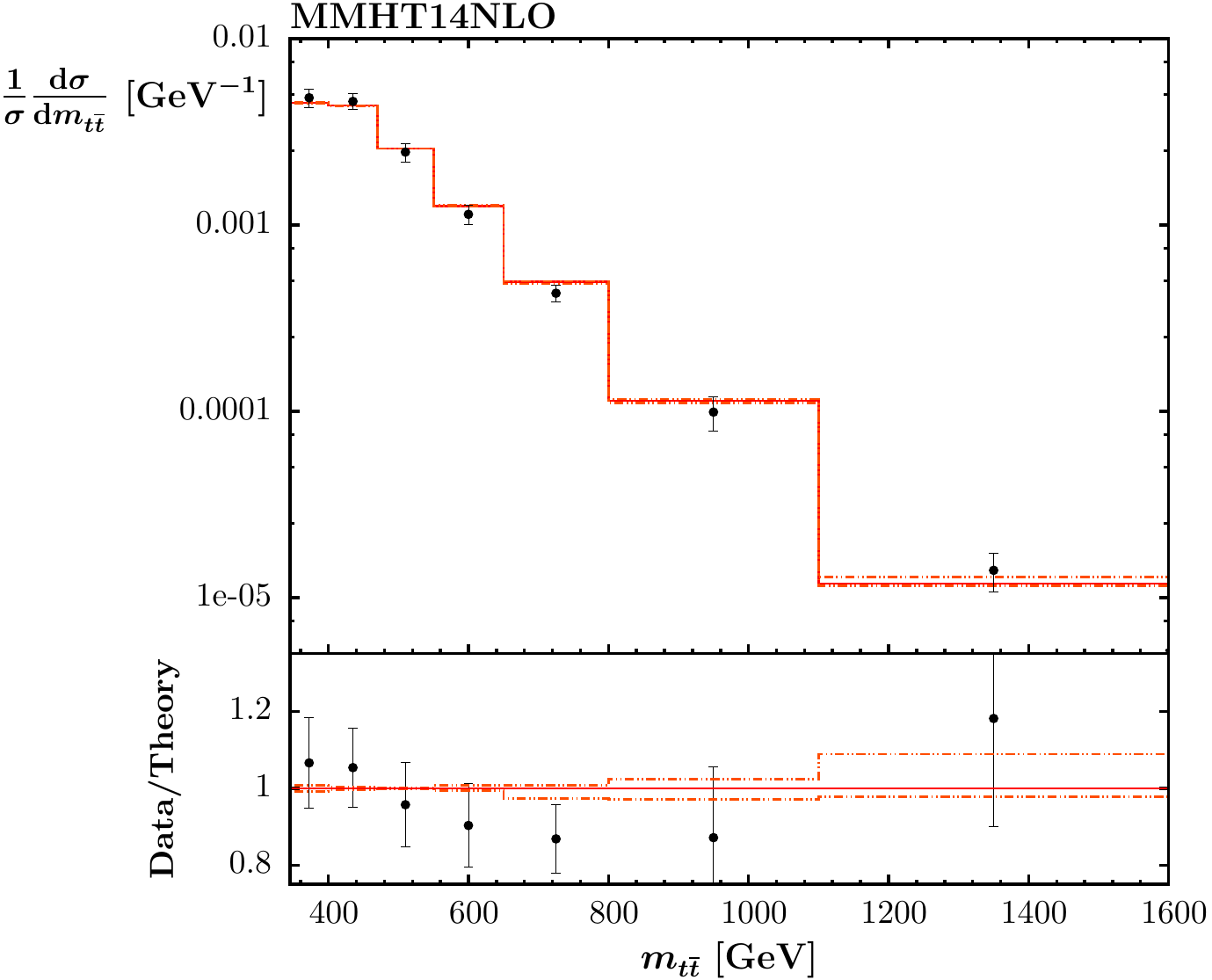}\\
\includegraphics[height=6.5cm]{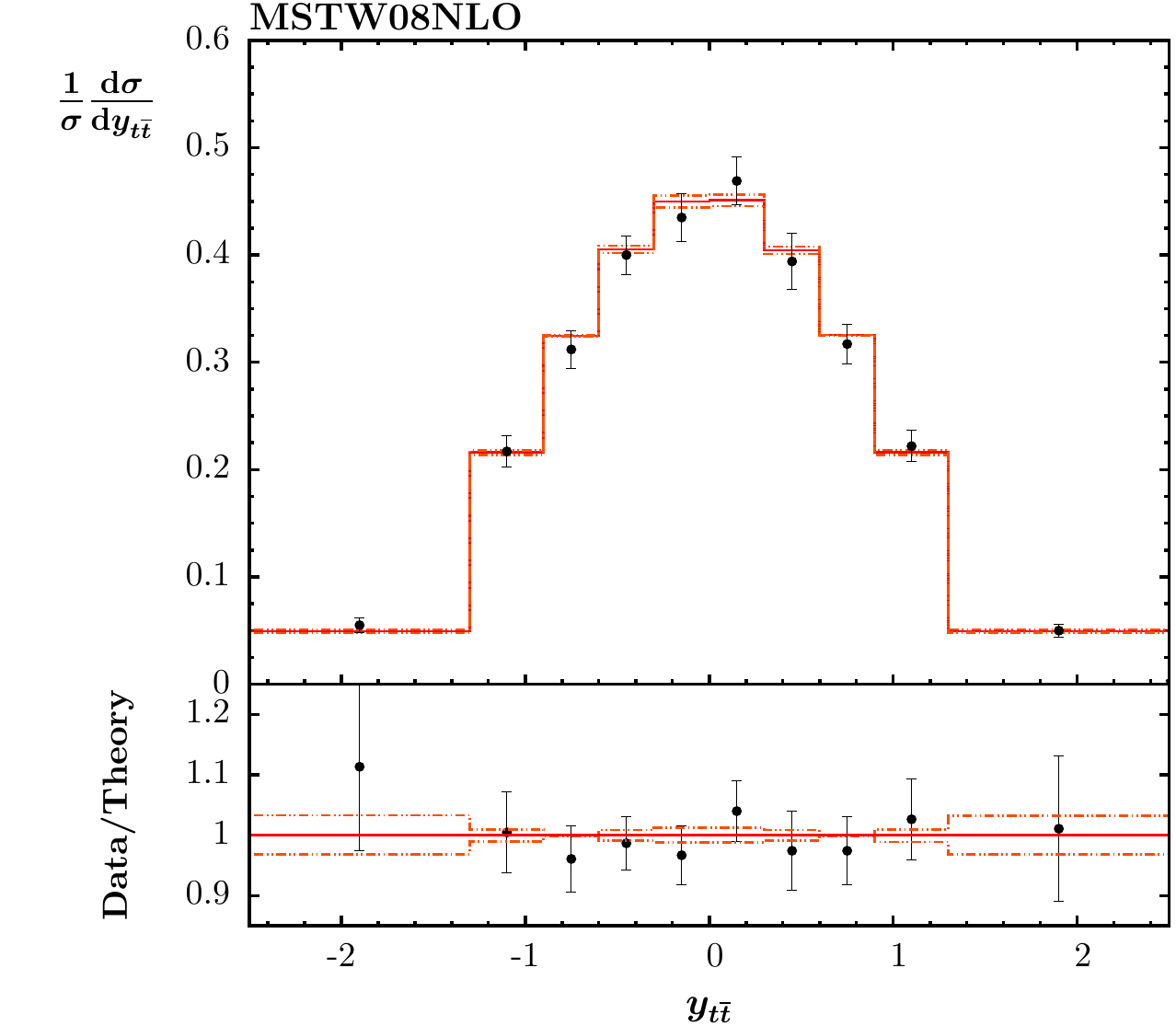}
\includegraphics[height=6.5cm]{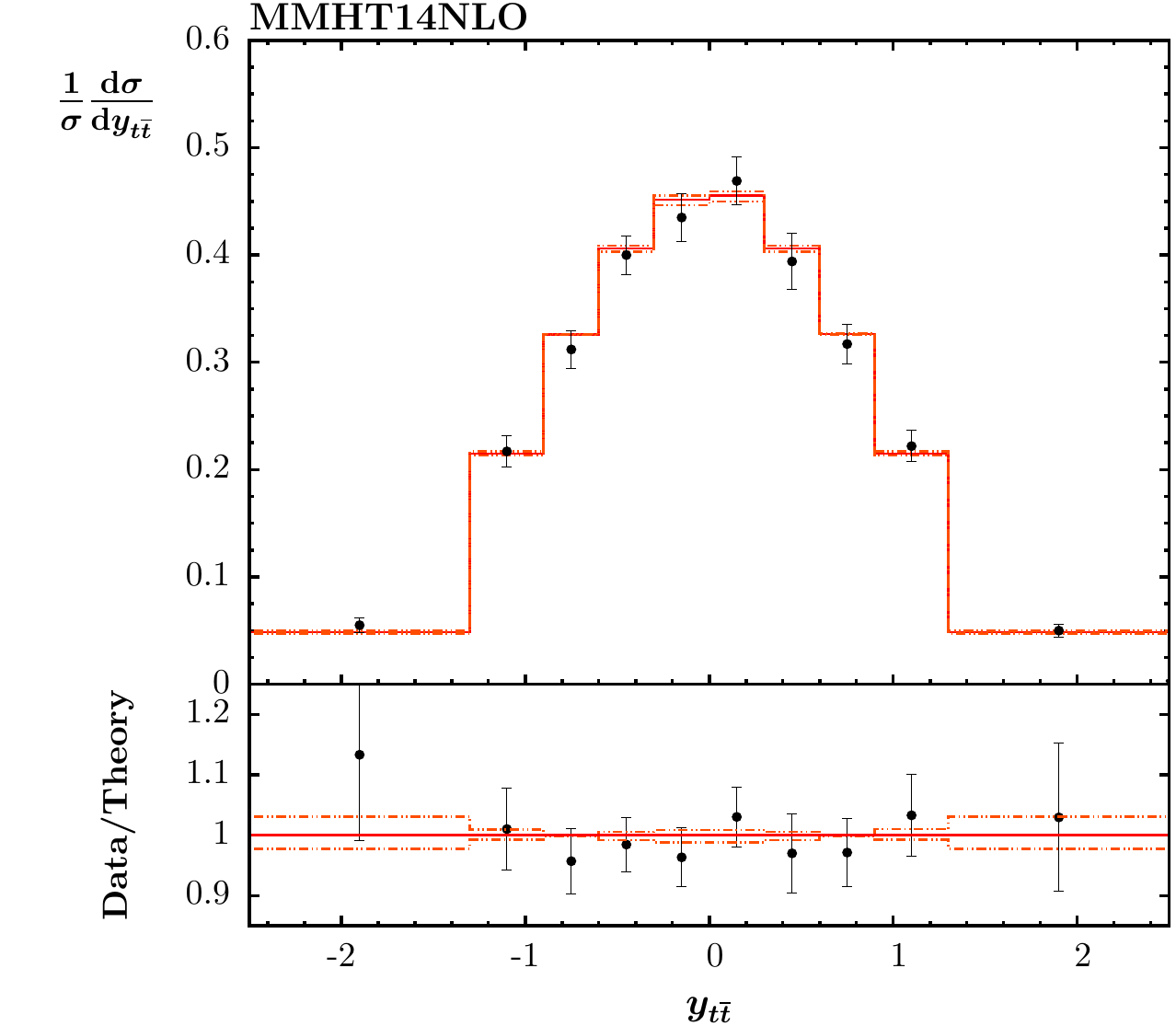}
\caption{\sf The CMS differential top quark data as functions of $p_T^t$ (top pair of plots), of $m_{t\bar t}$  (middle plots), and of $y_{t\bar t}$ (bottom plots), compared to the predictions of
 the MSTW2008 PDFs (left) and MMHT2014 PDFs (right). The dotted lines represent 
the PDF uncertainties.}
\label{fig:CMStopdiff}
\end{center}
\end{figure}

\subsection{Dijet production at the LHC}
The comparison to the dijet data in \cite{ATLAS-jet7,CMS-jet7} was studied 
in \cite{Wattjets}. It was clear that at high rapidity there was a 
significant difference in conclusions depending on which scale choice was 
used, i.e. one depending just on $p_T$ or one with rapidity dependence as 
well. There is also double counting between the events included in the 
inclusive and the dijet data. In \cite{ATLASdijet} the data are limited to 
relatively low rapidity, and full account of correlations between 
data sets is taken. The analysis in \cite{ATLASdijet} shows that for 
the full data sample MSTW2008 PDFs fit extremely well, better than most 
alternatives, and, as seen in this article, there should be little change 
if the MMHT2014 PDFs are used. We will include appropriate dijet data samples 
in the future. However, we will probably wait for the complete NNLO formulae for the cross sections to 
become available, before including them in the NNLO analysis.  
We also note that MSTW2008 PDFs give an excellent description of the
higher luminosity $7~\TeV$ ATLAS jet data \cite{ATLAShighlumijet}, so 
presumably MMHT2014 PDFs will as well.

\subsection{$W+$charm jet production}
We also compare to the CMS  \cite{CMSWc}  $W$ plus charm jet data with total cross section on $W$ plus
charm jets, satisfying $p_T^{\rm jet} > 25$ GeV and $|\eta^{\rm jet}|<2.5$, for
two values of the cut on the $W$ decay lepton: $p_T^{\rm lep} > 25$ GeV and $p_T^{\rm lep} > 35$ GeV.
The results are shown in Table \ref{tab:Wc} for the total $W+c$ cross section and for the ratio $R_c^{\pm} \equiv \sigma(W^+ + \bar{c} +X)/\sigma(W^- +c+X)$.
 The predictions are calculated using
MCFM, and we get completely consistent results with the data in \cite{CMSWc} when 
using the NNLO MSTW 2008 PDFs and $m_c=1.5~\GeV$. However, since the cross section is calculated at 
NLO, we use NLO PDFs, and we take our default mass to be $m_c=1.4~\GeV$.  (This change in mass increases the cross
sections by about $1\%$, though a little more in the lower than the higher $p_T^{\rm lep}$ bin.) 
The cross sections are then slightly larger than quoted in
\cite{CMSWc}, but still below the data. The ratio of $c$ to $\bar c$ production is 
slightly lower than the data, but consistent. When using MMHT2014 the cross sections 
increase by a few percent, and are actually slightly larger than the data, though well within the data uncertainty. 
The PDF uncertainty  in the cross section is now very much larger, reflecting the increase in the uncertainty
on the total $s + \bar s$ production. The ratios are slightly lower, and the uncertainty is
very similar to that with MSTW2008, reflecting the fact that the uncertainty on $s -\bar s$ is
essentially unchanged. The ATLAS measurements \cite{ATLASWc} are 
not corrected to the parton level, so cannot be directly compared. However, they appear 
to be a few percent higher than the CMS measurements. This is in reasonable disagreement with MSTW2008 PDFs,
but appears very likely to be fully consistent with MMHT2014 PDFs. The ratio, where non-perturbative corrections
presumably largely cancel, is close to 0.90, so is again likely to be 
very compatible with the MMHT2014 prediction.

\begin{table} [h]
\begin{center}
\begin{tabular}{|c|c|c|c|c|}\hline
                                      &  GeV& data & MSTW2008 &  MMHT2014 \\ \hline
  $\sigma(W + c)$ & $p_T^{\rm lep}>25$ & $107.7 \pm 3.3 ({\rm stat.}) \pm 6.9 ({\rm sys.})$     & $102.8 \pm 1.7$ & $110.2 \pm 8.1$\\
  $\sigma(W + c)$&$  p_T^{\rm lep}>35$ & $84.1 \pm 2.0 ({\rm stat.}) \pm 4.9 ({\rm sys.})$      & $80.4 \pm 1.4$  & $86.5 \pm 6.5$\\
  $R^{\pm}_c $&$ p_T^{\rm lep}>25$     & $0.954 \pm 0.025 ({\rm stat.}) \pm 0.004 ({\rm sys.})$ & $0.937 \pm 0.029$ & $0.924 \pm 0.026$\\
  $R^{\pm}_c  $&$ p_T^{\rm lep}>35$    & $0.938 \pm 0.019 ({\rm stat.}) \pm 0.006 ({\rm sys.})$ & $0.932\pm 0.030$ & $0.904 \pm 0.027$\\
 \hline
\end{tabular}
\end{center}
\caption{\sf The values of the total $W+c$ cross section (in pb), and the $W^+/W^-$ ratio $R_c^{\pm}$, measured by CMS \cite{CMSWc}, compared with the predictions obtained using MSTW2008 and MMHT2014 NLO PDFs.  The charm jet is subject to the acceptance cuts $p_T^{\rm jet} > 25$ GeV and $|\eta^{\rm jet}|<2.5$.}
\label{tab:Wc}
\end{table}

\subsection{Differential top-quark-pair data from the LHC \label{sec:7.3}}
Finally we compare to some recent differential top quark data \cite{CMSdifftop}. 
The comparison between NLO theory, with the calculation performed using MCFM \cite{MCFMtop},  
and data are shown for both MSTW2008 and 
MMHT2014 in Fig.~\ref{fig:CMStopdiff} as functions of $p_T^t$, of $m_{t \bar t}$, and of
$y_{t\bar t}$. One can see that the $p_T^t$ distribution of the data falls more quickly than
the prediction. The same is arguably true, to a lesser extent, for the $m_{t \bar t}$ 
distribution, except for the last point, while the rapidity distribution matches 
data very well. The same trend is true for the other `top' data sets. However, there is an 
indication \cite{Difftop} that NNLO corrections soften the $p^t_T$ distribution in particular, so
the relatively poor comparison may be due mainly to the missing higher-order corrections.

\section{Comparison of MMHT with other available PDFs   \label{sec:8}}   
 
\begin{figure} [t]
\begin{center}
\includegraphics[height=5cm]{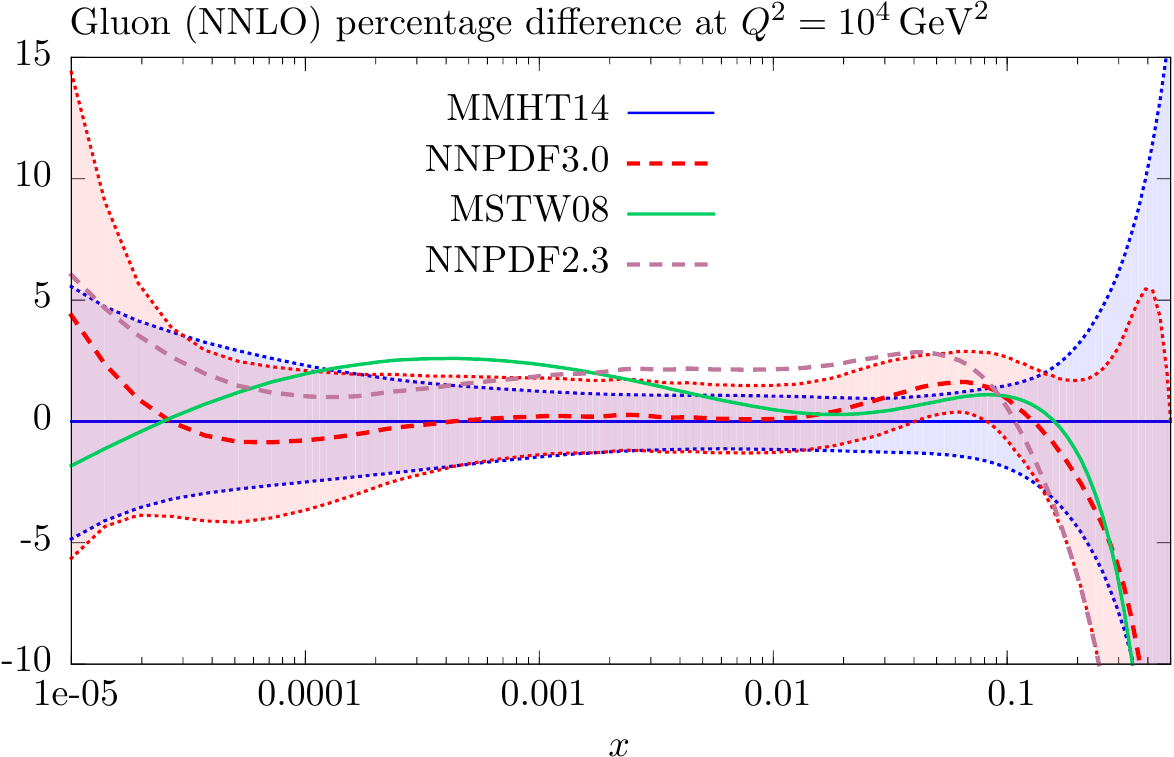}
\includegraphics[height=5cm]{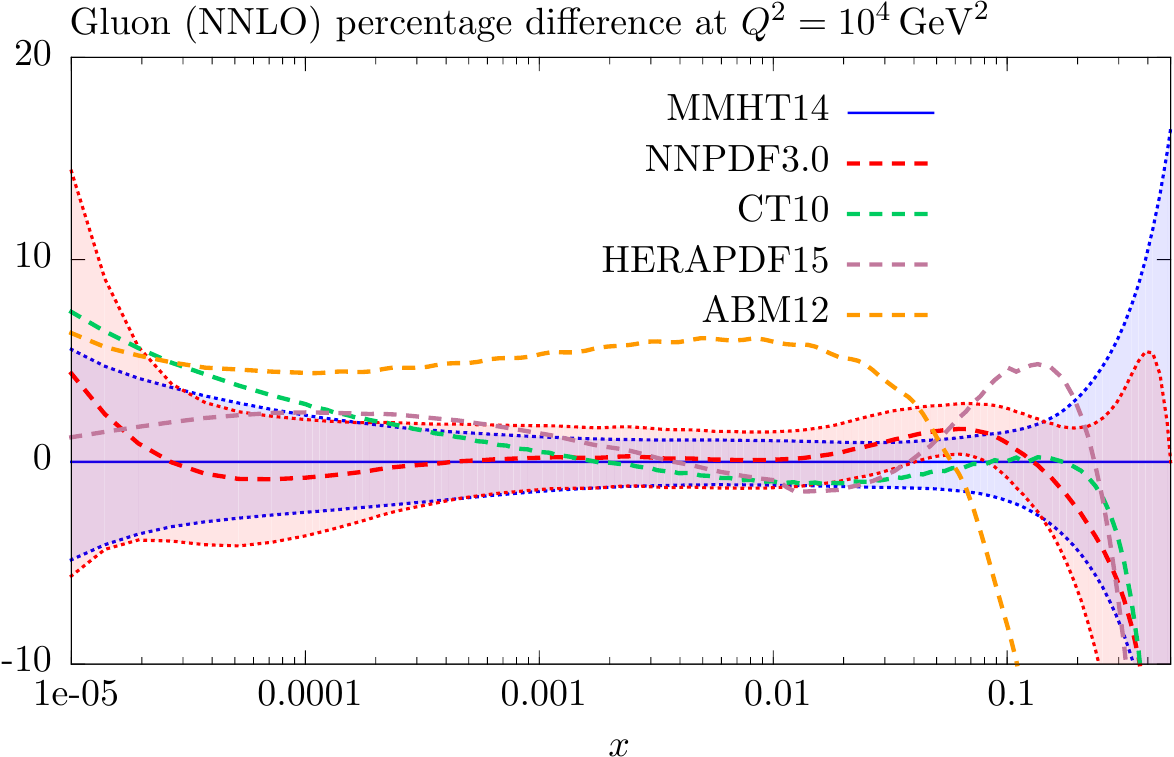}
\includegraphics[height=5cm]{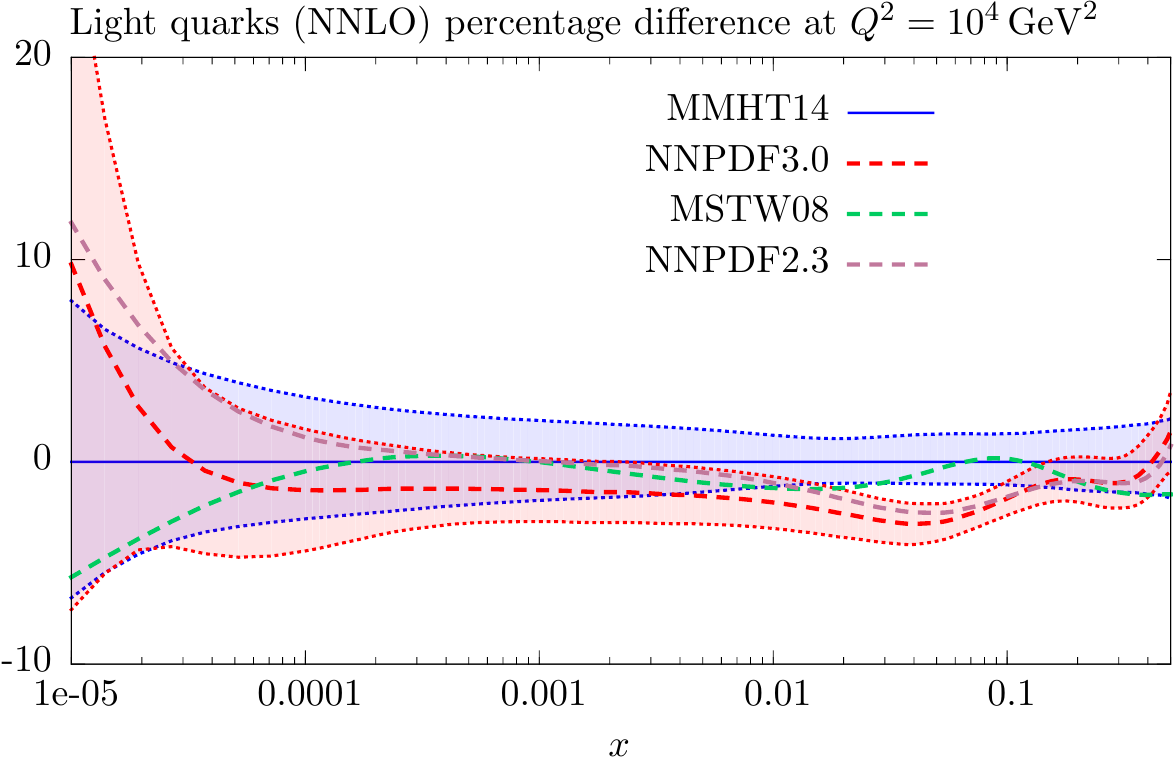}
\includegraphics[height=5cm]{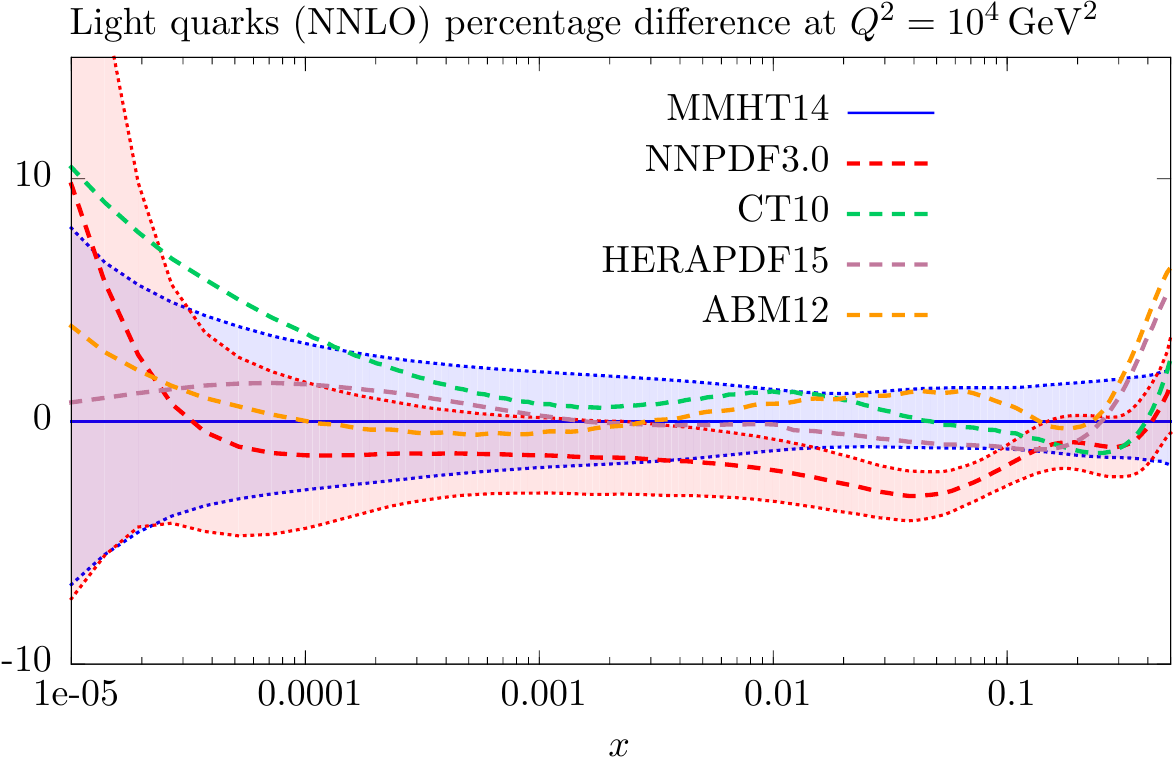}
\caption{\sf The comparison between NNLO NNPDF3.0 and MMHT14  PDFs at 
$Q^2=10^4~\GeV^2$ showing the  $g$ and light quark PDFs.  
Also shown (without error corridors, which would be similar to those of the 
newer sets in most cases) are the NNPDF2.3 and MSTW08 PDFs (left) 
which they supersede and (right) CT10 HERAPDF1.5 and ABM12 PDFs.}
\label{fig:NNPDFMMHTsing}
\end{center}
\end{figure}

\begin{figure} [t]
\begin{center}
\includegraphics[height=5cm]{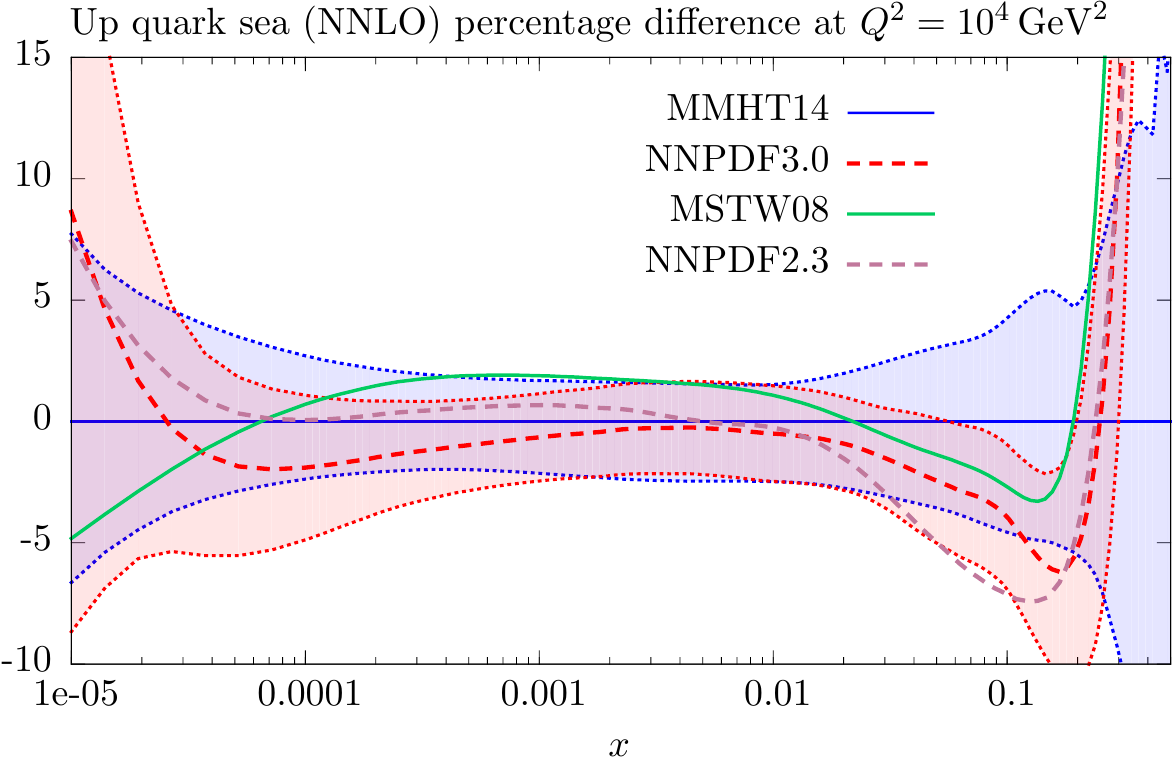}
\includegraphics[height=5cm]{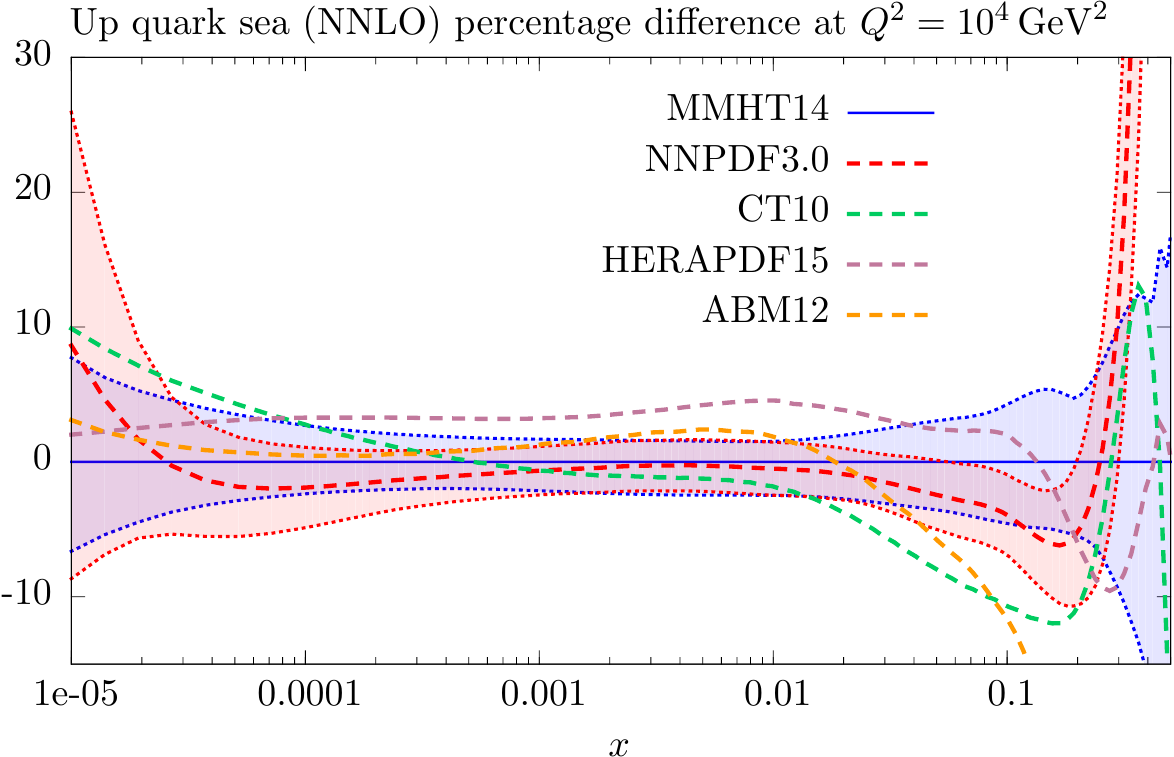}
\includegraphics[height=5cm]{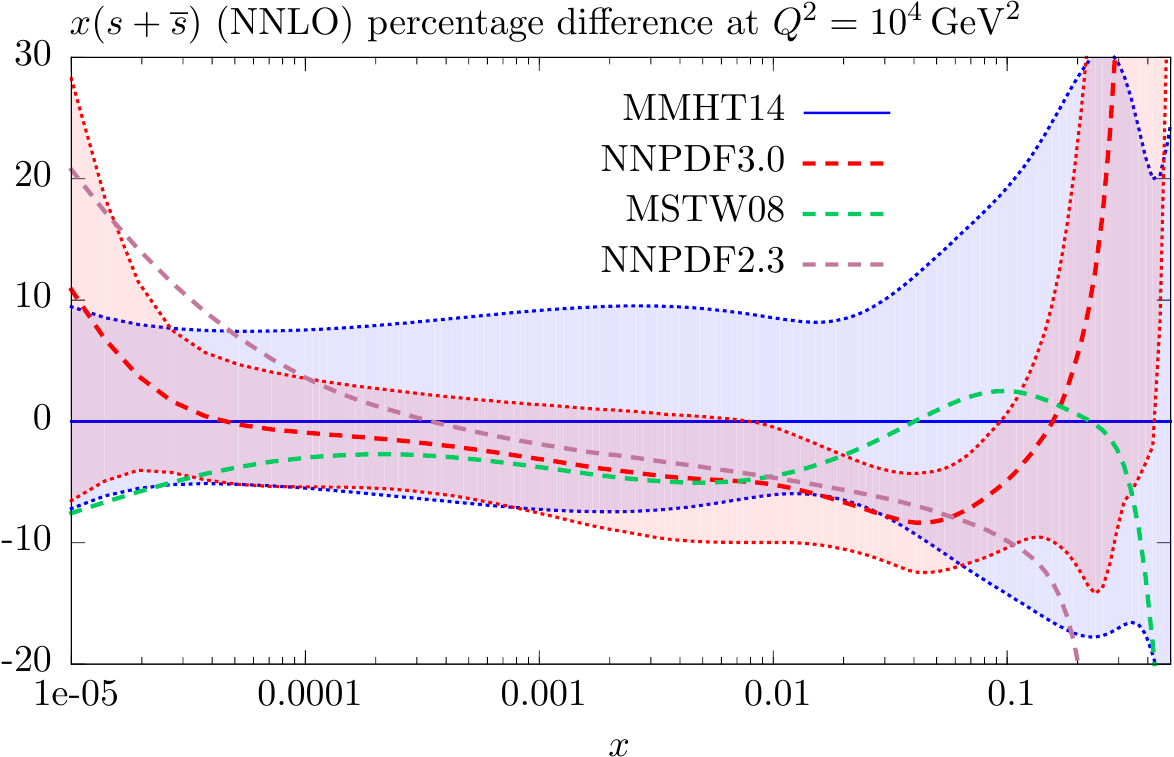}
\includegraphics[height=5cm]{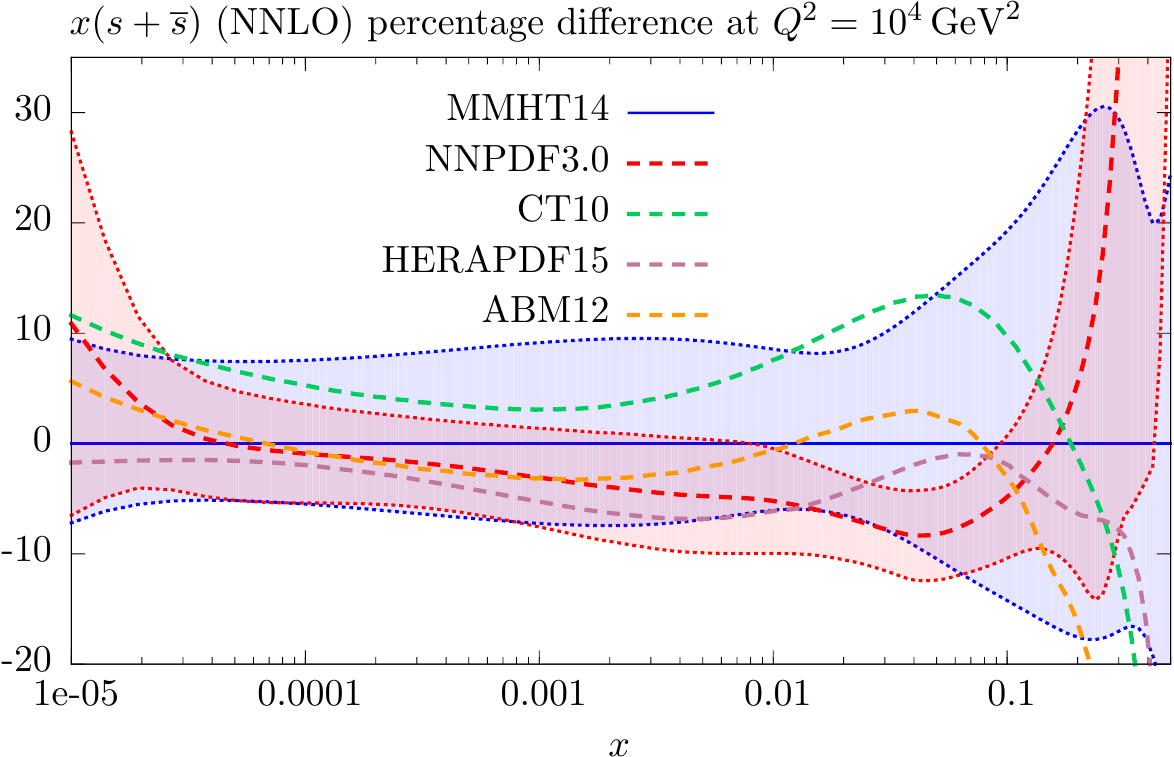}
\caption{\sf The comparison between NNLO NNPDF3.0 and MMHT14  PDFs at 
$Q^2=10^4~\GeV^2$ showing the  $\bar u$ and $s + \bar s$ quark PDFs.  
Also shown (without error corridors) are the NNPDF2.3 and MSTW08 PDFs (left) 
which they supersede and (right) CT10 HERAPDF1.5 and ABM12 PDFs..}
\label{fig:NNPDFMMHTsea}
\end{center}
\end{figure}

\begin{figure} [t]
\begin{center}
\includegraphics[height=5cm]{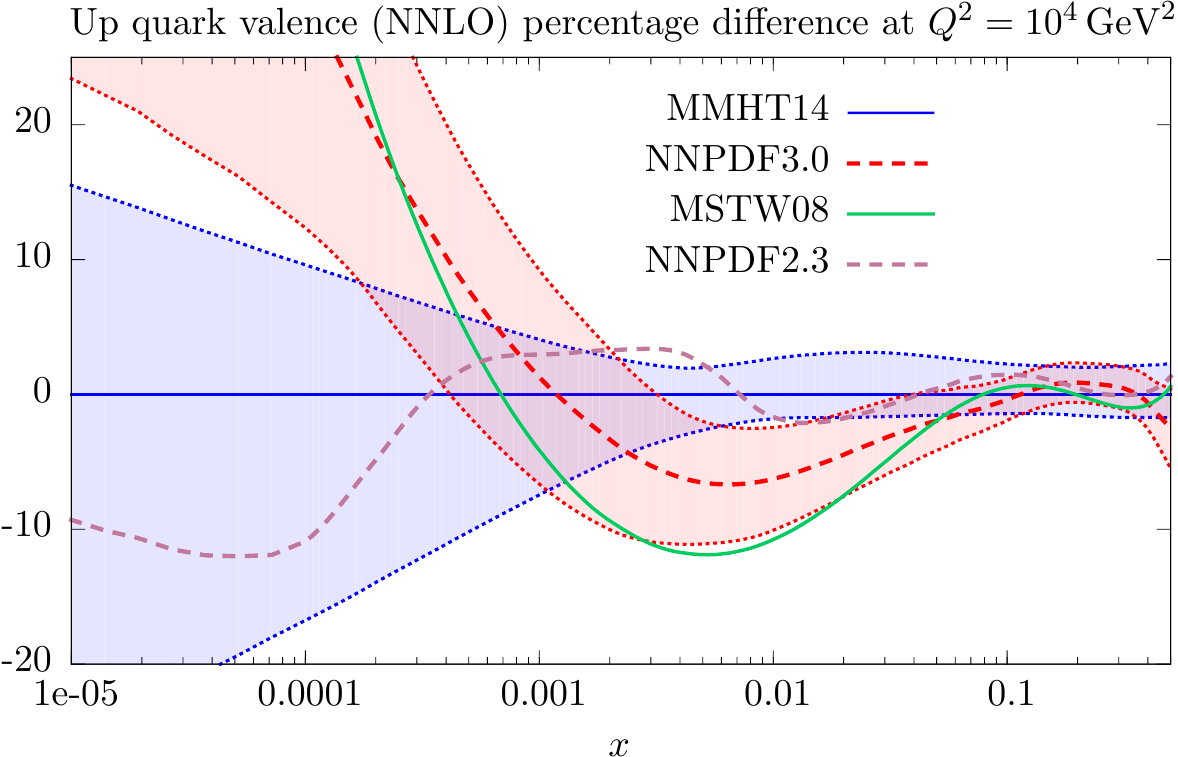}
\includegraphics[height=5cm]{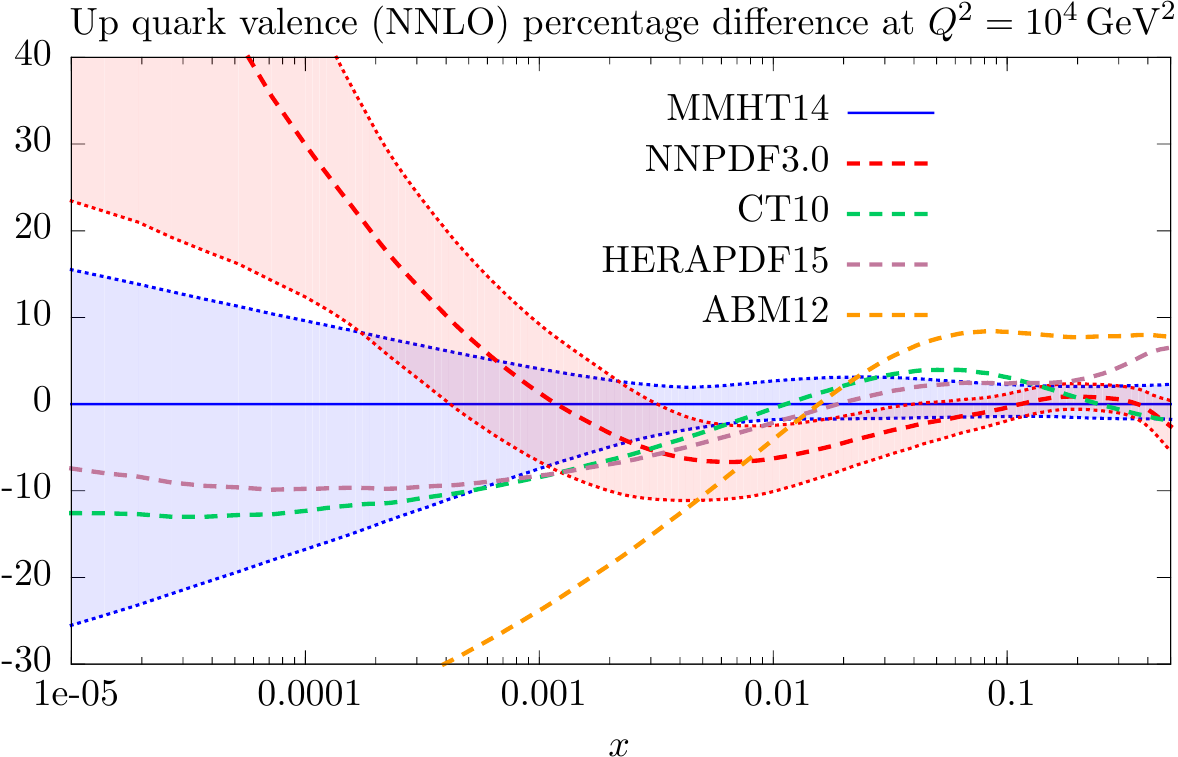}
\includegraphics[height=5cm]{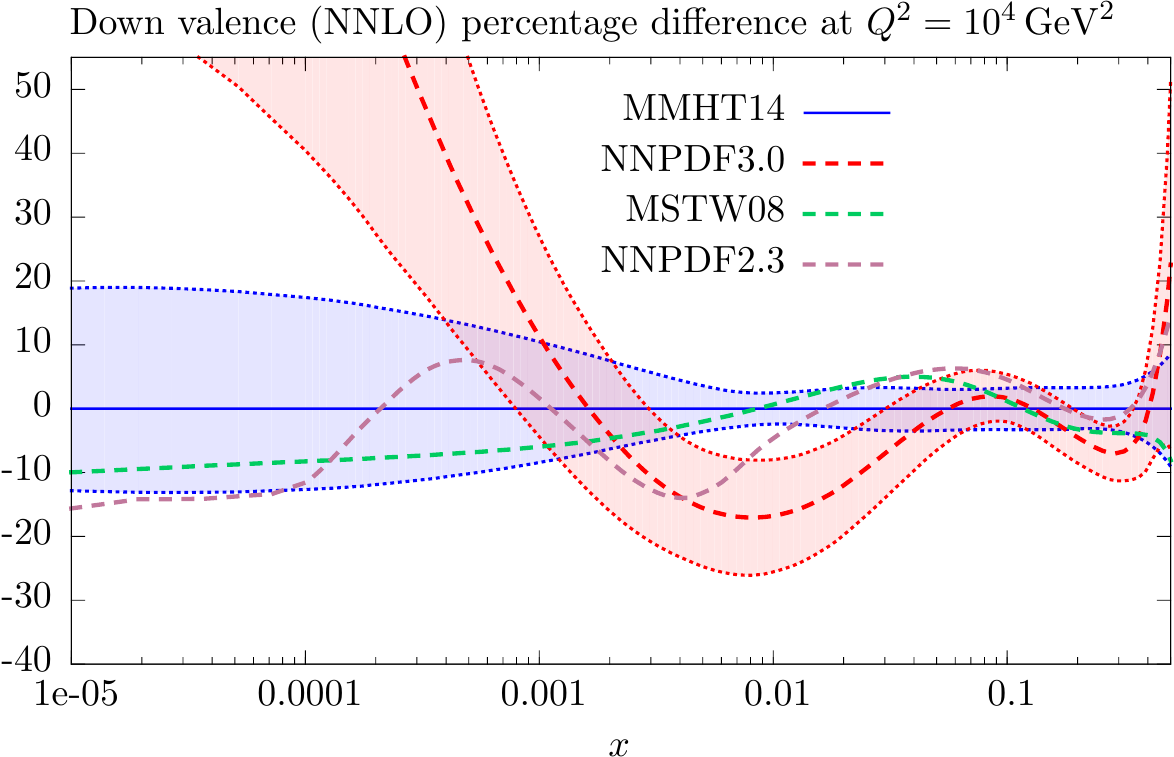}
\includegraphics[height=5cm]{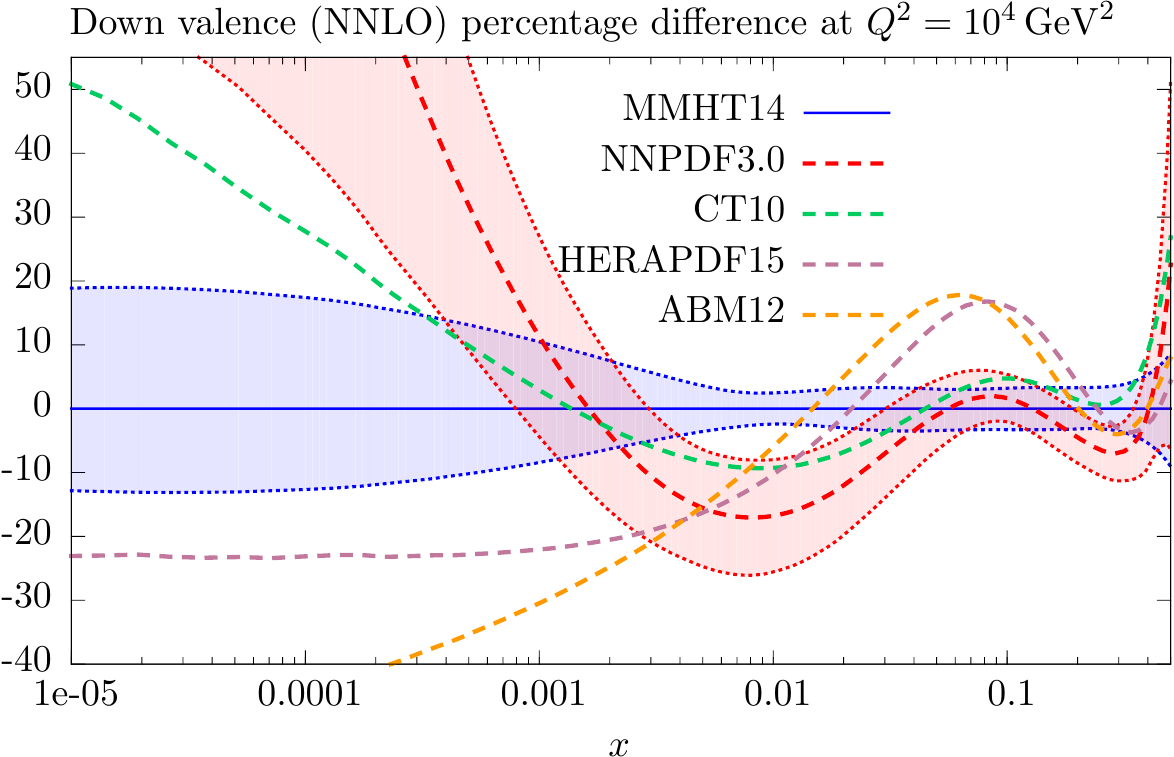}
\caption{\sf The comparison between NNLO NNPDF3.0 and MMHT14  PDFs at 
$Q^2=10^4~\GeV^2$ showing the  $u_V$ and $d_V$ quark PDFs.  
Also shown (without error corridors) are the NNPDF2.3 and MSTW08 PDFs (left) 
which they supersede and (right) CT10 HERAPDF1.5 and ABM12 PDFs..}
\label{fig:NNPDFMMHTval}
\end{center}
\end{figure}

Here we compare the MMHT14 PDFs to PDF sets obtained by other groups. The most
direct comparison is with the NNPDF3.0 PDFs which have
very recently been obtained in a new global analysis performed by the NNPDF 
collaboration \cite{NNPDF3}. This involves a fit to very largely the same data sets, 
including much of the available LHC data, and also uses a general mass variable flavour
number scheme which has been shown to converge with that used in our 
analysis as the order increases \cite{HERAbench}. There do, however, remain some 
significant differences in the two theoretical approaches. For example, NNPDF3.0 does not apply 
deuteron and heavy nuclear target corrections. Moreover, the MMHT and NNPDF 
collaborations use quite a different procedure for the analysis.  The NNPDF collaboration combine 
a Monte Carlo representation of the probability measure in the space of PDFs with the 
use of neural networks to give a set of unbiased input distributions. On the other 
hand, here, we use parameterisations of the input distributions based on Chebyshev 
polynomials where the optimum order of the polynomials for the various PDFs is 
explored in the fit.

Although the most direct comparison is between the MMHT14 and NNPDF3.0 sets of PDFs, we also compare to older PDF sets; i.e. the MSTW08 \cite{MSTW} 
and NNPDF2.3 \cite{NNPDF23} sets, which MMHT14 and NNPDF3.0 supersede, and with the 
ABM12 \cite{ABM14}, CT10 \cite{CT10} and HERAPDF1.5 \cite{HERAPDF15} 
sets which are obtained from a smaller selection of data.\footnote{The ABM12 
analysis does include some of the LHC $W,Z$ data.}

\subsection{Representative comparison plots of various PDF sets}

As a representative sample, we show in Figs. \ref{fig:NNPDFMMHTsing}, 
\ref{fig:NNPDFMMHTsea} and \ref{fig:NNPDFMMHTval} the comparison of MMHT14
and NNPDF3.0 for six PDFs: namely the $g$, light quark, $u_V$, $d_V$, $\bar u$ and $s + \bar s$,   at $Q^2=10^4~\GeV^2$ at NNLO.
All the plots show the MMHT14 and NNPDF3.0 PDFs with their error corridors.
The plots on the left of the Figures also show
  the MSTW08 and NNPDF2.3 PDFs (but now without their error corridors), which 
have been superseded by the MMHT14 and NNPDF3.0 sets, respectively, The  plots on the right of the Figures show the comparison with the central values of ABM12, CT10 and HERAPDF1.5 PDFs.  These representative plots of PDFs  
are sufficient to draw general conclusions concerning the comparisons, 
which we discuss in the subsections below.

As noted above, the treatment of the input distributions and the uncertainties are 
quite different in the NNPDF and MMHT analyses. However, remarkably, we see from 
Figs. \ref{fig:NNPDFMMHTsing}$-$\ref{fig:NNPDFMMHTval}, that in regions where 
the NNLO PDFs are tightly constrained by the data, with a few exceptions, 
the values, and also the error corridors, are very consistent between the two analyses.

\subsection{Comparison of gluon PDFs and sea quark PDFs}

We may conclude (at  $Q^2=10^4~\GeV^2$) that to within 
2\% accuracy, the NNLO gluon is determined in the domain 
$3\times 10^{-4} \lapproxeq x \lapproxeq  5\times 10^{-2}$.
There is much better agreement between MMHT14 and NNPDF3.0 for the gluon
than between MSTW08 and NNPDF2.3.\footnote{We note that NNPDF3.0 uses a charm pole
mass of $m_c=1.275~\GeV$ rather than the value $m_c=\sqrt{2}~\GeV$ used for NNPDF2.3.
As noted in \cite{MSTWmass} (see Fig.~4), and \cite{NNPDFmass} (see Fig.~40) 
this type of change has some effect on the gluon, potentially
of order $1\%$ at $Q^2=10^4~\GeV^2$ (except at very high and low $x$), 
but very little change near $x=0.01$. The value of $m_b$ is also changed, but this should 
have negligible change on the PDFs, except for the $b$ distribution.}    
In the region $x \sim 0.01$ NNPDF2.3 is
outside the combined error band of the two newer sets (leading to the reduced 
cross section for Higgs production via gluon fusion for the NNPDF update noted 
in \cite{NNPDF3}). For $x \sim 0.0001-0.001$ MSTW08 is outside the combined 
error band (though quite close to NNPDF2.3). 

The CT10 and HERAPDF1.5 gluons are 
in good agreement with MMHT14/NNPDF3.0, except for HERAPDF near $x=0.1-0.2$, 
though at the edge of the error band precisely at the central Higgs rapidity 
$x$ values of $0.01-0.02$. ABM12 is much larger below $x \sim 0.05$ and much smaller
for $x>0.1$. Part of this is due to the much smaller strong coupling obtained by
ABM12, but the general effect persists even if $\alpha_S(M_Z^2)=0.118$ is used. 
It was argued in \cite{ThorneFFNS} that this difference with ABM12 is primarily 
due to their use of a fixed-flavour number scheme (FFNS).

The very good agreement in the MMHT14 and NNPDF3.0 gluon distributions is 
responsible for the comparably good agreement in the small-$x$ ($x<0.01$) light quark, 
$\bar u$ and $s + \bar s$ distributions, which are driven at small $x$ by evolution 
mainly from the gluon. For these values of $x$ the superseded MSTW08 and NNPDF2.3 distributions
for these PDFs also show good agreement, although there has been a noticeable transfer 
from $\bar u$ to $s +\bar s$ quarks in going from MSTW08 to MMHT14. It would be 
surprising to see much change in the sea quarks in this region, as a linear 
combination of them is very tightly constrained by HERA structure function data. 
Indeed, there is also generally good agreement with ABM12, CT10 and HERAPDF1.5
distributions. CT10 lies a little higher at very small $x$, consistent with the similar 
feature for the gluon distribution. HERAPDF has a distinctly higher $\bar u$ distribution 
at lower $x$, but this is compensated, to some extent, by a smaller $s + \bar s$ 
distribution.

Perhaps the most surprising discrepancy between MMHT14 and NNPDF3.0 is in the total 
light quark distribution at $x\sim 0.05$, see Fig. \ref{fig:NNPDFMMHTsing}. This seems to be a particular feature of NNPDF, with 
NNPDF2.3 and NNPDF3.0 being very similar, while all the other PDF sets are very similar to 
MMHT14 in this region. The difference is $\sim 3\%$, but the PDF uncertainty is 
only $\sim 1\%$ here. The main reason for this difference seems to be that NNPDF have the smallest 
strange quark in this region, as well as smaller valence quarks than other PDF sets. 
NNPDF are the only sets of PDFs which have used HERA-II data, which constrain this $x$ range, 
so this may have some effect. Also, the singlet-quark distribution is probed in 
charged current neutrino DIS by $F_2(x,Q^2)$, and some difference may be due to whether 
nuclear corrections are included when fitting to these data.   
The smaller NNPDF light quark distribution for $x \sim 0.05$ 
is perhaps apparent in NNPDF3.0 having smaller quark-quark luminosity than CT10 and MSTW08 
in Fig.~59 of \cite{NNPDF3} for $M_X\sim 600~\GeV$ at the LHC with 13~TeV centre-of-mass energy. 
However, in the luminosity plot the error bands easily overlap due to sampling a range of $x$ 
values for each $M_X$.

\subsection{Comparison of $s+\bar s $ distributions}
The MMHT14 and NNPDF3.0 $s + \bar s$ distributions are fully compatible, 
but NNPDF3.0 has a lower distribution. The latter observation is due to the increase in the strange fraction 
in MMHT14 arising from the improved treatment of the $D\to \mu$ branching ratio $B_{\mu}$, whereas NNPDF3.0 is similar to NNPDF2.3 (and also to MSTW08, except at fairly high
$x$ values). The improved treatment of $B_{\mu}$ means MMHT14 has a rather larger uncertainty for $s + \bar s$
than previously, and this also seems to be larger than that for NNPDF3.0. 

MMHT14 also has a larger total strange distribution than HERAPDF (as already 
noted  at small $x$), but the two are compatible. There is quite good agreement with ABM12
except for $x>0.2$, where there is little constraint from data. CT10 has the largest 
$s+ \bar s$ distribution, and the central value is even outside the MMHT14 error band near 
$x=0.05$, though their uncertainty band is large. 
However, it was recently reported in \cite{PDF4LHC} that a sign error was discovered in 
the CT10 heavy flavour contribution to charged-current DIS. This led to a considerable 
underestimate of the dimuon cross section, and hence a larger strange distribution.
A significant reduction of $s+ \bar s$ is therefore expected in future CT PDF sets.

\subsection{Comparison of valence quark distributions}

There is, perhaps unsurprisingly, more difference in the PDFs for valence distributions, 
as seen in  Fig. \ref{fig:NNPDFMMHTval}, since there is less direct constraint from the data. MMHT14 and 
NNPDF3.0 agree well for both $u_V$ and $d_V$ at $x > 0.05$ where 
the valence quarks provide
the dominant contribution to 
the structure function data. However, 
at lower $x$ values. where sea quarks dominate, the PDFs start to differ significantly. Both the
$u_V$ and $d_V$ of NNPDF3.0 become smaller than those of MMHT14 for $x \sim 0.01$ (though more so for 
$d_V$), and then become larger at very small $x$ as a result of the quark number constraint. 

The same sign 
difference for both valence quarks for $x \sim 0.01$ allows $u_V-d_V$ to be similar for MMHT14
and NNPDF3.0, so both fit the LHC lepton asymmetry data at low rapidity, which is sensitive to 
$u_V-d_V$ at $x \sim 0.01$. It may be the case that the absence of deuteron corrections in 
NNPDF3.0 compared to the relatively large ones now used in the MMHT14 analysis leads to a 
difference in the $d_V$ distribution which also impacts on the $u_V$ distribution due to the 
constraint on the difference between them. Indeed, MSTW08 (which had a more restricted deuteron 
correction) and NNPDF2.3 agree quite well for $d_V$. However, there is also some direct constraint on 
valence distributions from nuclear target data, and also sensitivity to the $F_3(x,Q^2)$ structure 
function. Here MMHT apply nuclear correction factors, while NNPDF do not, and also 
employ a larger $Q^2$ cut for $F_3(x,Q^2)$ than for $F_2(x,Q^2)$ due to the probable large higher-twist corrections at lower $x$ values. As already commented on, the valence distributions in 
MMHT14 and MSTW08 are quite different due to the extended parameterisation and to the deuteron 
corrections -- the main features of the change are already present in \cite{MMSTWW}. 
Note that there are also some quite significant changes in going from NNPDF2.3 to NNPDF3.0 at smaller $x$. 

The MMHT14 $u_V$ distribution agrees quite well with that of both CT10 and HERAPDF1.5. The ABM12
$u_V$ distribution is very different in shape to all the rest, perhaps due to the approach of
fitting higher-twist corrections, rather than employing a {\it conservative} kinematic cut as 
the other groups do. MMHT14 also exhibits reasonable agreement with the CT10 $d_V$ 
distribution, but both HERAPDF1.5 and ABM12 have quite different shapes (though similar to 
each other). HERAPDF has little constraint on $d_V$ and the uncertainty is large, 
though it is not influenced by
assumptions about deuteron corrections or by imposing isospin 
symmetry conservation. The reason for the
difference for ABM12 may be similar to that proposed for the difference in $u_V$.      
The valence quarks are very different as $x \to 0$, perhaps suggesting an underestimation 
of uncertainty here, even by NNPDF. However, it is not clear what experimental data would be sensitive
to the very small $x$ valence quark differences.

\subsection{Comparison at NLO}

\begin{figure}[t]
\begin{center}
\includegraphics[height=5cm]{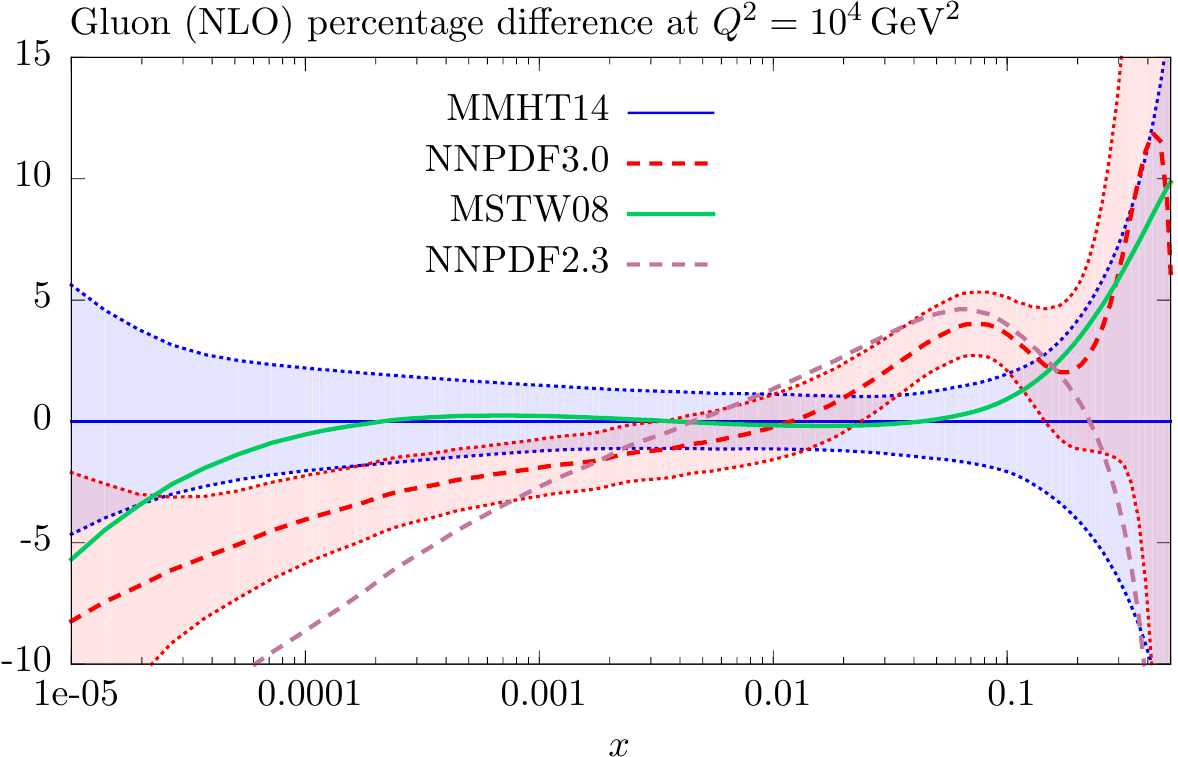}
\includegraphics[height=5cm]{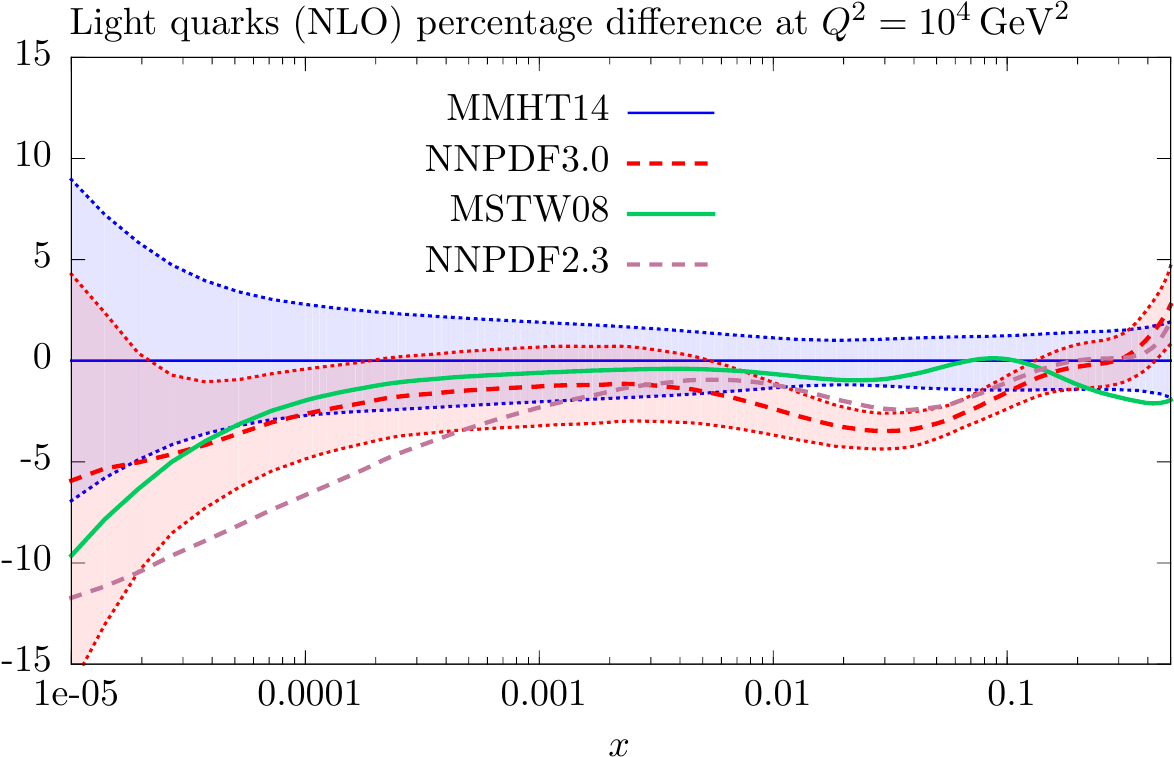}
\caption{\sf The comparison between NLO NNPDF3.0 and MMHT14 PDFs at $Q^2=10^4~\GeV^2$. The two plots 
show the  $g$ and light quark PDFs.  Also shown (without error corridors) are the NNPDF2.3 and 
MSTW08 PDFs which they supersede.}
\label{fig:NNPDFMMHTNLO}
\end{center}
\end{figure}

The same type of PDF comparison  is made between NNPDF3.0 and MMHT14 
at NLO in Fig.~\ref{fig:NNPDFMMHTNLO}. For the gluon (left-hand plot) this 
shows less agreement between the values of the MMHT14 and NNPDF3.0 PDFs than the comparison at NNLO, 
though the width of the error corridors are still comparable. 
NNPDF3.0 is larger for $x \sim 0.1$ but becomes considerably smaller  at
very low $x$. Even so, the 
plots show that there is now closer agreement than between the 
MSTW08 \cite{MSTW} and NNPDF2.3 \cite{NNPDF23}  PDFs that they supersede, though 
the form of the difference is the same.
For the quarks the differences between PDF sets are largely similar at NLO as at 
NNLO (an exception being that HERAPDF1.5 has a smaller high-$x$ gluon at NLO and larger high-$x$ sea quarks
compared to its NNLO comparison to other sets). The main additional difference between 
NNPDF3.0 and MMHT14 (and between NNPDF2.3 and MSTW08) is simply that inherited from 
the gluon difference, i.e. the smaller NNPDF gluon at low $x$ leads to smaller 
low $x$ sea quarks. This is illustrated in the NLO comparison of the light quark
distributions shown in the right-hand-side plot of Fig.~\ref{fig:NNPDFMMHTNLO}, and is similar for all sea 
quarks at low $x$.

So far we have compared the PDF sets at $Q^2=10^4 ~\GeV^2$.
The comparison of MMHT14 and NNPDF3.0 (and other) PDFs at lower $Q^2$, say 
$Q^2=10~\GeV^2$, shows the same general trends, but now the error corridors 
are wider, particularly at very small $x$, as illustrated for MMHT2014 PDFs 
in Figs. \ref{fig:NNLOpdfs} and \ref{fig:NLOpdfs} respectively.

\section{Conclusions \label{sec:9}}

We have performed fits to the available global hard scattering data to determine the PDFs of the 
proton at NLO and NNLO, as well as at LO. These PDF sets, denoted MMHT2014, supersede the MSTW2008 
sets, that were obtained using a similar framework, since we have made improvements in the 
theoretical procedure and since more data have become available in the intervening period.  
The resulting MMHT2014 PDF sets may be accessed, as functions of $x,Q^2$ in computer retrievable 
form, as described in Section \ref{sec:access}.

How has the theoretical framework been improved?  This was the subject of Section \ref{sec:theory}. 
First, we now base the parameterisation of the input distributions on Chebyshev polynomials. It was 
shown in \cite{MMSTWW} that this provided a more stable determination of the parameters. We now also
use more free parameters than previously, i.e. an additional two for each valence quark, for the overall 
sea distribution and the strange sea. However, we only use 5 more in determining PDF eigenvectors as 
there is some still some redundancy in parameters. 
Next, note that even with the advent of LHC data, we find we still need the fixed-target nuclear data 
to determine the flavour separation of the PDFs. So our second improvement is to use a 
physically-motivated parametric form for the deuteron correction, and to allow the data to determine 
the parameters with the uncertainties determined by the quality of the fit. The first step in this 
direction was taken in \cite{MMSTWW}, but now we find that the global fit results in a correction 
factor even more in line with theoretical expectations, see Fig. \ref{fig:deutunc}.  
There are similar improvements for 
the heavy-nuclear corrections for the deep inelastic neutrino scattering data, with an update of 
the corrections used, and again allowing some freedom to modify these corrections and for the 
fit to choose the final form. The third improvement concerns the treatment of the heavy 
$(c,b)$ quark thresholds. We use an optimal GM-VFNS to give improved smoothness in the transition 
region where the number of active flavours increases by one.  The fourth improvement is to use 
the multiplicative, rather than the additive, definition of correlated uncertainties. Another 
important change in our procedure is the treatment of the $D\to \mu$ branching 
ratio, $B_{\mu}$, needed in the analysis of (anti)neutrino produced dimuon data. These data give 
the primary constraints on the $s$ and $\bar s$ PDFs. In the present analysis we avoid using the 
determination of $B_{\mu}$ obtained independently from the same dimuon data, but instead, in the global fit, we 
include the value, and its uncertainty, obtained from direct measurements.
It turns out that the global fit determines a consistent value of $B_{\mu}$, but with a larger 
uncertainty than the direct measurement, leading to a much larger uncertainty on the strange quark 
PDFs than that in the MSTW2008 PDFs, see Figs. \ref{fig:NLO3} and \ref{fig:gsNNLO}.

What data are now included, that were not available for the MSTW08 analysis? This was the subject 
of Sections \ref{sec:preLHC} and \ref{sec:4}. First, we are now able to use the combined H1 and 
ZEUS run  I HERA data for the neutral and charged current, and for the charm structure functions. 
Then we have $W$ charge asymmetry data updated from the Tevatron experiments and new from the 
LHC experiments. We also have LHC data for $W,Z$, top-quark-pair and jet production. It is 
interesting to see which data sets most constrain the PDFs. This is discussed in Section 
\ref{sec:5.3.3}; and displayed in Tables \ref{chi2eigNLO} and \ref{chi2eigNNLO} for the NLO and 
NNLO PDF sets respectively. It is still the case that the constraints come from  a very wide 
variety of data sets, both old and new, with LHC data providing some important constraints, 
particularly on quark flavour decomposition.
 
Some LHC data are not included in the present fits; namely dijet production, $W$+charm jet data and 
the differential top-quark-pair distributions. However, as shown in Section \ref{sec:7}, these data 
seem to be well predicted by MMHT14 partons, except for the behaviour of $t\bar t$ production at 
large $p_T^t$ (using NLO QCD), see Section \ref{sec:7.3}. In all these cases full NNLO corrections 
are still awaited, and it will be interesting to see how they change the predictions we have at NLO.

The new MMHT14 PDFs only significantly differ from the MSTW08 PDF sets for $u_V-d_V$ for 
$x\sim 0.01$, see Fig. \ref{fig:NLO3}. The only data probing valence quarks in this region are 
the $W$ charge asymmetry measurements at the Tevatron and the LHC. The MSTW08 partons gave a poor 
description of these data. This was cured by changing to a Chebyshev polynomial parameterisation 
of the input distributions, with more free parameters, and by a better treatment of the form of 
the deuteron corrections, as first noted in \cite{MMSTWW}, and further improved here. It is 
therefore not surprising that the MSTW08 PDFs still give reliable predictions for all other data, 
see Tables \ref{tab:sigmaNLO} and \ref{tab:sigmaNNLO} for some NLO and NNLO predictions respectively.
The only other significant change is in the total strange quark distribution, with a moderate increase 
in magnitude (larger than the MSTW2008 uncertainty) for the best fit value, but a very significant 
increase in uncertainty. Thus, we may conclude that one is unlikely to obtain an inaccurate prediction 
for the vast majority of processes using MSTW08 PDFs, but we recommend the use of MMHT14 PDFs for the optimum
accuracy for both the central value and uncertainty.    

As we enter an era of precision physics at the LHC, it is crucial to have PDFs determined as 
precisely as possible. So improvements to the MSTW08 PDFs are valuable. In this respect, it is 
important to notice that the values and error corridors of the two very recent sets of PDFs 
(the MMHT14 and NNPDF3.0 sets, obtained with very different methodologies) are consistent with 
each other at NNLO, with only a few differences of more than one standard deviation, and that 
the values are closer together than hitherto, 
see Figs. \ref{fig:NNPDFMMHTsing}-\ref{fig:NNPDFMMHTval}. Hence,  although it appears that
the intrinsic uncertainties from individual PDF sets are not shrinking at present, with new data
being balanced by better means of estimating full PDF uncertainty, the PDF uncertainties from 
combinations of PDFs, for example as in \cite{PDF4LHC1}, are very likely to decrease in the future.

We note that the current strategy is to upgrade and to run the LHC at $\sqrt{s}=$ 14 TeV, with 
increasing integrated luminosity from 30 fb$^{-1}$ (already taken at $\sqrt{s}=$ 8 TeV) to 300 fb$^{-1}$ at the first stage, and 
eventually, in the High Luminosity LHC (HL-LHC), to 3000 fb$^{-1}$ \cite{ECFAreport}.  The increase in luminosity 
means that we can increase the mass reach for the direct search of new particles. For example, 
the last factor of 10 gain in luminosity means the centre-of-mass energy reach goes from about 
7.5 to 8.5 TeV \cite{ECFAreport}, while HL-LHC continues to operate at $\sqrt{s}=$ 14 TeV.
 However the knowledge of the PDFs at large $x$ will also 
have to improve. From the present study, we see that gluon PDF at NNLO at $Q^2=10^4~\GeV^2$ is 
known to within a small number of $\%$ 
for $0.001\lapproxeq x \lapproxeq 0.2$, but that, at the moment, we have 
little constraint from the data in the larger $x$ domain.  For the two processes which constrain 
the high $x$ gluon PDF, that is jet production 
and the differential distributions for top-quark-pair production, it will be important to complete the NNLO formalism. There are already 
some results for the former process in \cite{GGGP1,GGGP2,GGGP3} and for the latter process in \cite{NNLOtopTev}.  On the experimental side 
it will be important to reliably measure the distributions for these processes, particularly for values of $p^t_T$, 
and rapidity $y_t$, that are as large as possible.

\section*{Acknowledgements}

We particularly thank W. J. Stirling  and G. Watt for numerous discussions on PDFs
and for previous work without which this study would not be possible. We would like 
to thank Richard Ball, Jon Butterworth, Mandy Cooper-Sarkar, Albert de Roeck, Stefano Forte, Jun Gao,    
Joey Huston, Misha Ryskin, Pavel Nadolsky, Voica Radescu, Juan Rojo and Maria Ubiali for various discussions on PDFs
and related issues. We would also like to thank Jon Butterworth and Mandy Cooper-Sarkar for helpful 
information on ATLAS data, Klaus Rabbertz and Ping Tan for help with CMS data and 
Ronan McNulty, Tara Shears and David Ward for LHCb data.  We would also like to 
thank Andrey Sapranov, Pavel Starovoitov, Mark Sutton for help with 
APPLgrid, and Ben Watt for playing an instrumental role in interfacing this to the fitting 
code. We would also like to thank Alberto Accardi for providing the numbers for the CJ12 deuteron corrections and for discussions about the comparison. This work is supported partly by the London Centre for Terauniverse Studies (LCTS),
using funding from the European Research Council via the Advanced 
Investigator Grant 267352. RST would also like to thank the IPPP, Durham, for 
the award of a Research Associateship held while most of this
work was performed. We thank the 
Science and Technology Facilities Council (STFC) for support via grant
awards ST/J000515/1 and ST/L000377/1.

\newpage

\bibliography{references}{}
\bibliographystyle{h-physrev}

\end{document}